\documentclass[12pt]{article}

\usepackage[margin={1.0in}]{geometry}
\usepackage{rotating}
\usepackage{lscape}
\usepackage{graphicx}
\usepackage{color}
\usepackage{venndiagram}

\usepackage{subcaption}

\usepackage{comment}
\usepackage{threeparttable}
\usepackage{booktabs}
\usepackage{longtable}
\usepackage{threeparttablex}
\usepackage{tabularx}
\usepackage{pdflscape} 
\usepackage{natbib}

\usepackage{authblk}
\usepackage{mathtools}
\usepackage{dsfont}
\usepackage{lmodern} \usepackage{anyfontsize} \usepackage[sc]{mathpazo}
\usepackage{courier}
\usepackage[utf8]{inputenc}
\usepackage[T1]{fontenc}
\usepackage{setspace}
\usepackage{amssymb}
\usepackage{amsmath}
\usepackage{mathrsfs}
\usepackage{amsthm}
\usepackage{bbm}

\usepackage{silence}
\usepackage{microtype}

\usepackage{float} \usepackage{multirow}
\usepackage{enumerate}
\usepackage{hyperref}

\usepackage{tikz}
\usetikzlibrary{positioning, arrows.meta, calc}

\usepackage{tcolorbox} 

\usepackage{listings} 
\lstdefinestyle{llmpromptstyle}{
    backgroundcolor=\color{gray!5},
    commentstyle=\color{blue},
    keywordstyle=\color{magenta},
    numberstyle=\tiny\color{gray},
    stringstyle=\color{purple},
    basicstyle=\ttfamily\footnotesize,     breakatwhitespace=false,
    breaklines=true,                     breakindent=0pt,
    captionpos=b,                        keepspaces=true,
    numbers=left,                        numbersep=5pt,
    showspaces=false,
    showstringspaces=false,
    showtabs=false,
    tabsize=2,
    frame=single,                        rulecolor=\color{black}
}
\usepackage[nameinlink]{cleveref} \crefname{appendix}{Appendix}{Appendices}

\usepackage{afterpage}

\newcommand{\currentappendix}{A}
\makeatletter
\def\ps@appendixstyle{  \let\@oddhead\@empty
  \let\@evenhead\@empty
  \def\@oddfoot{\hfil\currentappendix.\arabic{page}\hfil}  \def\@evenfoot{\hfil\currentappendix.\arabic{page}\hfil}}
\makeatother

\providecommand{\sym}[1]{\ifmmode^{#1}\else\(^{#1}\)\fi}

\usepackage{array}
\usepackage{adjustbox}

\usepackage{xfrac}

\hypersetup{
	colorlinks=true, 	linktoc=all,        	linkcolor=black,  	urlcolor  = blue,
	citecolor = black,
    pdftitle={Bank Runs},
    pdfauthor={Sergio Correia, Stephan Luck, Emil Verner}
}

\usepackage[nottoc,notlot,notlof]{tocbibind}

\hypersetup{pdfstartview={Fit}} \RequirePackage[font=small,format=plain,labelfont=bf,textfont=it]{caption}

\def\eatcell#1\unskip{}
\newcolumntype{E}{>{\eatcell}c@{}}
\newcolumntype{H}{>{\lrbox0}c<{\endlrbox}@{}}

\newtheorem{prediction}{Prediction}

\newcommand{\E}{\mathbb{E}}

\begin{document}

\title{\vspace{0cm} \LARGE \bf Bank Runs With and Without Bank Failure}

\author{\vspace{.5cm} Sergio Correia, Stephan Luck, and Emil Verner\textsuperscript{*} \\
}

\date{ First Version: January 29, 2026 \\
This Version: \today } 

\pagenumbering{gobble}
\maketitle

\begin{abstract}
\fontsize{12.0pt}{15.0pt}\selectfont

We study the causes and consequences of bank runs. By applying large language models to historical newspapers, we create a comprehensive database of  bank runs in U.S. history with information on 3,984
runs on individual banks from 1863 to 1934. Our novel data allow us to establish that runs are considerably more likely in  weak banks but also occur in strong banks, especially in response to negative news about the real economy or the broader banking system. However, runs typically only result in failure for banks with poor fundamentals. Strong banks survive runs through various mechanisms, including signaling strength, interbank cooperation, and temporary suspension. At the local level, runs on banks with poor fundamentals translate into substantially larger declines in deposits, lending, and manufacturing activity than runs on strong banks. Our findings imply that poor fundamentals are central to explaining both when runs occur and when they have severe economic effects, tempering the view that small shocks can generate discontinuous jumps to bad equilibria through self-fulfilling run dynamics.
\noindent

\end{abstract}

\let\oldthefootnote\thefootnote
\renewcommand{\thefootnote}{\fnsymbol{footnote}}
\footnotetext[1]{Correia: Federal Reserve Bank of Richmond, \href{mailto:sergio.correia@rich.frb.org}{sergio.correia@rich.frb.org}; Luck: Federal Reserve Bank of New York, \href{mailto:stephan.luck@ny.frb.org}{stephan.luck@ny.frb.org};
Verner: MIT Sloan and NBER, \href{mailto:everner@mit.edu}{everner@mit.edu}.
We thank Sriya Anbil, Matt Baron, Carlos Burga, Gabriel Chodorow-Reich, Melissa Dell, Thomas Eisenbach, Carola Frydman, Claudia Goldin, Itay Goldstein, Andrew Jalil, Matt Jaremski, Sam Hanson, Todd Keister, Stephen Morris, Sarah Quincy, Jeremy Stein, Richard Thakor, and participants at the Barcelona Summer Forum, Catholic University of Chile International Conference on Finance, Chicago Booth, DC Finance Conference, ECB, Empirical Macro Discussion Group, Federal Reserve Board of Governors, Federal Reserve Bank of New York, Frankfurt School, Georgetown University, Goethe University/SAFE, Harvard University, Harvard/MIT Financial Economics Workshop, IMF Macrofinancial Seminar, Minnesota Corporate Finance Conference, NBER Development of the American Economy (Spring 2026), NYU Financial History Workshop, Princeton University, University of Georgia, and YPFS Fighting a Financial Crisis Conference for helpful comments and suggestions. Special thanks to Matt Jaremski for sharing datasets on bank names and clearinghouse existence across time and cities.
We also thank Moxie Clifford, Trang Do, and Camilla MacMichael for excellent research assistance.
The views expressed in this paper are those of the authors and do not necessarily reflect the position of the Federal Reserve Bank of New York, the Federal Reserve Bank of Richmond, or the Federal Reserve System.}

\let\thefootnote\oldthefootnote

\onehalfspacing

\clearpage
\pagenumbering{arabic}

\section{Introduction}
\label{sec:intro}

Bank runs are a salient feature of many financial crises. While it is well understood that financing illiquid assets with demandable liabilities makes banks vulnerable to a sudden withdrawal of funding \citep{Diamond1983}, the role of runs in financial crises remains controversial. Runs can be seen as the key turning point, whereby even small shocks can generate severe crises with widespread bank failures. Alternatively, runs can be viewed as a consequence of deeper fundamental problems in the financial system,  exacerbating crises rather than being their primary cause.

This paper uses novel microdata on bank runs in the U.S. from 1863 to 1934 to study their causes and consequences. We address the following questions: What are the determinants of runs? When do bank runs result in bank failure?  What are the real economic consequences of bank runs? Can runs trigger the failure of healthy banks and amplify small shocks into large crises?

A key challenge for understanding the causes and consequences of runs is the absence of systematic microdata on the occurrence of bank runs.   Existing micro-level studies of bank distress typically focus on bank failures, as these are objectively recorded in historical records \citep{Calomiris2003a,PuriJF,CLV2026}. However, this approach misses the broader universe of runs that do not involve failure. At the same time, studies of bank runs typically focus on granular data from specific episodes \citep{Iyer2012,Iyer2016,CiprianiEtAl2024} or consider aggregate data on banking panics \citep{Calomiris1991,Baron2021,JKMS2024}. Due to the absence of systematic micro data on bank runs, a consensus on how fundamentals map into the likelihood of bank runs and panics has yet to emerge \citep[see][]{Blanchard2018Comment}.

We overcome this gap by studying the near-universe of runs in the U.S. banking system from 1863 to 1934 as recorded in contemporary newspapers. Our approach leverages advances in large language models (LLMs) and the digital accessibility of historical newspaper articles \citep{AmericanStories}. This allows  us to construct a novel database of 3,984
runs on individual banks during 1863--1934. We also identify 13,772
bank suspensions and 10,341
failures. Each event we identify is backed by newspaper accounts, which we publicly document at \href{https://finhist.com/bank-runs}{finhist.com/bank-runs}. Our approach allows us to study bank runs that result in failure, runs that result in temporary suspension but not failure, and runs that have no further consequences. We combine these data on bank runs with information on bank balance sheets, data on macroeconomic conditions, and newly digitized data on local business failures and manufacturing activity.

We provide three main findings. First, runs are considerably more likely in weak banks, but can also occur in strong banks. Second, runs typically only result in failure for banks with weak fundamentals. Third, runs on weak banks lead to substantially larger local declines in lending and real activity than runs on strong banks. Taken together, our findings suggest that poor fundamentals are key for whether runs pass through into failure and have severe consequences for the broader economy.

The historical U.S. banking system provides an appealing environment to study runs. Our sample covers the numerous panics of the National Banking Era and the Great Depression. Bank runs are less common in contemporary banking systems due to a variety of government interventions that affect depositor behavior, such as deposit insurance \citep{Iyer2019} and lending of last resort \citep{MetrickSchmelzing2021}. However, government interventions in the financial sector were more limited before 1934, making runs much more common in this time period. This historical laboratory has shaped economists' understanding of bank runs and banking crises \citep[e.g.,][]{Sprague1910,FriedmanSchwartz,Bernanke1983,Calomiris1991,Wicker2006,Richardson2007}, though this work has largely relied on aggregate data or episode-specific analysis of narrative evidence. Our paper, in contrast, combines both comprehensive narrative accounts and exhaustive bank-level microdata.

We start by validating our novel bank runs database. Spikes in the rate of bank runs recorded in newspapers line up closely with existing narrative chronologies of banking crises in the United States \citep{Jalil2015,Baron2021}, both at the aggregate and regional level. For national banks, the annual unconditional probability of a run is 0.39\%. This number rises to 1.4\% during crisis years. Moreover, national bank failures for which newspapers recorded a bank run involve a 5-percentage-point larger deposit outflow right before failure than those without mention of a run. For banks that survive bank runs, we also find large deposit declines in their annual financial statement around the event. In addition, our dataset achieves a match rate of about 95\% when cross-validating against four ground-truth samples.

With these novel data on runs, we address three fundamental questions on the role of bank runs in bank failures and financial fragility.

\textit{First, what are the determinants of runs?} We empirically examine the role of bank fundamentals and negative news about the economy in triggering bank runs. We establish that runs are substantially more likely in banks with weak observable fundamentals. We measure fundamentals using balance sheet measures capturing bank capitalization and the ability of a bank to finance itself with cheap deposits relative to expensive noncore funding. Existing research shows that these metrics capture bank health and failure risk \citep{White1984,Calomiris2003a,CLV2026}. We document that runs are considerably more likely in poorly-capitalized banks, banks with higher leverage and less liquid assets, and banks that rely more on noncore funding.

While runs are more likely in banks with weak fundamentals, the ability to predict runs with fundamentals alone is somewhat low---especially compared to the predictability of bank failures. However, the incidence of runs is well explained when combining information on bank fundamentals with information on the local and national economic and banking conditions. For instance, we find that runs are more likely after a downturn in the stock market, a rise in local non-financial business failures, or runs on other banks in the same city. News shocks that are not bank-specific help explain runs on healthier banks.

\textit{Second, when do bank runs result in bank failure?}  We find that a bank's annual probability of failure increases by 38 percentage points when subject to a run. This is a large effect, given that the annual unconditional probability of bank failure is 0.89\%. However, it also implies that a run is not a death sentence. There are more runs without bank failure than runs with failure, highlighting a key novel insight from our approach.

We further establish that whether a run results in failure depends crucially on a bank's financial health. Runs are significantly more likely to result in failure for banks with weak fundamentals. Conditional on a run, a bank with very weak fundamentals---defined as being in the lowest decile of our preferred metric that summarizes a bank's financial health---has a 59\% probability of failing. In contrast, the \textit{ex ante} strongest banks, measured as being in the top decile of the same metric, seldom fail when subject to a run. Thus, weak bank fundamentals are necessary for a run to be associated with bank failure.

To better understand how banks with sound fundamentals survive runs, we leverage additional context provided in newspaper articles. We find that newspapers often report that surviving banks accommodate withdrawals and attempt to signal strength to depositors. A common signal of strength is an equity or deposit injection by owners or other investors. In some cases, banks weathered liquidity shocks by borrowing from other banks. Further, 28
\unskip\%  of runs that do not result in failure nevertheless lead to a temporary suspension of convertibility. Suspension was a common way to successfully resolve runs, often alongside bank examination to signal solvency to depositors. In 11\% of runs, newspapers mention that the local bank clearinghouse is involved by providing liquidity or conducting an examination.

The finding that runs can occur for both weak and strong banks, but that only weak banks fail from runs, has important implications for theories of bank runs. First, the finding that runs happen predominantly in weak banks supports the notion that fundamentals are important \citep{MorrisShin2000,Goldstein2005}. Second, the fact that even healthy banks can be subject to runs following adverse news shocks is consistent with information-based theories of bank runs, where depositor confusion leads to withdrawals from both weak and strong banks because depositors struggle to distinguish which banks are solvent \citep{Gorton1988,Chari1988,Calomiris1991,Dang2020information}. Finally, the fact that both weak and strong banks can be subject to a run even though they can be separated based on simple public balance sheet metrics suggests that depositors are at times inattentive. The predictability of both runs and failures suggests that depositors tend to be sleepy and are slow in responding to weak bank fundamentals \citep{HANSON2015449}. However, the importance of news shocks suggests that salient information can cause depositors to wake up and run.

In line with this interpretation, we further find that runs following negative public signals have contrasting implications for bank failure, depending on the nature of the signal. Runs following runs on other local banks are associated with a lower probability of failure. This suggests news of other runs makes depositors ``jittery,'' leading to a rise in runs on healthy banks that do not result in failure. In contrast, runs following a rise in the local business failure rate are associated with a higher probability of failure, consistent with these runs responding to a rise in bank losses that erodes solvency.

We also explicitly examine the scope for non-fundamental runs to trigger failure. We define non-fundamental runs as those that, according to the newspapers, are caused by misinformation or depositor confusion and are not related to bank or economic conditions. These runs are the  most plausible empirical equivalent to the pure liquidity-driven runs studied theoretically in \citet{Diamond1983}. With the assistance of LLMs, we identify a sample of 53
non-fundamental runs on national banks.   This exercise also provides sharper identification by zooming in on a set of runs that are relatively random and thus free of confounding factors. The probability of failure following a non-fundamental run is 11\%, substantially lower than for the average run. Moreover, we find that these non-fundamental runs trigger failure when they happen to occur in weak or fraudulent banks, reinforcing the key role of poor fundamentals for the pass-through of runs to failure. While strong fundamentals reduce the probability of failure after a run generally, this is especially true for non-fundamental runs for which only banks with the weakest fundamentals fail. This pattern casts doubt on a strong form of the view that liquidity problems alone can trigger severe financial distress.

\textit{Third, what are the economic consequences of bank runs with and without failure?}
At the bank level, we find that banks that are subject to runs and survive see a contraction in deposits and loans of about 7\% of pre-run assets. The contraction is persistent, lasting for over four years. To refine the identification, we study the consequences of runs we classify as non-fundamental. Such runs also have modest negative effects on deposits and lending. Further, bank-level deposit and lending contractions following runs are much stronger for banks with \textit{ex ante} weak fundamentals, while the effects are small and not significant for strong banks. Thus, bank fundamentals shape the severity of the consequences of runs, even when runs do not translate into failure.

Finally, we study the consequences of bank runs for local city-level deposits, lending, and manufacturing activity. We first show that runs only lead to declines in deposits, lending, and manufacturing activity when they occur in \textit{ex ante} weak banks. Runs on weak banks translate into a 5\% decline in a proxy of city-level manufacturing production. Non-fundamental runs entail no significant adverse local effects. Further, we use the richness of our bank distress episodes data to consider the city-level consequences of runs \textit{without} failure versus runs \textit{with} failure. We find that spikes in the local rate of runs \textit{without} failure lead to modest declines in deposits and loans at the city level and no effect on manufacturing activity. In contrast, runs with failure lead to more severe contractions in local deposits, credit, and manufacturing activity.

Taken together, our findings have two broad implications. First, while runs can occur in both weak and strong banks, poor fundamentals are necessary for runs to result in bank failures. Second, failures, and their corresponding weak fundamentals, are a key mechanism for banking sector distress to translate into substantial contractions in credit and real activity. Pure liquidity events, in which depositors run due to fear or confusion but bank fundamentals are sound, are not associated with severe economic outcomes at the local level. These patterns, in turn, temper the view that small shocks can result in large jumps to bad equilibria via self-fulfilling runs on demandable debt. Rather, bank runs are only associated with major economic distress when they take place in the context---and as a consequence---of already weak fundamentals. At the same time, because even insolvent banks can remain liquid for some time, a run may be an important determinant for the timing and economic costs of bank distress.

\paragraph{Related Literature}

Our paper relates to a rich empirical literature on bank runs, bank failures, and financial crises (see, e.g., \citet{SufiTaylor2021FinancialCrisesSurvey} and \citet{FrydmanXu2023} for overviews).

The main contribution of our paper is to provide insights from a novel comprehensive database that covers nearly four thousand bank runs in the historical U.S. banking system. Our novel data allows us to provide new systematic evidence on which banks become subject to bank runs, which shocks trigger runs, which banks fail in runs, how banks survive runs, and the consequences of runs with and without failure for broader credit and real activity. Hence, our study relates to micro-level studies of bank runs. Existing evidence suggests that deposit growth is lower in relatively weaker banks \citep[see, e.g.,][]{chen2024liquidity} and especially when solvency concerns are salient during crisis episodes \citep{Schumacher2000}. \citet{Jiang2023}, \citet{Metrick2023}, and \citet{CiprianiEtAl2024} provide evidence that the banks most exposed to interest rate risk were those most affected by the banking stress during the March 2023 banking turmoil. Furthermore, previous work has shown that banks can withstand runs when they have strong ties to their depositors \citep{Iyer2012}. Runs by relatively less well-informed depositors are also less likely to result in failure \citep[see, e.g.,][]{OgradaWhite,Saunders1996,Calomiris1997,Iyer2016,Blickle2022,JKMMS_Wholesale_2026}. Moreover, existing work suggests that interbank and wholesale markets may be crucial for how runs play out \citep{afonso2011stressed,Perignon2018,Mitchener2019}. Finally, \citet{Richardson2009} also show that central bank interventions can be important in influencing bank survival rates during times of heightened financial distress.

Our paper also relates to studies of banking panics using aggregate or regional data. \citet{Gorton1988}, \citet{Calomiris1991}, \citet{Alston1994}, \citet{Wicker1996}, and \citet{Wicker2006} present either time-series or regional evidence suggesting that U.S. banking panics from the National Banking Era through the Great Depression were largely predictable responses to fundamental shocks that altered depositors’ perceptions of bank solvency.  \citet{Baron2021} argue that widespread panic runs are not a necessary feature of banking crises, showing that panics are typically preceded by declines in bank equity prices that reflect the recognition of underlying losses. \citet{JKMS2024} identify systemic bank runs  with narrative evidence and observed deposit outflows and find that such runs occur infrequently, but when they do, they are associated with large, persistent output losses.

Our paper also relates to a large literature on understanding bank failures \citep[e.g.,][]{White1984,Calomiris2003a,CLV2026}. These studies find that failed banks displayed weak balance sheets before failure and argue that runs usually accelerated the failure of already insolvent banks, rather than causing widespread collapse of otherwise solvent institutions; see \cite{CLV_AR_2026} and references therein for a comprehensive review. Our paper makes a distinct contribution to this literature by expanding the scope of bank distress to systematically include runs and suspensions without bank failures. We provide the novel insight that runs can occur in strong banks, but they typically only result in failure for weak banks.

Further, our paper contributes to the literature that shows that financial disruptions can have persistent consequences for the real economy by linking the consequences of both bank failures and bank runs without failures to real economic outcomes at the bank and local level.  \citet{Bernanke1983} emphasizes the role of bank failures and the resulting credit market impairments in deepening the Great Depression. \citet{ChodorowReich2014} provides micro-level evidence that credit supply shocks during the 2008--09 financial crisis led to substantial employment losses among firms dependent on weak lenders. Similarly, \citet{Frydman2015} document that runs on trust companies during the Panic of 1907 curtailed investment and dividend payments by connected firms, illustrating how financial distress transmits to the real economy even outside traditional banks.

Finally, our methodology provides an innovation to the narrative approach to identification in macroeconomics and finance \cite[e.g.,][]{RomerRomer2004}. By providing historical classifications of banking crises, the narrative approach has been fruitful for understanding crises \citep{Bordo2001crisis,Reinhart2009,Jorda2013,Baron2021}. The narrative approach has generally been used in a macroeconomic context with a relatively limited set of events. Our methodology using textual analysis and LLMs allows us to apply the narrative approach at a scale that would previously have been insurmountable without a squadron of researchers. Furthermore, the narrative approach to classifying banking crises involves researcher judgment, leading to some disagreement about which episodes constitute widespread bank runs \citep[e.g.,][]{Jalil2015,Baron2021,Hoon2025costs}. Our newspaper-based approach provides a systematic method to resolve these disagreements based on large-scale and consistent analysis of a given corpus of text. It is particularly well-suited to understanding events such as runs when such events are recorded in the text, as it usually does not require judgment to determine whether there is a discussion of a bank run.

\paragraph{Roadmap} The next section outlines testable predictions from theories of bank runs. Section \ref{sec:data}  describes the methodology behind the construction of our novel bank runs dataset. Section \ref{sec:data_overview} validates our database and documents the incidence of bank distress in the U.S. from 1863 to 1934. Section \ref{sec:determinants} studies the empirical determinants of bank runs. Section \ref{sec:runs_failure} studies the pass-through of runs to failure. Section \ref{sec:consequences} estimates the consequences of runs for affected banks and for local deposits, lending, and manufacturing activity. Section \ref{sec:conclusion} concludes.

\section{Conceptual Framework}
\label{sec:theory}

To guide our empirical analysis, we outline different prominent theories of the causes of bank runs and their predictions for whether runs result in bank failure. We then discuss the predicted real economic consequences of bank runs with and without failure.

Our emphasis on bank failure as a central outcome is motivated by a long tradition in macro-finance arguing that bank failure is particularly costly for real activity because it disrupts credit intermediation \citep[e.g.,][]{Bernanke1983}. Bank lending is ``special'' because it mitigates informational and agency frictions through relationship-based screening and monitoring. When a bank fails, this informational capital is destroyed, leading to persistent reductions in credit. As a result, bank failure provides a natural mechanism through which runs translate into broader real economic costs. Nevertheless, as we discuss below, runs that do not result in failure may still impose economic costs through alternative channels.

\subsection{Causes of Runs and Implications for Bank Failure}

To fix ideas, we use a simple framework to outline the empirical predictions from three different theories of runs: non-fundamental bank runs \citep[e.g.,][]{Diamond1983}, fundamental-based panic runs \citep[e.g.,][]{MorrisShin2000,Goldstein2005}, and information-based runs \citep[e.g.,][]{Gorton1988,Chari1988}. We summarize the main insights from this framework in the main text, leaving the full model details and derivations to \Cref{appendix:model}.

\paragraph{Setup} There are two dates, $t\in \{1,2\}$. There is a bank that initially holds liquid cash $C$ and illiquid loans $L$, financed by deposits $D$ and equity $E$. There is a continuum of risk-neutral depositors, each holding one unit of deposits. Deposits are demandable at $t=1$. However, the bank lacks the liquid funds to pay all depositors in the initial period, $D>C$.  In $t=2$, the loan portfolio pays off $\theta L$. We refer to the gross return $\theta$ as the fundamental of the economy. There is no discounting, deposits pay zero interest, and the return on cash is zero.

At time $t=1$, depositors observe $\theta$ and decide whether to keep their deposits in the bank or withdraw. Depositors who keep their deposits in the bank until $t=2$ receive a convenience benefit of $c$ per unit of deposit. If total withdrawals exceed cash, the bank is illiquid, and it is forced into failure (receivership). In failure, the loan portfolio value is reduced to $(1-\rho)\theta L$. The reduction in value $\rho$ can reflect that the receiver is less skilled at managing bank assets than the banker or a fire sale discount. We assume sequential service of withdrawals. Depositors who withdraw while the bank is liquid receive full repayment, whereas depositors remaining in the bank in failure receive a pro-rata share of the remaining assets.

\paragraph{Non-fundamental Runs}
The bank is fundamentally insolvent when the fundamental is sufficiently low that it cannot fully repay depositors, $\theta < \theta^S\equiv \frac{D-C}{L}$. However, the bank may be subject to a run that renders it illiquid and forces it into failure if $\theta<\theta^L\equiv \frac{D-C}{(1-\rho)L}$. The threshold for run-induced failure exceeds the solvency threshold, $\theta^L>\theta^S$. Thus, for $\theta \in [\theta^S,\theta^L)$, a self-fulfilling run constitutes an equilibrium. Because depositors are served on a first-come, first-served basis, they run if they expect others to run, leading to a self-fulfilling panic. A run that renders a bank illiquid will also render it insolvent by forcing it into value-destroying failure.

Thus, the bank is exposed to the possibility of a self-fulfilling panic run that is not strongly tied to bank or broader economic fundamentals. We refer to these as \textit{non-fundamental runs} \citep{Diamond1983}.\footnote{Many papers explore non-fundamental runs in a Diamond-Dybvig framework \citep[e.g.,][]{PeckShell2003}. \cite{He2016} model the dynamics of information-acquisition in response to rumor-based runs, showing that socially inefficient information acquisition can increase the likelihood of runs. Their model also captures the notion of gradual withdrawals.}   Non-fundamental runs can occur in relatively healthy banks and following a small shock that brings the fundamental $\theta$ just below $\theta^L$. These runs are thus unpredictable jumps to a bad equilibrium that are not necessarily linked to particularly weak fundamentals.

\paragraph{Fundamental-based Panic Runs}
Models of non-fundamental runs do not predict a sharp relation between fundamentals and runs. However, runs may be more likely following bad realizations of fundamentals. Models of \textit{fundamental-based panic runs} capture this idea \citep{MorrisShin2000,Goldstein2005}.\footnote{See also \citet{Allen1998} and \citet{Rochet2004} for models where runs occur after the realization of poor fundamentals. To capture the idea that runs should occur in relatively weaker banks, \cite{Ennis2003} and \cite{Gertler2015} assume that non-fundamental runs are more likely in banks with lower $\theta$.} In \citet{Goldstein2005}, bank fundamentals are stochastic, but agents do not have common knowledge about fundamentals. Instead, each depositor receives a slightly noisy private signal of the bank's asset value $\theta$. This global games set-up can feature a unique equilibrium as a function of fundamentals. A bank run occurs when bank fundamentals $\theta$ fall below a specific threshold $\theta^* \in (\theta^S,\theta^L)$. Importantly, the model generates panic runs, in the sense that, absent the run, the bank would have been solvent and survived for $\theta \in [\theta^S, \theta^*)$.\footnote{In practice, depositors may be sleepy and not attentive to bank fundamentals \citep[e.g.,][]{Egan2025}. Therefore, depositors may react slowly to weak fundamentals, leading to runs on weak banks once these banks are already insolvent \citep{CLV2026}.} Runs thus amplify bad fundamentals. The empirical prediction is that fundamental-based panic runs should occur in banks with weak fundamentals.\footnote{In \cite{CalomirisKahn1991} and \cite{Diamond2001}, runs are triggered by negative information about bank assets. In these frameworks, demandable deposits are the optimal contract to discipline bankers.}

\paragraph{Information-based Runs} In another class of theories, runs are modeled as a response to aggregate public signals that lead depositors to revise their assessment of the riskiness of deposits. We refer to these as \textit{information-based runs}. These aggregate signals are not bank-specific. This class of models assumes that bank assets are opaque and depositors cannot tell apart good and bad banks \citep{Dang2017}. In particular, suppose at time $t=1$ depositors receive a noisy public signal $y=\theta + \epsilon$, where $\epsilon$ is a mean-zero shock. In this setting, if a signal is sufficiently bad, $y < \overline y$, a run constitutes an equilibrium. Such runs away from bank debt can occur even in response to small shocks \citep{Gorton2012book,Gorton2014}.\footnote{In recent work by \cite{Dang2020information}, runs arise when an adverse signal about the collateral backing bank debt leads agents to produce information about bank debt. As a result, there is a regime switch, and bank debt moves from being information-insensitive to information-sensitive. Agents flee bank debt out of fear that others have private information about the value of bank debt.} Information-based theories also capture the idea that depositors can misinterpret public signals and mistakenly run on healthy banks. If the realization of the shock $\epsilon$ is sufficiently low, then a bank with a high $\theta>\theta^L$ can be run.

Depositors can react to a range of signals. For example, depositors might run in response to a signal that a recession is imminent, such as following a decline in the stock market, a rise in business failures, or runs on other banks  \citep{Gorton1988,Calomiris1991}. Depositors might misinterpret a long line at a bank as a signal that the bank is insolvent, even though the line was caused by a liquidity demand shock \citep{Chari1988}. Related to this, runs can be driven by herding \citep{Gu2011}.\footnote{ \cite{Gu2011} writes a model of herding by depositors, where there can be both efficient and inefficient runs. Inefficient runs occur when depositors are misled by past observed withdrawals through an informational cascade.} The bank run in the film \textit{Mary Poppins} is an evocative example. Runs on other banks may lead depositors to revise their beliefs that their own bank is insolvent and run, an information-based contagion \citep{ChenJPE1999}. Behavioral theories of flawed belief formation can also lead to runs on healthy banks, as suggested by the original connotation of ``panic.'' Information-based theories predict that panics should occur in response to adverse public signals, leading to runs on both weak and strong banks.

 \begin{prediction}
\textbf{Causes of bank runs:}
Non-fundamental runs can happen both in banks with weak and strong fundamentals. Fundamental-based panics happen in weak banks. Information-based runs are more likely in weak banks, but can also happen in strong banks after negative news that is not necessarily bank-specific.
\end{prediction}

\paragraph{Bank Runs and Bank Failure} What do theories of bank runs predict about bank failure? In standard theories of non-fundamental and fundamental-based panic runs, in equilibrium, runs necessarily trigger failure. For example, in \cite{Diamond1983} and \cite{Goldstein2005}, the run equilibrium leads to the liquidation of the bank. If the liquidation can be avoided, then there is no run in the first place. Thus, these models predict that, in equilibrium, runs cannot occur without failure.\footnote{This statement is based on a version of global games models in which all depositors receive almost the same signal about the state of the world, and which is the most commonly considered version. Note that partial runs that do not result in failure are possible in models such as \citet{Goldstein2005} when the common knowledge assumption is more broadly relaxed and when depositors do not receive the exact same signal, i.e., the noise does not go to zero.}

In contrast, information-based theories allow for runs without failure, as depositors may run on strong banks following overly negative signals. Several mechanisms can prevent the failure of a solvent bank in response to a run, even in the absence of a public lender of last resort.  Runs might be managed by accommodating withdrawals or attempting to signal that the bank is solvent, such as through information provision or equity injection. For instance, the run in the celebrated and commonly invoked \citep[see, e.g.,][]{NobelCommittee2022} film \textit{It's a Wonderful Life} is resolved without failure because George Bailey is able to restore depositor confidence through a mix of persuasion and use of his  personal wealth.\footnote{Note that George Bailey operates a Building \& Loan rather than a commercial bank. The run in the movie is fended off after George Bailey uses both his ability as a gifted speaker and his honeymoon savings to end the run.}  A run on a healthy bank by uninformed depositors can also be managed by interbank cooperation. In the National Banking Era, bank clearinghouses would regularly inspect member banks and provide liquidity in response to runs \citep{Timberlake1984,Gorton1985}. Heavy runs can be resolved by suspension of convertibility, whereby banks refuse to convert deposits into currency for a time. \cite{Gorton1985suspension} models suspension of convertibility as a costly way for a solvent bank to signal its solvency and avoid an inefficient liquidation. If these mechanisms are empirically relevant, then they should imply that even if runs on solvent banks do occur, they should rarely lead to bank failures.\footnote{Note that information-based runs where agents make mistakes can be resolved with suspension, but they still occur in equilibrium. However, in full-information models, such as classic models of non-fundamental runs, suspension of convertibility would rule out inefficient runs on solvent banks in the first place.}

In our simple framework, suppose that if withdrawals exceed cash, the bank can choose one of two actions. Either it can enter failure, where assets are liquidated at $(1-\rho)\theta L$, as above. Or, it can suspend and pay a proportional cost $\tau$ to have its books examined and certify that it is solvent. Supposing that the cost of examination is lower than that of failure, $\tau < \rho$, the bank chooses to temporarily suspend and pay the cost $\tau$ if $\theta> \theta^{Susp} \equiv \frac{D-C}{(1-\tau)L},$ where $\theta^{Susp} \in (\theta^S, \theta^L)$. Thus, if depositors see an overly negative public signal of fundamentals, but true fundamentals are relatively strong, $\theta \geq \theta^{Susp}$, then the bank is subject to a run, suspends, is examined, and reopens, thereby avoiding failure. A mechanism such as suspension and examination thus reduces the set of solvent banks that can be pushed into failure by a run. Moreover, because survival is more likely, the run threshold itself falls, making runs less likely, especially on strong banks.

\begin{prediction}
\textbf{Bank runs and failures:}
    In canonical non-fundamental and fundamental-based run models, run equilibria typically imply failure. In information-based models, runs can occur on solvent banks and need not end in failure when mechanisms such as suspension and examination, or interbank support are available.
\end{prediction}

\subsection{Broader Economic Consequences of Runs and Failures}

Our analysis speaks to two broad conceptual views about the consequences of runs for the banking system and the real economy. One view is that the run itself is a key discontinuous turning point whereby a small shock can translate into high economic costs through deposit outflows,  bank failures, and credit contraction. This view plays a key role in prominent accounts of the Great Depression and Great Recession \citep{FriedmanSchwartz,Gorton2012book,Bernanke1983,Bernanke2018}. Theories of non-fundamental runs and fundamental-based panic runs most directly embody this mechanism \citep{Diamond1983,Goldstein2005}. Information-based theories also allow runs to lead to discontinuous jumps, but the consequences of runs depend more on underlying fundamentals and on whether banks can survive the run.

Another view is that runs themselves are not the root cause of severe downturns, but are instead primarily a reflection of deeper fundamental solvency problems \citep{AdmatiHellwig2014,Gennaioli2018,Baron2021}. Solvency problems are the cause of severe economic contraction, as in models of the bank lending channel \citep{Holmstrom1997,GertlerKiyotaki2010}. Bank capital is the key state variable capturing the ability of the banking system to intermediate funds. If runs often arise in response to these fundamental weaknesses, they may be the trigger that results in inevitable bank failures and economic costs, but not their root cause. In this view, avoiding runs would not prevent the costs of poor fundamentals. At the same time, even if runs are not the root cause of banking sector distress, runs may be an important amplification mechanism for bad fundamentals, leading to worse outcomes than if the run were averted \citep[e.g.,][]{Goldstein2005,Gertler2015}.

While these two views are challenging to fully distinguish, they entail different empirical implications.
On the one hand, if runs themselves are key turning points and the primary cause of distress, then runs should unconditionally be associated with severe outcomes, irrespective of a bank's initial fundamentals.\footnote{Of course, observing severe outcomes after runs does not imply that the run was the \textit{cause} of these outcomes, so this is a necessary but not sufficient condition.} On the other hand, if runs themselves are not the root cause of distress, runs should only be associated with severe consequences when they result in failure. In that case, runs should result in bank failure and large economic contractions when they occur in banks with weak fundamentals, but they should be less costly for banks with sound fundamentals.

Finally, we note that, even absent outright bank failure, runs may have adverse real effects by reducing the quantity and increasing the cost of bank funding, thereby increasing the cost of credit intermediation. Theoretical models of the \textit{bank lending channel} capture the idea that financing constraints among borrowers and banks imply real effects of shocks to bank funding or capital \citep{Holmstrom1997}.\footnote{A large empirical literature finds evidence for this channel \citep[e.g.,][]{PeekRosengren2000,KhwajaMian2008,ChodorowReich2014,Frydman2015,Huber2018}.}  Again, in cases where banks survive a run, bank-lending channel models predict that the cost of these runs is higher in weaker, less well-capitalized banks. Our analysis will thus compare runs with and without failure in weak and strong banks to test this idea.

\begin{prediction}
    \textbf{Real consequences of runs and failures:}

    (a) If runs themselves are key turning points and the primary cause of distress, then runs should typically be associated with severe outcomes, irrespective of initial bank fundamentals.

    (b) If runs are symptoms or amplifiers of deeper fundamental banking sector distress, they should result in severe economic consequences when they occur in banks with weak fundamentals, but be less costly if they occur in banks with sound fundamentals.
\end{prediction}

\section{Data}

\label{sec:data}

\subsection{Novel Bank Distress Events Database}

We construct a novel and comprehensive dataset of bank run events recorded in newspapers for all U.S. banks. The dataset is available at \href{https://finhist.com/bank-runs}{finhist.com/bank-runs} and is searchable by year, state, and city. The full bank distress events database comprises events between 1800 and 1963. For this paper, we focus on the 1863--1934 sample to span the period from the start of the National Banking Era to the introduction of federal deposit insurance.

\paragraph{Concepts and Definitions}  Our conceptual definition of a bank run is an episode in which a substantial number of depositors suddenly demand that their bank convert claims into cash or the liabilities of another bank. We also collect newspaper mentions of other key bank events: suspensions, reopenings, and failures.  Our definition of a suspension is when a bank ``closes its doors.'' This includes both temporary and permanent closures of a bank.\footnote{Our definition of suspensions does not include partial suspensions of convertibility, such as when a bank imposes maximum withdrawal limits for each depositor or legally invokes 30/60 day notice rules for savings or time deposits. We found that defining suspension as a bank closing its doors was the most objective and empirically verifiable definition. However, below we separately study whether banks partially suspended payments in response to runs.} A reopening occurs when a bank temporarily suspends and subsequently resumes business. A bank failure is a suspension that becomes permanent.\footnote{In rare cases, banks are placed in receivership but restored to solvency later. We nonetheless classify these cases as failures.} This includes receiverships for national banks and some state banks, and assignments for private and other banks. We exclude orderly voluntary liquidations from failures. For our analysis focusing on national banks, we define failures as national bank receiverships.

\paragraph{Sources} We rely on several databases of historical newspapers. Our main source is \textit{Chronicling America} \citep{ChroniclingAmerica}, a large-scale database of publicly available historical newspaper scans maintained by the U.S. National Digital Newspaper Program. To conduct our analysis at the article level, we draw on the \textit{American Stories} dataset from \citet{AmericanStories} to separate pages into articles.\footnote{\textit{Chronicling America} provides scanned newspaper images together with metadata for each newspaper issue, as well as text obtained by applying reduced-quality optical character recognition (OCR) techniques. However, its key limitation for our purposes is that it provides content only at the page level and cannot distinguish between different articles. Moreover, the low quality of the text makes it challenging to apply textual analysis or even keyword matching methods. To address this, we use the \textit{American Stories} dataset \citep{AmericanStories}. This dataset provides, for most of the scanned pages in \textit{Chronicling America}, layout information indicating the coordinates of each article, as well as text with an OCR of sufficiently high quality for keyword matching.}
Because copyright limitations restrict the number of articles available in the \textit{Chronicling America} database after the mid-1920s,\footnote{The number of articles in \textit{American Stories} drops from 7.3 million articles in 1922 to 2.6 million articles in 1923 and only 1.3 million articles by 1930 (see \Cref{fig:ca-total-articles}).} for the 1923-1935 period we further supplement it with articles from commercial sources.\footnote{In particular, we use Newspapers.com and NewspaperArchive.com, two major commercial newspaper databases, as well as the New York Times' TimesMachine digital archive.} \Cref{fig:ca-selected-articles} shows the number of articles used in our final analysis across time and source.

\paragraph{Methodology} We produce our dataset of bank distress events in two stages. The procedure is summarized in \Cref{fig:diagram}. In the first stage, we process scanned newspaper images and obtain an intermediate dataset of articles related to bank-distress events. This stage has four components.
In the first component, illustrated in \Cref{fig:example-layout-ocr}, we perform a layout analysis of each page, segment pages into articles, and obtain each article's text through inexpensive but relatively imprecise OCR techniques.
Next, we search each article with keywords and regular expressions (see \Cref{tab:keywords}), and create a tentative list of articles related to bank distress events. Third, we filter out obvious false positives with an off-the-shelf inexpensive LLM.\footnote{See \Cref{prompt1} for the LLM prompt used and \Cref{fig:example-false-positives} for some examples of articles filtered out.} At this point, we have reduced the total number of newspaper articles by 99.8 percent (from 372 million to 801 thousand), so we are able to apply a more precise analysis. In the fourth and last step, we apply state-of-the-art OCR to obtain high-quality article text, and employ a more sophisticated LLM to (i) determine if the article is really about a bank run or other form of distress, and (ii) extract structured information from each article, including every date, event, and bank discussed. This results in an intermediate dataset of 269 thousand articles discussing 314 thousand bank-distress events (see \Cref{fig:examples} for some examples).

At this stage, the dataset has the drawback of being at the article-bank-event level rather than at the bank-event level. Thus, a given bank run might be discussed multiple times throughout many articles. For instance, the failure of the Knickerbocker Trust Company in 1907 is discussed in over 1,000 articles. Further, a single article might not be enough to understand the nature of a bank distress episode, as it might assume some degree of knowledge from earlier articles.

To address this, in the second stage,  we transform the intermediate article-bank-event dataset into our final dataset at the bank-episode level. This is done in roughly two steps. First, we validate and standardize state, city, and bank names. This allows us to group all articles for a given bank and to combine this data with datasets containing bank balance sheet information, receivership records, and city- or state-level economic indicators. \Cref{appendix:validation:city} discusses how we standardize state and city names. This step is relatively straightforward and correctly matches more than 99 percent of non-missing records against existing city-level datasets. However, standardizing bank names is considerably more challenging, as there is no comprehensive list of bank names across bank types (national, state, private) or historical periods (Antebellum, National Banking Era, Federal Reserve). To overcome this challenge, we build a new near-exhaustive bank list based on existing bank-level datasets and several newly digitized sources.\footnote{This new list of banks augments those of \citet{WeberJEH}, \citet{Jaremski_Fishback_2018}, and \citet{CLV2026}, with additional banks manually compiled for this paper, based on \textit{Rand McNally Bankers Directory} (multiple years),  \textit{Trust Companies of the United States} (multiple years), and other sources.} We then write a Python package that flexibly matches bank names while being robust to typos and variations on how these names are written.\footnote{\Cref{appendix:validation:bank} discusses this package in more detail.} This allows us to match 75\% of all bank names between 1863 and 1934, and up to 94\% if we focus only on national banks.\footnote{We do not truly know if a bank is a national bank unless we can match it, so as a workaround we proxy for national banks with banks that contain the string ``nat'', which captures ``National,'' ``Natl.'', etc.}

In the final step, for each bank, we group all the events that take place within a period of time of up to 365 days. These collections of \textit{events} capture what we call ``bank distress \textit{episodes},'' which we analyze by feeding the corresponding newspaper articles to an ensemble of more advanced LLMs.\footnote{We query three LLMs: \texttt{gpt-5-mini}, \texttt{gemini-3-flash-preview}, and \texttt{deepseek-v4-pro}. If all three models agree on the episode type, we select the answer from the first one in this list. Otherwise, we apply a majority vote with priority-based tie breaking in the order listed above.} An example of a bank distress episode involving three bank events is a bank run, suspension, and reopening. We ask the LLMs to jointly analyze all newspaper articles that describe a given bank distress episode, reclassify all the details of this episode, including event types and dates, and finally classify the type of the overall episode depending on its internal dynamics.\footnote{Due to cost and context-window limitations of current LLMs, we only include up to 25 articles for each bank episode queried. For episodes that exceed this count, we try to preserve a representative sample of articles discussing different events (runs, suspensions, etc.) while selecting for high-quality articles within each sample.} To minimize classification errors, we combine queries to independent LLMs, and apply a final manual review.

\paragraph{Episode Types} Our bank distress episodes can be interpreted as a self-consistent list of events that affected a specific bank. As mentioned above, we classify bank run, suspension, failure, and reopening events. We then further classify bank distress episodes into the following categories based on their internal dynamics.

First, we can thus study episodes with runs but no failure, distinguishing between those resolved with and without suspension:

\begin{itemize}
    \item \textit{Run only.} A bank run takes place, but the bank remains open throughout the run, and the run resolves on its own. The bank does not suspend or fail.
    \item \textit{Run-suspension-reopening.} A bank run takes place. As a consequence of the bank run, the bank suspends and stops paying depositors. However, the bank eventually reopens and does not fail. \end{itemize}
Next, we can consider runs with failure:
\begin{itemize}
    \item \textit{Run-suspension-failure.}  A bank run takes place. The bank suspends all payments. The bank remains closed permanently and has a receiver assigned (or similar resolution process for some non-national bank types).
\end{itemize}
Further, we also consider episodes without a run but where the bank suspends temporarily or permanently:
\begin{itemize}
\item \textit{Suspension-failure}. A bank suspends. There is no recorded bank run. The bank remains closed permanently and has either a receiver assigned or is taken over by another bank.
\item \textit{Suspension-reopening}. A bank suspends, but there is no run on this specific bank. This can, for example, occur as part of a clearinghouse decision to suspend following runs on other banks. However, the bank reopens.
\end{itemize}
Note that by construction a bank must suspend before or at failure. Thus, any episode with failure will also involve a suspension. For our analysis using national banks, we rely on information on whether a bank failed directly from the OCC Annual Report to Congress digitized in \cite{CLV2026}.

\paragraph{Measurement Issues} We note three measurement issues that are important to keep in mind when using information on bank distress recorded in newspapers. First, our database employs a binary classification of whether a run occurs according to newspapers. However, the severity of runs can vary, as we will show using data on bank deposits.

Second, newspapers may not always report runs. This concern applies especially to smaller banks. As with any narrative classification, there is always a concern that an event is not discussed in the narrative accounts under consideration. There can be episodes of ``quiet'' distress. Nevertheless, we do find that newspapers contain a massive amount of valuable information when validating our measure against existing narrative chronologies of banking crises or existing information on bank failures for national banks, as we show in the next section. Moreover, the typical run on a national bank in our dataset is discussed by 9
separate underlying newspaper articles, indicating that runs were generally newsworthy and widely reported.\footnote{\Cref{tab:num_articles_per_episode} provides statistics on the number of articles used to construct each episode. The episode with the most newspaper articles is the run, suspension, and reopening of the \href{https://finhist.com/bank-runs/episodes/1019971294.html}{Knickerbocker Trust Company} during the Panic of 1907.}

Third, during crises, newspapers sometimes speak more generally about several banks in an area being subject to runs or suspension, without explicitly listing all banks. Our baseline classification of a bank run requires a newspaper article to explicitly mention that a specific bank was run. However, when a newspaper clearly and explicitly reports that all banks in a city were subject to a run (or all banks of a given type, such as all savings banks), we augment our classification to include runs on these banks. We conduct robustness tests when including these ``all banks were run'' cases and find similar results.

\subsection{Additional Data Sources}

We combine the list of bank distress episodes with bank financial data for national banks. We use balance sheets of national banks from 1863-1941 from \citet{CLV2026}, based on national bank call reports submitted in the OCC's Annual Report to Congress.\footnote{The data in \citet{CLV2026} build on and extend the data in \citet{Carlson2022} and are digitized using the OCR techniques developed in \citet{Correia2022}.} These data contain detailed balance sheet line items, including total assets, loans, securities, deposits, equity capital, and noncore (non-deposit and non-equity) funding.

For national banks, we use the precise information on closures of national banks (receivership, mergers, assignment, and voluntary liquidations) from \citet{vanBelkum1968}, \citet{KeyBankData},  and \citet{CLV2026}. We define a national bank as failed when the OCC appoints a receiver. This definition of failure includes banks that eventually exit receivership, restore solvency, and continue operating, as well as banks that exit receivership and wind down their operations in an orderly voluntary liquidation that imposes no losses to creditors. However, this definition explicitly does not include temporary suspensions. To measure the extent of the deposit outflow in a bank failure, we further use deposits at suspension based on data on OCC receiverships digitized by \cite{CLV2026}. Given that information on bank fundamentals and closures is consistently available only for national banks, most of our analysis focuses on the sample of national banks.

To capture local defaults of non-financial firms that can trigger bank asset losses,  we create a novel state-level dataset of quarterly non-financial business failures covering 1886--1935 by digitizing various issues of \textit{Bradstreet's}. To obtain a business failure rate, we scale non-financial business failures by the number of manufacturing establishments from the census. To proxy real economic activity at the city level, we also create an index of manufacturing activity from \textit{Bradstreet's} Trade at a Glance tables (1917-33) and descriptions of Business Conditions (1933-35). The index is consistently available every week for 30
cities from 1917 through 1933 and every month from 1933 to 1935. \Cref{appendix:other_data} provides additional details and shows that the index is highly correlated with aggregate industrial production ($R^2=49\%$). Finally, we use information on banking crisis years and the month of banking panics from \cite{Baron2021}, as well as information on regional banking panics from  \citet{Wicker1996} and \cite{Jalil2015}.

\section{Bank Runs in the US, 1863-1934}
\label{sec:data_overview}

This section provides an initial look at the incidence of bank runs, suspensions, and failures in the US from 1863 to 1934. In addition to providing basic descriptive statistics, we also validate the quality of our newly constructed database, a key step to ensure the reliability of our findings.

\subsection{Descriptive Statistics on Bank Distress Episodes}

\begin{table}[ht!]
\begin{center}
\begin{threeparttable}
\caption{\textbf{Bank Runs, Failures, and Suspensions in Newspapers, 1863-1934}  \label{tab:event_table} }
\centering
\footnotesize
   \begin{tabular}{lcccccc}
\toprule
\multicolumn{7}{c}{\textbf{Panel A}: Total number of events, all bank types} \\ \hline  \\
Sample  &
\multicolumn{4}{c}{Bank run} & Failure & Suspension  \\ \cmidrule(lr){2-5}
\cmidrule(lr){6-6} \cmidrule(lr){7-7}
 & All & Run w/o  & Run with  & Run with &  &  \\
  & & suspension & sus. + reopen. & failure &  &  \\
\midrule

All&3,984&1,435&639&1,910&10,341&13,772\\
 \midrule
No crisis&2,130&823&232&1,075&6,198&7,245\\
Banking crisis&1,854&612&407&835&4,143&6,527\\
 \midrule
1863-1913 (NB Era)&2,042&991&341&710&3,717&4,932\\
1914-1928 (Early Fed)&797&230&87&480&2,655&3,082\\
1929-March 6, 1933 &1,132&208&210&714&3,367&4,825\\
After March 6, 1933&13&6&1&6&602&933\\
 \midrule
\\ \midrule
\multicolumn{7}{c}{\textbf{Panel B}: Number of events as a share of total number of banks (in \%), national banks} \\ \midrule
Sample  &
\multicolumn{4}{c}{Bank run} & Failure & Suspension  \\ \cmidrule(lr){2-5}
\cmidrule(lr){6-6} \cmidrule(lr){7-7}
 & All & Run w/o  & Run with  & Run with &  &  \\
  & & suspension & sus. + reopen. & failure &  &  \\ \midrule
All&0.39&0.15&0.06&0.18&0.89&1.17\\
  \midrule
No crisis&0.23&0.09&0.02&0.11&0.49&0.55\\
Banking crisis&1.42&0.49&0.29&0.65&3.45&5.07\\
 \midrule
1863-1913 (NB Era)&0.39&0.18&0.08&0.13&0.32&0.53\\
1914-1928 (Early Fed)&0.21&0.07&0.01&0.12&0.61&0.64\\
1929-March 6, 1933 &1.03&0.24&0.12&0.68&3.68&4.95\\
After March 6, 1933&0.03&0.01&0.00&0.03&3.50&4.09\\
 \bottomrule
\end{tabular}

\begin{tablenotes} \item
Notes: Panel A reports the counts of all runs, suspensions, and failures we identify in newspapers from 1863 to 1934. The sample covers bank distress episodes for all bank types: national banks, state banks, trust and private banks. Panel B reports the average annual rate of runs, suspensions, and failures for national banks, defined as the number of distress events indicated in the table header divided by the number of national banks in the previous year. For this sample, a bank failure is defined as an OCC receivership. The number of national banks and OCC receiverships per year comes from \citet{CLV2026}. Both panels distinguish between runs without suspension (``run only''), runs with suspension and reopening, and runs with failure. We also split the sample by different eras (National Banking Era, Early Federal Reserve, and Great Depression before and after the 1933 banking holiday) and whether a year involved a major banking crisis as identified in \cite{Baron2021}.

\end{tablenotes}
\normalsize
\end{threeparttable}
\end{center}
\end{table}

\begin{figure}[ht!]
    \centering
    \caption{\textbf{Bank Run, Suspension, and Failure Episodes} \label{fig:venn} }
    \includegraphics[width=1.0\linewidth]{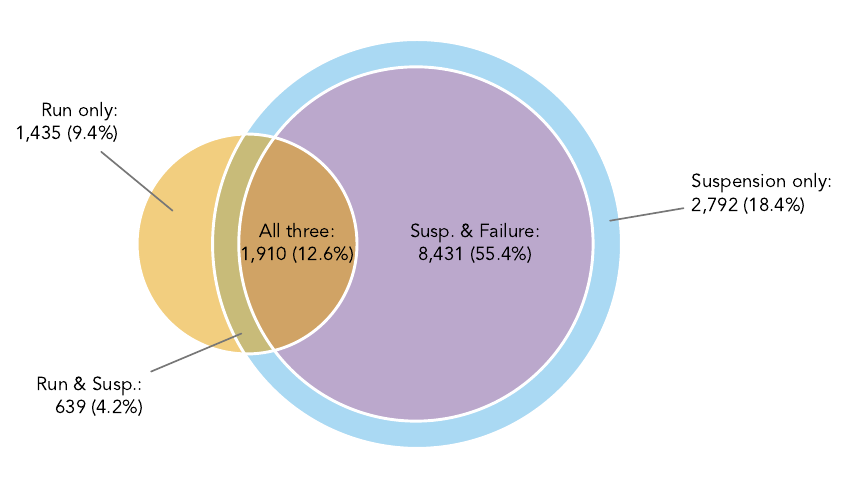}

    \begin{minipage}{\textwidth}
    \footnotesize
    Notes: The sample is all bank distress episodes involving a run, suspension, and/or bank failure over the period 1863-1934 for all bank types. By definition, failure requires suspension, so there are no cases with failure but without suspension.

    \end{minipage}
\end{figure}

Panel (a) of \Cref{tab:event_table} reports the raw counts of bank runs, failures and suspensions identified in our new database for all bank types. \Cref{fig:venn} shows a Venn diagram of episodes with a run, suspension, and/or failure. As noted above, each observation represents an episode of bank distress for a particular bank, with the same bank possibly being subject to distress at different points in time over the sample period.

The majority of runs do not involve bank failure, and the vast majority of failures do not involve runs.  During the 1863--1934 period, our dataset contains 3,984
\unskip bank run episodes.  The first key insight is that the majority of runs do not involve failure: There are 2,074
\unskip runs without failure and 1,910
\unskip runs with failure. We also find  13,772
\unskip episodes involving a suspension, and 10,341
\unskip episodes involving a failure.\footnote{As noted above, a bank always suspends before or at failure. Hence, by definition, there are no failures without suspensions.} A second key insight is thus that runs are substantially less common than bank suspensions and failures, and most bank failures do not involve runs, as illustrated in the Venn diagram (\Cref{fig:venn}).

Panel (b) in \Cref{tab:event_table} shows the unconditional probability of bank distress episodes in our 1863-1934 panel of national banks. To calculate probabilities of bank distress, we focus on national banks, where we have a precise measure of the total number of banks for the entire sample. Of the 3,984
\unskip bank run episodes, 1,280
\unskip  are runs on national banks. For national banks, we also have clear records that determine whether a bank failed, as the OCC recorded all national bank receiverships. Such precise records are not available for state banks.\footnote{There is no census on the universe of non-national banks. Moreover, there is no systematic information on failures of state banks or trusts. Therefore, much of our analysis relies on the merged sample of national banks.} A national bank's annual unconditional probability of failing between 1864 and 1934 is 0.89
\unskip\%, while the unconditional probability of being subject to a run is 0.39
\unskip\%. Again, the annual probability of a run without failure is 0.20
\unskip\%, so more than half of runs on national banks are not associated with failure. About 28
\unskip\% of runs without failure involve temporary suspension.

\subsection{Bank Runs and Narrative Banking Crises}

\begin{figure}[ht!]
\centering
\caption{\textbf{Bank Runs With and Without Failure, National Banks, 1863-1934 } \label{fig:events}}

{\includegraphics[width=1.0\textwidth]{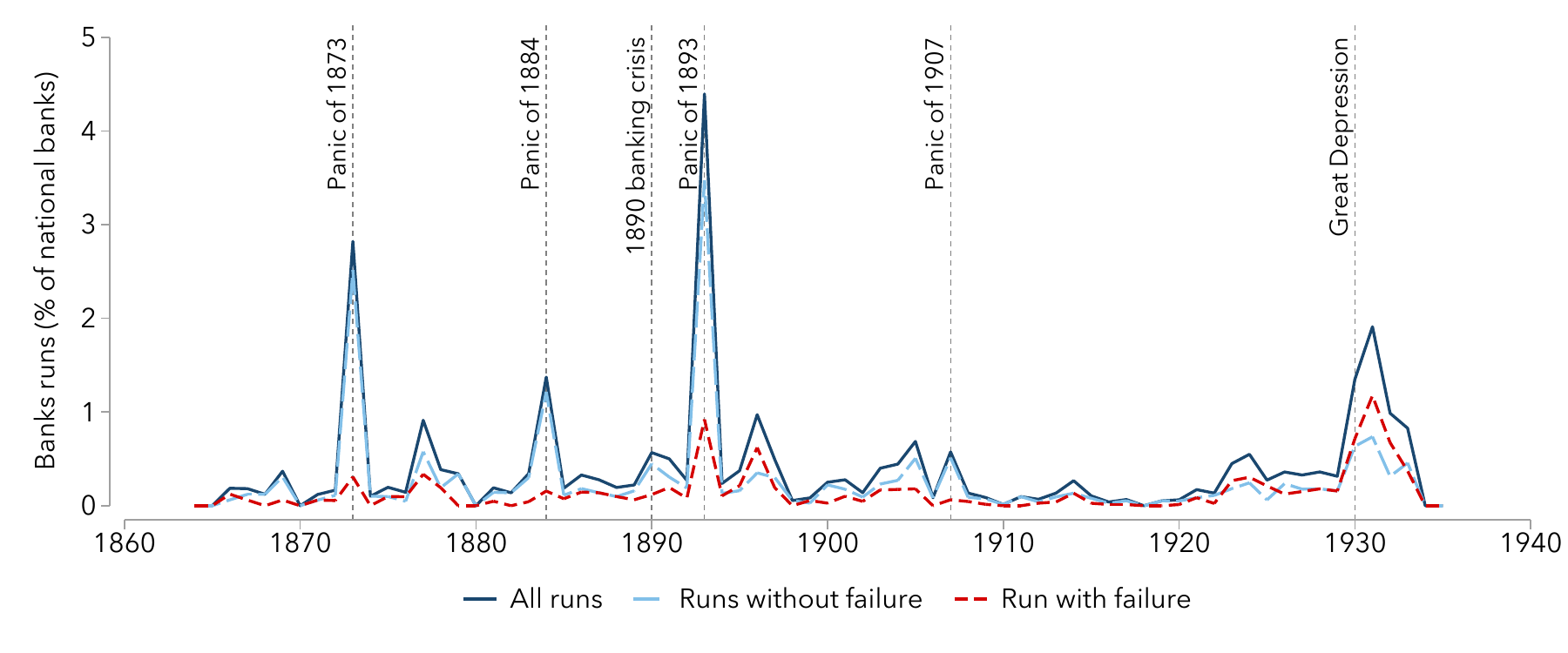}}

\begin{minipage}{\textwidth}
\footnotesize
Notes: This figure plots the ratio of the number of national banks subject to a run over the total number of national banks in each year. The number of national banks each year is from \citet{CLV2026}. The figure also separates runs into runs without failure and runs with failure.
\end{minipage}
\end{figure}

\Cref{fig:events} presents the rate of runs in the time series by plotting the number of runs as a fraction of the total number of banks in each year.\footnote{Appendix \Cref{fig:events_app} plots the fraction of national banks subject to a distress episode involving a run, suspension, or failure. Runs and suspensions spike during all major banking crisis years. Failures spike during some of the crisis years, such as 1893 and the Great Depression. We note that we ``only'' identify around 700 suspensions after the Banking Holiday in 1933 when all banks suspended \citep[see, e.g.,][]{Jaremski2023}. This is explained by the fact that newspapers may not have seen a need to report individual bank suspensions right around the time the entire banking system suspended.} We again focus on national banks so that we can calculate rates relative to the total number of banks. The rate of runs spikes during the major crisis years: 1873, 1884, 1893, 1907 (to a lesser extent), and the Great Depression. This incidence of runs confirms prominent narratives of crises in the pre-FDIC era \citep[e.g.,][]{Calomiris1991,Wicker1996,Wicker2006}. The highest rate of runs occurs in 1873, 1893, and 1930-33, which involved the most severe crises in the commercial banking system. The spike in runs is smaller for the milder crises of 1884 and 1890, as well as for 1907, where distress was more pronounced among non-banks (trusts) and in the money market. \Cref{appendix:case_studies} provides case studies of the major runs and failures in these crises, illustrating that our database captures the well-known bank-level events discussed in prominent narratives of U.S. banking crises \citep[e.g.,][]{Sprague1910,FriedmanSchwartz,Wicker1996,Wicker2006,Rockoff2021,Conti2025}.

\Cref{fig:events} also distinguishes between bank runs with and without failure. Runs without failure are especially pronounced during major banking crises, especially the ``panics'' of the national banking era. The Panics of 1873 and 1893 featured a particularly large number of runs without failure (but with temporary suspensions), as noted by previous work \citep[e.g.,][]{Carlson2005}. Runs with failure spiked during the Panic of 1893 and the Great Depression. Interestingly, most failures do not involve discussions of runs. Failures without runs were elevated during the agricultural downturns of the 1890s and 1920s; see Appendix \Cref{fig:events_app}.

\begin{figure}[ht]
\centering
 \caption{\textbf{Probability of Bank Runs With and Without Failure for National Banks During and Outside of Banking Crises} }
\includegraphics[width=0.9\linewidth]{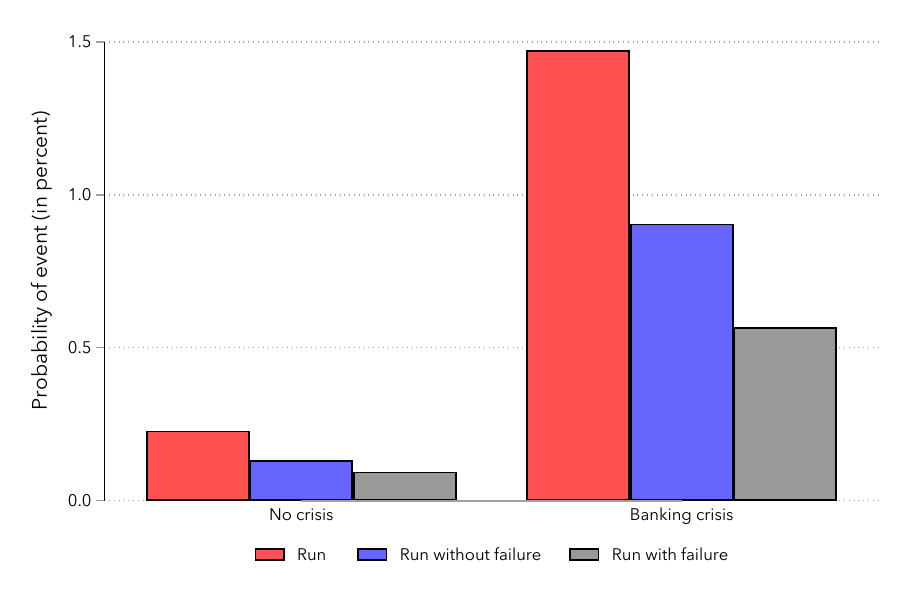}

\label{fig:runs_rate_crisis}

\begin{minipage}{\textwidth}
\footnotesize
Notes: The figure plots the probability of bank runs, runs without failure, and runs with failure during and outside of years identified as banking crises in \cite{Baron2021}. The sample is all national banks during 1863-1934. \Cref{fig:probability_eventS_by_crisisBVX} plots the probability of runs, suspensions, and failures during and outside of crises.
\end{minipage}
\end{figure}

In the time series, runs are much more likely during major banking crisis years. \Cref{fig:runs_rate_crisis} plots the rate of runs during and outside of banking crisis years and confirms this finding. Crisis years are based on the narrative chronology by \citet{Baron2021}.\footnote{\cite{Baron2021} define panics as episodes of severe and sudden withdrawals of funding by bank creditors from a significant part of the banking system.} The probability of a run increases by roughly a factor of six during banking crisis years compared to other years. The probability of a national bank being subject to a run is around 0.2\% outside of banking crisis years but increases to more than 1.4\% during a crisis; see also panel (b) of \Cref{tab:event_table}. The probability of a failure increases from 0.49\% to 3.5\%, and the probability of a suspension from 0.55\% to 5.1\%.

\Cref{tab:Jalil} confirms that bank runs and other distress events recorded in newspapers are substantially more likely in years and U.S.\ states that have been classified as having a banking crisis/panic by narrative chronologies of crises in a regression framework. Specifically, we test whether our list of bank distress events lines up with the state-level narrative banking panic dates from \citet{Jalil2015} (pre-1930) and \citet{Wicker1996} (1930-33), and with the aggregate banking crisis dates from \cite{Baron2021}.\footnote{\citet{Jalil2015} provides information on state-level banking panics. \citet{Jalil2015} also reports city-level panics, which we aggregate to the state level. \citet{Wicker1996} provides a detailed descriptive account of the regional panics during the Great Depression, which we hand-code from his text. \cite{Baron2021}, as noted above,  provide years of the onset of banking crises and the starting month of major banking panics.}  With this narrative information, we estimate the following regression at the state-month level:
\begin{align}
\text{Number of bank distress events}_{st} = \beta_0 & + \beta_1 \times \text{Major banking crisis}_t  \label{eq:jalil} \\ &+ \beta_2 \times \text{Major panic month}_t  \notag \\ &+ \beta_3 \times \text{Regional banking panic}_{st} + \epsilon_{st}.  \notag
\end{align}
The dependent variable is the number of runs, suspensions, or failures at the state-month level. $\text{Major banking crisis}_t$ is a dummy that takes a value of one in years of major banking crises, and  $\text{Major panic month}_t $ is a dummy that is one in months of major banking panics. $\text{Regional banking panic}_{st}$ is a dummy that takes the value one if a state is subject to a banking panic in a given month.
We compute robust standard errors clustered by state.

\begin{table}[!ht]
\caption{\textbf{Bank Runs, Suspensions, and Failures Correlate with Narrative Banking Crisis Classifications} } \label{tab:Jalil}
  \centering
       \begin{minipage}{1.0\textwidth}
        \begin{center}
        {\begin{tabular}{l*{3}{c}}   \toprule  Dependent variable & Run & Failure & Suspension \\   \cmidrule(lr){2-2} \cmidrule(lr){3-3}  \cmidrule(lr){4-4}
            &\multicolumn{1}{c}{(1)}         &\multicolumn{1}{c}{(2)}         &\multicolumn{1}{c}{(3)}         \\
\hline
Major banking crisis year&        0.20\sym{***}&        0.38\sym{***}&        0.68\sym{***}\\
            &     (0.030)         &     (0.074)         &      (0.11)         \\
Major banking panic month&        0.89\sym{**} &       -0.34\sym{***}&        0.90\sym{**} \\
            &      (0.43)         &     (0.095)         &      (0.45)         \\
Regional banking panic&        2.17\sym{***}&        2.34\sym{***}&        5.37\sym{***}\\
            &      (0.47)         &      (0.43)         &      (0.95)         \\
\hline
\(N\)       &       41609         &       41609         &       41609         \\
Mean dep. var.&        0.10         &        0.19         &        0.28         \\
\hline\hline
\end{tabular}
}

   \end{center}
    {\footnotesize Notes: This table reports estimates of \Cref{eq:jalil} using a monthly state-level panel with the number of distress events as the outcome variable. Major banking crises and panic months are taken from \citet{Baron2021}. This chronology captures the start year of banking crises; therefore, for the Great Depression, we also include 1931--1933 as crisis years. Regional banking crises  are from \citet{Jalil2015} (1884-1929) and \citet{Wicker1996} (1930-1933). \Cref{fig:Jalil_granular} presents estimates separately including a dummy for each regional crisis.  The sample period is 1863-1934.
    We compute robust standard errors clustered by state.     *,**, and *** indicate significance at the 10\%, 5\%, and 1\% level, respectively.}
    \end{minipage}
\end{table}

Our database of bank distress events correlates strongly with existing narrative chronologies of banking crises in U.S.\ history. On average, there are 0.1 bank run events per state-month. During years with a banking crisis, the number of runs in a state-month triples, increasing by 0.2 on average.

Moreover, bank run spikes also line up with the month of banking panics from narrative accounts. During panic months, the number of runs increases by another 0.9. This result is also confirmed by \Cref{fig:monthly_ts} in the Appendix, which plots the rate of bank runs at the monthly frequency. With vertical lines, we plot the monthly start date of banking panics according to \citet{Baron2021}.  During the National Banking Era, newspapers also pick up several additional borderline cases of banking panics, such as 1896.  For the Great Depression, we capture the various waves of runs discussed in narrative accounts. In particular, we capture the rise in runs in late 1930, the rise in runs from March to August 1931, the spike in runs after the U.K. leaves gold in September 1931, and another rise from the end of 1932 to the banking holiday in March 1933, consistent with the aggregate narrative of \cite{FriedmanSchwartz}. We also capture a small spike in runs in early 1930 and during the Chicago panic in the summer of 1932 \citep{Calomiris1997}. Further, while bank runs were more likely before the banking holiday of 1933, failures and suspensions were mostly reported during or after the banking holiday, in line with \citet{Jaremski2023}.

\Cref{tab:Jalil} further shows that during months of regional banking panics identified by \citet{Jalil2015} and \citet{Wicker1996}, the number of runs increases by 2.2 in a state. Bank suspensions and failures display a similar pattern. Further, in \Cref{fig:Jalil_granular}, we include dummies for individual regional crises identified by \citet{Jalil2015} and \citet{Wicker1996} in the regression. Most individual crises see considerable increases in the number of runs and failures. For instance, the number of runs increases by 9 during the Florida ``Fruit Fly Panic'' in  July 1929 \citep{Carlson1929Florida} and by 15 in the June 1932 Chicago panic \citep{Calomiris1997}.

Taken together, this evidence indicates that our database of runs lines up quantitatively with existing qualitative narrative chronologies of banking crises at the aggregate and regional level. The detailed case studies of bank-level events during major banking crises in \Cref{appendix:case_studies} support this statistical evidence. These case studies, with links to original news articles, show that we capture the famous bank runs and failures of all major crises in our sample period.

\subsection{Data Quality Checks}

Beyond showing that our bank distress events dataset lines up with aggregate and regional banking crisis chronologies, we provide several additional formal validation exercises.

\paragraph{Cross-Validation Exercises} To quantify the accuracy of our data construction pipeline, we conduct four formal cross-validation exercises. These exercises are detailed in  \Cref{app:cross-validation}. Overall, the match rate for national bank events is about 95\% across the four exercises.

First, we build a ground truth dataset of 287
\unskip runs, suspensions, and reopening events based on reading \citet{Wicker1996,Wicker2006}. These two studies provide detailed accounts of bank-specific events from historical newspapers.  For national banks, the key sample for most of our analysis, our dataset matches 94.1
\unskip\% of runs and 94.9
\unskip\% of all events in this ground truth. For all banks, including state and private banks, we match 85.0
\unskip\% of runs. Importantly, relative to the descriptive account in \citet{Wicker1996,Wicker2006}, we identify nearly 40 times more bank runs, and we can match these runs to bank fundamentals.

Second, we match all but one event from a list of 19
\unskip prominent runs and failures discussed in \cite{Rockoff2021}, yielding a match rate of 94.7
\unskip\%. We match all national bank events in this list. The only missing event is the failure of a trust that is not in our bank census.

Third, we match 93.5
\unskip\% of suspensions from a list of 123
\unskip  national bank suspensions published by the trade journal \textit{Bradstreet's} following the Panic of 1893.

Finally, we hired eight undergraduate students for one month to search \url{newspapers.com} for runs across our full sample period. This provided us with a sample of 271
\unskip bank runs from 1863 through 1934 that we manually verified. Compared to the previous three ground truth datasets, this last dataset provides a sample of somewhat more random runs discussed in newspapers, as it is not based on prominent runs or events during major crises.  Our dataset matches 94.3
\unskip\% of the runs on national banks in this ground truth. Extrapolating linearly, a heroic assumption, constructing our new distress events database through hand collection would require roughly 9,400 human hours, in addition to the untold cost of dissuading a group of promising research assistants from becoming economists.

We find that a bank event is not matched for one of two main reasons: either the bank is not in our bank census (not a concern for national banks) or we do not have a news article on the particular event in our corpus. Reassuringly for our LLM-based approach, we found that false positives and false negatives caused by LLM errors are very rare. Conditional on high-quality news article inputs, the ensemble of LLMs is highly accurate in classifying runs and other distress events.

\paragraph{Additional Data Quality Checks} We provide four additional data quality checks.  First, we correlate the true failure rate for national banks based on official OCC receiverships with the failure rate according to our classification from newspapers. The aim of this exercise is to assess how informative newspapers are about bank distress events whose occurrence we can verify with certainty. We construct an alternative failure rate from our bank distress episodes database by dividing the number of national bank failures in newspapers by the total number of national banks. \Cref{fig:scatter_rates} shows that the overall correlation between the true failure rate and the failure rate based on newspapers in annual data is good, but not perfect, as one might expect. This indicates that newspapers are informative about important bank distress events.

Second, we study whether deposit outflows in failing banks are larger for bank failures involving a run. \Cref{fig:deposit_outflows} plots the distribution of actual deposit outflows in national bank failures for failures with and without a bank run according to our bank distress episodes database. Deposit outflows are measured as the growth in deposits from the last call report before suspension to the time of suspension. The figure shows that failures with narrative evidence of a run in newspapers have larger deposit outflows before failure. On average, failures with narrative evidence of a run exhibit a 5
\unskip percentage points larger deposit outflow before failure. This finding provides evidence that narrative accounts of runs in newspapers indeed capture many failures with large deposit outflows.\footnote{Note that the deposit outflow from the last call report to failure can capture a steady deposit drain over a period of up to a year. Therefore, there are failures with larger deposit outflows that do not necessarily involve a run based on our conceptual definition of a run.}

The previous exercise focuses on deposit outflows conditional on bank failure. As a third exercise, \Cref{fig:events_conditional_on_dep_growth} plots the probability of a bank distress episode as a function of annual deposit growth, conditional on survival. Bank runs and suspensions that do not result in failure are all significantly more common for banks in the bottom 1\% of the deposit growth distribution. Thus, our classification of runs correlates with statistical evidence of deposit outflows. 

Finally, \Cref{fig:events_conditional_on_size} plots the probability of a run or failure reported in newspapers as well as the actual annual rate of bank failures by bank size, measured using total assets. It reveals that our bank distress events database from newspaper accounts is more likely to report events for larger banks, in line with events for larger banks being more newsworthy. Thus, the incidence of bank runs is more likely to be under-reported for smaller banks.  To address this potential shortcoming of our data, we show below that our main findings are robust to restricting the sample to large banks.

\section{Determinants of Bank Runs}

\label{sec:determinants}

What are the characteristics of banks subject to runs? Which shocks trigger runs? Are the determinants of runs the same as the determinants of bank failure? Guided by the predictions of the theories discussed in \Cref{sec:theory}, this section presents evidence on the determinants of bank runs.

\subsection{Empirical Framework}

We start by assessing the likelihood of a bank run as a function of bank-specific and broader economic conditions. As discussed in \Cref{sec:theory}, theories of bank runs have different predictions for what can explain the incidence of runs. Theories of non-fundamental runs \citep[e.g.,][]{Diamond1983} suggest that even healthy banks can be subject to a run, thus reducing the explanatory power of bank fundamentals. In contrast, theories of fundamental-based panic runs \citep[e.g.,][]{Goldstein2005} suggest that runs should predominantly happen in banks with weak fundamentals. Finally, theories of information-based runs \citep[e.g.,][]{Gorton1988} suggest that runs are more likely after adverse public signals about the state of the banking sector or the macroeconomy.

Motivated by these predictions, we examine the ability of various sets of variables to capture an elevated risk of runs and compare the determinants of bank runs to those of bank failure. Our goal is to describe the likelihood of runs as a function of bank-specific fundamentals and public signals that may trigger runs. To do this, we estimate predictive regressions of the following form:
\begin{align}
\text{Run}_{bt+1} &= \beta_0 +X^{Bank}_{bt} \beta  + X^{Macro}_{t} \gamma + X_{bt}^{Local} \delta + \epsilon_{bt+1}, \label{eq:run_failure_fundamentals}
\end{align}
where $\text{Run}_{bt+1}$ is an indicator variable for whether bank $b$ is subject to a bank run within the next year. In addition to examining bank runs, we also present results for several other outcomes: bank runs that do not result in failure, runs that result in failure, and all failures. The model allows us to test whether runs are predictable, and whether factors that imply a higher probability of a run are similar to those indicating a higher risk of failure. We use the sample of all national banks from 1863 to 1934, for which we have information on both bank runs and bank-level fundamentals captured in bank balance sheets.

On the right-hand side, we include a range of fundamental bank balance-sheet metrics in $X^{Bank}_{bt}$ to proxy for bank health and risk-taking. These measures have been shown to capture the risk of bank failures in prior work \citep[e.g.,][]{White1984,Calomiris1997,CLV2026}. First, we proxy for bank solvency with the surplus-to-equity ratio. Surplus profit captures past retained earnings and is thus a measure of past bank profitability and capitalization. Second, we include a measure of noncore funding. Noncore funding is defined as the ratio of non-deposit and non-equity funding to total assets. This measure includes wholesale funding, such as bills payable and rediscounts. Noncore funding tends to be more expensive and risk-sensitive than deposit funding. Existing work shows that reliance on noncore funding was often a signal of troubled banks suffering losses \citep[e.g.,][]{White1984,Calomiris1997}. Therefore, we note that, while noncore funding captures bank funding structure, it is both a signal of weak solvency and of potential liquidity risk. We also interact the solvency and noncore funding measures, as banks with low solvency and high reliance on noncore funding might be especially likely to experience future distress. Third, we include the liquid assets ratio as a proxy for both a bank's available liquid funds to accommodate withdrawals and the riskiness of a bank's investment mix.  Fourth, we include deposits-to-assets to capture the reliance on deposit funding. Finally, we include recent asset growth. Troubled banks might contract shortly before failure, so asset growth may contain information about the future likelihood of failure.

We also examine whether adverse public signals that are salient to depositors imply an increase in the likelihood of a run. In $X^{Macro}_t$, we include aggregate information that is not specific to bank $b$ that might lead depositors to revise their probability that banks are troubled. Specifically, we include the following time series variables: annual total return on the stock market, real GDP growth, the aggregate national bank failure rate, the aggregate rate of runs on national banks, and the aggregate non-financial business failure rate. In $X^{Local}_{bt}$, we include local public signals: an indicator for whether there is a run on another bank in the same city and the state-level non-financial business failure rate. All variables are measured as of year $t$. For example, the stock market return used to predict a run in year $t+1$ is measured as the return from December $t-1$ to December of $t$.

In addition to the estimated coefficients, we are interested in the extent to which bank-specific fundamentals and economic conditions can explain bank runs and other bank distress episodes. To assess the explanatory power, we calculate the in-sample area under the receiver-operating characteristic curve (AUC) for various predictive models. The receiver-operating characteristic (ROC) curve traces the true-positive rate against the false-positive rate across classification thresholds. The AUC is the integral under the ROC curve. An AUC of 1 indicates perfect classification performance, while an AUC of 0.5 indicates an uninformative predictor.

\subsection{Determinants of Runs}

\Cref{tab:pred_run} presents the results from predicting bank runs as a function of bank fundamentals, macro conditions, and local economic conditions. Column 1 uses only bank fundamentals. It shows that banks with low solvency (proxied by surplus-to-equity) and a high reliance on noncore funding are more likely to experience a run. The interaction between the solvency and noncore funding measures is also predictive of runs. Further, banks with a low share of liquid assets are more likely to be subject to a run. A higher deposits-to-assets ratio, indicating higher leverage, is also positively related to runs. These patterns are consistent with the notion that runs are more likely in banks that are closer to insolvency and less liquid, consistent with theories of fundamental-based runs where the threshold for whether a run occurs is a function of bank leverage and illiquidity.

\begin{table}[!ht]
\caption{\textbf{Predictability of Bank Runs Based on Bank Fundamentals and Economic Conditions}  } \label{tab:pred_run}
  \centering
\footnotesize
       \begin{minipage}{1.0\textwidth}
        \begin{center}
        \begin{tabular}{l*{4}{c}}
        \toprule
        Dependent variable & \multicolumn{4}{c}{Run in $t+1$} \\ \cmidrule(lr){2-5}
                            &         (1)   &         (2)   &         (3)   &         (4)   \\
\cmidrule(lr){1-5} \textbf{\textit{Bank fundamentals:}}&               &               &               &               \\
\cmidrule(lr){1-5} Surplus / Equity&       -0.30** &               &               &       -0.51***\\
                    &      (0.13)   &               &               &      (0.19)   \\
Noncore funding / Assets&       11.06***&               &               &       10.23***\\
                    &      (1.78)   &               &               &      (1.83)   \\
Surplus $\times$ Noncore funding&      -11.05***&               &               &      -11.75***\\
                    &      (2.60)   &               &               &      (2.35)   \\
Liquid Assets / Assets&       -0.65***&               &               &       -0.67*  \\
                    &      (0.24)   &               &               &      (0.34)   \\
Deposits / Assets   &        0.74*  &               &               &        0.67*  \\
                    &      (0.40)   &               &               &      (0.34)   \\
Asset Growth (3 years)&       -0.15   &               &               &       -0.05   \\
                    &      (0.14)   &               &               &      (0.09)   \\
\cmidrule(lr){1-5} \textbf{\textit{Macro conditions:}}&               &               &               &               \\
\cmidrule(lr){1-5} NB failure rate&               &       -0.79   &               &       -0.29   \\
                    &               &      (3.96)   &               &      (3.73)   \\
NB run rate         &               &        5.40   &               &       -5.30   \\
                    &               &     (17.17)   &               &     (14.63)   \\
Business failure rate&               &       21.34*  &               &       19.25   \\
                    &               &     (11.78)   &               &     (13.53)   \\
Real GDP growth     &               &        0.93   &               &        0.33   \\
                    &               &      (0.89)   &               &      (0.72)   \\
Stock market return &               &       -0.68** &               &       -0.65***\\
                    &               &      (0.26)   &               &      (0.22)   \\
\cmidrule(lr){1-5} \textbf{\textit{Local conditions:}}&               &               &               &               \\
\cmidrule(lr){1-5}  Run on other bank in same city&               &               &        3.47***&        3.45***\\
                    &               &               &      (0.95)   &      (0.95)   \\
Local business failure rate&               &               &       15.16***&        1.95   \\
                    &               &               &      (3.60)   &      (6.28)   \\
\cmidrule(lr){1-5} N&      290437   &      277086   &      273481   &      243908   \\
Mean dep. var       &         .39   &         .38   &         .38   &         .38   \\
  \bottomrule
        \end{tabular}
   \end{center}
    {\footnotesize Notes: This table presents estimates of \Cref{eq:run_failure_fundamentals}. \citet{Driscoll1998} standard errors are reported in parentheses, with a bandwidth of three years to allow for residual correlation within and across banks in proximate years.   *,**, and *** indicate significance at the 10\%, 5\%, and 1\% level, respectively. }
        \end{minipage}
\end{table}

Column 2 in \Cref{tab:pred_run} focuses on the information content of adverse public signals about the  economy and the banking system that are not bank-specific. We find that runs are more likely following years with low stock returns. They are also more likely when business failures are high.

Column 3 focuses on the predictive content of local economic conditions. Runs are significantly more likely following a run on another bank in the same city. The likelihood of a run is also elevated when the state-level business failure rate is higher. These findings are consistent with information-based runs, whereby agents observing adverse news about the economy or banking system revise their probability that other local banks are troubled, increasing the likelihood of runs on other banks. It can also be explained by deteriorating local economic conditions weakening many local banks.

\subsection{Predictability of Runs With and Without Failure}

\begin{table}[ht]
   \caption{\textbf{AUC Metric for Predictability of Bank Runs and Bank Failures}  \label{tab:AUC} }
        \begin{minipage}{1\textwidth}
        \begin{center}
        \small
        \begin{tabular}{lcccccc}
        \toprule

\cmidrule(lr){2-7}
Prediction horizon $h$  & \multicolumn{6}{c}{1 year}  \\             \cmidrule(lr){2-7}   \\
\midrule
 & (1) & (2) & (3) & (4) & (5) & (6) \\ \midrule
\multicolumn{7}{c}{\textbf{Panel A: Bank run}} \\ \midrule
AUC                 &       0.692&       0.817&       0.749&       0.788&       0.799&       0.840 \\
 \midrule
\multicolumn{7}{c}{\textbf{Panel B: Bank run without failure}} \\ \midrule
AUC                 &       0.617&       0.798&       0.687&       0.734&       0.755&       0.818\\
 \midrule
\multicolumn{7}{c}{\textbf{Panel C: Bank run with failure}} \\ \midrule
AUC                 &       0.838&       0.878&       0.865&       0.878&       0.883&       0.895\\
 \midrule
\multicolumn{7}{c}{\textbf{Panel D: Bank failure}} \\ \midrule
AUC                 &       0.874&       0.903&       0.897&       0.892&       0.904&       0.911\\
 \midrule
\multicolumn{7}{c}{\textbf{Specification details}} \\ \midrule
Bank fundamentals & \checkmark &  \checkmark &  \checkmark  &  \checkmark   & \checkmark  & \checkmark  \\
Macro conditions&  &  & \checkmark  &  & \checkmark   &  \\
Local conditions &  &   &    & \checkmark & \checkmark &\checkmark  \\
Year FE & & \checkmark  & &  & &  \checkmark  \\
        \bottomrule
        \end{tabular}
        \end{center}
        {\footnotesize Notes: This table reports the area under the receiver operating characteristic curve (AUC) across different outcome variables (rows) and specifications (columns). The corresponding regression coefficients underlying the models are reported in \Cref{tab:pred_run_run} (for Panel A), \Cref{tab:pred_run_run_no_fail} (for Panel B), \Cref{tab:pred_fail_w_run} (for Panel C), and \Cref{tab:pred_fail} (for Panel D).

        }
        \end{minipage}
 \end{table}

\Cref{tab:AUC} studies how well the three sets of variables can classify bank runs and other bank distress episodes based on the AUC metric. Panel A focuses on the predictability of bank runs. Panel A, column 1 shows that bank runs are only moderately predictable based on bank fundamentals, with an AUC of 0.69.

\Cref{tab:AUC} column 2 includes year fixed effects in the predictive regression. This specification asks what is the highest AUC that can be obtained from annual time series variation. Including year fixed effects substantially increases the AUC for predicting runs to 0.82. Thus, time series variation is an important source of variation in runs. This is consistent with runs clustering in particular ``panic'' years.

In columns 3-5, we attempt to ``interpret'' the year fixed effects by including the proxies of economic conditions. Adding information on local conditions substantially increases the AUC for bank runs, moving the AUC closer to the specification with year fixed effects. The AUC of 0.80 in column 5 indicates that runs are substantially, though not extremely strongly, predictable based on bank-level and aggregate information.

We next compare the predictability of bank runs with and without failure (\Cref{tab:AUC} panels B and C). We also compare the predictability of runs to the predictability of bank failures (panel D). Column 1 in \Cref{tab:AUC} reveals that bank fundamentals are much more strongly related to bank runs \textit{with} failure compared to bank runs without failure. Runs without failure are weakly predictable based on bank fundamentals, with an AUC of only 0.62. In contrast, bank failures are tightly linked to bank fundamentals with an AUC of 0.87 (panel D).

Column (2) shows that including year fixed effects substantially increases the AUC for predicting runs \textit{without} failure. The relative increase is more modest for predicting runs \textit{with} failure. This suggests that time series variation is especially important for understanding runs \textit{without} failure, while bank-specific fundamentals are more important for classifying runs with failure. As a result, the inclusion of variables capturing macro and local conditions leads to a substantial increase in the AUC for bank runs without failure, from 0.62 to 0.76. However, it leads to a more moderate improvement for predicting runs with failure, which are already well-captured by bank-specific fundamentals.

Overall, the evidence indicates that, to understand bank runs, both bank fundamentals and adverse local and macro news play an important role. For runs without failure, adverse macro news is substantially more important than bank-specific information. This is in line with banking panics occurring after bad news about the economy or the banking system, leading to runs even on healthy banks. While balance sheet metrics contain substantial information for distinguishing strong and weak banks, confused depositors do not fully incorporate this information. On the other hand, bank fundamentals are much more important for understanding bank runs with failure, and bank failures more broadly. Bank failures, including those with runs, tend to occur in banks with the weakest balance sheets, consistent with the evidence in \citet{CLV2026}. Thus, fundamentals-based theories are promising for explaining bank runs and failures, while information-based theories are important for explaining runs without failure.

\section{Bank Runs and Bank Failure}

\label{sec:runs_failure}

We next investigate the relation between bank runs and bank failures in more detail. Which runs lead to bank failure? Can a run lead to the failure of a healthy bank? Or do runs only lead to failures for banks with weak fundamentals?

The previous section showed that runs are more likely in banks with weak fundamentals. Yet even strong banks can be subject to runs, especially in response to adverse public signals. In this section, we show that a run is much more likely to translate into failure for weak banks. Runs essentially never result in failure for banks with strong fundamentals. Moreover, failing banks display weak and deteriorating fundamentals, irrespective of whether they fail with or without a run.

\subsection{Bank Runs and Failure across Weak and Strong Banks}

\paragraph{Summarizing Fundamentals} We start by constructing a simple bank-level measure of fundamentals, denoted $\text{Fundamentals}_{bt},$ based on the regression of bank failure on bank fundamentals (i.e., in column 1 of \Cref{tab:AUC} with only bank fundamentals, reported in column 1 of \Cref{tab:pred_fail}). The idea is to summarize fundamentals that are predictive of failure into one number, akin to an Altman Z-score \citep{Altman1968}. To avoid look-ahead bias, we estimate the regression iteratively using data up to $t-1$ to obtain predicted failure probabilities for time $t+1$. $\text{Fundamentals}_{bt}$ is then defined as the negative of the predicted value from the regression that predicts failure in year $t+1$ based on bank fundamentals, so a higher value indicates a bank with stronger fundamentals.

\begin{figure}[!ht]
\centering

\caption{\textbf{Probability of Bank Runs by Bank Fundamentals} \label{fig:cond_prob} }

\includegraphics[width=0.85\textwidth]{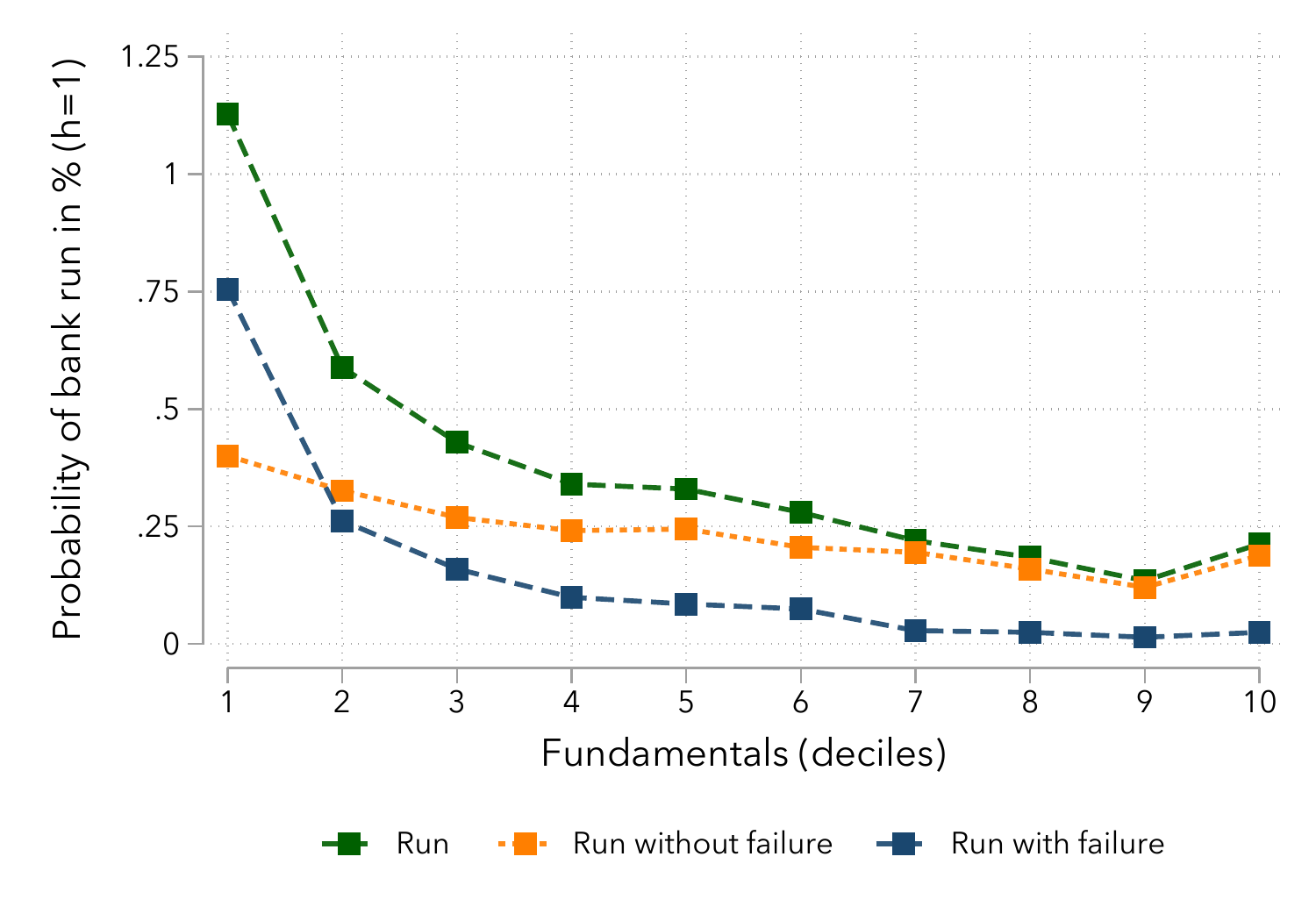}

\begin{minipage}{\textwidth}
\footnotesize
Notes: This figure plots the probability of a bank run, a bank run that does not result in failure, and a bank run that does result in failure across deciles of a proxy of bank-specific fundamentals (see text for definition).

\end{minipage}
\end{figure}

\paragraph{Fundamentals and Bank Runs With and Without Failure} \Cref{fig:cond_prob} provides a simple visualization of the relationship between bank fundamentals and the likelihood of a run, a run with failure, and a run without failure. It shows the conditional probability of a bank run over the next year across deciles of our proxy for fundamentals. In line with the patterns documented in \Cref{tab:pred_run}, runs are more likely in banks with weak fundamentals. A bank in the lowest decile of fundamentals has a roughly 110 basis point probability of being exposed to a run in a given year. The probability drops for stronger banks to below 25 basis points for banks with the strongest fundamentals. Thus, while much less likely, strong banks can be subject to runs. Moreover, we note that, while the probability of a run is higher for weak banks, both the level and the difference relative to strong banks are still relatively low in absolute terms, in line with the relatively modest predictability of runs based on fundamentals alone; see \Cref{tab:AUC}.

\Cref{fig:cond_prob} also plots the probability of runs with and without failure across the distribution of bank fundamentals. The figure reveals two distinctive patterns. Runs that do not result in failure are only weakly related to fundamentals. These runs are only slightly more likely to happen in weak banks than in strong banks.

Runs with failure, in contrast, are concentrated among banks with weak fundamentals. Their incidence is highest in the lowest decile and declines sharply for banks with stronger fundamentals. A bank that is above the 7th decile of the fundamentals distribution will very rarely be subject to a run and fail. Hence, a look at the relationship of runs, failures, and fundamentals in the raw data suggests that even though strong banks can be subject to a bank run, for a run to be associated with failure, a bank must have weak fundamentals.

\paragraph{Pass-through of Bank Runs to Bank Failure} To more formally study the importance of bank fundamentals for whether a bank can withstand a run, we estimate the following specification for the pass-through of a bank run into failure:
\begin{align}
    \text{Failure}_{bt} = \beta_0 &+ \beta_1 \text{Run}_{bt} + \beta_2 \text{Fundamentals}_{bt-1}  \label{eq:failure_run} \\
    &+ \beta_3 \text{Run}_{bt} \times \text{Fundamentals}_{bt-1} + \epsilon_{bt},  \nonumber
\end{align}
where $\text{Failure}_{bt}$ is an indicator for whether bank $b$ fails in year $t$, $\text{Run}_{bt}$ is an indicator for whether the bank is exposed to a run in year $t$, and $\text{Fundamentals}_{bt-1}$ is our proxy for bank health measured in year $t-1$, defined above. This specification asks whether a run is associated with an increased likelihood of failure over the next year and how this varies across banks with weak and strong fundamentals.

\begin{table}[!ht]
\caption{\textbf{Pass-Through of Runs to Failure: The Role of Bank-Specific Fundamentals} } \label{tab:fail_run_fundamentals}
  \centering
   \footnotesize
       \begin{minipage}{1.0\textwidth}
        \begin{center}
        {\begin{tabular}{l*{6}{c}}   \toprule  Dependent variable & \multicolumn{6}{c}{Failure in t} \\  \cmidrule(lr){2-7}
                &\multicolumn{1}{c}{(1)}         &\multicolumn{1}{c}{(2)}         &\multicolumn{1}{c}{(3)}         &\multicolumn{1}{c}{(4)}         &\multicolumn{1}{c}{(5)}         &\multicolumn{1}{c}{(6)}         \\
\midrule
Run             &     0.38\sym{***}&     0.25\sym{***}&     0.43\sym{***}&     0.40\sym{***}&     0.33\sym{***}&     0.29\sym{***}\\
                &  (0.051)         &  (0.039)         &  (0.053)         &  (0.051)         &  (0.053)         &  (0.053)         \\
Fundamentals    &                  &    -3.40\sym{***}&                  &                  &                  &                  \\
                &                  &   (0.82)         &                  &                  &                  &                  \\
Fundamentals $\times$ run&                  &    -12.7\sym{***}&                  &                  &                  &                  \\
                &                  &   (1.82)         &                  &                  &                  &                  \\
Strong fundamentals&                  &                  &  -0.0094\sym{**} &                  &  -0.0054\sym{**} &                  \\
                &                  &                  & (0.0038)         &                  & (0.0025)         &                  \\
Strong fundamentals $\times$ run&                  &                  &    -0.32\sym{***}&                  &    -0.27\sym{***}&                  \\
                &                  &                  &  (0.049)         &                  &  (0.048)         &                  \\
Very strong fundamentals&                  &                  &                  &  -0.0072\sym{**} &                  &  -0.0034\sym{**} \\
                &                  &                  &                  & (0.0029)         &                  & (0.0017)         \\
Very strong fundamentals $\times$ run&                  &                  &                  &    -0.30\sym{***}&                  &    -0.23\sym{***}\\
                &                  &                  &                  &  (0.067)         &                  &  (0.066)         \\
\midrule
Observations    &   272535         &   272535         &   272535         &   272535         &    73859         &    73859         \\
Mean dep. var.  &   0.0084         &   0.0084         &   0.0084         &   0.0084         &   0.0050         &   0.0050         \\
Sample          &                  &                  &                  &                  &Large Banks         &Large Banks         \\
\bottomrule
\end{tabular}
}

   \end{center}
    {\footnotesize Notes: This table presents estimates of \Cref{eq:failure_run}. We define a bank to have  ``strong'' fundamentals when  $\text{Fundamentals}_{bt-1}$ is in the upper tercile of its historical distribution. ``Very strong fundamentals'' indicates when  $\text{Fundamentals}_{bt-1}$  is in the highest decile. \citet{Driscoll1998} standard errors are reported in parentheses with a bandwidth of three years to allow for residual correlation within and across banks in proximate years.   *,**, and *** indicate significance at the 10\%, 5\%, and 1\% level, respectively. }
        \end{minipage}
\end{table}

\Cref{tab:fail_run_fundamentals} presents the results from estimating \Cref{eq:failure_run} for the sample of national banks from 1863 to 1934. Column 1 shows that the probability of bank failure increases by 38 percentage points when a bank is subject to a bank run, roughly in line with \Cref{fig:venn}. This increase in the probability of failure is considerable given that the unconditional probability of failure is 0.84\% in this sample. Thus, a bank subject to a run is more than 40 times more likely to fail than the average bank, and runs often result in failure. At the same time, a run is not a death sentence, and runs without failure are more common than runs with failure.

Column 2 in \Cref{tab:fail_run_fundamentals} further shows that banks with strong fundamentals have a significantly lower probability of failure, echoing the findings in \citet{CLV2026}. Moreover, in line with the evidence from \Cref{fig:cond_prob}, the negative coefficient on $\text{Fundamentals}_{bt-1} \times \text{Run}_{bt}$ implies that strong fundamentals significantly reduce the probability of bank failure after a  run. 

In order to quantify how fundamentals matter for whether a bank fails after a run, we further interact the run indicator with an indicator for whether a bank's fundamentals are ``strong.'' We define a bank as having  ``strong'' fundamentals when  $\text{Fundamentals}_{bt-1}$ is in the upper tercile of its historical distribution.  Column 3  shows that banks in the two bottom terciles of the fundamentals distribution have a 43 pp higher chance of failing when subject to a run, but banks with strong fundamentals have a 32 pp lower probability of failure. Thus, having strong observable fundamentals substantially reduces the probability of failure when a bank is exposed to a run. Column 4 further shows that banks with very strong fundamentals, defined as being in the top decile of $\text{Fundamentals}_{bt-1}$, have a 30 pp lower probability of failure compared to all other banks, which have a 40\% chance of failing in this specification. Thus, very strong banks rarely fail when subject to a run. While the fact that very strong banks can be subject to runs is in line with both theories of non-fundamental runs and information-based runs (Prediction 1), the fact that strong banks do not fail after a run is only predicted by theories of information-based bank runs (Prediction 2).

\begin{figure}[!ht]
\centering
 \caption{\textbf{Pass-Through of Runs to Failure across Deciles of Bank-Specific Fundamentals} \label{fig:failure_run_by_fundamentals}}
\includegraphics[width=0.9\linewidth]{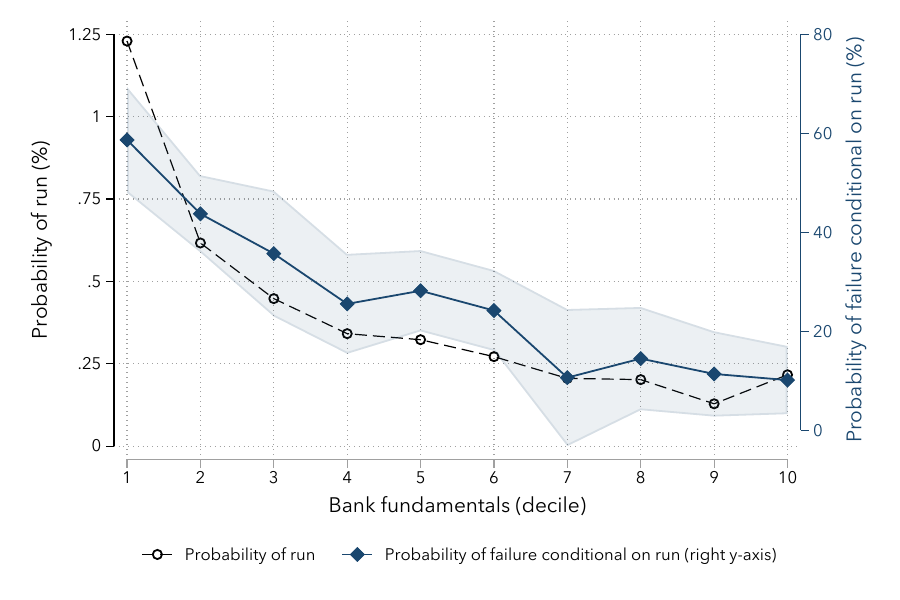}

\begin{minipage}{\textwidth}
\footnotesize
Notes: This figure plots the likelihood of a run (left y-axis) and the probability of failure conditional on a run (right y-axis) across deciles of bank fundamentals. The right y-axis plots the coefficients from estimating \Cref{eq:failure_run_deciles} separately for each decile of the distribution of $\text{Fundamentals}_{bt-1}$. $\text{Fundamentals}_{bt-1}$ is a proxy of bank health (see text). The sample consists of all national banks from 1863 through 1934.  The shaded area represents 95\% confidence intervals based on Driscoll-Kraay standard errors with a bandwidth of three years to allow for residual correlation within and across banks in proximate years.
\end{minipage}
\end{figure}

\Cref{fig:failure_run_by_fundamentals} presents a visualization of the likelihood that a run leads to failure across the full distribution of bank fundamentals. For each decile $d$ of $\text{Fundamentals}_{bt-1}$, we estimate
\begin{align}
\text{Failure}_{bt} = \beta^d_0 + \beta^d_1 \text{Run}_{bt} + \epsilon_{bt}, \quad d=1,...,10,    \label{eq:failure_run_deciles}
\end{align}
and plot the estimated coefficients $\{\hat \beta^d_1\}$. The figure shows that a run is associated with a 59 percentage point increase in failure probability for banks in the lowest decile of observable fundamentals. The increase in failure probability associated with a run falls nearly monotonically for banks with stronger fundamentals. For banks in the strongest decile of our proxy for fundamentals, the increase is only about 10 percentage points. Overall, this suggests that bank-specific fundamentals are central to understanding the pass-through of bank runs to failure.

Before proceeding, we note that these estimates provide upper bounds on the likelihood that a run causes a healthy bank to fail. The reason is that $\text{Fundamentals}_{b,t-1}$ may, in some cases, be an imprecise proxy of bank health. For instance, publicly available financial statements cannot capture episodes where a bank was insolvent but had not yet recognized losses (e.g., due to an unexpected shock, fraud, or hidden losses). In these cases, news about a bank's trouble that is not yet reflected in bank balance sheets would trigger a run and failure, but the run is a response to a signal of insolvency.\footnote{\cite{CLV2026} document that failed national banks had both low recovery rates and low asset quality according to OCC bank examiners. That study argues this suggests that the vast majority of failed banks were fundamentally insolvent, unless one assumes large value destruction from receivership itself.}

\begin{figure}[!ht]
\centering
\caption{\textbf{Daily Dynamics of Failure and Suspension Probabilities around Bank Runs} \label{fig:cond_prob_daily} }
\subfloat[Suspensions and bank failures around bank runs]{\includegraphics[width=0.5\textwidth]{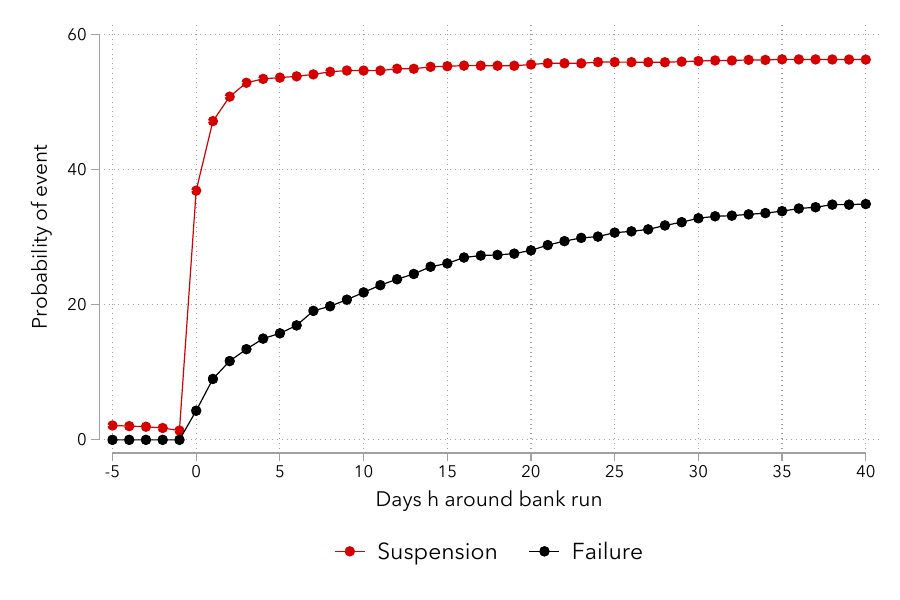}}
\subfloat[Bank failure around runs by bank fundamentals]{\includegraphics[width=0.5\textwidth]{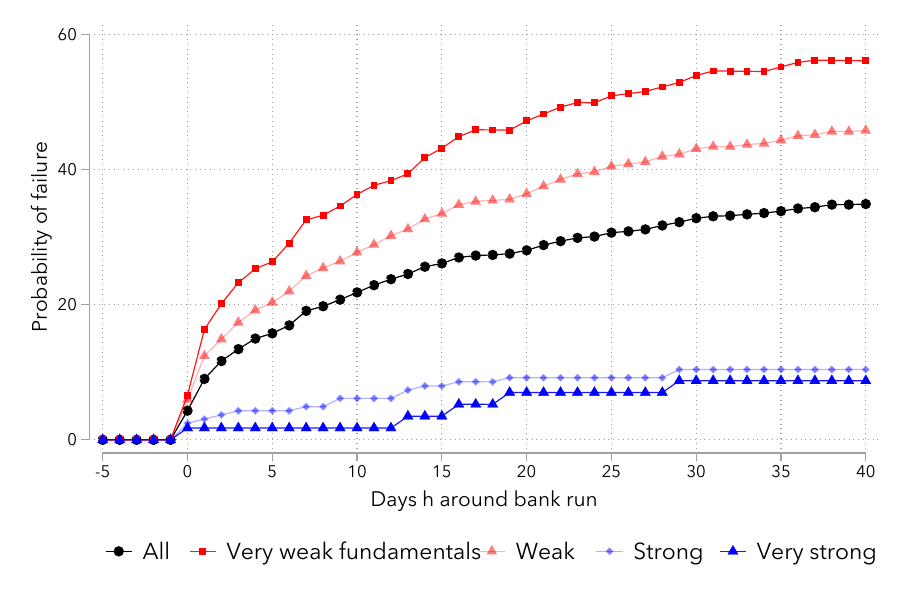}}

\begin{minipage}{\textwidth}
\footnotesize
Notes: Both panels plot coefficients from estimating \Cref{eq:lp_fail}. In panel (a), we report results with failure and suspension as the outcome variables. In panel (b), we estimate the responses separately based on bank fundamentals, measured with $\text{Fundamentals}_{bt-1}$. Specifically, we define bank fundamentals categories as: ``very weak'' (bottom decile), ``weak'' (bottom tercile), ``strong'' (top tercile), and ``very strong'' (top decile).
\end{minipage}
\end{figure}

\paragraph{High Frequency Dynamics} \Cref{fig:cond_prob_daily} zooms in on the daily dynamics of the probability of failure following a run. The figure reports the cumulative probabilities of bank suspension and bank failure around a run for all national banks, estimated from local projections of the form:
\begin{align}
\text{Failure}_{bt+h}= \beta^{h} \text{Run}_{bt} + \epsilon_{bt + h},  \hspace{1cm}  h = -5,...,40. \label{eq:lp_fail}
\end{align}

\Cref{fig:cond_prob_daily} panel (a)  plots the cumulative probability of bank suspension and bank failure following a run. We find that the probability of bank failure rises gradually to about 35\% in the 40 days following a run. Thus, essentially all failures following a run happen roughly within a month of the run.  The dynamics of suspension display two differences compared to the dynamics of failure. First, suspension is more likely than failure, occurring in more than 50\% of runs. Second, suspension typically occurs immediately, within the same day or within a few days of the run. Nearly 40 percent of runs involve suspension on the same day.

\Cref{fig:cond_prob_daily} panel (b) further visualizes the cumulative probability of bank failure following a run across banks by their fundamentals. For each run occurring in year $t$, we split banks into the bottom decile (very weak), bottom tercile (weak), top tercile (strong), and top decile (very strong) of $\text{Fundamentals}_{bt-1}$, measured as of the most recent call report. In line with the evidence in \Cref{tab:fail_run_fundamentals}, we find a striking difference in the likelihood of bank failure across weak and strong banks following a bank run. Very weak banks see a relatively quick spike in failures after runs and a cumulative probability of failure of about 56\% within the 40 days following a run. In contrast, the probability of failure is about 9\% for very strong banks.

\paragraph{Event Study of Fundamentals around Runs and Failures} Appendix \ref{app:results}  provides further illustration of the marked differences in bank fundamentals across banks subject to runs that fail and those that survive. We conduct simple event studies of bank fundamentals around runs without failure and runs with failure (see \Cref{fig:dynamics_characteristics}). Banks subject to runs that do not fail look similar to ordinary banks not subject to any form of distress. In contrast, banks that are run and fail have worse fundamentals five years before failure. Moreover, these banks have deteriorating profitability, rising non-performing loans, a declining deposit base, and an increasing reliance on expensive noncore funding in the three-to-five years before failure. This evidence reinforces the finding that poor fundamentals are key to determining whether a bank will fail. Runs that do not result in failure tend to occur in healthy banks, while runs that result in failure occur in observably weak banks.

In \Cref{fig:dynamics_characteristics}, we also compare banks that fail with a run to banks that fail without a narrative indication of a run. Banks that fail without a run are slightly worse across a range of fundamentals compared to banks that fail with a run, but the differences are modest. This suggests that runs may accelerate the failure of banks with deteriorating fundamentals, triggering the failure of marginally healthier banks. However, given that these banks are on very similar trajectories, many failures with runs were likely inevitable.

\paragraph{Robustness and Subsamples} As noted above, bank distress episodes are more likely to be accurately reported in newspapers for the largest banks. Runs might be underreported for small banks, especially if they do not result in failure, possibly leading to an upward-biased estimate of the likelihood that a run results in failure for small banks. To address this concern, it is useful to verify that our findings hold in the sample of larger banks.  Column 5 of \Cref{tab:fail_run_fundamentals} zooms in on  banks in the top quartile of total assets. In the sample of large banks, we find that banks have a 33 pp higher chance of failing when subject to a run, but banks with strong fundamentals have a 27 pp lower probability of failure. These numbers are similar to the full sample estimates.

To ensure that the results are not sensitive to the construction of the summary fundamentals measure, in \Cref{tab:fail_run_fundamentals_granular}, we estimate a variant of \Cref{eq:failure_run} where we interact $\text{Run}_{bt}$ with the underlying bank fundamental metrics, rather than using the summary $\text{Fundamentals}_{bt-1}$ measure. We find that the pass-through from runs to failure is lower for banks with higher surplus-to-equity, lower reliance on noncore funding, and a lower share of a proxy for nonperforming loans.

In \Cref{tab:fail_run_fundamentals_by_era}, we estimate the pass-through of runs to failure across subsamples. The pass-through of runs to failure is higher in weaker banks across all sample periods. Interestingly, runs without failure are significantly more common in the pre-Federal Reserve sample. After the introduction of the Fed, most runs are associated with failure. This holds both in the 1920s sample and the Great Depression. Two changes could account for this shift.  First, the introduction of the Fed reduced the incidence of panic runs on relatively healthy banks that generally did not translate into failure, but it could not prevent runs from closing insolvent banks. Second, the Fed's founding may have displaced the private-sector arrangements that allowed banks to survive runs.\footnote{Bank failures are especially highly predictable in the post-1913 period. Moreover, bank asset quality and recovery rates are very low in this sample \citep{CLV2026}. This suggests that the Fed may have prevented runs on solvent banks and even delayed runs on insolvent banks via discount window access \citep{Schwartz1992misuse}, implying that once runs occurred, they usually happened in deeply insolvent banks, making failure inevitable.}

\vspace{.5cm}

Taking stock, the evidence indicates that, while runs can occur for strong and weak banks, they are substantially more likely to occur in weak banks, consistent with theories of fundamental-based panic runs. Moreover, conditional on a run, a weak bank is much less likely to survive the run. Thus, the combination of these two findings implies that runs almost never trigger the failure of strong banks.

\subsection{Bank Responses to Runs}

Banks with strong fundamentals typically do not fail when subject to a run. What measures do banks take in response to runs? To answer this question, we exploit additional information in  newspapers. We identify seven common responses described in newspapers:
\begin{itemize}
    \item[(i)] the bank accommodated withdrawals by paying depositors;
    \item[(ii)] the bank borrowed from another bank;
    \item[(iii)] the bank sent a public signal to reassure depositors, such as conspicuously delivering gold or cash to the bank;
    \item[(iv)] bank owners or investors injected equity or provided a loan;
    \item[(v)] partial suspension of convertibility;
    \item[(vi)] full suspension of convertibility; and
    \item[(vii)] the bank underwent an examination by state supervisors, federal examiners, or by the local clearinghouse to ascertain its solvency.
\end{itemize}
These responses are not mutually exclusive. We query an LLM to identify whether the newspaper description of a bank run episode mentions any of the seven potential bank responses.  Appendix \ref{appendix:data} provides the exact prompts we use.

\begin{figure}[ht!]
\centering

\caption{\textbf{Bank Responses to Runs Discussed in Newspapers} \label{fig:responses} }

{\includegraphics[width=1.0\textwidth]{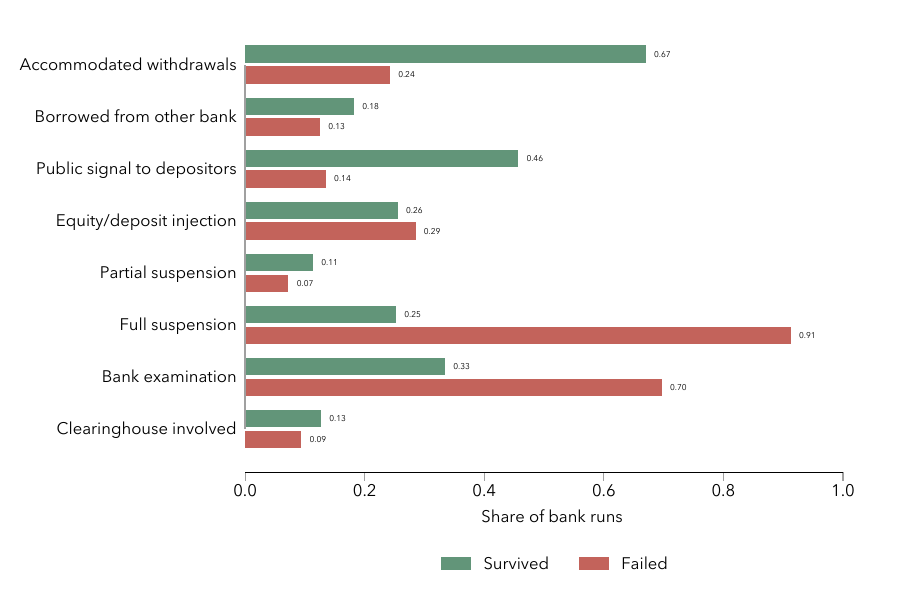}}

\begin{minipage}{\textwidth}
\footnotesize
Notes: This figure plots the share of episodes involving a bank run in which the newspapers mention one of the listed actions. The shares are reported separately for runs without and with failure.
\end{minipage}
\end{figure}

\Cref{fig:responses} reports the share of run episodes mentioning each of these responses. We distinguish between runs with and without failure. Banks that survive runs are more likely to be described as accommodating withdrawals compared to banks that fail in the run.
They are also  more likely to send a public signal to reassure depositors. Just like the commonly invoked George Bailey, who as noted above fended off the run on his bank by a mix of persuasion and using his own private wealth, owners of solvent banks commonly undertook successful measures to calm down worried depositors, including persuasion, other public signals such  as delivering a ``truckload of cash'', or injecting additional equity into the bank (although the latter is equally likely in both failing and surviving banks).
Surviving banks are also more likely to borrow from other banks.\footnote{We note that borrowing from other banks is likely to be underreported in newspapers, as banks may not want to make this public to avoid the stigma of needing assistance from other banks.}  Surviving banks often take several of these actions in conjunction to ward off a run (see \Cref{fig:responses_cond_prob}). Surviving banks are also more likely to partially suspend convertibility, though this is much less common than full suspension of convertibility.

In contrast, newspapers are substantially less likely to report that banks that fail in a run take these actions. Instead, newspapers emphasize that these banks are likely to fully suspend convertibility of deposits into currency. Moreover, banks that fail are also more likely to be subject to an examination by state supervisors, federal examiners, or the local clearinghouse to evaluate if the bank is solvent. When these banks are deemed to be insolvent, they are closed. Finally, in about 10\% of runs, newspapers explicitly mention clearinghouse involvement, either through the provision of liquidity (loan certificates), examination, or another measure.\footnote{\Cref{tab:fail_run_Clearinghouses} examines whether the pass-through of runs to failure is lower for banks that are likely to be part of a local clearinghouse. We identify cities with clearinghouses using data from \citet{Jaremski2018}. National banks in cities with a clearinghouse were more likely to survive a run, suggesting that clearinghouses mitigated the effect of runs. This effect may be explained by the ability of clearinghouses to provide emergency liquidity as well as to credibly examine and transmit information about member banks' health. \Cref{appendix:case_studies} discusses several cases where clearinghouse intervention prevented runs from translating into failures through examination and liquidity provision, such as \href{https://finhist.com/bank-runs/episodes/29000885.html}{Fourth National Bank of New York} in 1873 and \href{https://finhist.com/bank-runs/episodes/112101013.html}{Metropolitan National Bank} in 1884.}

\subsection{Non-Fundamental Runs and Bank Failures}

\label{sec:nf_run}

Can non-fundamental runs trigger bank failures? To answer this question, we use additional information from the context provided in newspapers around bank runs to classify a subset of the runs in our dataset as ``non-fundamental'' runs. We define a run as ``non-fundamental'' when newspapers affirmatively describe that the run was triggered by a specific, discrete episode of misinformation or confusion that triggered withdrawals and the source explicitly indicates that the information reported was false. These non-fundamental runs are \textit{not} triggered by adverse news about the bank, failures or distress of other banks, systemic panic, or local economic shocks. See the prompt in \Cref{appendix:prompts} for the exact definition used in the classification.

We identify  53
runs on national banks that are classified as non-fundamental, indicating that such runs are uncommon. \Cref{fig:run_examples_theories} provides examples of non-fundamental runs. For instance, the \href{https://finhist.com/bank-runs/episodes/71000886.html}{First National Bank of Minneapolis} was subject to a run after newspapers mistakenly reported it had suspended, confusing it with the First National Bank of Memphis. In Los Angeles in 1910, a large crowd gathered to see a famous boxer, who was depositing his money at the \href{https://www.finhist.com/bank-runs/episodes/353801322.html}{Merchants National Bank}. This led to a run on the bank when others misinterpreted the crowd as a run.\footnote{This episode could be classified as both a non-fundamental run and an information-based run.}

\begin{table}[!ht]
\caption{\textbf{Pass-Through of Runs to Failure: Narrative Classification of Non-fundamental Runs} } \label{tab:fail_run_fundamentals2}
  \centering
  \footnotesize
       \begin{minipage}{1.0\textwidth}
        \begin{center}
        {\begin{tabular}{l*{3}{c}}   \toprule  Dependent variable & \multicolumn{3}{c}{Failure in t} \\  \cmidrule(lr){2-4}
                &\multicolumn{1}{c}{(1)}         &\multicolumn{1}{c}{(2)}         &\multicolumn{1}{c}{(3)}         \\
\midrule
Run             &     0.38\sym{***}&                  &                  \\
                &  (0.051)         &                  &                  \\
Non-fundamental run&                  &     0.11\sym{***}&     0.14\sym{***}\\
                &                  &  (0.036)         &  (0.045)         \\
Other run       &                  &     0.40\sym{***}&     0.44\sym{***}\\
                &                  &  (0.053)         &  (0.054)         \\
Strong fundamentals&                  &                  &  -0.0094\sym{**} \\
                &                  &                  & (0.0038)         \\
Strong fundamentals $\times$ Non-fundamental run&                  &                  &    -0.14\sym{***}\\
                &                  &                  &  (0.045)         \\
Strong fundamentals $\times$ Other run&                  &                  &    -0.32\sym{***}\\
                &                  &                  &  (0.051)         \\
\midrule
Observations    &   272535         &   272535         &   272535         \\
Mean dep. var.  &   0.0084         &   0.0084         &   0.0084         \\
\bottomrule
\end{tabular}
}

   \end{center}
    {\footnotesize Notes: This table presents estimates of \Cref{eq:failure_nonfund_run}. We define a bank to have  ``strong'' fundamentals when  $\text{Fundamentals}_{bt-1}$ is in the upper tercile of the historical distribution.  \citet{Driscoll1998} standard errors are reported in parentheses with a bandwidth of three years to allow for residual correlation within and across banks in proximate years.    *,**, and *** indicate significance at the 10\%, 5\%, and 1\% level, respectively. }
        \end{minipage}
\end{table}

\Cref{tab:fail_run_fundamentals2} studies the pass-through of non-fundamental runs to bank failure based on estimating
\begin{align}
    \text{Failure}_{bt} &= \beta_0 + \beta_1 \text{Non-Fundamental Run}_{bt}  + \beta_2 \text{Other Run}_{bt}   +  \epsilon_{bt},   \label{eq:failure_nonfund_run}
\end{align}
While the typical run in our sample is associated with a 38 pp higher probability of failure (column 1), this drops to 11 pp for non-fundamental runs. Thus, non-fundamental runs are much less likely to result in bank failure. Moreover, column 3 shows that strong banks (fundamentals in the upper tercile) subject to non-fundamental runs have a zero probability of failure (0\%=14\%-14\%). Thus, non-fundamental runs, while rare, can sometimes trigger failure if they happen to affect weak banks and uncover deeper asset quality issues.\footnote{For instance, several newspapers reported that a ``heavy run'' on the \href{https://finhist.com/bank-runs/episodes/878401489.html}{First National Bank (Clovis, NM)} in January 1924 was caused by a ``false rumor.'' However, the run led to suspension and receivership of the bank after it revealed that the bank had doubtful assets. The OCC assessed that all its assets were of ``doubtful'' quality at suspension, and the  depositor recovery rate was only 29\%, indicating deep insolvency. In this sense, even a random run can be the trigger for closing an insolvent bank.}

\subsection{Adverse Public Signals, Runs, and Failures}

In \Cref{sec:determinants}, we found that runs, especially those that do not result in failure, often happen in the context of negative news about the macro economy and the banking sector. Motivated by this insight, we next study whether runs result in failure when they happen in the context of adverse public signals. \Cref{tab:fail_run_signal} presents results from estimating variants of the following specification
\begin{align}
    \text{Failure}_{bt} &= \beta_0 + \beta_1 \text{Run}_{bt} + \beta_2 \text{Adverse signal}_{rt}  \label{eq:failure_run_signal} \\
    &+ \beta_3 \text{Run}_{bt} \times \text{Adverse signal}_{rt} + \epsilon_{bt},  \nonumber
\end{align}
where $\text{Adverse signal}_{rt}$ is an aggregate or regional adverse public signal, such as a run on another bank in the same city. This specification asks whether the pass-through from runs to failure differs for runs occurring after adverse public signals that are not bank-specific.

\begin{table}[!ht]
\caption{\textbf{The Pass-Through of Runs to Failure: The Role of Adverse Public Signals} } \label{tab:fail_run_signal}
  \centering
\footnotesize
       \begin{minipage}{1.0\textwidth}
        \begin{center}
        {\begin{tabular}{l*{5}{c}}   \toprule  Dependent variable & \multicolumn{5}{c}{Failure in t} \\  \cmidrule(lr){2-6}
                &\multicolumn{1}{c}{(1)}         &\multicolumn{1}{c}{(2)}         &\multicolumn{1}{c}{(3)}         &\multicolumn{1}{c}{(4)}         &\multicolumn{1}{c}{(5)}         \\
\midrule
Run             &     0.38\sym{***}&     0.44\sym{***}&     0.27\sym{***}&     0.20\sym{***}&     0.25\sym{***}\\
                &  (0.051)         &  (0.052)         &  (0.057)         &  (0.065)         &  (0.056)         \\
Run on other bank in same city&                  &    0.010\sym{***}&                  &                  &    0.016\sym{***}\\
                &                  & (0.0038)         &                  &                  & (0.0044)         \\
Run on other bank in same city $\times$ Run&                  &    -0.14\sym{***}&                  &                  &    -0.15\sym{***}\\
                &                  &  (0.035)         &                  &                  &  (0.031)         \\
Local business failure rate&                  &                  &     0.78\sym{***}&                  &     0.28\sym{**} \\
                &                  &                  &   (0.16)         &                  &   (0.13)         \\
Local business failure rate $\times$ Run&                  &                  &     7.03\sym{***}&                  &     3.76\sym{***}\\
                &                  &                  &   (1.36)         &                  &   (1.10)         \\
Business failure rate&                  &                  &                  &     1.28\sym{***}&     1.00\sym{**} \\
                &                  &                  &                  &   (0.34)         &   (0.43)         \\
Business failure rate $\times$ Run&                  &                  &                  &     10.6\sym{***}&     7.11\sym{***}\\
                &                  &                  &                  &   (2.53)         &   (2.26)         \\
\midrule
Observations    &   282082         &   282082         &   243908         &   246960         &   243908         \\
Mean dep. var.  &   0.0082         &   0.0082         &   0.0091         &   0.0090         &   0.0091         \\
\bottomrule
\end{tabular}
}

   \end{center}
    {\footnotesize Notes: This table presents results from estimating variants of \Cref{eq:failure_run_signal}. \citet{Driscoll1998} standard errors are reported in parentheses with a bandwidth of three years to allow for residual correlation within and across banks in proximate years.   *,**, and *** indicate significance at the 10\%, 5\%, and 1\% level, respectively.
  }
        \end{minipage}
\end{table}

\Cref{tab:pred_run} showed that the occurrence of a run in the same city is one of the strongest non-bank-specific antecedents of a run. Column 2 in \Cref{tab:fail_run_signal} shows that runs that coincide with a run on another bank in the same city are 14 pp \textit{less} likely to result in failure compared to other runs. The lower probability of failure for such runs is consistent with the notion that a run on another bank leads depositors to become ``jittery.'' This finding is in line with depositors revising upward their belief that their own bank is insolvent. In some cases, this signal leads depositors to mistakenly run on healthy banks. As a consequence, runs within the context of runs on other banks are less likely to lead to failure. However, the response is not necessarily irrational, as these runs still involve a probability of failure of 30\%, suggesting many of these banks that are run are indeed troubled.

In column 3 of \Cref{tab:fail_run_signal}, we use the state-level nonfinancial business failure rates as the proxy of an adverse signal from a deterioration in economic conditions. In contrast to the finding for runs on other banks, we find that runs that occur when the state-level business failure rate is elevated are \textit{more} likely to be associated with a bank failure. In terms of magnitudes, a 1 pp rise in the state-level business failure rate---roughly a one standard-deviation increase---raises the pass-through from runs to failure by 7 pp. A rise in state-level business failures signals an increase in bank asset losses and thus an increased risk of bank insolvency. Runs in response to this deterioration of economic fundamentals are more likely to translate to failure, as these are precisely the times when banks are more likely to be insolvent. Columns 4 and 5 reveal a similar pattern when we use the aggregate nonfinancial business failure rate, rather than the local failure rate.

To further illustrate that bank runs and failures often follow a rise in non-financial business failures, \Cref{fig:failures_dynamics} shows the dynamics of the state-level non-financial business failure rate around innovations in the state-level rate of bank runs and bank failures. State-level non-financial business failures are from \textit{Bradstreet's} and are available quarterly. Both bank runs and bank failures are preceded by and coincide with an elevated rate of business failures, which lead to asset losses for banks. The relation is especially strong for bank failures. This evidence supports the key role of deteriorating fundamentals for understanding bank failures and runs. The small spike in business failures around runs without failure further suggests that increases in business failures can lead to runs on healthy banks that do not ultimately fail, consistent with information-based theories of runs.

\section{Consequences of Runs for Lending and Real Activity}

\label{sec:consequences}

In this final section, we study the consequences of runs and failures on financial intermediation and real economic activity. We first study the impact of bank runs on surviving banks at the bank level. We then turn to the impact of bank runs and failures on local aggregate lending, deposits, and manufacturing activity.

\subsection{Bank-Level Consequences of Bank Runs Without Failure}

What are the consequences of bank runs without failure at the bank level? To answer this question, we estimate local projections of the following form:
\begin{align}
\frac{Y_{bt+h}-Y_{bt-1}}{\text{Assets}_{bt-1}}
= \alpha_b + \gamma_t + \beta^{h} \text{Run}_{bt}^{Without \; failure} + \epsilon_{bt+h} \label{eq:loc_proj_pre_post}, \quad h=-5,\, \dots,\, 4
\end{align}
with the prediction horizon $h$ running from five years before to five years after the distress episode. $\text{Run}_{bt}^{Without \; failure}$ is an indicator variable equal to one for a bank run without failure. We focus on runs without failure to understand the consequences of runs for surviving banks; for failing banks, we cannot observe bank-level outcomes after the run. We measure the change in a given bank balance sheet item ($Y_{b,t+h}-Y_{b,t-1}$) relative to total assets in the year before the event, $\text{Assets}_{b,t-1}$, so magnitudes are relative to pre-event total assets. $\alpha_b$ is a bank fixed effect, and $\gamma_t$ is a year fixed effect.

\begin{figure}[!ht]
\centering

\caption{\textbf{Bank-Level Deposit and Loan Dynamics around Bank Runs Without Failure} \label{fig:LP_banklevel} }

\subfloat[Deposits and loans: bank runs without failure \label{fig:dynamics_run_loans_deposits}]{
{\includegraphics[width=0.49\textwidth]{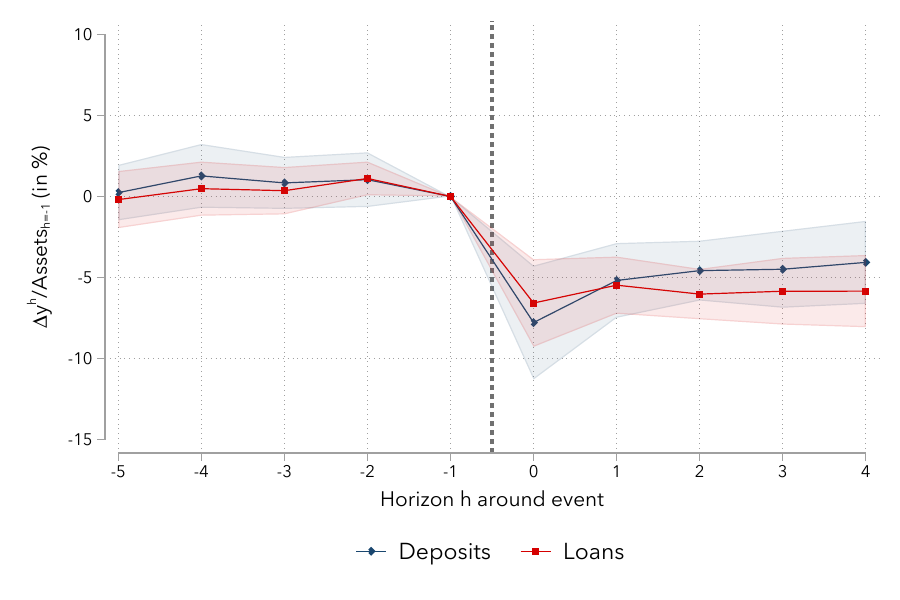}}} \hfill
\subfloat[Deposits: runs with and without temporary suspension\label{fig:dynamics_run_deposits}]{
{\includegraphics[width=0.49\textwidth]{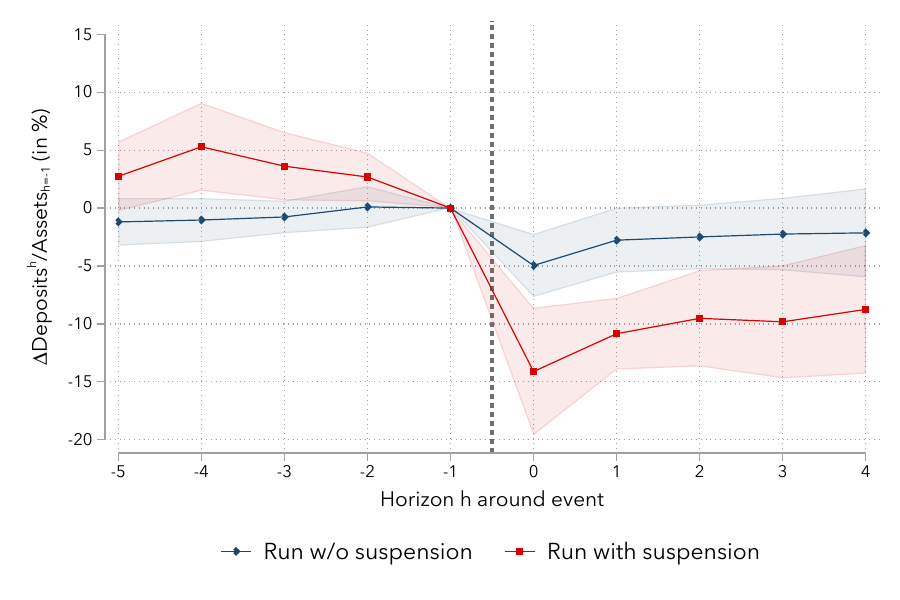}}}

\subfloat[Deposits: weak vs strong banks \label{fig:dynamics_dep_fund}]{\includegraphics[width=0.49\textwidth]{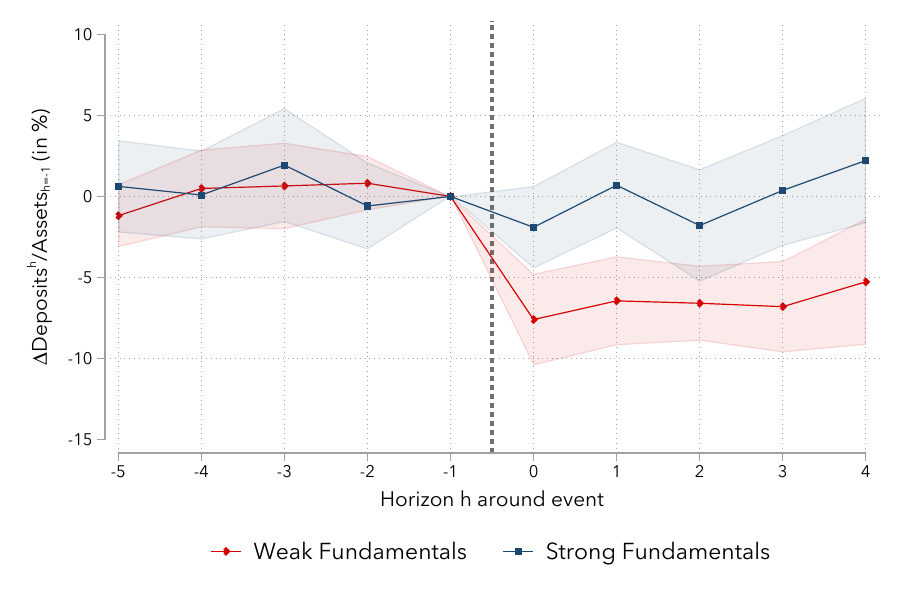}} \hfill
\subfloat[Loans: weak vs strong banks \label{fig:dynamics_loans_fund}]{
\includegraphics[width=0.49\textwidth]{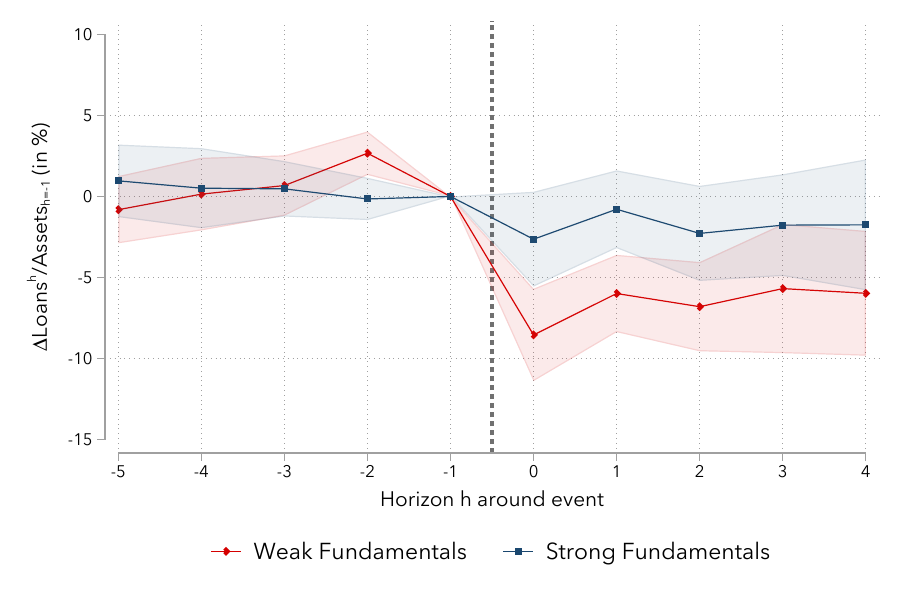}}

\subfloat[Deposits: non-fundamental runs and other runs \label{fig:dynamics_nf_run_deposits}]{
{\includegraphics[width=0.49\textwidth]{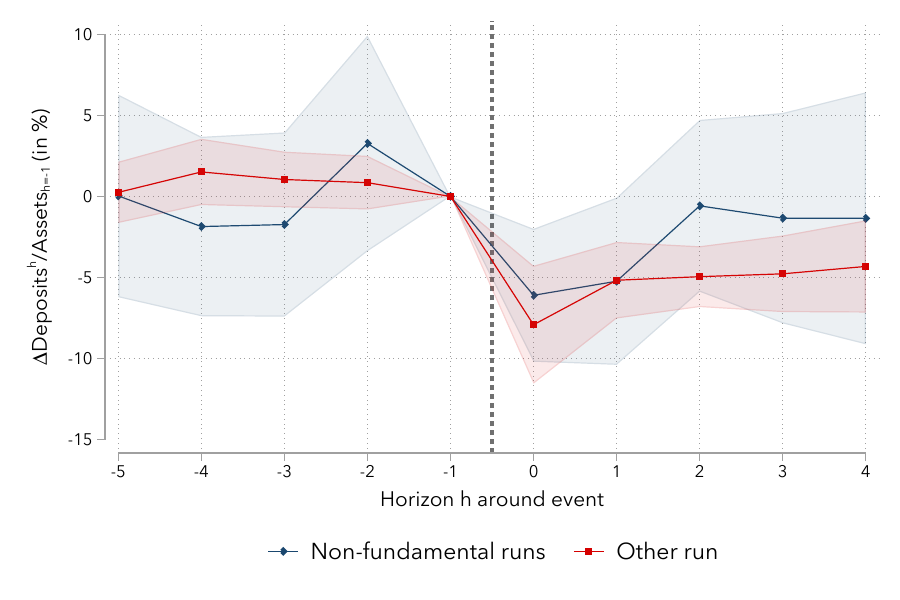}}}
\subfloat[Loans: non-fundamental runs and other runs \label{fig:dynamics_nf_run_loans}]{
{\includegraphics[width=0.49\textwidth]{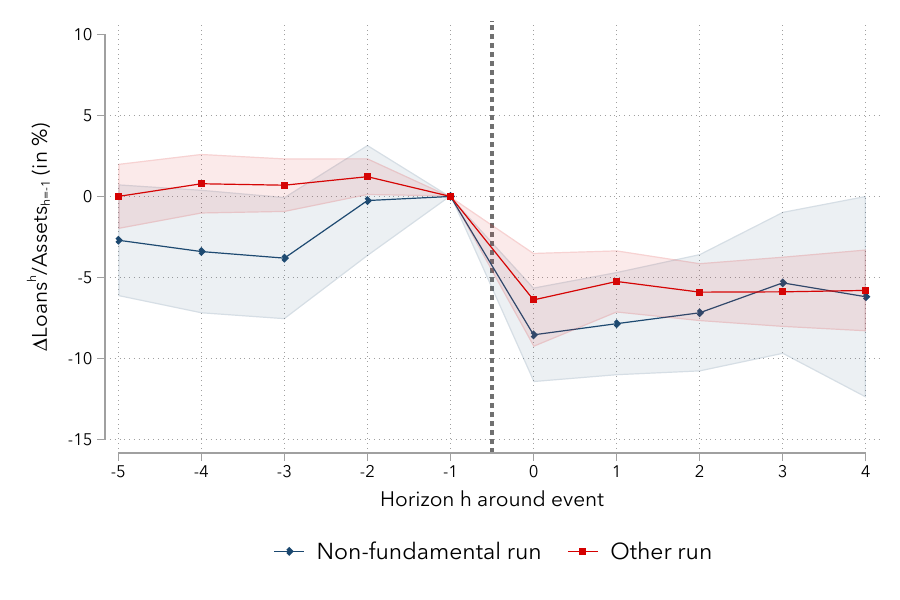}}}

\begin{minipage}{\textwidth}
\footnotesize
Notes: This figure reports coefficients from estimating Equations \eqref{eq:loc_proj_pre_post} (panels a and b), \eqref{eq:loc_proj_pre_post2}  (panels c and d), and \eqref{eq:LP_run_no_fail} (panels e and f). All panels focus on episodes where there is a run without bank failure. Shaded areas represent 95\% confidence intervals based on \citet{Driscoll1998} standard errors with a bandwidth of three years to allow for residual correlation within and across banks in proximate years. Strong (weak) fundamentals are defined as $\text{Fundamental}_{b,t-1}$ being in the upper (lower) tercile.
\end{minipage}
\end{figure}

Panel (a) of \Cref{fig:LP_banklevel} shows the results from estimating \Cref{eq:loc_proj_pre_post} for deposits and loans on the 1865-1934 sample of national banks. There is no pre-trend before a run without failure in either deposits or loans.\footnote{As discussed above, for failing banks there is a gradual deterioration in loans and deposits before the run and failure (see \Cref{fig:dynamics_characteristics} and the discussion in \Cref{app:results}).} Once a bank is exposed to a run, deposits decline by around 7.5\% as a share of pre-event assets and remain depressed thereafter. Loans decline by a similar magnitude. While we have selected episodes of bank distress that do not result in failure, these are still associated with a decline in deposits and loans. Thus, runs have a lasting impact at the bank level even in the absence of bank failure.

Panel (b) of \Cref{fig:LP_banklevel}  further explores heterogeneity across runs with and without (temporary) suspension. Run episodes in which the bank is subject to a run but does not suspend at any point have relatively mild reductions in deposits of about 5\%. After these episodes, banks recover their deposits within five years. Episodes with a run, suspension, and reopening are more severe, leading to a 15 percent decline in deposits that persists for five years. The larger decline for runs with suspension is likely because these runs are more severe. In addition, suspension itself may reduce depositor confidence and lead depositors to switch banks.

\paragraph{Dynamics For Weak and Strong Banks}
As discussed in \Cref{sec:theory}, if runs themselves are the primary cause of distress, then runs should typically be associated with severe outcomes, irrespective of initial bank fundamentals. In contrast, if runs themselves are mainly a trigger of distress for weak banks, then runs should be less costly for banks with sound fundamentals. Thus, we next ask: What are the implications of a run without failure across weak and strong banks? To answer this question, we estimate local projections of the form:
\begin{align}
\frac{Y_{b,t+h}-Y_{b,t-1}}{\text{Assets}_{b,t-1}}
= \alpha_b + \gamma_t + &\beta_S^{h} \text{Run}^{Without \; failure}_{b,t} \times \text{Strong Fundamentals}_{b,t-1} \label{eq:loc_proj_pre_post2} \\
+ &\beta_W^{h} \text{Run}^{Without \; failure}_{b,t} \times \text{Weak Fundamentals}_{b,t-1} \nonumber \\ + &\epsilon_{b,t+h} , \quad \quad \quad h=-5,\, \dots,\, 4 \nonumber
\end{align}
where strong (weak) fundamentals are defined as being in the top (bottom) tercile of the fundamentals measure in $t-1$.

Panels (c) and (d) of \Cref{fig:LP_banklevel} present the results. Runs without failure lead to larger deposit declines for banks with weak fundamentals. Banks with strong fundamentals see essentially no decline in loans and deposits, whereas banks with weak fundamentals see a persistent decline in deposits and lending of about 7\%. Thus, bank fundamentals not only determine whether runs trigger bank failure; they also shape the consequences of bank runs when banks survive runs.

Taken together, the dynamics of deposits and loans around bank runs reinforce the notion that runs tend to be associated with severe distress when they happen in banks with weak fundamentals. Runs on banks with sound fundamentals have little impact on a bank's long-term viability.

\paragraph{Dynamics Around Non-Fundamental Runs}

A concern with studying the dynamics of deposits and loans around runs without failure is that such runs do not happen randomly but occur in response to bad news about the economy that leads depositors to run as a precautionary measure. These shocks can simultaneously reduce the demand for loans and deposits.  To make some progress on the causal effect of runs,  we replace the shock with our measure of non-fundamental runs discussed in \Cref{sec:nf_run}. These runs, caused by false rumors or misinformation, are more exogenous events not caused by bank or economic conditions. This allows us to trace the effects of runs that are closer in spirit to the notion of random runs in \cite{Diamond1983}.

\Cref{fig:LP_banklevel} panels (e) and (f) study the consequences of non-fundamental runs and other runs by estimating
\begin{align}
\frac{Y_{bt+h}-Y_{bt-1}}{\text{Assets}_{bt-1}}
= \alpha_b + \gamma_t &+ \beta_{NF}^{h} \text{Run}^{Without \; failure, \; non-fund}_{bt} \label{eq:LP_run_no_fail} \\ &+ \beta_{Other}^{h} \text{Run}^{Without \; failure, \; other}_{bt} + \epsilon_{bt+h} , \quad h=-5,\, \dots,\, 4.  \nonumber
\end{align}
Non-fundamental runs without failure lead to a 5\% drop in deposits and loans, which is slightly milder than for other runs without failure. Interestingly, deposits of banks subject to non-fundamental runs recover fully within two years of the run, similar to ``run w/o suspension'' events (see panel b). Loans remain roughly 5\% lower three years after the run. Thus, runs that we specifically identify as non-fundamental and that are thus exogenous to bank and local economic conditions do lead to moderate disruptions in bank-level deposits and credit.

\subsection{Local Consequences of Runs and Failures}

We next examine the consequences of bank runs at the local aggregate level. As we have seen, most banking crises from 1863 to 1934 had a strong regional component. We can use our detailed data to exploit cross-regional variation to understand the local consequences of banking sector distress. We first examine the impact on city-level aggregate bank deposits and lending. We then provide evidence for the effect of runs on city-level manufacturing activity.

\subsubsection{Deposits and Lending}

\paragraph{Runs on weak vs strong banks} We study the local economic consequences of bank runs on local deposits and lending, distinguishing between whether runs affect weak or strong banks. We estimate impulse responses from local projections of the following form:
\begin{align}
    g^y_{c,t-1,t+h}=  \alpha_c & + \beta^h_{S} \text{Run}^{Strong}_{c,t} + \beta^h_{I} \text{Run}^{Interm}_{c,t} + \beta^h_{W} \text{Run}^{Weak}_{c,t}  \label{LP1} \\ & + X_{c,t-1} \Gamma^h + \epsilon_{c,t+h}, \quad h=0,...,H. \notag
\end{align}
The outcome variable, $g^y_{c,t-1,t+h}$, is the symmetric growth rate of a variable such as total lending in city $c$ from year $t-1$ to $t+h$, defined as $g^y_{c,t-1,t+h}=100\frac{y_{c,t+h}-y_{c,t-1}}{0.5(y_{c,t+h}+y_{c,t-1})}$. The symmetric growth rate is bounded between -200 and 200. We employ this definition of growth to accommodate the possibility that an outcome in a local area goes to zero.\footnote{Results are similar if we use the log of $y_{c,t}$ as the outcome, but we prefer the symmetric growth rate to avoid exit and reentry of mostly smaller cities in the estimation sample.} Outcome variables are constructed based on aggregating national bank balance sheets to the city and year level. Note that if a bank fails and exits, its loans and deposits go to zero. In estimating \eqref{LP1}, we therefore weight by the number of banks in a city to mitigate the influence of large relative changes in small cities with few banks.\footnote{\Cref{fig:city_level_dynamics_unweighted} reports the unweighted version of this specification. The impulse responses are qualitatively similar but quantitatively larger for the unweighted regression.}

The shocks---$\text{Run}^{Strong}_{c,t}$, $\text{Run}^{Interm}_{c,t}$, and $\text{Run}^{Weak}_{c,t}$---capture local exposure to runs on banks with strong, intermediate, and weak fundamentals. These measures are defined as indicator variables that equal one if a bank with fundamentals in category $j$ is subject to a run in city $c$ and year $t$, and zero otherwise. Fundamentals are defined as the top (strong), middle (intermediate), and bottom (weak) terciles of the fundamentals measure, $Fundamentals_{b,t-1}$. By using indicator variables, we effectively exploit the extensive margin of the variation in whether a city is exposed to a run; results are qualitatively similar when using asset-weighted exposure to capture the share of banks subject to runs. In $X_{c,t-1}$, we include $L=3$ yearly lags of both the run shocks and the dependent variable, as is standard in the local projections framework \citep{Jorda2005}. The local projection impulse responses are the sequence of estimated coefficients $\{\hat \beta_S^h , \hat \beta_I^h, \hat \beta_W^h \}$.

\begin{figure}[ht!]
\centering

\caption{\textbf{Consequences of Runs for City-Level Deposits and Lending} \label{fig:city_level_dynamics} }

\subfloat[Impact of runs on deposits, weak vs strong banks]{
{\includegraphics[width=0.49\textwidth]{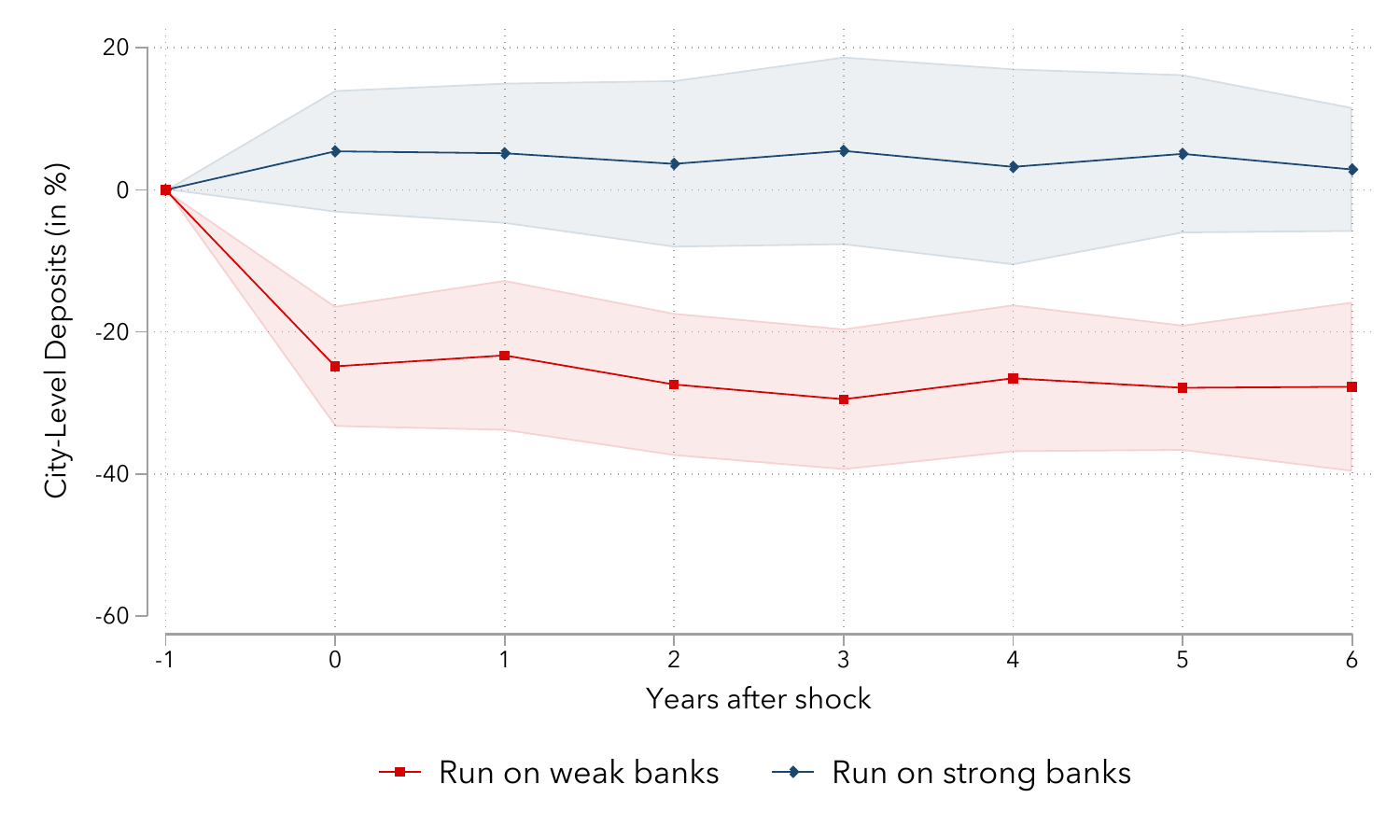}}}
\hfill
\subfloat[Impact of runs on loans, weak vs strong banks]{
{\includegraphics[width=0.49\textwidth]{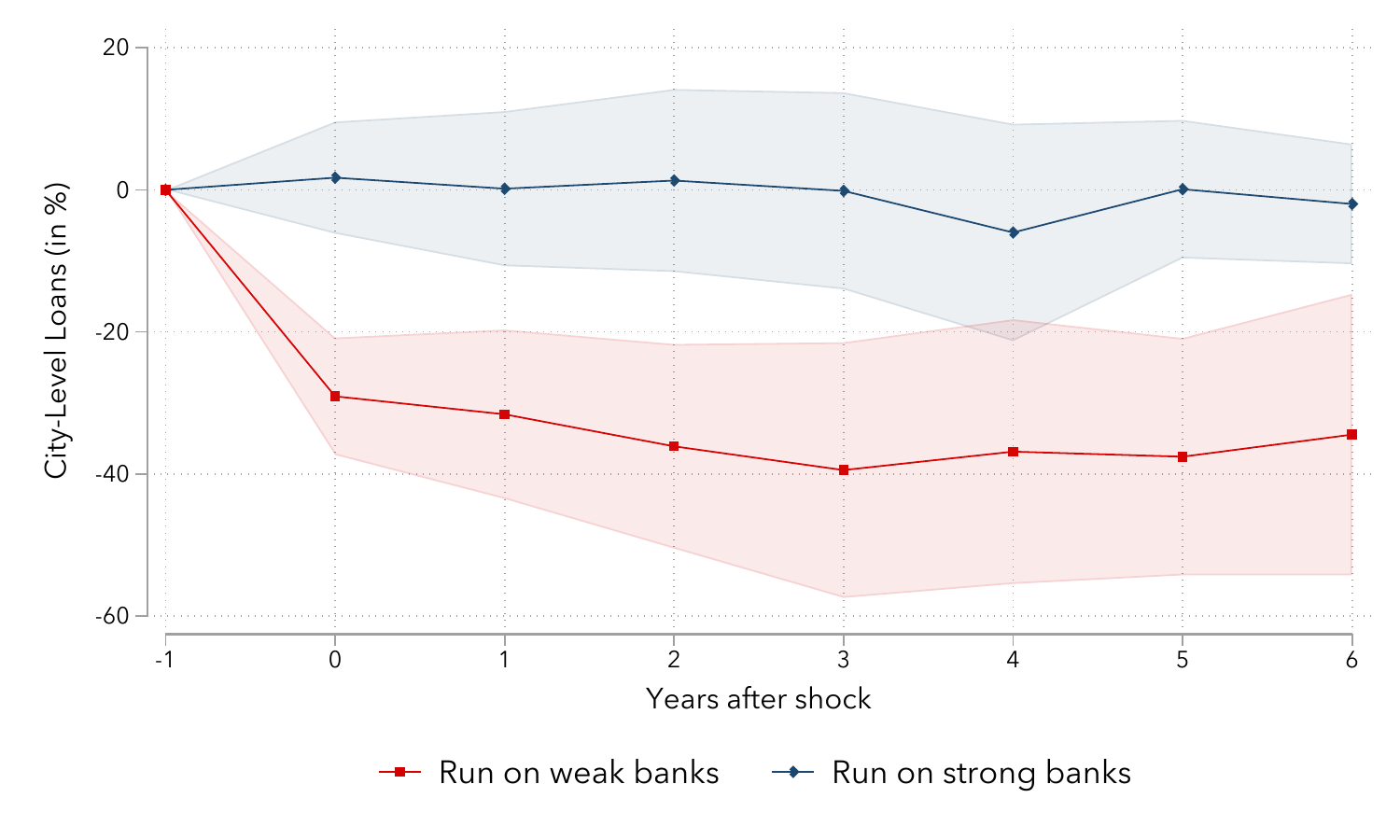}}}

\subfloat[Impact of runs on deposits, non-fundamental runs]{
{\includegraphics[width=0.49\textwidth]{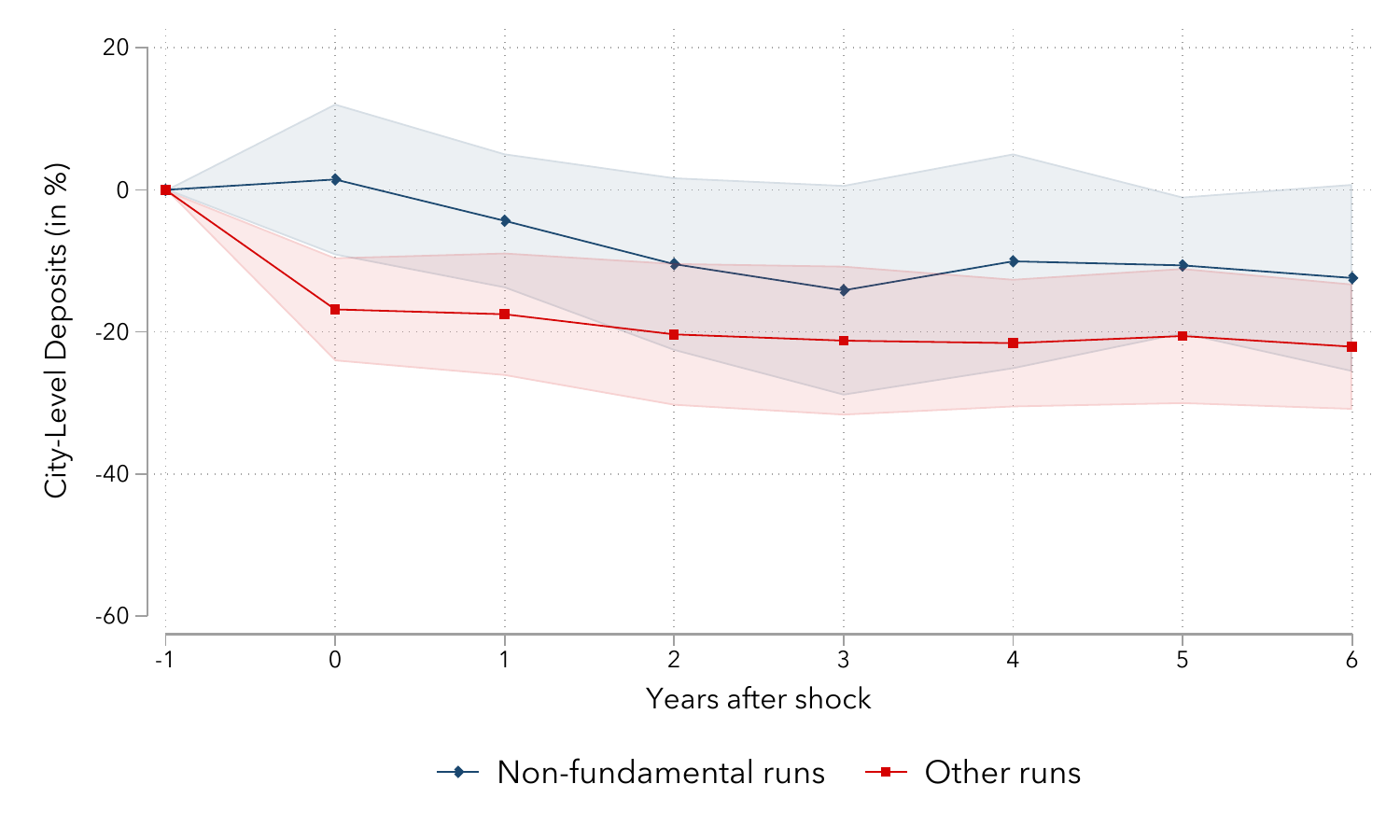}}}
\hfill
\subfloat[Impact of runs on loans, non-fundamental runs]{
{\includegraphics[width=0.49\textwidth]{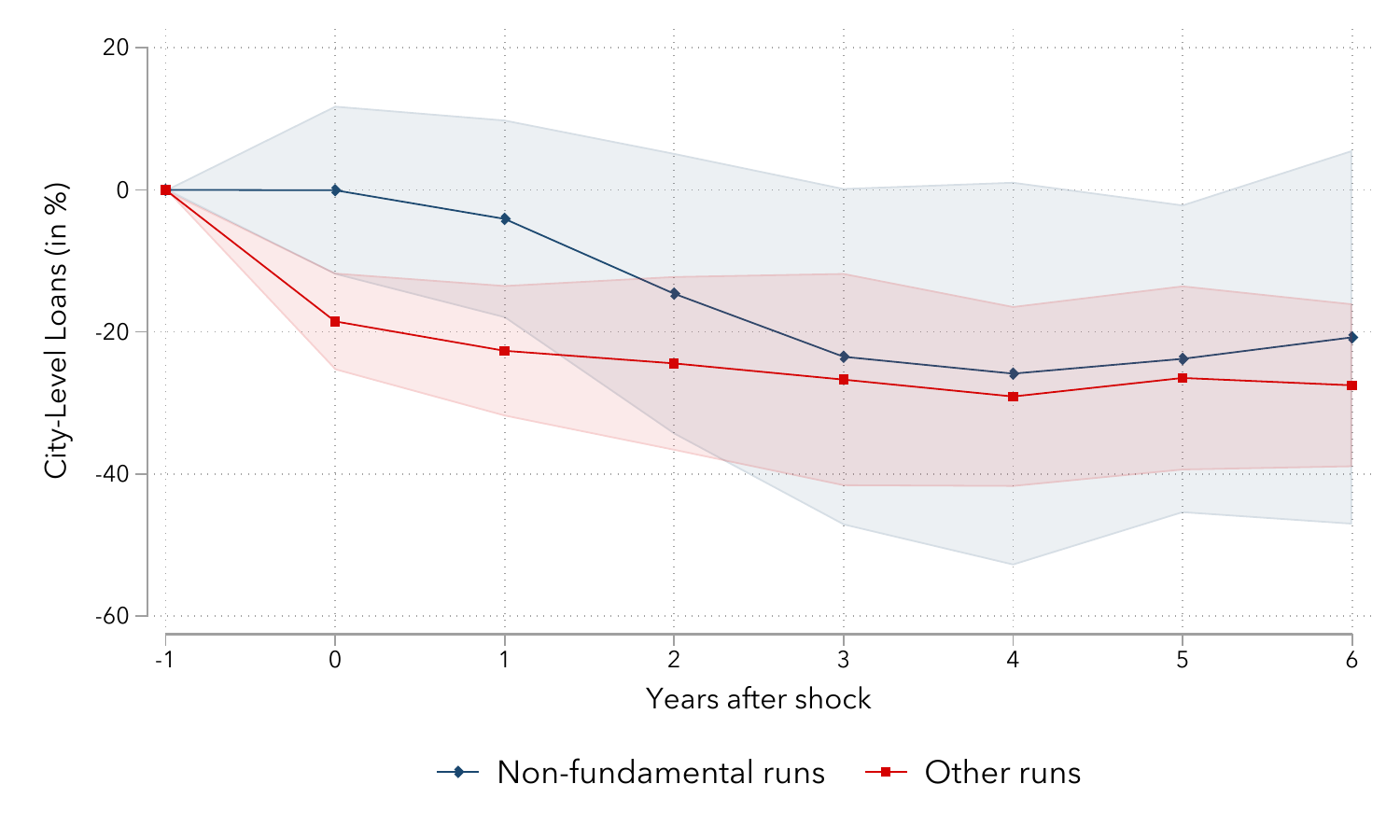}}}

\subfloat[Impact of runs with and without failure on deposits]{
{\includegraphics[width=0.49\textwidth]{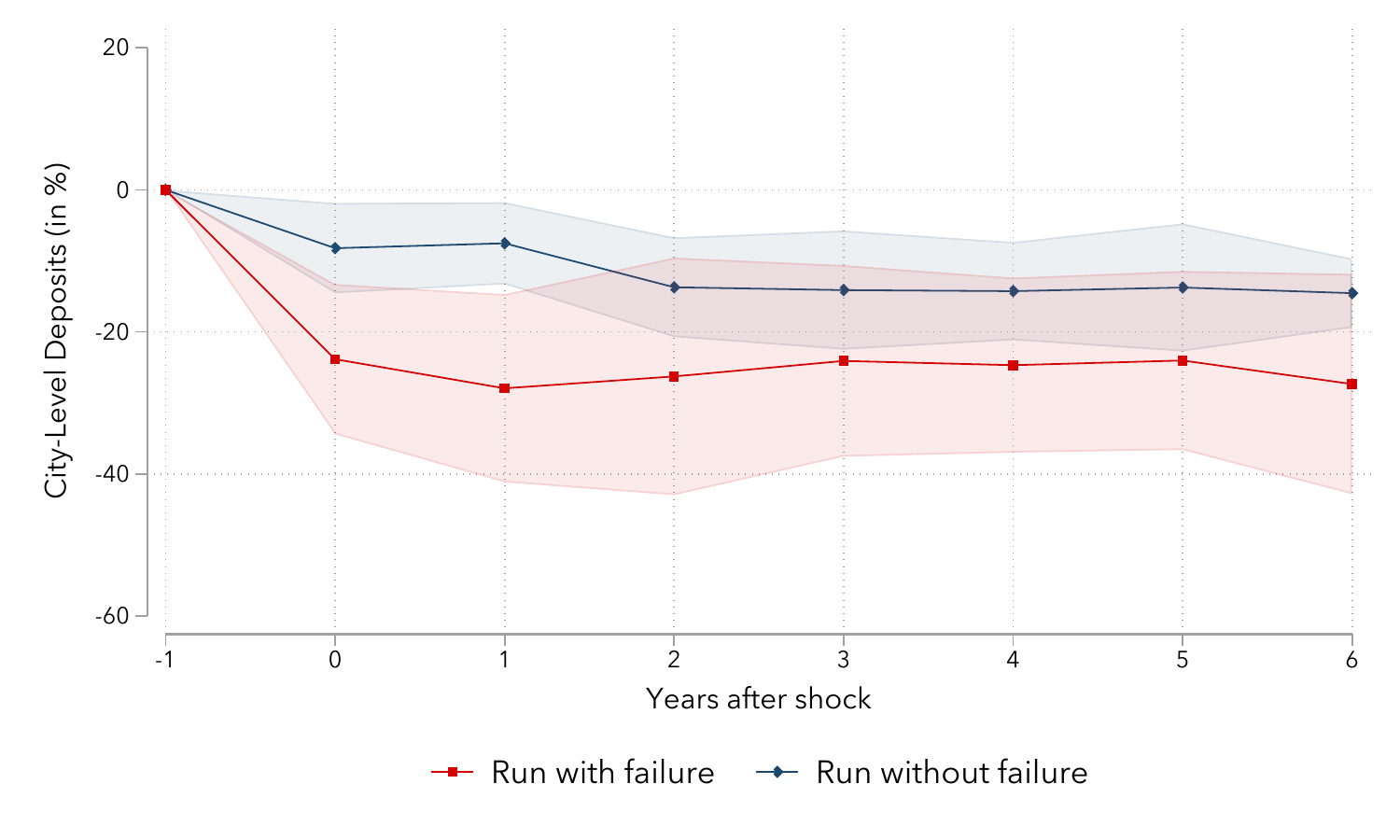}}}
\hfill
\subfloat[Impact of runs with and without failure on loans]{
{\includegraphics[width=0.49\textwidth]{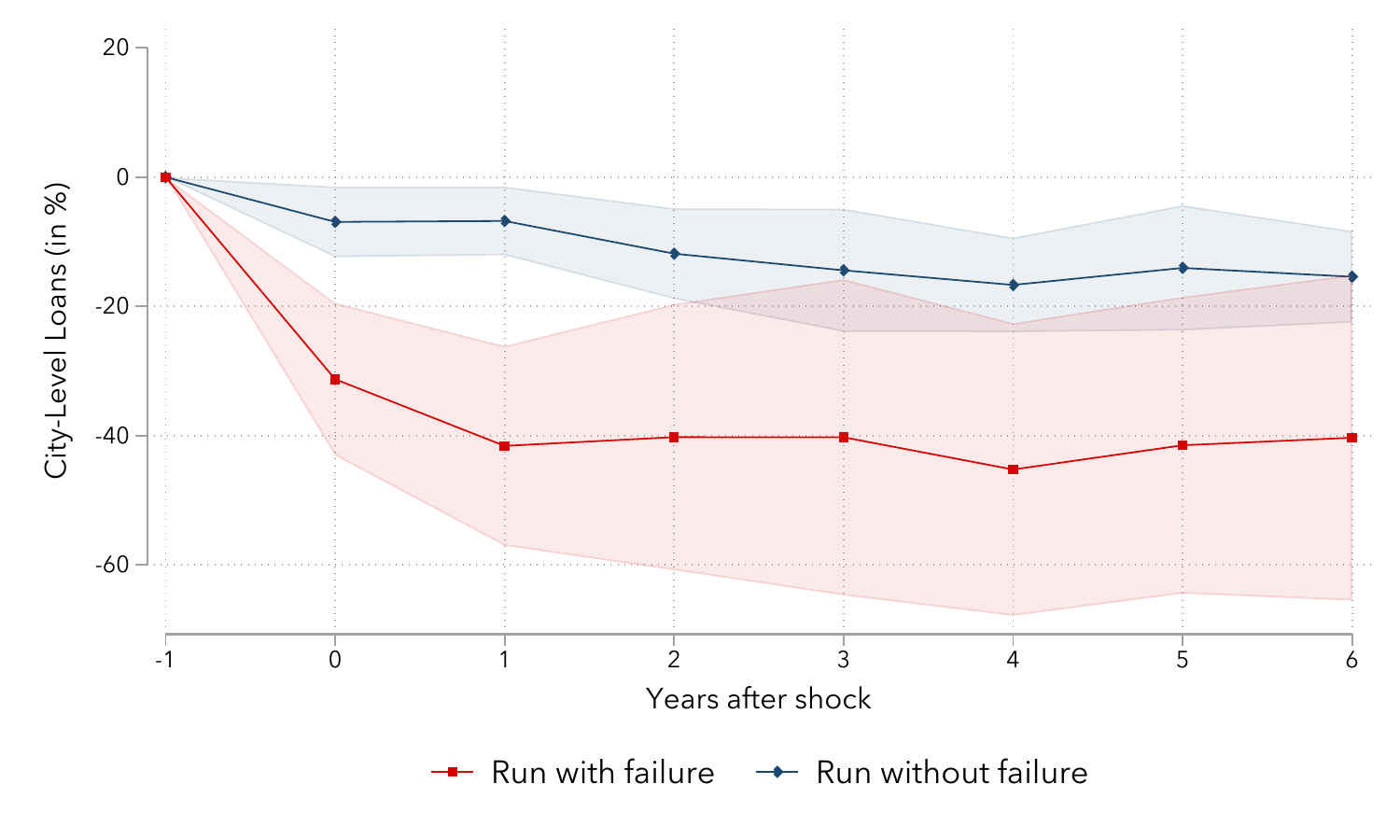}}}

\begin{minipage}{\textwidth}
\footnotesize
Notes: This figure presents impulse responses based on estimation of \Cref{LP1} (panels a and b), \Cref{LP_nf} (panels c and d), and \Cref{LP2} (panels e and f). The analysis is based on a city-level annual panel. Loans, deposits, and bank distress are based on aggregating the sample of national banks from 1863-1934 to the city level at an annual frequency. The regressions are weighted by the number of banks in a city; unweighted estimates are reported in \Cref{fig:city_level_dynamics_unweighted}. Error bands represent 95\% confidence intervals based on \citet{Driscoll1998} standard errors with a horizon-dependent bandwidth of $L_h = \lceil 1.5h \rceil$.
\end{minipage}
\end{figure}

\Cref{fig:city_level_dynamics} presents the impulse responses of city-level deposits (panel a) and loans (panel b) to innovations in the city-level runs on weak and strong banks. Runs predict a reduction in city-level deposits and loans, but only when they occur in weak banks. Runs on weak banks lead to a reduction in city-level deposits and lending of over 20\% in the years immediately after the shock. In contrast, runs on strong banks are not associated with significant declines in deposits and lending.  The difference between the responses for runs on strong and weak banks is statistically significant. This is in line with the view that runs themselves are not universally associated with severe crises but rather result in severe economic consequences when bank fundamentals are weak.

\paragraph{Non-fundamental and other runs} In \Cref{fig:LP_banklevel}, we saw that non-fundamental runs led to modest, but not negligible, reductions in bank-level deposits and lending. \Cref{fig:city_level_dynamics} panels (c) and (d) report the implications of such runs for city-level deposits and lending based on estimating:
\begin{align}
    g^y_{c,t-1,t+h}=  \alpha_c & + \beta^h_{NF} \text{Run}^{Non-fund}_{c,t} + \beta^h_{Other} \text{Run}^{Other}_{c,t} \label{LP_nf} \\ &+ X_{c,t-1} \Gamma^h + \epsilon_{c,t+h}. \quad h=0,...,H.  \notag
\end{align}
$\text{Run}^{Non-fund}_{c,t}$ is an indicator for whether there is a non-fundamental run in city $c$ and year $t$ as defined in \Cref{sec:nf_run} and $\text{Run}^{Other}_{c,t}$ is an indicator for the occurrence of other runs in a city. While non-fundamental runs reduce deposits and lending at affected banks, \Cref{fig:city_level_dynamics} shows they have essentially no effect on local deposits and lending.\footnote{Loans do exhibit a gradual decline, but the estimates are noisy and not statistically significant. In the unweighted specification (\Cref{fig:city_level_dynamics_unweighted}, panel d), the impact of non-fundamental runs on loans is more precise and close to zero.} The null aggregate effect is because deposits are reallocated seamlessly to other local banks following non-fundamental runs.

\paragraph{Runs with and without failure} Next, we study the consequences of runs with and without failure. This provides another test to address whether runs alone can be associated with severe economic consequences, or whether bank failures, indicative of bank insolvency, are necessary for runs to be associated with severe crises.

To do this, we estimate impulse responses from the following local projections specification:
\begin{align}
    g^y_{c,t-1,t+h}=  \alpha_c &+ \beta^h_{RNF} \text{Run}^{Without\;  failure}_{c,t} +  \beta^h_{RWF} \text{Run}^{With\; failure}_{c,t}  \label{LP2}  \\ & + X_{c,t-1} \Gamma^h + \epsilon_{c,t+h}, \quad h=0,...,H.  \notag
\end{align}
$g^y_{c,t-1,t+h}$ is again the symmetric growth rate of aggregate city-level loans or deposits from $t-1$ to $t+h$. We consider two mutually exclusive bank distress episodes: (i) run without failure, and (ii) run with failure. Each measure is defined as an indicator variable for whether a city is subject to a given distress episode in year $t$. This exercise allows us to distinguish the consequences of episodes that are closer to pure liquidity events from those involving bank insolvency. Given that the likelihood of failure conditional on a run is strongly predictable based on bank fundamentals, this exercise complements the previous analysis on runs across banks with strong and weak fundamentals. However, instead of using an \textit{ex ante} measure of bank health, here we use an \textit{ex post} measure of bank distress.

Panels (e) and (f) in \Cref{fig:city_level_dynamics} present the results from estimating \Cref{LP2}. Not all bank distress episodes are alike. Runs with failure translate into substantially larger contractions in city-level deposits and lending than runs without failure. Runs without failure lead to a modest decline in city-level deposits that is borderline statistically significant. The effect on loans is more persistent and significant, but also modest in size. Thus, bank distress involving bank failure is associated with substantially worse lending contractions than episodes with runs that do not result in failure. Failures, which are related to bank insolvency, are necessary for bank runs to involve severe local deposit and lending contractions.

In terms of magnitudes, \Cref{fig:city_level_dynamics} shows that if a city experiences a run that results in bank failure, deposits decline by around 30\% and loans by 40\%. The responses are persistent. These responses are large in part because of restrictions on branching and acquisitions, which meant that bank failure would lead to persistent reductions in banking services in a city \citep{QuincyXu2025}. This is especially true in smaller cities with fewer banks. In contrast, if a city is exposed to a run that does not result in bank failure, both deposits and loans decline by about 10\%.

\subsubsection{Manufacturing Activity}

As a final exercise, we study the real economic consequences of runs at the local city level. We use a newly collected dataset on weekly local economic activity constructed from \textit{Bradstreet's}, a business journal that provides city-level summaries of economic conditions in major cities for various sectors, including manufacturing, retail trade, and wholesale trade.\footnote{The underlying data are weekly for 1917 through March 11, 1933, and monthly for the remainder of 1933 through 1935. For the latter period, we fill values forward to retain a weekly panel structure. See \Cref{app:bradstreets} for details on the \textit{Bradstreet's} data and the construction of the manufacturing activity index.} The key benefit of this data is that it is weekly, so we can match to runs at the weekly level and study high-frequency real dynamics.  We focus on \textit{Bradstreet's} description of activity in the manufacturing sector. Using the short textual descriptions, we classify economic conditions into poor (1), fair (2), or good (3).  We normalize our index of manufacturing activity to have the same variance as the cyclical component of industrial production (IP), defined as the deviation of log IP from its one-sided three-year moving average. This analysis covers 30
major cities from 1917 through 1935, providing high-frequency evidence on the consequences of runs for the real economy.

Before proceeding, we note that the results based on our index of manufacturing activity from \textit{Bradstreet's} are mainly driven by larger cities during the Great Depression. Moreover, the index is based on qualitative information, so the transformation to a numeric index may introduce measurement error. Nevertheless, in the appendix, we show that the aggregated city-level manufacturing index constructed from \textit{Bradstreet's} is highly correlated with the cyclical component of industrial production. It captures the major 1920-21 recession, the mild recessions of 1923-24 and 1926-27, and the major downturn starting in August 1929, indicating that it contains valuable information at the business cycle frequency.

\paragraph{Runs on weak vs strong banks} We first study the city-level consequences of runs on weak versus strong banks. We estimate impulse responses to run events using the specification:
\begin{align}
Y^{Manuf}_{c,t+h}=  \alpha_c & + \beta^h_{S} \text{Run}^{Strong}_{c,t}   + \beta^h_{I} \text{Run}^{Interm}_{c,t} + \beta^h_{W} \text{Run}^{Weak}_{c,t}  \label{LP_TAAG1}  \\ & + X_{c,t-1} \Gamma^h + \epsilon_{c,t+h}, \quad h=0,...,H. \notag
\end{align}
$Y^{Manuf}_{c,t}$ is the city-level manufacturing index in city $c$ and week $t$. $\alpha_c$ is a city fixed effect. The shocks $\text{Run}^{Strong}_{c,t},$ $\text{Run}^{Interm}_{c,t},$ and $\text{Run}^{Weak}_{c,t}$ are indicator variables for whether a strong, intermediate, or weak bank was subject to a run in city $c$ and week $t$ based on the tercile of $Fundamentals_{b,t-1}$.  We include 24 weekly lags of both the dependent and the innovation variables in $X_{c,t-1}$ and study responses out to 80 weeks (1.5 years).

\begin{figure}[ht!]
\centering
\caption{\textbf{Consequences of Runs for City-Level Manufacturing Activity} \label{fig:LP_TAAG} }

\subfloat[Impact of runs on manufacturing index, weak vs strong banks \label{fig:LP_TAAG_fund} ]{
{\includegraphics[width=0.6\textwidth]{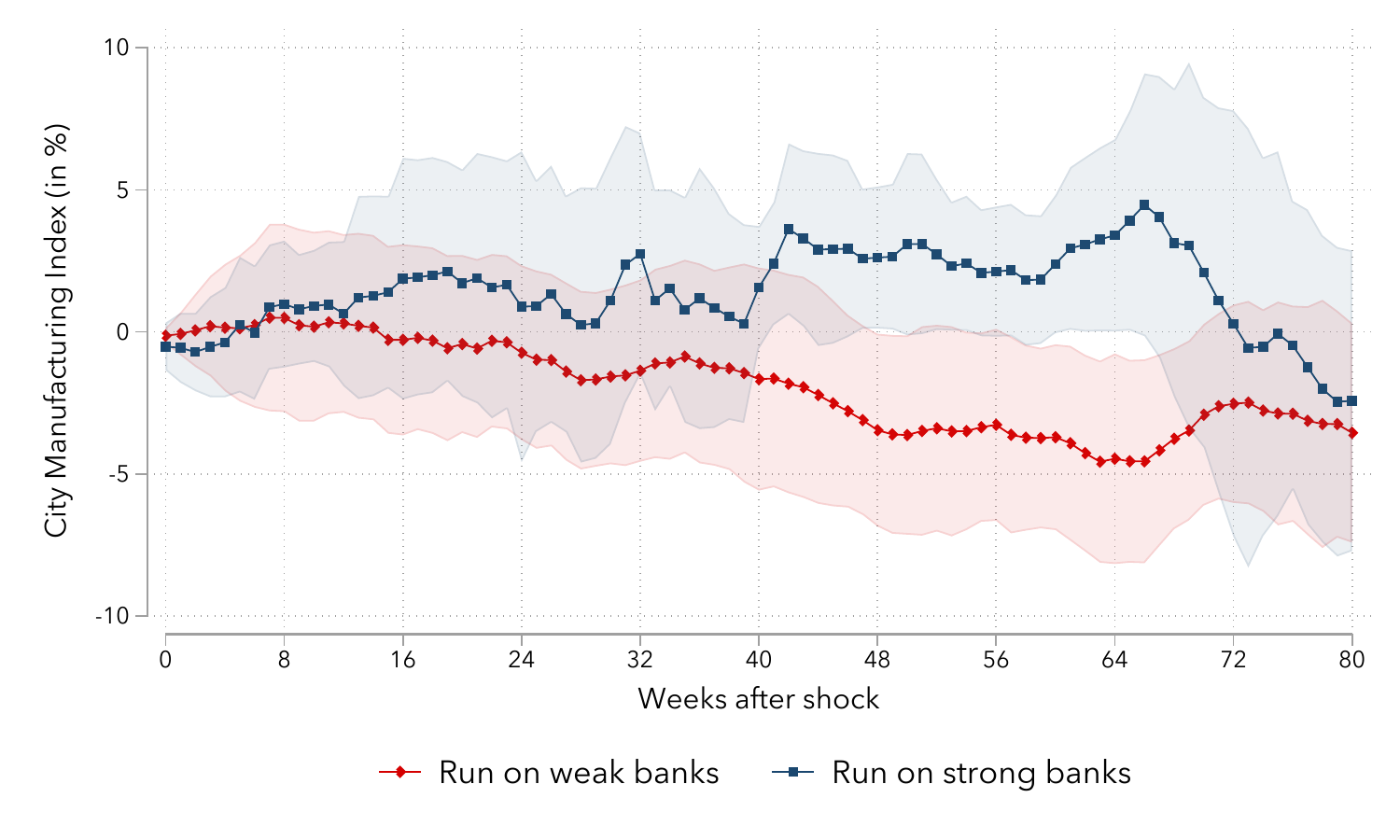}}}

\subfloat[Impact of runs with and without failure on manufacturing index \label{fig:LP_TAAG_fail} ]{
{\includegraphics[width=0.6\textwidth]{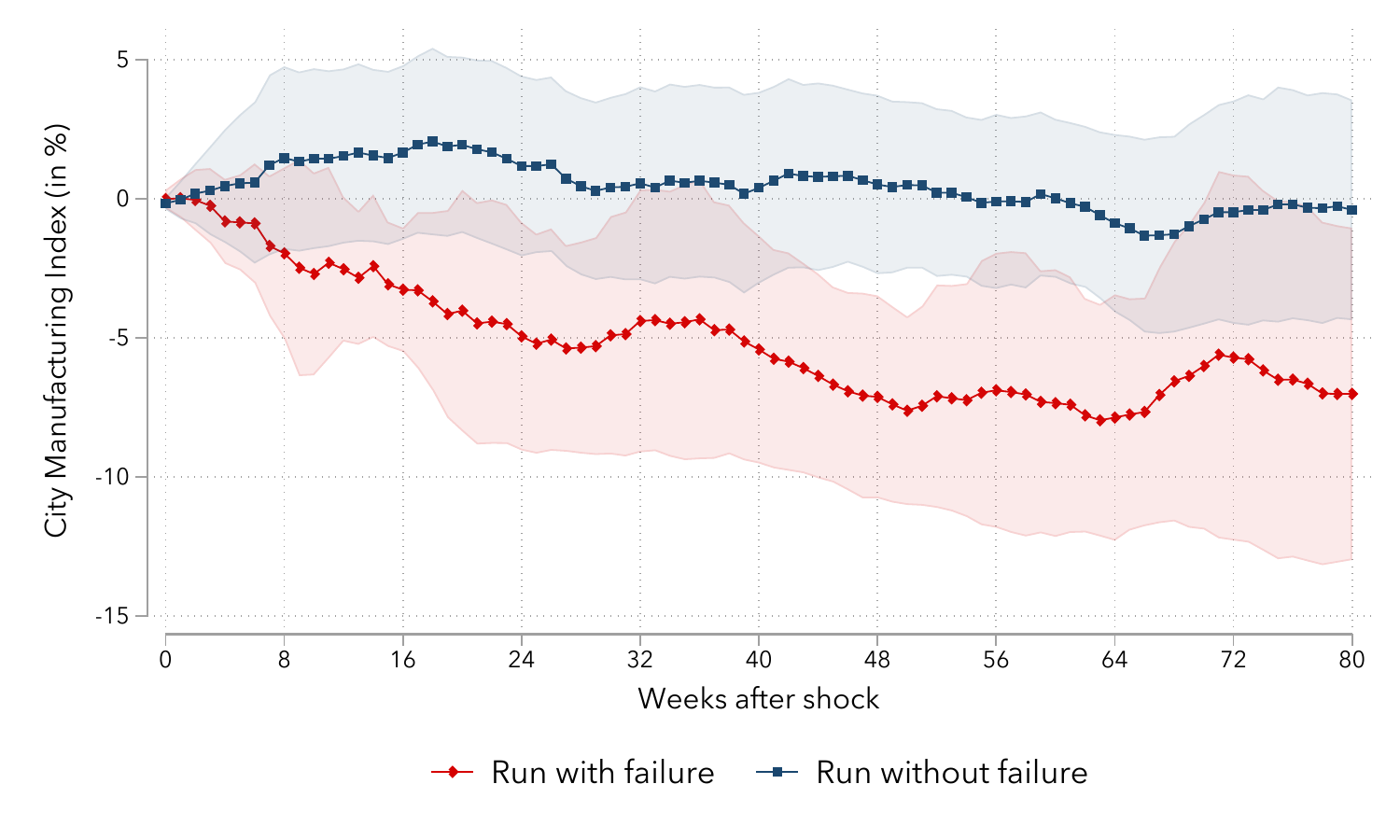}}}

\begin{minipage}{\textwidth}
\footnotesize
Notes: This figure presents impulse responses based on estimation of \Cref{LP_TAAG1} (panel a) and \Cref{LP_TAAG2} (panel b). The analysis is based on a city-level weekly panel from 1917 to 1935. The outcome variable is a city-level weekly index of manufacturing activity, which we normalize to have the same mean and variance as industrial production; see \Cref{app:bradstreets} for details. Error bands represent 95\% confidence intervals based on standard errors dually clustered on city and time (week).

\end{minipage}
\end{figure}

\Cref{fig:LP_TAAG_fund} shows that runs lead to substantially worse declines in city-level manufacturing when they occur in weak banks. Runs on strong banks do not predict a deterioration in city-level manufacturing. The estimates are slightly positive, close to zero, and not statistically significant at most horizons. In contrast, runs on weak banks lead to a gradual decline in manufacturing activity over the subsequent year. Recall that the manufacturing index is normalized to have the same mean and variance as the cyclical component of aggregate industrial production. Thus, in terms of magnitudes, runs on weak banks in a city are associated with a roughly 5\% decline in city industrial production after one year.

We stress one important point about identification that affects the interpretation of the estimates of \Cref{LP_TAAG1} in \Cref{fig:LP_TAAG}. The negative response of real activity following a run may be caused by the run itself, but it can also be caused by an underlying real economic shock that leads depositors to run. While the high-frequency nature of our analysis allows us to closely link the timing of runs and real activity, we cannot fully disentangle these stories. However, distinguishing between runs on weak versus strong banks (and runs with versus without failure below) allows us to understand which types of runs are associated with the most severe economic consequences. The findings imply that runs alone, without fundamental weaknesses or failure, do not lead to severe negative real effects.

\paragraph{Runs with and without failure} Finally, we study the city-level consequences of runs with and without failure for manufacturing activity. We estimate impulse responses from the following local projection at the weekly level.
\begin{align}
Y^{Manuf}_{c,t+h}=  \alpha_c & + \beta^h_{RNF} \text{Run}^{Without \; failure}_{c,t} \label{LP_TAAG2} + \beta^h_{RWF} \text{Run}^{With \; failure}_{c,t}
\\ & + X_{c,t-1} \Gamma^h + \epsilon_{c,t+h}, \quad h=0,...,H. \notag
\end{align}
In this specification, $\text{Run}^{Without \; failure}_{c,t}$ is an indicator for whether a city $c$ is subject to a run without failure in week $t$, and $\text{Run}^{With \; failure}_{c,t} $ is an indicator for whether there is a run with a bank failure. We define a run as involving failure if the bank fails within 30 days of the date of the run, based on the evidence that most failures occur within 30 days of the run (see \Cref{fig:cond_prob_daily}).

\Cref{fig:LP_TAAG_fail} reveals that runs with bank failures lead to substantially worse declines in local economic activity than runs without failure. The response of manufacturing activity to runs without failure (runs with failure) is similar to the response to runs on strong (weak) banks in panel (a). Runs without failure do not result in significantly lower manufacturing activity within the next 18 months. Runs with failure predict a decline in manufacturing activity of 5--8\%.

Overall, the evidence on local consequences of runs for credit markets and real activity tempers the view that runs that originate as pure liquidity events can generate severe disruptions in local credit markets and real activity. Runs on strong banks and non-fundamental runs carry limited adverse effects beyond the affected banks. Instead, the evidence indicates that poor fundamentals are necessary for runs to generate severe local financial disruptions. A key channel through which runs disrupt real activity is through bank failure, which, in turn, usually requires that a bank is fundamentally insolvent, or at least very weak in terms of fundamentals \citep{CLV2026}. Thus, runs should not be seen as universally carrying high economic costs; rather, runs involve large economic contractions when bank fundamentals are particularly weak.

\section{Conclusion}

\label{sec:conclusion}

In this paper, we apply textual analysis and large language models to historical newspapers to study bank runs recorded in digitally available newspapers in the United States from 1863 to 1934. Our novel database uncovers nearly four thousand bank runs. We combine these data with balance sheet information on banks, allowing for the first comprehensive micro-level evidence on the causes and consequences of bank runs. Our data allow us to derive several key insights about the nature of banking crises.

We provide three main findings. First, runs are more likely in banks with observably weaker fundamentals, broadly consistent with fundamental-based panic runs. But strong banks can also be run, especially in response to adverse news about the economy or the banking system, in line with information-based theories of runs. Second, most runs do not result in bank failure, and runs typically only trigger the failure of weak banks. Banks with strong fundamentals manage to survive runs through various mechanisms, including interbank cooperation, signaling strength, equity injections or loans, examination, and suspension, consistent with information-based theories of runs. Third, weak fundamentals are necessary for bank runs to result in substantial contractions in financial intermediation and real manufacturing activity. Moreover, bank failure is an important mechanism for capturing the local economic costs of runs.

Taken together, our findings lend little support to the view that small shocks can result in large jumps to bad equilibria via runs on demandable debt. In contrast, our findings suggest that absent  government interventions in the banking sector such as deposit insurance, runs can occur in both weak and strong banks, but poor fundamentals are necessary for runs to result in bank failures. Moreover, poor bank fundamentals are necessary for runs to result in severe real economic contractions. Failures are a key proxy for economic distress resulting from turmoil in the banking sector. Events that are primarily liquidity driven are less commonly associated with severe economic outcomes and should be less of a concern than solvency-driven crises.

{\singlespacing
\footnotesize
\bibliographystyle{chicagofixed}
\bibliography{literature}

\begin{thebibliography}{}

\bibitem[\protect\citeauthoryear{Admati and Hellwig}{Admati and
  Hellwig}{2014}]{AdmatiHellwig2014}
Admati, A. and M.~Hellwig (2014).
\newblock {\em The bankers' new clothes: What's wrong with banking and what to
  do about it}.
\newblock Princeton University Press.

\bibitem[\protect\citeauthoryear{Afonso, Kovner and Schoar}{Afonso, Kovner and
  Schoar}{2011}]{afonso2011stressed}
Afonso, G.~M., A.~Kovner and A.~Schoar (2011, August).
\newblock Stressed, not frozen: The federal funds market in the financial
  crisis.
\newblock {\em Journal of Finance\/}~{\em 66\/}(4), 1109--1139.

\bibitem[\protect\citeauthoryear{Allen and Gale}{Allen and
  Gale}{1998}]{Allen1998}
Allen, F. and D.~Gale (1998).
\newblock Optimal financial crises.
\newblock {\em Journal of Finance\/}~{\em 53\/}(4), 1245--1284.

\bibitem[\protect\citeauthoryear{Alston, Grove and Wheelock}{Alston, Grove and
  Wheelock}{1994}]{Alston1994}
Alston, L.~J., W.~A. Grove and D.~C. Wheelock (1994).
\newblock Why do banks fail? evidence from the 1920s.
\newblock {\em Explorations in Economic History\/}~{\em 31\/}(4), 409--431.

\bibitem[\protect\citeauthoryear{Altman}{Altman}{1968}]{Altman1968}
Altman, E.~I. (1968).
\newblock Financial ratios, discriminant analysis and the prediction of
  corporate bankruptcy.
\newblock {\em The Journal of Finance\/}~{\em 23\/}(4), 589--610.

\bibitem[\protect\citeauthoryear{Anderson, Choi and Rhee}{Anderson, Choi and
  Rhee}{2024}]{Anderson2024}
Anderson, H., J.~Choi and J.~Rhee (2024, April).
\newblock {CEO} ownership, risk management, and bank runs at unlimited
  liability banks during the 1890s.
\newblock FDIC CFR Working Paper 2024-03, Federal Deposit Insurance
  Corporation.

\bibitem[\protect\citeauthoryear{Baron, Verner and Xiong}{Baron, Verner and
  Xiong}{2021}]{Baron2021}
Baron, M., E.~Verner and W.~Xiong (2021).
\newblock Banking crises without panics.
\newblock {\em The Quarterly Journal of Economics\/}~{\em 136\/}(1), 51--113.

\bibitem[\protect\citeauthoryear{Bernanke}{Bernanke}{2018}]{Bernanke2018}
Bernanke, B. (2018).
\newblock The real effects of disrupted credit: Evidence from the global
  financial crisis.
\newblock {\em Brookings Papers on Economic Activity\/}~{\em 49\/}(2 (Fall)),
  251--342.

\bibitem[\protect\citeauthoryear{Bernanke}{Bernanke}{1983}]{Bernanke1983}
Bernanke, B.~S. (1983).
\newblock Nonmonetary effects of the financial crisis in the propagation of the
  {Great Depression}.
\newblock {\em The American Economic Review\/}~{\em 73\/}(3), 257--276.

\bibitem[\protect\citeauthoryear{Blanchard}{Blanchard}{2018}]{Blanchard2018Comment}
Blanchard, O. (2018).
\newblock Comment on ``the real effects of disrupted credit: Evidence from the
  global financial crisis''.
\newblock {\em Brookings Papers on Economic Activity\/}~{\em 49\/}(2),
  323--331.
\newblock Fall.

\bibitem[\protect\citeauthoryear{Blickle, Brunnermeier and Luck}{Blickle,
  Brunnermeier and Luck}{2024}]{Blickle2022}
Blickle, K., M.~Brunnermeier and S.~Luck (2024, 05).
\newblock {Who Can Tell Which Banks Will Fail?}
\newblock {\em The Review of Financial Studies\/}~{\em 37\/}(9), 2685--2731.

\bibitem[\protect\citeauthoryear{Bordo, Eichengreen, Klingebiel, and
  Martinez-Peria}{Bordo et~al.}{2001}]{Bordo2001crisis}
Bordo, M., B.~Eichengreen, D.~Klingebiel, and M.~S. Martinez-Peria (2001).
\newblock Is the crisis problem growing more severe?
\newblock {\em Economic policy\/}~{\em 16\/}(32), 52--82.

\bibitem[\protect\citeauthoryear{Calomiris and Gorton}{Calomiris and
  Gorton}{1991}]{Calomiris1991}
Calomiris, C.~W. and G.~Gorton (1991).
\newblock The origins of banking panics: models, facts, and bank regulation.
\newblock In {\em Financial markets and financial crises}, pp.\  109--174.
  University of Chicago Press.

\bibitem[\protect\citeauthoryear{Calomiris and Kahn}{Calomiris and
  Kahn}{1991}]{CalomirisKahn1991}
Calomiris, C.~W. and C.~M. Kahn (1991).
\newblock The role of demandable debt in structuring optimal banking
  arrangements.
\newblock {\em American Economic Review\/}~{\em 81\/}(3), 497--513.

\bibitem[\protect\citeauthoryear{Calomiris and Mason}{Calomiris and
  Mason}{1997}]{Calomiris1997}
Calomiris, C.~W. and J.~R. Mason (1997).
\newblock Contagion and bank failures during the {Great Depression}: The {June
  1932 Chicago Banking Panic}.
\newblock {\em American Economic Review\/}~{\em 87\/}(5), 863--883.

\bibitem[\protect\citeauthoryear{Calomiris and Mason}{Calomiris and
  Mason}{2003}]{Calomiris2003a}
Calomiris, C.~W. and J.~R. Mason (2003).
\newblock Fundamentals, panics, and bank distress during the depression.
\newblock {\em American Economic Review\/}~{\em 93\/}(5), 1615--1647.

\bibitem[\protect\citeauthoryear{Carlson}{Carlson}{2005}]{Carlson2005}
Carlson, M. (2005).
\newblock Causes of bank suspensions in the panic of 1893.
\newblock {\em Explorations in Economic History\/}~{\em 42\/}(1), 56--80.

\bibitem[\protect\citeauthoryear{Carlson, Correia and Luck}{Carlson, Correia
  and Luck}{2022}]{Carlson2022}
Carlson, M., S.~Correia and S.~Luck (2022).
\newblock The effects of banking competition on growth and financial stability:
  Evidence from the {National Banking Era}.
\newblock {\em Journal of Political Economy\/}~{\em 130\/}(2), 462--520.

\bibitem[\protect\citeauthoryear{Carlson, Mitchener and Richardson}{Carlson,
  Mitchener and Richardson}{2011}]{Carlson1929Florida}
Carlson, M., K.~J. Mitchener and G.~Richardson (2011).
\newblock Arresting banking panics: {Federal Reserve} liquidity provision and
  the forgotten panic of 1929.
\newblock {\em Journal of Political Economy\/}~{\em 119\/}(5), 889--924.

\bibitem[\protect\citeauthoryear{Chari and Jagannathan}{Chari and
  Jagannathan}{1988}]{Chari1988}
Chari, V.~V. and R.~Jagannathan (1988).
\newblock Banking panics, information, and rational expectations equilibrium.
\newblock {\em Journal of Finance\/}~{\em 43\/}(3), 749--761.

\bibitem[\protect\citeauthoryear{Chen, Goldstein, Huang, and Vashishtha}{Chen
  et~al.}{2024}]{chen2024liquidity}
Chen, Q., I.~Goldstein, Z.~Huang, and R.~Vashishtha (2024).
\newblock Liquidity transformation and fragility in the {U.S.} banking sector.
\newblock {\em The Journal of Finance\/}~{\em 79\/}(4), 3985--4036.

\bibitem[\protect\citeauthoryear{Chen}{Chen}{1999}]{ChenJPE1999}
Chen, Y. (1999).
\newblock Banking panics: The role of the first-come, first-served rule and
  information externalities.
\newblock {\em Journal of political economy\/}~{\em 107\/}(5), 946--968.

\bibitem[\protect\citeauthoryear{Chernow}{Chernow}{2017}]{Chernow2017}
Chernow, R. (2017).
\newblock {\em Grant}.
\newblock Bloomsbury Publishing.

\bibitem[\protect\citeauthoryear{Chodorow-Reich}{Chodorow-Reich}{2014}]{ChodorowReich2014}
Chodorow-Reich, G. (2014).
\newblock The employment effects of credit market disruptions: Firm‐level
  evidence from the 2008–9 financial crisis.
\newblock {\em The Quarterly Journal of Economics\/}~{\em 129\/}(1), 1--59.

\bibitem[\protect\citeauthoryear{Cipriani, Eisenbach and Kovner}{Cipriani,
  Eisenbach and Kovner}{2024}]{CiprianiEtAl2024}
Cipriani, M., T.~M. Eisenbach and A.~Kovner (2024, May).
\newblock Tracing bank runs in real time.
\newblock Staff Report 1104, Federal Reserve Bank of New York.
\newblock Revised December 2024.

\bibitem[\protect\citeauthoryear{{Commercial and Financial
  Chronicle}}{{Commercial and Financial Chronicle}}{1908}]{CFC8Feb1908}
{Commercial and Financial Chronicle} (1908, February).
\newblock {Commercial and Financial Chronicle}: February 8, 1908: Bank and
  quotation section.
\newblock Vol. 86, no. 2224. New York: William B. Dana Company.
\newblock Digitized by FRASER, Federal Reserve Bank of St. Louis. Accessed June
  15, 2026.

\bibitem[\protect\citeauthoryear{Conti-Brown and Vanatta}{Conti-Brown and
  Vanatta}{2025}]{Conti2025}
Conti-Brown, P. and S.~H. Vanatta (2025).
\newblock {\em Private Finance, Public Power: A History of Bank Supervision in
  {America}}.
\newblock Princeton University Press.

\bibitem[\protect\citeauthoryear{Correia and Luck}{Correia and
  Luck}{2023}]{Correia2022}
Correia, S. and S.~Luck (2023).
\newblock Digitizing historical balance sheet data: A practitioner's guide.
\newblock {\em Explorations in Economic History\/}~{\em 87}, 101475.

\bibitem[\protect\citeauthoryear{Correia, Luck and Verner}{Correia, Luck and
  Verner}{2022}]{CLV_Pandemic_2022}
Correia, S., S.~Luck and E.~Verner (2022).
\newblock Pandemics depress the economy, public health interventions do not:
  Evidence from the 1918 flu.
\newblock {\em The Journal of Economic History\/}~{\em 82\/}(4), 917--957.

\bibitem[\protect\citeauthoryear{Correia, Luck and Verner}{Correia, Luck and
  Verner}{2025}]{CLV2026}
Correia, S., S.~Luck and E.~Verner (2025, 09).
\newblock Failing banks.
\newblock {\em The Quarterly Journal of Economics\/}~{\em 141\/}(1), 147--204.

\bibitem[\protect\citeauthoryear{Correia, Luck and Verner}{Correia, Luck and
  Verner}{2026}]{CLV_AR_2026}
Correia, S., S.~Luck and E.~Verner (2026).
\newblock {Bank Failures: The Roles of Solvency and Liquidity}.
\newblock Staff Report 1181, Federal Reserve Bank of New York.

\bibitem[\protect\citeauthoryear{Dang, Gorton and Holmstr{\"o}m}{Dang, Gorton
  and Holmstr{\"o}m}{2020}]{Dang2020information}
Dang, T.~V., G.~Gorton and B.~Holmstr{\"o}m (2020).
\newblock The information view of financial crises.
\newblock {\em Annual Review of Financial Economics\/}~{\em 12\/}(1), 39--65.

\bibitem[\protect\citeauthoryear{Dang, Gorton, Holmström, and Ordoñez}{Dang
  et~al.}{2017}]{Dang2017}
Dang, T.~V., G.~Gorton, B.~Holmström, and G.~Ordoñez (2017, April).
\newblock Banks as secret keepers.
\newblock {\em American Economic Review\/}~{\em 107\/}(4), 1005–29.

\bibitem[\protect\citeauthoryear{Dell, Carlson, Bryan, Silcock, Arora, Shen,
  D'Amico-Wong, Le, Querubin, and Heldring}{Dell
  et~al.}{2023}]{AmericanStories}
Dell, M., J.~Carlson, T.~Bryan, E.~Silcock, A.~Arora, Z.~Shen, L.~D'Amico-Wong,
  Q.~Le, P.~Querubin, and L.~Heldring (2023).
\newblock American stories: A large-scale structured text dataset of historical
  {U.S.} newspapers.

\bibitem[\protect\citeauthoryear{Diamond and Dybvig}{Diamond and
  Dybvig}{1983}]{Diamond1983}
Diamond, D.~W. and P.~H. Dybvig (1983).
\newblock Bank runs, deposit insurance, and liquidity.
\newblock {\em Journal of Political Economy\/}~{\em 91\/}(3), 401--419.

\bibitem[\protect\citeauthoryear{Diamond and Rajan}{Diamond and
  Rajan}{2001}]{Diamond2001}
Diamond, D.~W. and R.~G. Rajan (2001).
\newblock Liquidity risk, liquidity creation, and financial fragility: A theory
  of banking.
\newblock {\em Journal of Political Economy\/}~{\em 109\/}(2), 287--327.

\bibitem[\protect\citeauthoryear{Driscoll and Kraay}{Driscoll and
  Kraay}{1998}]{Driscoll1998}
Driscoll, J.~C. and A.~C. Kraay (1998).
\newblock Consistent covariance matrix estimation with spatially dependent
  panel data.
\newblock {\em The Review of Economics and Statistics\/}~{\em 80\/}(4),
  549--560.

\bibitem[\protect\citeauthoryear{Egan, Horta{\c{c}}su, Kaplan, Sunderam, and
  Yao}{Egan et~al.}{2025}]{Egan2025}
Egan, M.~L., A.~Horta{\c{c}}su, N.~A. Kaplan, A.~Sunderam, and V.~Yao (2025).
\newblock Dynamic competition for sleepy deposits.
\newblock Technical report, National Bureau of Economic Research.

\bibitem[\protect\citeauthoryear{Ennis and Keister}{Ennis and
  Keister}{2003}]{Ennis2003}
Ennis, H.~M. and T.~Keister (2003).
\newblock Economic growth, liquidity, and bank runs.
\newblock {\em Journal of Economic Theory\/}~{\em 109\/}(2), 220 -- 245.
\newblock Festschrift for Karl Shell.

\bibitem[\protect\citeauthoryear{Friedman and Schwartz}{Friedman and
  Schwartz}{1963}]{FriedmanSchwartz}
Friedman, M. and A.~J. Schwartz (1963).
\newblock {\em A monetary history of the {United States}, 1867-1960}, Volume~9.
\newblock Princeton University Press.

\bibitem[\protect\citeauthoryear{Frydman, Hilt and Zhou}{Frydman, Hilt and
  Zhou}{2015}]{Frydman2015}
Frydman, C., E.~Hilt and L.~Y. Zhou (2015).
\newblock Economic effects of runs on early “shadow banks”: Trust companies
  and the impact of the panic of 1907.
\newblock {\em Journal of Political Economy\/}~{\em 123\/}(4), 902--940.

\bibitem[\protect\citeauthoryear{Frydman and Xu}{Frydman and
  Xu}{2023}]{FrydmanXu2023}
Frydman, C. and C.~Xu (2023).
\newblock Banking crises in historical perspective.
\newblock {\em Annual Review of Financial Economics\/}~{\em 15\/}(1), 265--290.

\bibitem[\protect\citeauthoryear{Gennaioli and Shleifer}{Gennaioli and
  Shleifer}{2018}]{Gennaioli2018}
Gennaioli, N. and A.~Shleifer (2018).
\newblock {\em A Crisis of Beliefs: Investor Psychology and Financial
  Fragility}.
\newblock Princeton University Press.

\bibitem[\protect\citeauthoryear{Gertler and Kiyotaki}{Gertler and
  Kiyotaki}{2010}]{GertlerKiyotaki2010}
Gertler, M. and N.~Kiyotaki (2010).
\newblock Financial intermediation and credit policy in business cycle
  analysis.
\newblock In {\em Handbook of monetary economics}, Volume~3, pp.\  547--599.
  Elsevier.

\bibitem[\protect\citeauthoryear{Gertler and Kiyotaki}{Gertler and
  Kiyotaki}{2015}]{Gertler2015}
Gertler, M. and N.~Kiyotaki (2015).
\newblock Banking, liquidity, and bank runs in an infinite horizon economy.
\newblock {\em American Economic Review\/}~{\em 105\/}(7), 2011--43.

\bibitem[\protect\citeauthoryear{Goldstein and Pauzner}{Goldstein and
  Pauzner}{2005}]{Goldstein2005}
Goldstein, I. and A.~Pauzner (2005).
\newblock Demand–deposit contracts and the probability of bank runs.
\newblock {\em Journal of Finance\/}~{\em 60\/}(3), 1293--1327.

\bibitem[\protect\citeauthoryear{Gorton}{Gorton}{1985a}]{Gorton1985suspension}
Gorton, G. (1985a).
\newblock Bank suspension of convertibility.
\newblock {\em Journal of monetary Economics\/}~{\em 15\/}(2), 177--193.

\bibitem[\protect\citeauthoryear{Gorton}{Gorton}{1985b}]{Gorton1985}
Gorton, G. (1985b).
\newblock Clearinghouses and the origin of central banking in the {United
  States}.
\newblock {\em The Journal of Economic History\/}~{\em 45\/}(2), 277--283.

\bibitem[\protect\citeauthoryear{Gorton}{Gorton}{1988}]{Gorton1988}
Gorton, G. (1988).
\newblock Banking panics and business cycles.
\newblock {\em Oxford Economic Papers\/}~{\em 40\/}(4), 751--781.

\bibitem[\protect\citeauthoryear{Gorton and Ordoñez}{Gorton and
  Ordoñez}{2014}]{Gorton2014}
Gorton, G. and G.~Ordoñez (2014, February).
\newblock Collateral crises.
\newblock {\em American Economic Review\/}~{\em 104\/}(2), 343–78.

\bibitem[\protect\citeauthoryear{Gorton and Tallman}{Gorton and
  Tallman}{2016}]{GortonTallman2016}
Gorton, G. and E.~W. Tallman (2016).
\newblock How did pre-{Fed} banking panics end?
\newblock Technical report, National Bureau of Economic Research.

\bibitem[\protect\citeauthoryear{Gorton}{Gorton}{2012}]{Gorton2012book}
Gorton, G.~B. (2012).
\newblock {\em {Misunderstanding Financial Crises: Why We Don't See Them
  Coming}}.
\newblock Number 9780199922901 in OUP Catalogue. Oxford University Press.

\bibitem[\protect\citeauthoryear{Gu}{Gu}{2011}]{Gu2011}
Gu, C. (2011).
\newblock Herding and bank runs.
\newblock {\em Journal of Economic Theory\/}~{\em 146\/}(1), 163--188.

\bibitem[\protect\citeauthoryear{Hanson, Shleifer, Stein, and Vishny}{Hanson
  et~al.}{2015}]{HANSON2015449}
Hanson, S.~G., A.~Shleifer, J.~C. Stein, and R.~W. Vishny (2015).
\newblock Banks as patient fixed-income investors.
\newblock {\em Journal of Financial Economics\/}~{\em 117\/}(3), 449--469.

\bibitem[\protect\citeauthoryear{He and Manela}{He and Manela}{2016}]{He2016}
He, Z. and A.~Manela (2016).
\newblock Information acquisition in rumor-based bank runs.
\newblock {\em Journal of Finance\/}~{\em 71\/}(3), 1113--1158.

\bibitem[\protect\citeauthoryear{Holmstrom and Tirole}{Holmstrom and
  Tirole}{1997}]{Holmstrom1997}
Holmstrom, B. and J.~Tirole (1997).
\newblock {Financial Intermediation, Loanable Funds, and The Real Sector*}.
\newblock {\em Quarterly Journal of Economics\/}~{\em 112\/}(3), 663--691.

\bibitem[\protect\citeauthoryear{Hoon, Liu, M{\"u}ller, Payne, and Zheng}{Hoon
  et~al.}{2025}]{Hoon2025costs}
Hoon, J., C.~Liu, K.~M{\"u}ller, J.~Payne, and Z.~Zheng (2025).
\newblock The costs of financial crises in the {United States}.
\newblock Available at SSRN 5350262.

\bibitem[\protect\citeauthoryear{Huber}{Huber}{2018}]{Huber2018}
Huber, K. (2018).
\newblock Disentangling the effects of a banking crisis: Evidence from {German}
  firms and counties.
\newblock {\em American Economic Review\/}~{\em 108\/}(3), 868--898.

\bibitem[\protect\citeauthoryear{Huntoon}{Huntoon}{2023}]{KeyBankData}
Huntoon, P. (2023, July).
\newblock {Key bank data, 1863-1935, charters 1-14348: Society of Paper Money
  Collectors, National Bank \& National Bank Note Summary Data}.

\bibitem[\protect\citeauthoryear{Iyer, Jensen, Johannesen, and Sheridan}{Iyer
  et~al.}{2019}]{Iyer2019}
Iyer, R., T.~L. Jensen, N.~Johannesen, and A.~Sheridan (2019, None).
\newblock The distortive effects of too big to fail: Evidence from the {Danish}
  market for retail deposits.
\newblock {\em The Review of Financial Studies\/}~{\em 32\/}(12), 4653--4695.

\bibitem[\protect\citeauthoryear{Iyer and Puri}{Iyer and Puri}{2012}]{Iyer2012}
Iyer, R. and M.~Puri (2012).
\newblock Understanding bank runs: The importance of depositor-bank
  relationships and networks.
\newblock {\em American Economic Review\/}~{\em 102\/}(4), 1414--45.

\bibitem[\protect\citeauthoryear{Iyer, Puri and Ryan}{Iyer, Puri and
  Ryan}{2016}]{Iyer2016}
Iyer, R., M.~Puri and N.~Ryan (2016).
\newblock A tale of two runs: Depositor responses to bank solvency risk.
\newblock {\em Journal of Finance\/}~{\em 71\/}(6), 2687--2726.

\bibitem[\protect\citeauthoryear{Jalil}{Jalil}{2015}]{Jalil2015}
Jalil, A.~J. (2015).
\newblock A new history of banking panics in the {United States}, 1825--1929:
  construction and implications.
\newblock {\em American Economic Journal: Macroeconomics\/}~{\em 7\/}(3),
  295--330.

\bibitem[\protect\citeauthoryear{Jamilov, K\"onig, Mahmood, M\"uller, and
  Saidi}{Jamilov et~al.}{2026}]{JKMMS_Wholesale_2026}
Jamilov, R., T.~K\"onig, S.~K. Mahmood, K.~M\"uller, and F.~Saidi (2026,
  January).
\newblock {Wholesale Funding Crises since 1800}.
\newblock Working paper.

\bibitem[\protect\citeauthoryear{Jamilov, K\"onig, M\"uller, and Saidi}{Jamilov
  et~al.}{2024}]{JKMS2024}
Jamilov, R., T.~K\"onig, K.~M\"uller, and F.~Saidi (2024, August).
\newblock {Two Centuries of Systemic Bank Runs}.
\newblock CEPR Discussion Papers 19382, C.E.P.R. Discussion Papers.

\bibitem[\protect\citeauthoryear{Jaremski}{Jaremski}{2018}]{Jaremski2018}
Jaremski, M. (2018).
\newblock The (dis) advantages of clearinghouses before the fed.
\newblock {\em Journal of Financial Economics\/}~{\em 127\/}(3), 435--458.

\bibitem[\protect\citeauthoryear{Jaremski and Fishback}{Jaremski and
  Fishback}{2018}]{Jaremski_Fishback_2018}
Jaremski, M. and P.~V. Fishback (2018).
\newblock Did inequality in farm sizes lead to suppression of banking and
  credit in the late nineteenth century?
\newblock {\em The Journal of Economic History\/}~{\em 78\/}(1), 155–195.

\bibitem[\protect\citeauthoryear{Jaremski, Richardson and Vossmeyer}{Jaremski,
  Richardson and Vossmeyer}{2023}]{Jaremski2023}
Jaremski, M.~S., G.~Richardson and A.~Vossmeyer (2023, March).
\newblock Signals and stigmas from banking interventions: Lessons from the bank
  holiday in 1933.
\newblock Working Paper 31088, National Bureau of Economic Research.

\bibitem[\protect\citeauthoryear{Jiang, Matvos, Piskorski, and Seru}{Jiang
  et~al.}{2023}]{Jiang2023}
Jiang, E.~X., G.~Matvos, T.~Piskorski, and A.~Seru (2023).
\newblock Monetary tightening and {US} bank fragility in 2023: Mark-to-market
  losses and uninsured depositor runs?
\newblock Technical report, National Bureau of Economic Research.

\bibitem[\protect\citeauthoryear{Jord{\`a}}{Jord{\`a}}{2005}]{Jorda2005}
Jord{\`a}, {\`O}. (2005).
\newblock Estimation and inference of impulse responses by local projections.
\newblock {\em American economic review\/}~{\em 95\/}(1), 161--182.

\bibitem[\protect\citeauthoryear{Jord{\`a}, Schularick and Taylor}{Jord{\`a},
  Schularick and Taylor}{2013}]{Jorda2013}
Jord{\`a}, {\`O}., M.~Schularick and A.~M. Taylor (2013).
\newblock When credit bites back.
\newblock {\em Journal of money, credit and banking\/}~{\em 45\/}(s2), 3--28.

\bibitem[\protect\citeauthoryear{Khwaja and Mian}{Khwaja and
  Mian}{2008}]{KhwajaMian2008}
Khwaja, A.~I. and A.~Mian (2008).
\newblock Tracing the impact of bank liquidity shocks: Evidence from an
  emerging market.
\newblock {\em American Economic Review\/}~{\em 98\/}(4), 1413--1442.

\bibitem[\protect\citeauthoryear{{Library of Congress}}{{Library of
  Congress}}{2025}]{ChroniclingAmerica}
{Library of Congress} (2025).
\newblock {Chronicling America}: Historic american newspapers.
\newblock \url{https://chroniclingamerica.loc.gov/}.
\newblock National Digital Newspaper Program (Library of Congress and the
  National Endowment for the Humanities).

\bibitem[\protect\citeauthoryear{Martin, Puri and Ufier}{Martin, Puri and
  Ufier}{2026}]{PuriJF}
Martin, C., M.~Puri and A.~Ufier (2026, April).
\newblock Deposit inflows and outflows in failing banks: The role of deposit
  insurance.
\newblock {\em The Journal of Finance\/}~{\em 81\/}(2), 643--685.

\bibitem[\protect\citeauthoryear{Metrick}{Metrick}{2024}]{Metrick2023}
Metrick, A. (2024, February).
\newblock The failure of {Silicon Valley Bank} and the panic of 2023.
\newblock {\em Journal of Economic Perspectives\/}~{\em 38\/}(1), 133–52.

\bibitem[\protect\citeauthoryear{Metrick and Schmelzing}{Metrick and
  Schmelzing}{2021}]{MetrickSchmelzing2021}
Metrick, A. and P.~Schmelzing (2021).
\newblock Banking-crisis interventions, 1257-2019.
\newblock Technical report, National Bureau of Economic Research.

\bibitem[\protect\citeauthoryear{Miron and Romer}{Miron and
  Romer}{1990}]{MironRomer1990}
Miron, J.~A. and C.~D. Romer (1990).
\newblock A new monthly index of industrial production, 1884--1940.
\newblock {\em The Journal of Economic History\/}~{\em 50\/}(2), 321--337.

\bibitem[\protect\citeauthoryear{Mishkin and White}{Mishkin and
  White}{2002}]{MishkinWhite2002}
Mishkin, F.~S. and E.~N. White (2002).
\newblock {US} stock market crashes and their aftermath: implications for
  monetary policy.

\bibitem[\protect\citeauthoryear{Mitchener and Richardson}{Mitchener and
  Richardson}{2019}]{Mitchener2019}
Mitchener, K.~J. and G.~Richardson (2019).
\newblock Network contagion and interbank amplification during the {Great
  Depression}.
\newblock {\em Journal of Political Economy\/}~{\em 127\/}(2), 465--507.

\bibitem[\protect\citeauthoryear{Moen and Tallman}{Moen and
  Tallman}{2000}]{MoenTallman2000}
Moen, J.~R. and E.~W. Tallman (2000).
\newblock Clearinghouse membership and deposit contraction during the panic of
  1907.
\newblock {\em The Journal of Economic History\/}~{\em 60\/}(1), 145--163.

\bibitem[\protect\citeauthoryear{Morris and Shin}{Morris and
  Shin}{2000}]{MorrisShin2000}
Morris, S. and H.~S. Shin (2000).
\newblock Rethinking multiple equilibria in macroeconomic modeling.
\newblock {\em NBER macroeconomics Annual\/}~{\em 15}, 139--161.

\bibitem[\protect\citeauthoryear{{New York State Department of Financial
  Services}}{{New York State Department of Financial Services}}{2024}]{nysdfs}
{New York State Department of Financial Services} (2024).
\newblock {Banking Institution History: Beginning 1784}.
\newblock NYS Open Data.
\newblock Accessed: 2025-10-04.

\bibitem[\protect\citeauthoryear{{Office of the Comptroller of the
  Currency}}{{Office of the Comptroller of the
  Currency}}{1873}]{OCCAnnualReport1873}
{Office of the Comptroller of the Currency} (1873).
\newblock Annual report of the {Comptroller of the Currency}, 1873.
\newblock Technical report, U.S. Treasury Department, Washington, DC.
\newblock Available via FRASER, Federal Reserve Bank of St. Louis.

\bibitem[\protect\citeauthoryear{{Office of the Comptroller of the
  Currency}}{{Office of the Comptroller of the
  Currency}}{1884}]{OCCAnnualReport1884}
{Office of the Comptroller of the Currency} (1884).
\newblock Annual report of the {Comptroller of the Currency}, 1884.
\newblock Technical report, U.S. Government Printing Office, Washington, DC.
\newblock Digitized by FRASER, Federal Reserve Bank of St. Louis. Accessed May
  21, 2026.

\bibitem[\protect\citeauthoryear{{Office of the Comptroller of the
  Currency}}{{Office of the Comptroller of the
  Currency}}{1890}]{OCCAnnualReport1890}
{Office of the Comptroller of the Currency} (1890).
\newblock {\em Annual Report of the {Comptroller of the Currency}, 1890}.
\newblock Washington, DC: Government Printing Office.

\bibitem[\protect\citeauthoryear{OGrada and White}{OGrada and
  White}{2003}]{OgradaWhite}
OGrada, C. and E.~N. White (2003).
\newblock The panics of 1854 and 1857: A view from the {Emigrant Industrial
  Savings Bank}.
\newblock {\em The Journal of Economic History\/}~{\em 63\/}(1), 213--240.

\bibitem[\protect\citeauthoryear{Peck and Shell}{Peck and
  Shell}{2003}]{PeckShell2003}
Peck, J. and K.~Shell (2003).
\newblock Equilibrium bank runs.
\newblock {\em Journal of political Economy\/}~{\em 111\/}(1), 103--123.

\bibitem[\protect\citeauthoryear{Peek and Rosengren}{Peek and
  Rosengren}{2000}]{PeekRosengren2000}
Peek, J. and E.~S. Rosengren (2000).
\newblock Collateral damage: Effects of the {Japanese} bank crisis on real
  activity in the {United States}.
\newblock {\em American economic review\/}~{\em 91\/}(1), 30--45.

\bibitem[\protect\citeauthoryear{Perignon, Thesmar and Vuillemey}{Perignon,
  Thesmar and Vuillemey}{2018}]{Perignon2018}
Perignon, C., D.~Thesmar and G.~Vuillemey (2018).
\newblock Wholesale funding dry-ups.
\newblock {\em Journal of Finance\/}~{\em 73\/}(2), 575--617.

\bibitem[\protect\citeauthoryear{Quincy and Xu}{Quincy and
  Xu}{2025}]{QuincyXu2025}
Quincy, S. and C.~Xu (2025).
\newblock Branching out: Capital mobility and long-run growth.
\newblock Technical report, National Bureau of Economic Research.

\bibitem[\protect\citeauthoryear{Reinhart and Rogoff}{Reinhart and
  Rogoff}{2009}]{Reinhart2009}
Reinhart, C.~M. and K.~S. Rogoff (2009).
\newblock The aftermath of financial crises.
\newblock {\em American Economic Review\/}~{\em 99\/}(2), 466--472.

\bibitem[\protect\citeauthoryear{Richardson}{Richardson}{2007}]{Richardson2007}
Richardson, G. (2007).
\newblock Categories and causes of bank distress during the {Great Depression},
  1929–1933: The illiquidity versus insolvency debate revisited.
\newblock {\em Explorations in Economic History\/}~{\em 44\/}(4), 588--607.

\bibitem[\protect\citeauthoryear{Richardson and Sablik}{Richardson and
  Sablik}{2015}]{Richardson2015}
Richardson, G. and T.~Sablik (2015, December).
\newblock Banking panics of the {Gilded Age}.
\newblock Federal Reserve History.

\bibitem[\protect\citeauthoryear{Richardson and Troost}{Richardson and
  Troost}{2009}]{Richardson2009}
Richardson, G. and W.~Troost (2009).
\newblock Monetary intervention mitigated banking panics during the {Great
  Depression}: Quasi‐experimental evidence from a {Federal Reserve} district
  border, 1929–1933.
\newblock {\em Journal of Political Economy\/}~{\em 117\/}(6), 1031--1073.

\bibitem[\protect\citeauthoryear{Rochet and Vives}{Rochet and
  Vives}{2004}]{Rochet2004}
Rochet, J.-C. and X.~Vives (2004).
\newblock {Coordination Failures and the Lender of Last Resort: Was Bagehot
  Right after All?}
\newblock {\em Journal of the European Economic Association\/}~{\em 2\/}(6),
  1116--1147.

\bibitem[\protect\citeauthoryear{Rockoff}{Rockoff}{2021}]{Rockoff2021}
Rockoff, H. (2021).
\newblock Oh, how the mighty have fallen: The bank failures and near failures
  that started {America}’s greatest financial panics.
\newblock {\em The Journal of Economic History\/}~{\em 81\/}(2), 331--358.

\bibitem[\protect\citeauthoryear{Romer and Romer}{Romer and
  Romer}{2004}]{RomerRomer2004}
Romer, C.~D. and D.~H. Romer (2004).
\newblock A new measure of monetary shocks: Derivation and implications.
\newblock {\em American economic review\/}~{\em 94\/}(4), 1055--1084.

\bibitem[\protect\citeauthoryear{{Royal Swedish Academy of Science}}{{Royal
  Swedish Academy of Science}}{2022}]{NobelCommittee2022}
{Royal Swedish Academy of Science} (2022).
\newblock Financial intermediation and the economy: Scientific background on
  the sveriges riksbank prize in economic sciences in memory of alfred nobel
  2022.
\newblock Technical report, {The Royal Swedish Academy of Sciences}, Stockholm,
  Sweden.

\bibitem[\protect\citeauthoryear{Saunders and Wilson}{Saunders and
  Wilson}{1996}]{Saunders1996}
Saunders, A. and B.~Wilson (1996).
\newblock Contagious bank runs: Evidence from the 1929–1933 period.
\newblock {\em Journal of Financial Intermediation\/}~{\em 5\/}(4), 409 -- 423.

\bibitem[\protect\citeauthoryear{Schroeder, Van~Riper, Manson, Knowles, Kugler,
  Roberts, and Ruggles}{Schroeder et~al.}{2025}]{ipums}
Schroeder, J., D.~Van~Riper, S.~Manson, K.~Knowles, T.~Kugler, F.~Roberts, and
  S.~Ruggles (2025).
\newblock {IPUMS National Historical Geographic Information System: Version
  20.0 [dataset]}.

\bibitem[\protect\citeauthoryear{Schumacher}{Schumacher}{2000}]{Schumacher2000}
Schumacher, L. (2000).
\newblock Bank runs and currency run in a system without a safety net:
  Argentina and the `tequila’ shock.
\newblock {\em Journal of Monetary Economics\/}~{\em 46\/}(1), 257 -- 277.

\bibitem[\protect\citeauthoryear{Schwartz}{Schwartz}{1992}]{Schwartz1992misuse}
Schwartz, A.~J. (1992, September/October).
\newblock The misuse of the {Fed}'s discount window.
\newblock {\em Federal Reserve Bank of St. Louis Review\/}~{\em 74\/}(5),
  58--69.

\bibitem[\protect\citeauthoryear{Sprague}{Sprague}{1910}]{Sprague1910}
Sprague, O. M.~W. (1910).
\newblock {\em History of Crises under the {National Banking System}}.
\newblock US Government Printing Office.

\bibitem[\protect\citeauthoryear{Sufi and Taylor}{Sufi and
  Taylor}{2021}]{SufiTaylor2021FinancialCrisesSurvey}
Sufi, A. and A.~M. Taylor (2021, August).
\newblock Financial crises: A survey.
\newblock Technical Report Working Paper 29155, National Bureau of Economic
  Research.
\newblock NBER Working Paper Series.

\bibitem[\protect\citeauthoryear{Tallman and Moen}{Tallman and
  Moen}{1990}]{TallmanMoen1990}
Tallman, E.~W. and J.~R. Moen (1990).
\newblock Lessons from the panic of 1907.
\newblock {\em Economic Review\/}~{\em 75\/}(3), 2--13.

\bibitem[\protect\citeauthoryear{Timberlake}{Timberlake}{1984}]{Timberlake1984}
Timberlake, R.~H. (1984).
\newblock The central banking role of clearinghouse associations.
\newblock {\em Journal of Money, Credit and Banking\/}~{\em 16\/}(1), 1--15.

\bibitem[\protect\citeauthoryear{{U.S. Census Bureau} and Steiner}{{U.S. Census
  Bureau} and Steiner}{2017}]{CESTA2017}
{U.S. Census Bureau} and E.~Steiner (2017).
\newblock {\em {Spatial History Project}}.
\newblock Center for Spatial and Textual Analysis, Stanford University.

\bibitem[\protect\citeauthoryear{van Belkum}{van Belkum}{1968}]{vanBelkum1968}
van Belkum, L. (1968).
\newblock {\em {National Banks of the Note Issuing Period, 1863-1935}}.
\newblock Hewitt's Numismatic Books.

\bibitem[\protect\citeauthoryear{Weber}{Weber}{2006}]{WeberJEH}
Weber, W.~E. (2006).
\newblock Early state banks in the {United States}: How many were there and
  when did they exist?
\newblock {\em The Journal of Economic History\/}~{\em 66\/}(2), 433--455.

\bibitem[\protect\citeauthoryear{White}{White}{1984}]{White1984}
White, E.~N. (1984).
\newblock A reinterpretation of the banking crisis of 1930.
\newblock {\em The Journal of Economic History\/}~{\em 44\/}(1), 119--138.

\bibitem[\protect\citeauthoryear{Wicker}{Wicker}{1996}]{Wicker1996}
Wicker, E. (1996).
\newblock {\em The banking panics of the {Great Depression}}.
\newblock Cambridge University Press.

\bibitem[\protect\citeauthoryear{Wicker}{Wicker}{2006}]{Wicker2006}
Wicker, E. (2006).
\newblock {\em Banking panics of the {Gilded Age}}.
\newblock Cambridge University Press.

\end{thebibliography}
}

\clearpage
\appendix

\crefalias{section}{appendix}
\crefalias{subsection}{appendix}

\pagestyle{appendixstyle}
\makeatletter
\def\ps@plain{\ps@appendixstyle}
\makeatother

\renewcommand{\theequation}{\thesection.\arabic{equation}}
\renewcommand{\thefigure}{\thesection.\arabic{figure}}
\renewcommand{\thetable}{\thesection.\arabic{table}}
\renewcommand{\thelstlisting}{\thesection.\arabic{lstlisting}}

\makeatletter
\@addtoreset{equation}{section}
\@addtoreset{figure}{section}
\@addtoreset{table}{section}
\@addtoreset{lstlisting}{section}
\@addtoreset{footnote}{section}
\makeatother

\thispagestyle{empty} \begin{center}
    \LARGE { \vspace{1cm} {\singlespacing \bf Bank Runs With and Without Bank Failure \\ \vspace{1cm}
    \textit{Online Appendix} }\\ \vspace{1cm}
    \large Sergio Correia, Stephan Luck, and Emil Verner\textsuperscript{*} \\ \vspace{0.7cm}
    }
\end{center}

\let\oldthefootnote\thefootnote
\renewcommand{\thefootnote}{\fnsymbol{footnote}}
\footnotetext[1]{Correia: Federal Reserve Bank of Richmond, \href{mailto:sergio.correia@rich.frb.org}{sergio.correia@rich.frb.org}; Luck: Federal Reserve Bank of New York, \href{mailto:stephan.luck@ny.frb.org}{stephan.luck@ny.frb.org}; Verner: MIT Sloan and NBER, \href{mailto:everner@mit.edu}{everner@mit.edu}.}
\let\thefootnote\oldthefootnote

\begin{itemize}
    \item Appendix \ref{appendix:additional_results}: Additional results, figures, and tables.
    \item Appendix \ref{appendix:model}: Model details.
    \item Appendix \ref{appendix:data}: Bank distress episodes dataset construction.
    \item Appendix \ref{appendix:other_data}: Details on other data.
    \item Appendix \ref{appendix:case_studies}: Case studies of bank runs.
\end{itemize}

\clearpage

\setcounter{page}{1}
\section{Additional Results}

\label{appendix:additional_results}

\subsection{Discussion of Selected Appendix Results}

\label{app:results}

\paragraph{Event Study of Bank Fundamentals around Runs and Failures} To further understand the differences between banks subject to runs without failure, banks subject to runs followed by failure, and banks that fail without a run, we perform a simple event study comparing the evolution of average bank characteristics around various event types.  Specifically, we estimate specifications of the following form:
\begin{align}
y_{bt+h} &=  \beta^h_1 \times \mathbf{1}[\text{Run w/o failure}_{bt}] \label{eq:dynamic_characteristics} \\
&+ \beta^h_2 \times \mathbf{1}[\text{Failure with run}_{bt}] \nonumber \\
&+\beta^h_3 \times \mathbf{1}[\text{Failure w/o run}_{bt}] \nonumber \\
&+ \beta^h_4 \times \mathbf{1}[\text{No Event}_{bt}] \nonumber \\ &+ \gamma^h_t + \epsilon_{bt+h}, \nonumber \hspace{3cm} h=-5,...,5.
\end{align}
The dependent variable $y_{bt+h}$ is a bank-level variable in year $t+h$. The right-hand-side variables are four mutually exclusive indicator variables: (i) bank is subject to a run in $t$ but does not fail, (ii) bank fails with a run mentioned in newspaper accounts, (iii) bank fails without a run in $t$, and (iv) no event for bank $b$ in year $t$. $\gamma_t$ is a year fixed effect. We exclude the constant term, so the estimated coefficient is the average level of $y_{bt+h}$ around the event. Because a bank that fails does not report bank fundamentals after failure, coefficients for failure are only estimated for $h=-5,..,-1$. Again, we focus on the sample of all national banks from 1863 to 1934.

\begin{figure}[ht]
\centering

\caption{\textbf{Characteristics of Banks Subject to Runs With and Without Failure} \label{fig:dynamics_characteristics} } 

\subfloat[Surplus/Equity]{\includegraphics[width=0.49\textwidth]{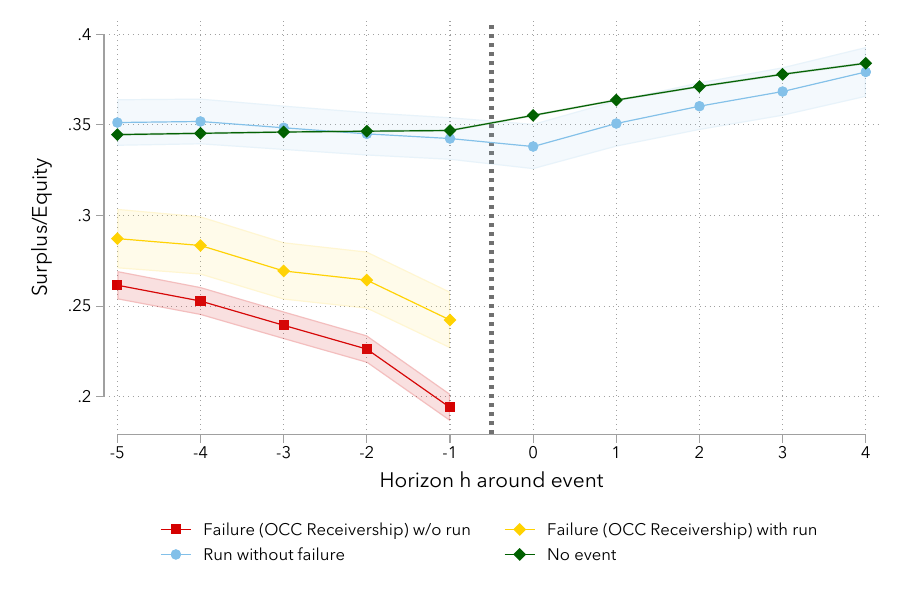}}
\hfill
\subfloat[Non-performing loans/Loans ]{\includegraphics[width=0.49\textwidth]{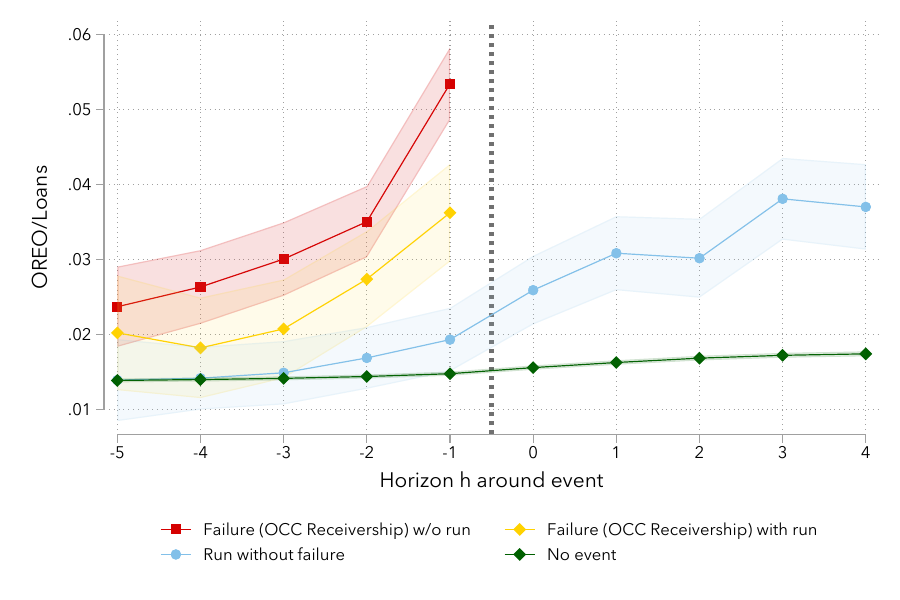}}

\subfloat[Deposit/Assets \label{fig:dep_a}]
{\includegraphics[width=0.49\textwidth]{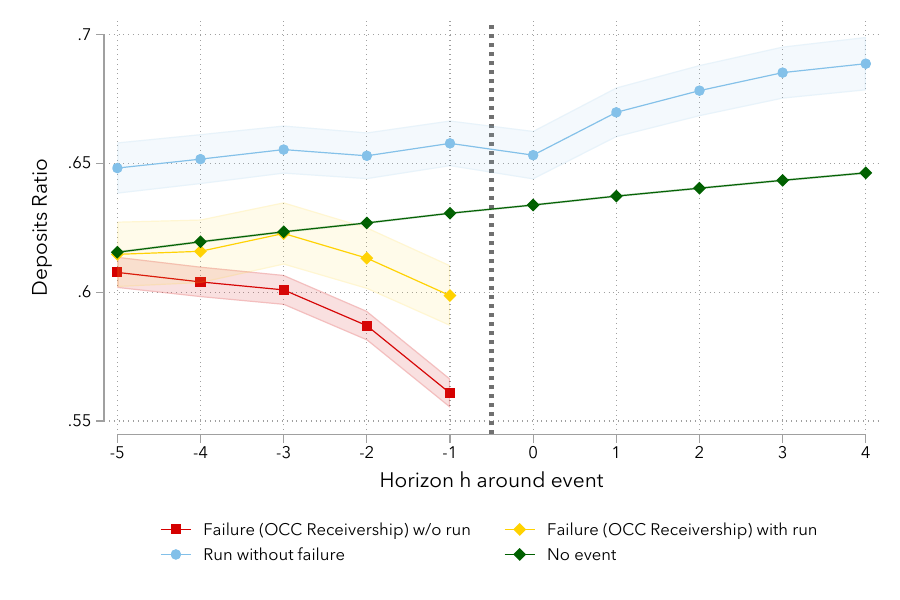}}
\hfill
\subfloat[Noncore Funding/Assets]{\includegraphics[width=0.49\textwidth]{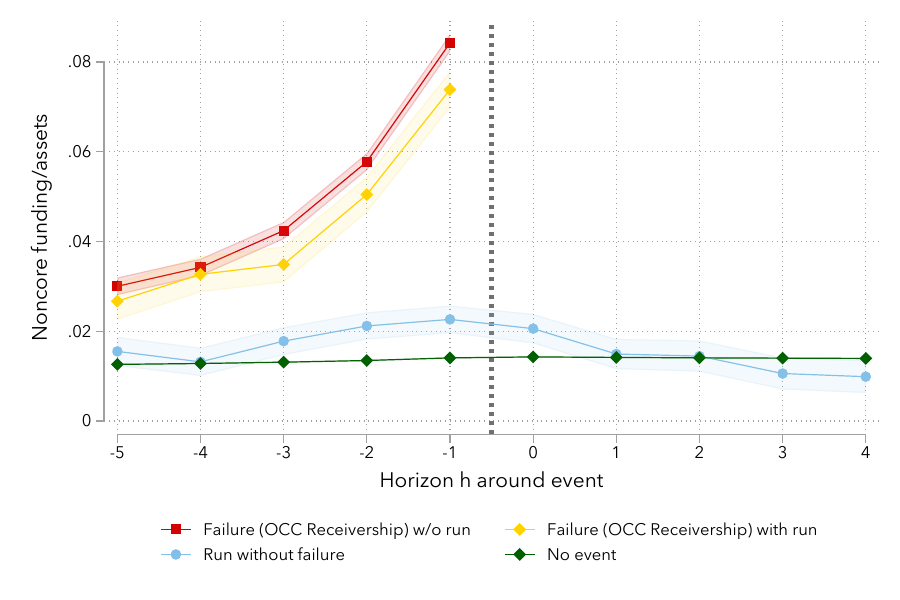}}

\subfloat[Liquid/Assets]
{\includegraphics[width=0.49\textwidth]{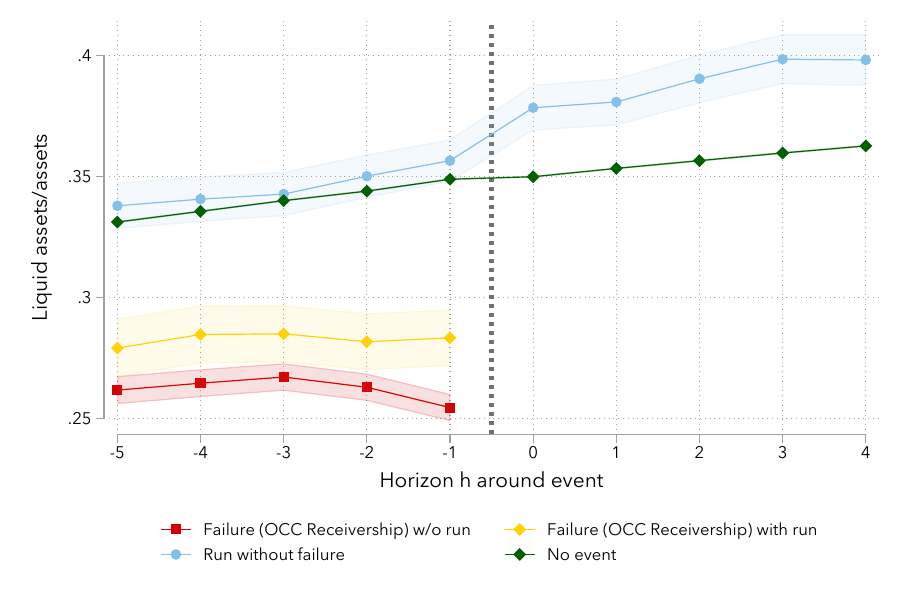}}
\hfill
\subfloat[Loans/Assets]
{\includegraphics[width=0.49\textwidth]{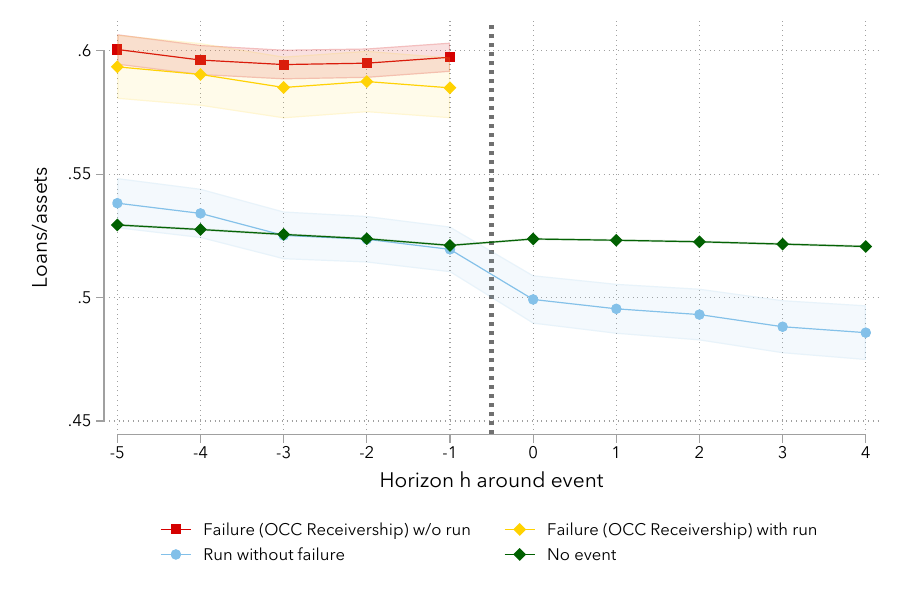}}

\begin{minipage}{\textwidth}
\footnotesize
Notes: This figure plots the coefficients from estimating \Cref{eq:dynamic_characteristics}. Non-performing loans are proxied by the line item ``other real estate owned'' (OREO). Shaded areas represent 95\% confidence bands based on \citet{Driscoll1998} standard errors with a bandwidth of three years to allow for residual correlation within and across banks in proximate years. \Cref{fig:dynamics_characteristics_growth} shows the corresponding plots using deposit growth, loan growth, asset growth, and liquid-asset growth as the outcome variables.
\end{minipage}
\end{figure}

\Cref{fig:dynamics_characteristics} plots the coefficients from estimating \Cref{eq:dynamic_characteristics} using different metrics of bank fundamentals, $y_{bt}$. The figure shows that the strongest differences in fundamentals are across failing and surviving banks, rather than whether a bank is subject to runs. Panel (a) of \Cref{fig:dynamics_characteristics} uses surplus-to-equity, our proxy of bank capitalization, as the outcome. Failing banks have a substantially lower level of surplus-to-equity before failure compared to banks subject to a run without failure and banks not subject to any distress event. Moreover, failing banks see a significant deterioration in this capitalization metric in the five years before failure. The difference holds whether or not failure involves a run. Banks that fail with a run are slightly stronger on this observable than banks that fail without a run. This helps understand why failures with runs are nearly as predictable as failures without runs, as we saw in \Cref{tab:AUC}. In contrast, banks that experience a run without failing have a similar level and evolution of capitalization as other banks not subject to a distress event.

Panel (b) of \Cref{fig:dynamics_characteristics} shows a similar pattern for non-performing assets. We proxy non-performing assets by the line item ``other real estate owned,'' which typically reflects seized real estate collateral following borrower default. Failing banks experience a rise in non-performing assets before failure. The rise in non-performing assets holds for banks that fail both with and without a run. Credit losses thus often play an important role in bank failures, both with and without runs.

In panel (b), we also see that banks subject to a run without failing do not exhibit this pre-trend in non-performing assets. These banks look similar to banks without exposure to any form of distress event. Banks subject to runs do, however, see an increase in non-performing assets after the run. There are two potential explanations for this finding. One is that a run occurs when there is a public signal that banks will face future losses, so runs predict, but do not cause, future losses \citep{Calomiris1991}. Another possibility is that the run causally increases non-performing assets by causing a deterioration of the local economy. 

\begin{figure}[ht]
    \centering
     \caption{Data Quality Validation}
     \subfloat[Annual National Bank Failure Rate \label{fig:scatter_rates}]{
    \includegraphics[width=0.49\linewidth]{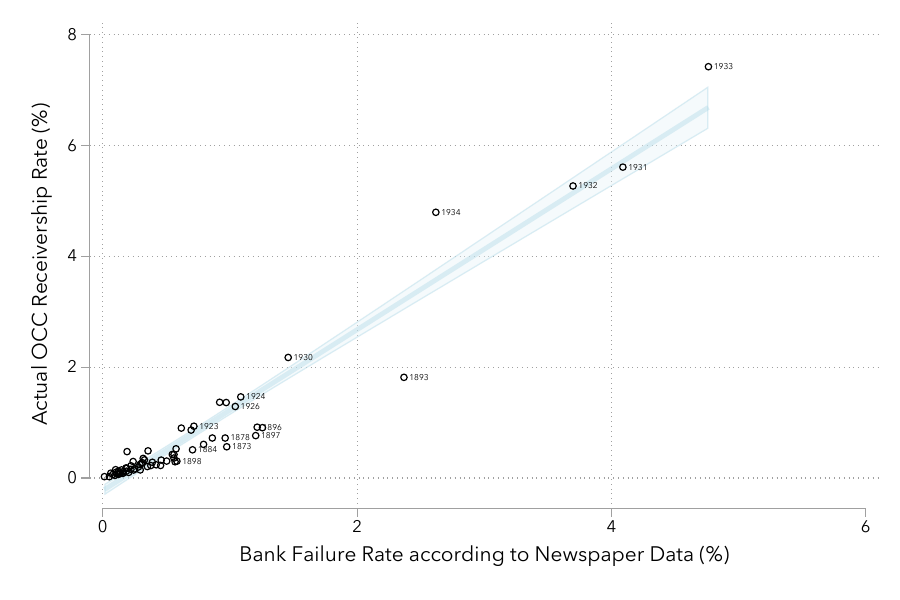}}
    \subfloat[Deposit Outflows  \label{fig:deposit_outflows}]{
    \includegraphics[width=0.49\linewidth]{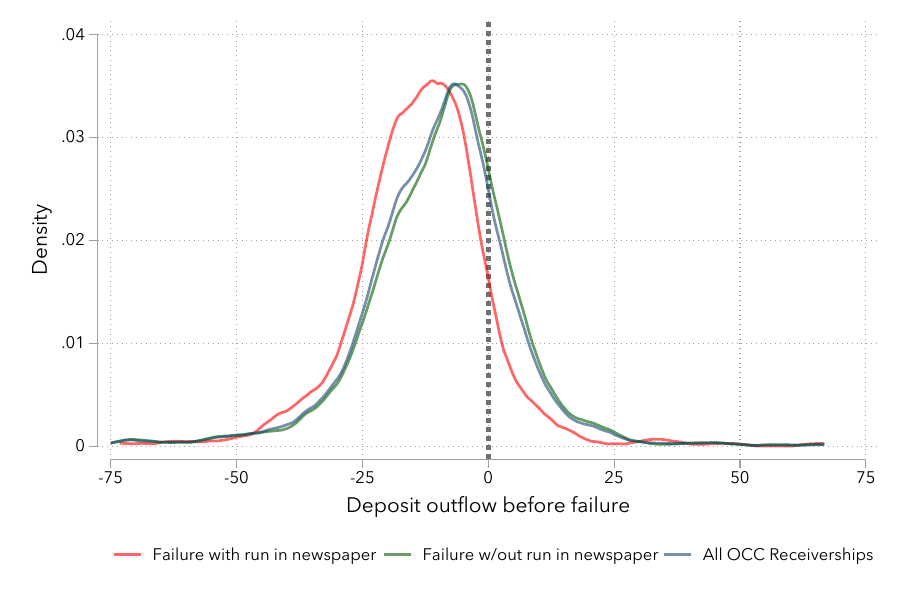}}

    \subfloat[Probability of distress event as a function of deposit growth \label{fig:events_conditional_on_dep_growth}]{
    \includegraphics[width=0.49\linewidth]{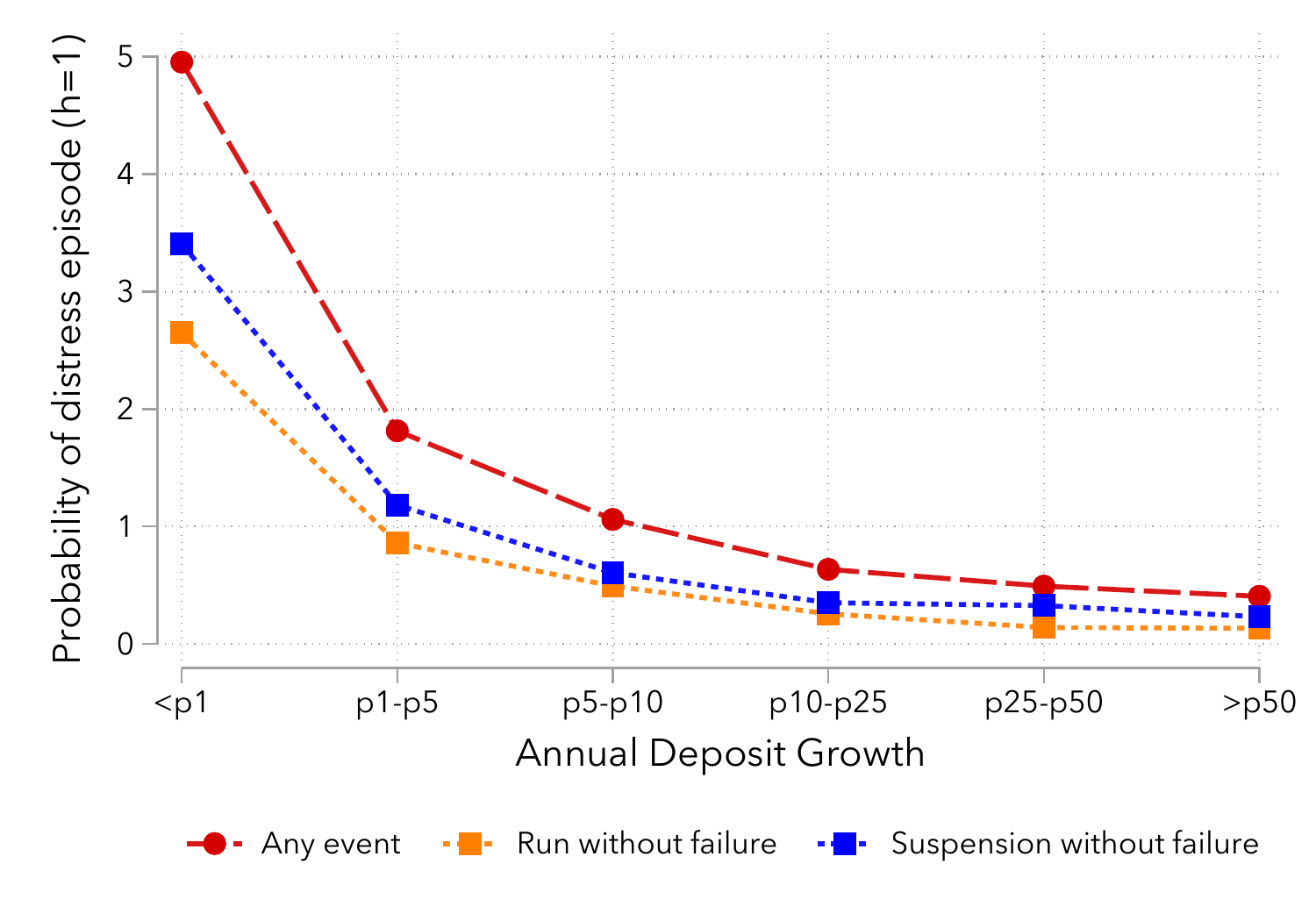}}
   \subfloat[Events by bank size \label{fig:events_conditional_on_size}]{
    \includegraphics[width=0.49\linewidth]{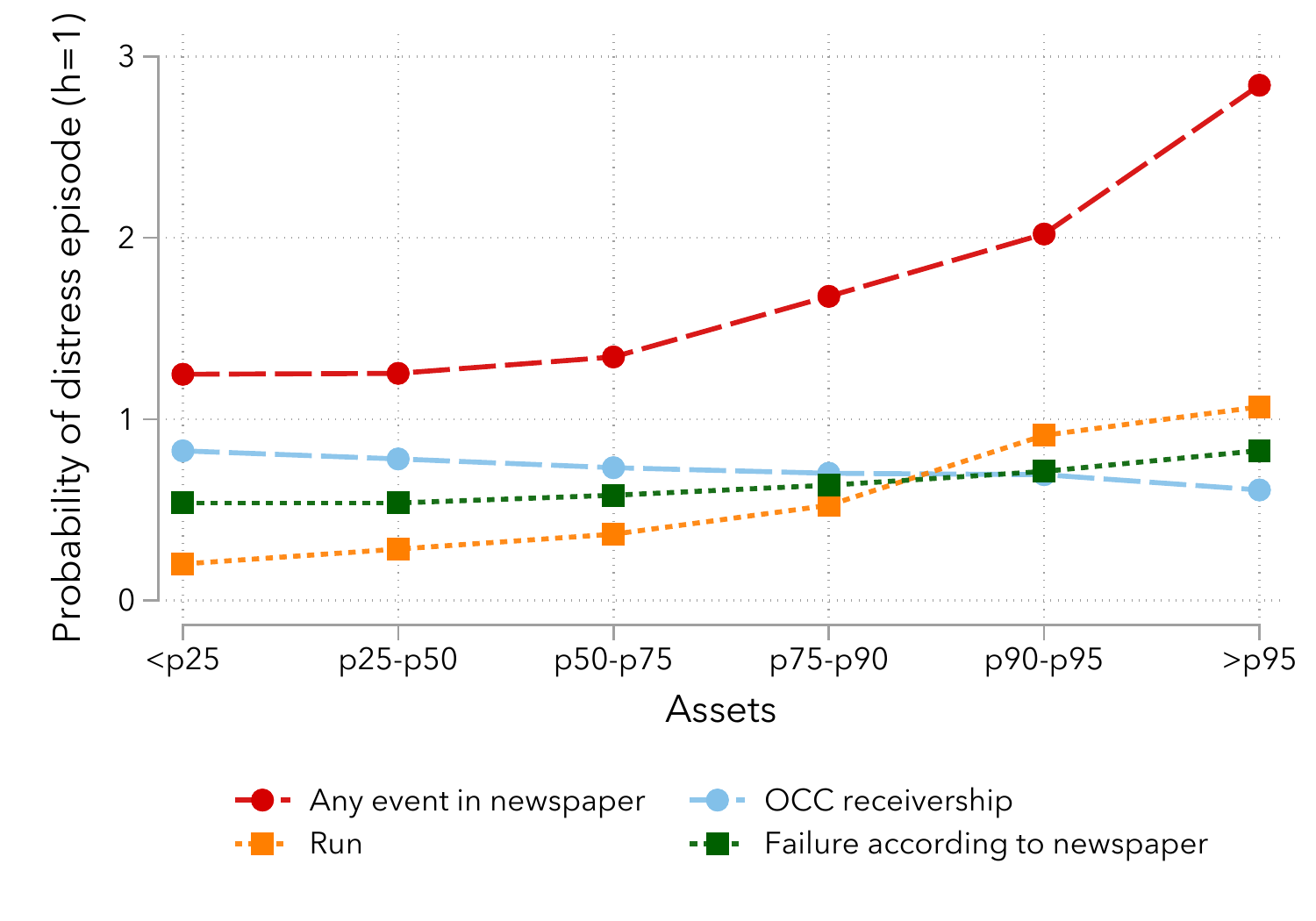}}

\begin{minipage}{\textwidth}
\footnotesize
Notes: The sample in all panels is based on national banks during 1863-1934. Deposit outflows before failure are calculated using the total deposits reported in the last call report before failure and deposits at failure.
\end{minipage}
\end{figure}

Panel (c) in \Cref{fig:dynamics_characteristics} shows that failing banks face a decline in their deposit-to-assets ratio of about 4 points before failure.\footnote{Note that banks that fail with a discussion of a run in newspapers see a larger deposit outflow from time -1 to 0 (\Cref{fig:deposit_outflows}). This suggests that the run discussed in newspaper accounts reflects an acute episode of depositor withdrawals, which is precisely the type of episode that would be likely to garner media attention. This decline in deposits for failing banks subject to runs is not shown here.}  Panel (d) shows that this deposit outflow is mirrored by an increased reliance on noncore funding in the years before failure. The decline in deposit funding and the rise in noncore funding in the years before failure are similar across banks that fail with and without a run.

Panels (c) and (d) further show that, in contrast to failing banks, banks that are subject to a run without failing do not exhibit a decline in deposits or an increased reliance on noncore funding in the years before the episode. This is another indication that these banks do not appear substantially worse than the average bank on observables. These episodes do, however, lead to a decline in deposits/assets after the run, as expected and explored further below.

Finally, panels (e) and (f) reveal striking differences in liquid assets and loan-to-asset ratios across failing and surviving banks. Failing banks have significantly lower liquid asset ratios and significantly higher loans-to-assets, indicating a riskier investment mix. Before the event, banks that are run without failing look similar to the average bank that is not exposed to a distress event. After a run, surviving banks reduce their loans-to-assets and increase liquid assets-to-assets.

Taken together, failing banks have very different observable characteristics than banks subject to runs that do not fail. Banks subject to runs that do not fail look similar to ordinary banks not subject to any form of distress. Moreover, failing banks have weak bank observables, irrespective of whether they are subject to a run before failure.

This evidence has three interesting implications. First, it reinforces the result that runs that do not result in failure tend to occur in healthy banks. Second, poor fundamentals are key to determining whether a bank will fail. Third, focusing on bank failures, there is suggestive evidence that runs trigger the failure of slightly healthier banks, on the margin, as observables are slightly better for banks that fail with a run compared to those that fail without a run. This suggests that runs may slightly accelerate the failure of weak banks. However, given that these banks are on similar trajectories, banks that fail with a run might have failed even in the absence of the run, but the run accelerated their demise. At the same time, we do not want to overstress this finding. When we classify failures with and without runs based on the size of deposit outflows in the run-up to failure, the differences across the two types of failure become even smaller and are no longer statistically significant \citep[see, e.g.,][]{CLV_AR_2026}.

\clearpage

\subsection{Additional Figures}

\begin{figure}[ht!]
\centering

\caption{\textbf{Bank Runs, Suspensions, and Failures for National Banks, 1863-1934} \label{fig:events_app}}

\subfloat[Runs, suspensions, and failures]{\includegraphics[width=1.0\textwidth]{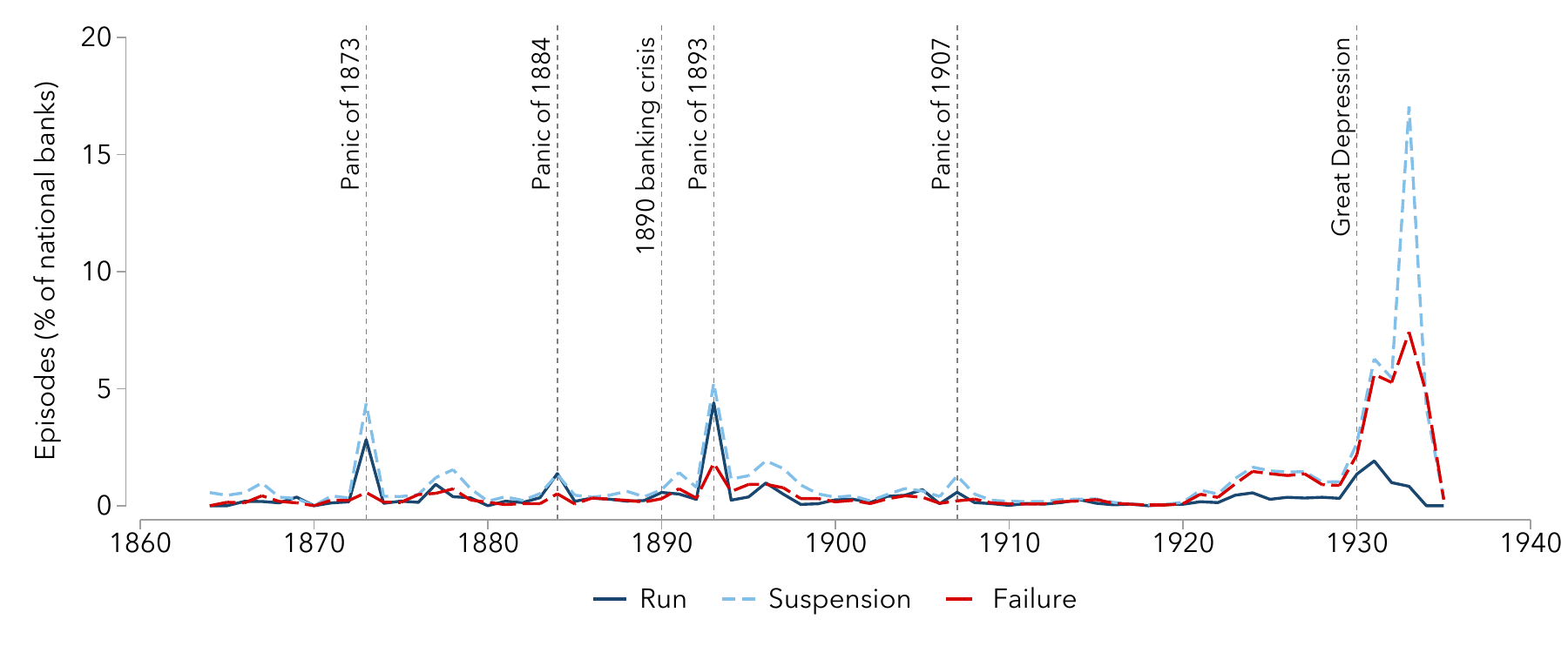}}

\subfloat[Failures with and without runs]{\includegraphics[width=1.0\textwidth]{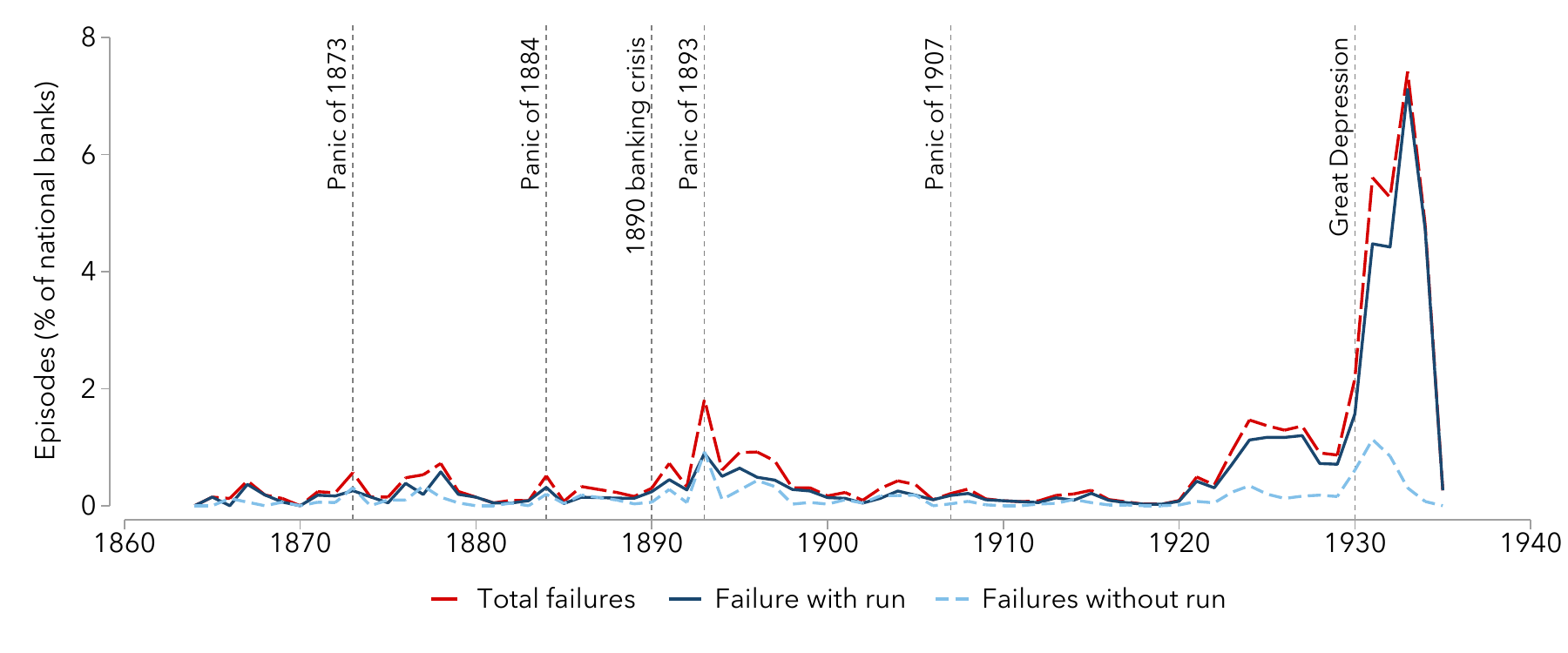}}

\begin{minipage}{\textwidth}
\footnotesize
Notes: This figure plots the ratio of the number of national banks subject to a distress episode over the total number of national banks in each year. In panel (a), we report the rate of bank runs, suspensions, and failures.
In panel (b), we show the number of bank failures (OCC receiverships) that are associated with runs reported in newspapers and those that are not associated with runs.
\end{minipage}
\end{figure}

\begin{figure}[ht]
    \centering
     \caption{ \textbf{Bank Run Rate, Monthly Time Series for National Banks} \label{fig:monthly_ts}}

    \subfloat[Full sample:1863-1934]{
    \includegraphics[width=1.0\linewidth]{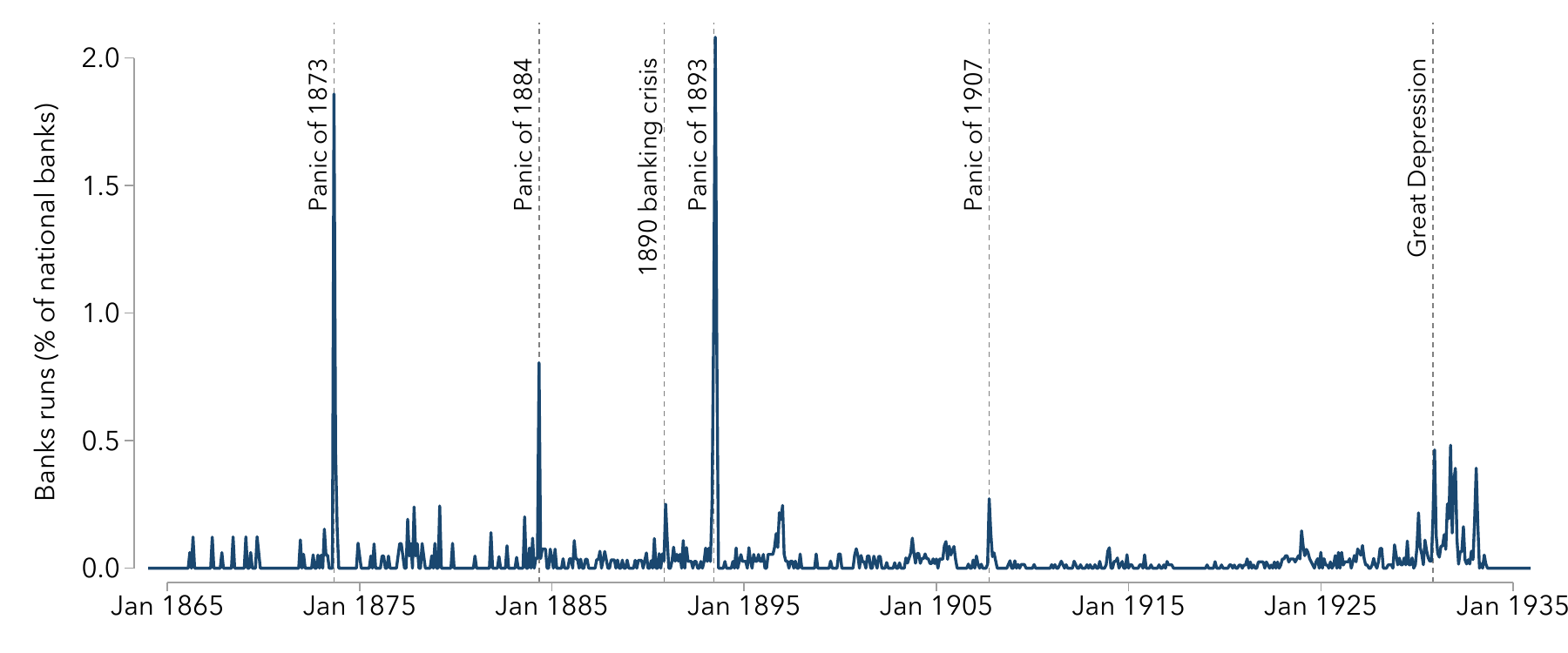}}

    \subfloat[Great Depression]{
    \includegraphics[width=1.0\linewidth]{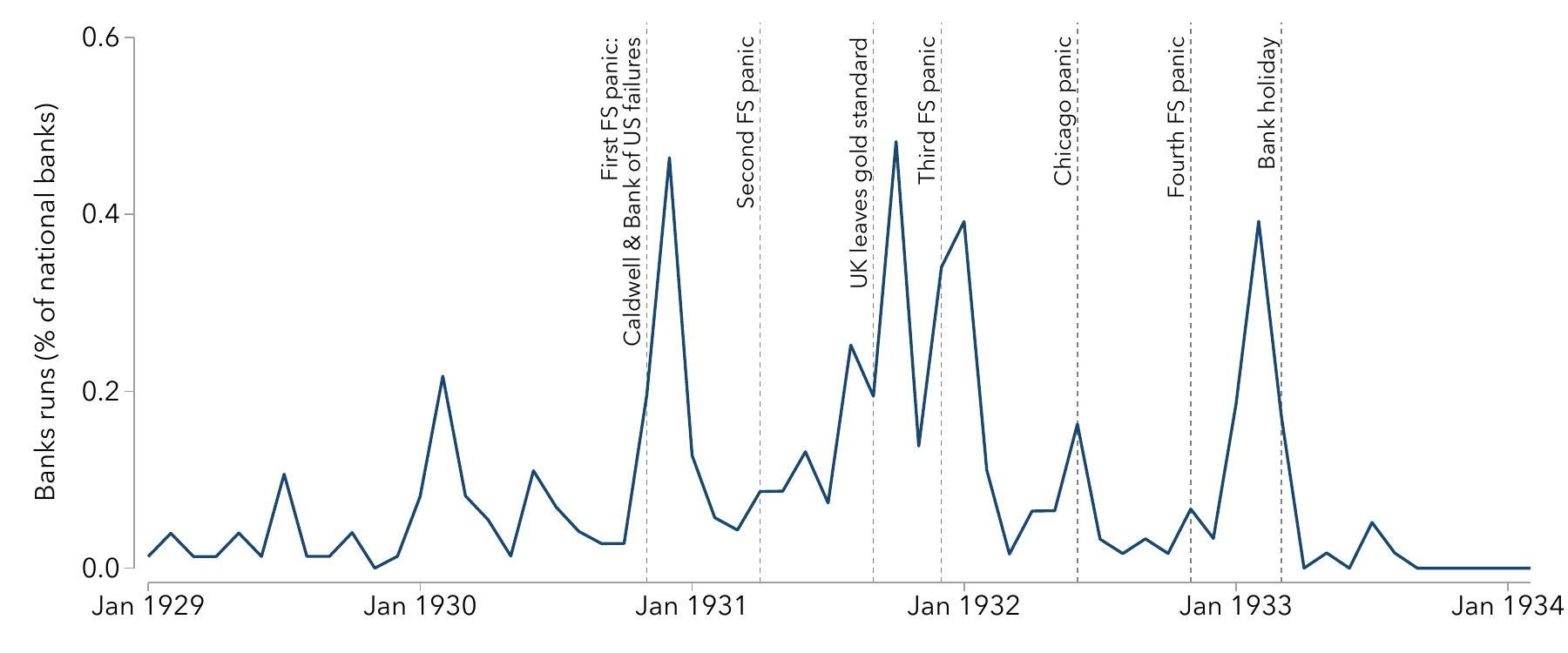}}

\begin{minipage}{\textwidth}
\footnotesize
Notes: Vertical lines in panel (a) represent the month of a banking panic from \cite{Baron2021}. Vertical lines in panel (b) represent key events during the Great Depression. ``FS panic'' refers to the start of banking panics as classified in \cite{FriedmanSchwartz}.
\end{minipage}
\end{figure}

\begin{figure}[ht]
\centering
 \caption{\textbf{Probability of Runs, Suspensions, and Failures for National Banks During and Outside of Banking Crises} }
\includegraphics[width=0.9\linewidth]{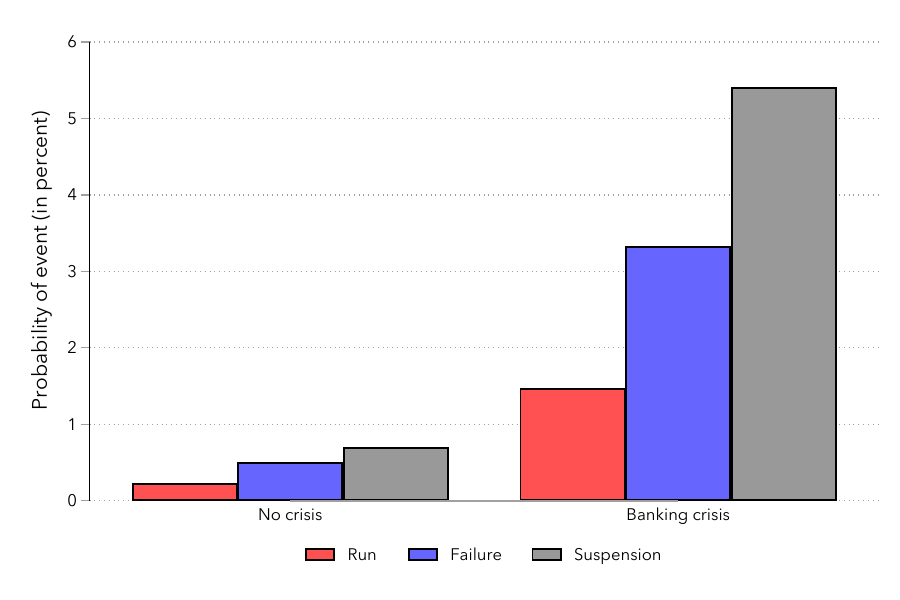}

\label{fig:probability_eventS_by_crisisBVX}

\begin{minipage}{\textwidth}
\footnotesize
Notes: The figure plots the probability of bank runs (both with and without bank failure), bank failures (defined as OCC receiverships), and suspensions (including both temporary suspensions and permanent closures) during and outside of years identified as banking crises in \cite{Baron2021}. The sample is all national banks during 1863-1934.
\end{minipage}
\end{figure}

\begin{figure}[ht]
\centering

\caption{\textbf{Characteristics of Banks Subject to Runs With and Without Failure: Growth Rates} \label{fig:dynamics_characteristics_growth} }

\subfloat[Deposit growth]{\includegraphics[width=0.49\textwidth]{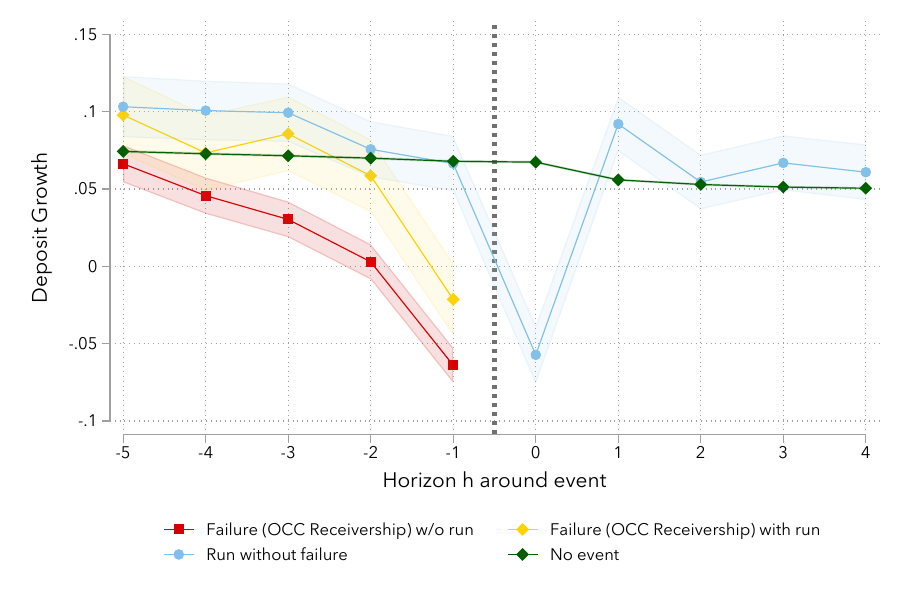}}
\hfill
\subfloat[Loan growth]{\includegraphics[width=0.49\textwidth]{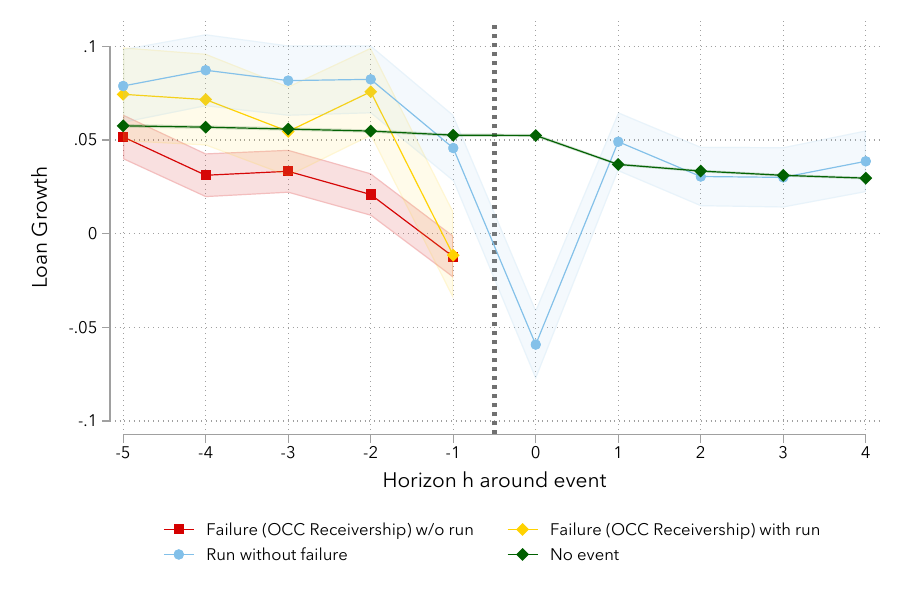}}

\subfloat[Asset growth]
{\includegraphics[width=0.49\textwidth]{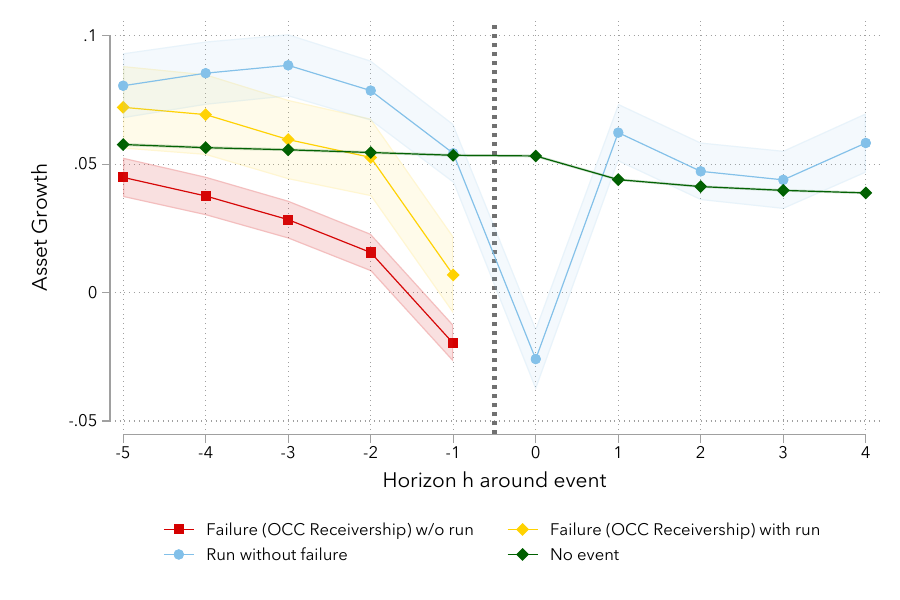}}
\hfill
\subfloat[Liquid asset growth]{\includegraphics[width=0.49\textwidth]{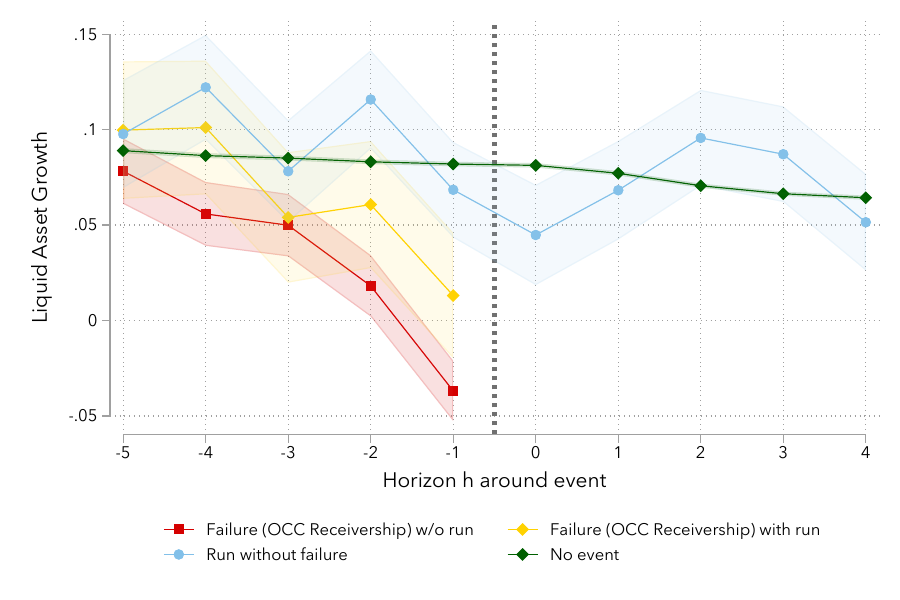}}

\begin{minipage}{\textwidth}
\footnotesize
Notes: This figure plots the coefficients from estimating \Cref{eq:dynamic_characteristics}.  Shaded areas represent 95\% confidence bands based on \citet{Driscoll1998} standard errors with a bandwidth of three years to allow for residual correlation within and across banks in proximate years.
\end{minipage}
\end{figure}

\begin{figure}[ht]
\centering

\caption{\textbf{Responses to Bank Runs: Conditional Probabilities} \label{fig:responses_cond_prob} }

{\includegraphics[width=0.95\textwidth]{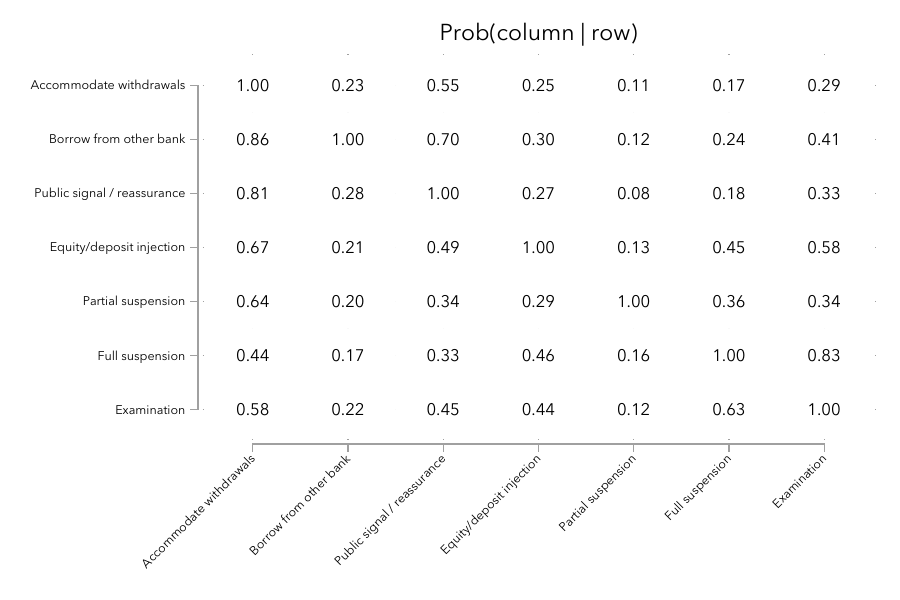}}

\begin{minipage}{\textwidth}
\footnotesize
Notes: This figure reports the probability of a response listed in each column, conditional on the bank taking the response in a given row. See \Cref{fig:responses} for further details on the classification of responses.
\end{minipage}
\end{figure}

\begin{figure}[ht!]
\centering

\caption{\textbf{Dynamics of State-level Non-financial Business Failure Rate around Innovations in Bank Runs and Failures} \label{fig:failures_dynamics} }

\subfloat[Runs versus failures]{
{\includegraphics[width=0.65\textwidth]{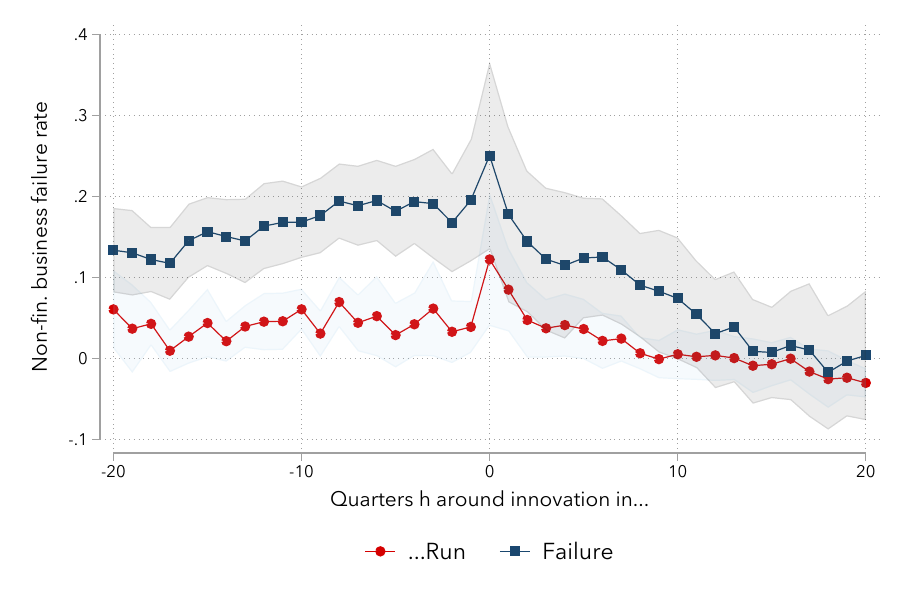}}}

\subfloat[Runs with failure, runs without failure, and failure without runs]{
{\includegraphics[width=0.65\textwidth]{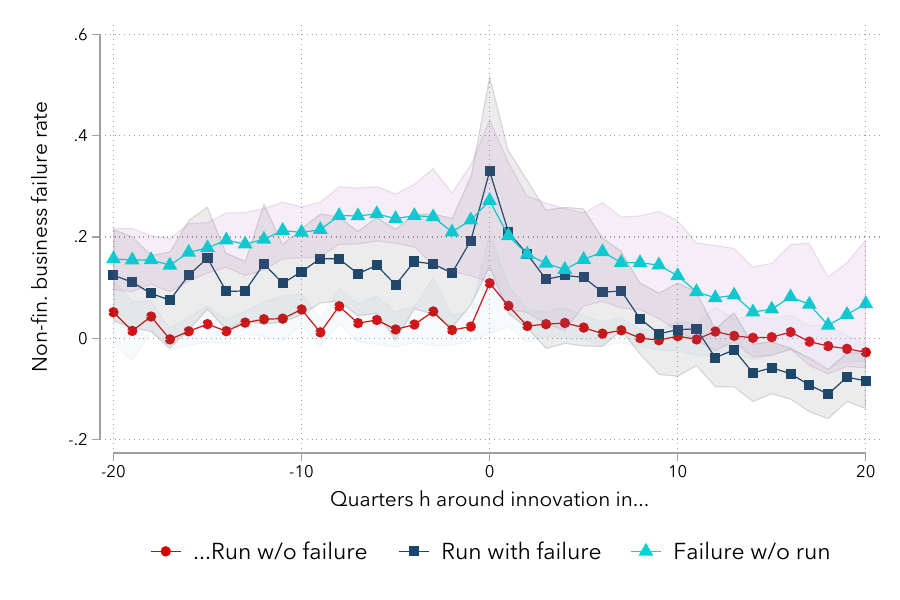}}}

\begin{minipage}{\textwidth}
\footnotesize
Notes: This figure plots the dynamics of the state-level non-financial business failure rate around innovations in the rate of bank runs and bank failures. Coefficients are based on estimation of \begin{align}
    \text{Business failure rate}_{st+h} = \alpha_s + \delta_{q(t+h)} + \beta^h \text{Bank distress}_{st} + u_{st+h}, \quad h=-20,...,0,...,20. \label{eq:LP_simple}
\end{align}
$\text{Business failure rate}_{st+h}$ is the state-level non-financial business failure rate in state $s$ for quarter $t+h$, and $\text{Bank Distress}_{st}$ is the share of banks exposed to a run or failure in state $s$ and quarter $t$. The non-financial business failure rate is defined as the number of business failures relative to the total number of manufacturing establishments in a state. Error bands represent 95\% confidence intervals based on \citet{Driscoll1998} standard errors  with a bandwidth of three years. The specification also controls for quarter fixed effects to capture seasonality in the failure rate.
\end{minipage}
\end{figure}

\begin{figure}[ht!]
\centering

\caption{\textbf{Consequences of Runs for City-Level Deposits and Lending: Unweighted Estimates} \label{fig:city_level_dynamics_unweighted} }

\subfloat[Impact of runs on deposits, weak vs strong banks]{
{\includegraphics[width=0.49\textwidth]{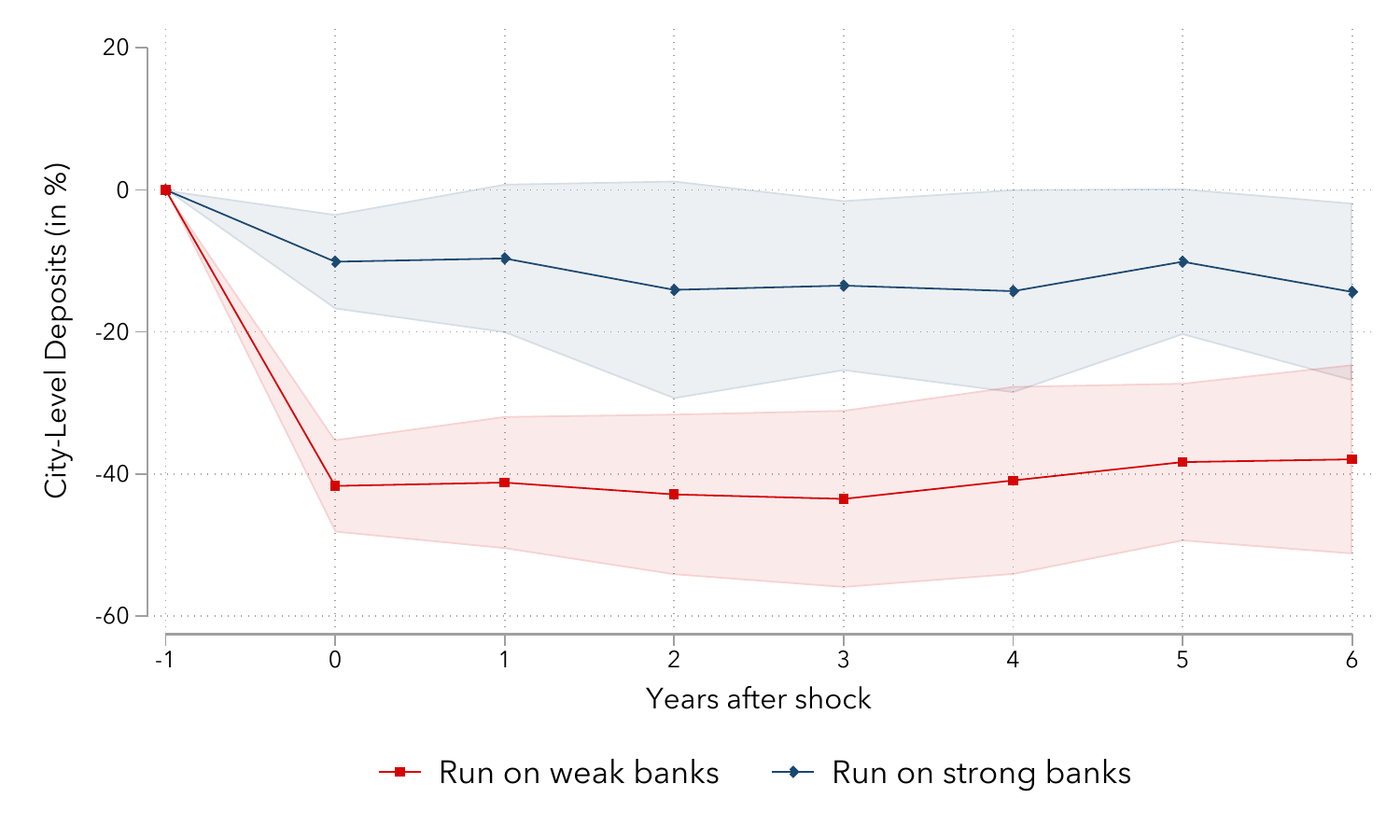}}}
\hfill
\subfloat[Impact of runs on loans, weak vs strong banks]{
{\includegraphics[width=0.49\textwidth]{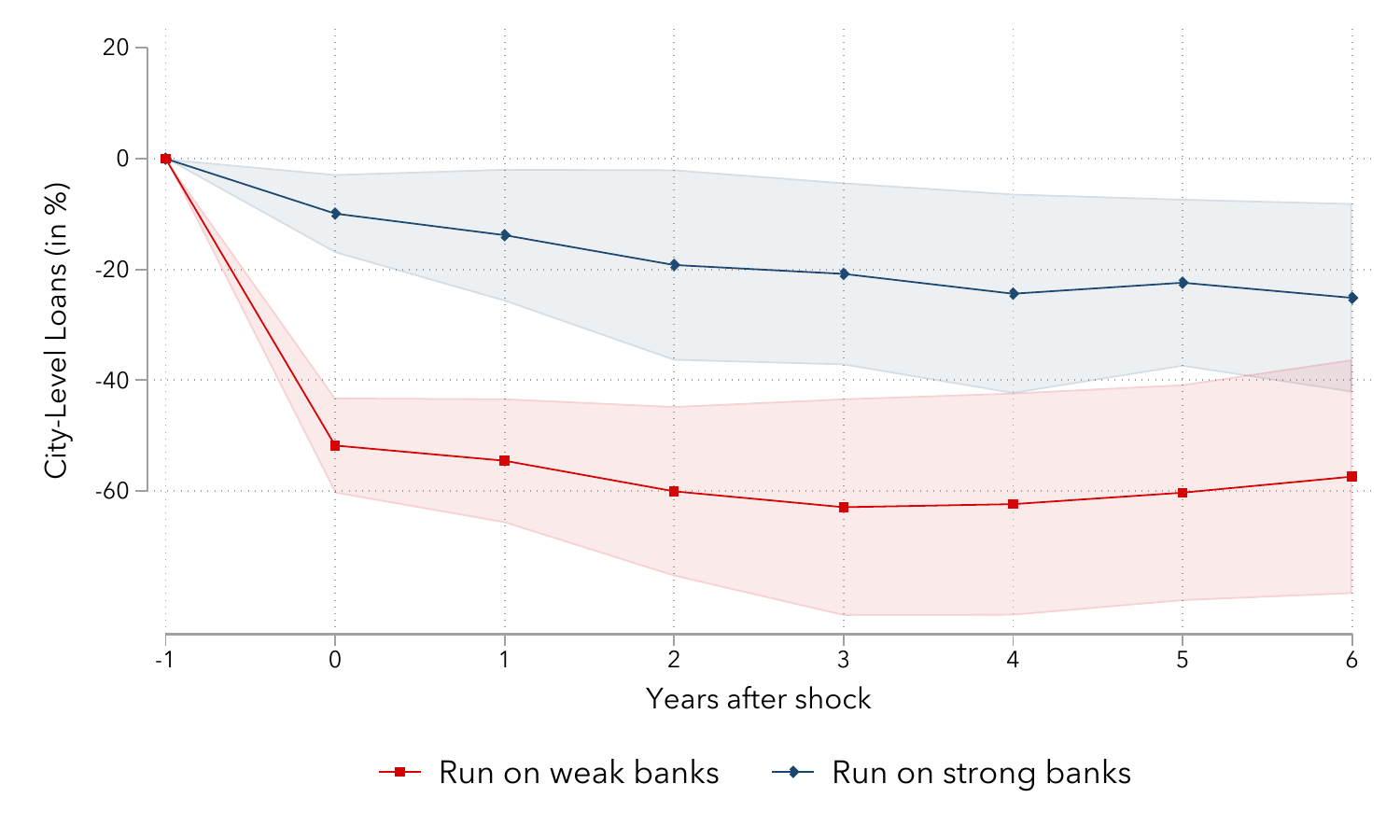}}}

\subfloat[Impact of runs on deposits, non-fundamental runs]{
{\includegraphics[width=0.49\textwidth]{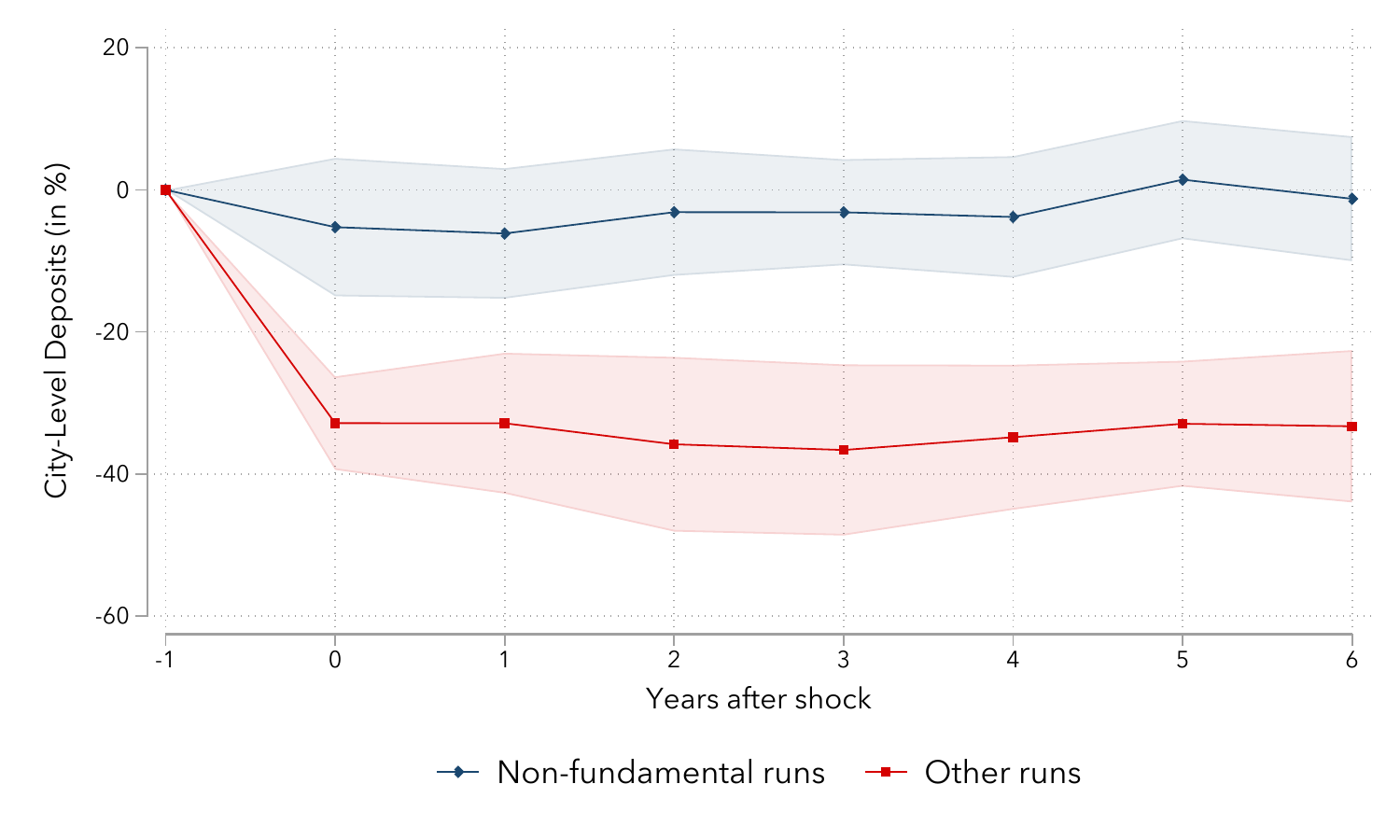}}}
\hfill
\subfloat[Impact of runs on loans, non-fundamental runs]{
{\includegraphics[width=0.49\textwidth]{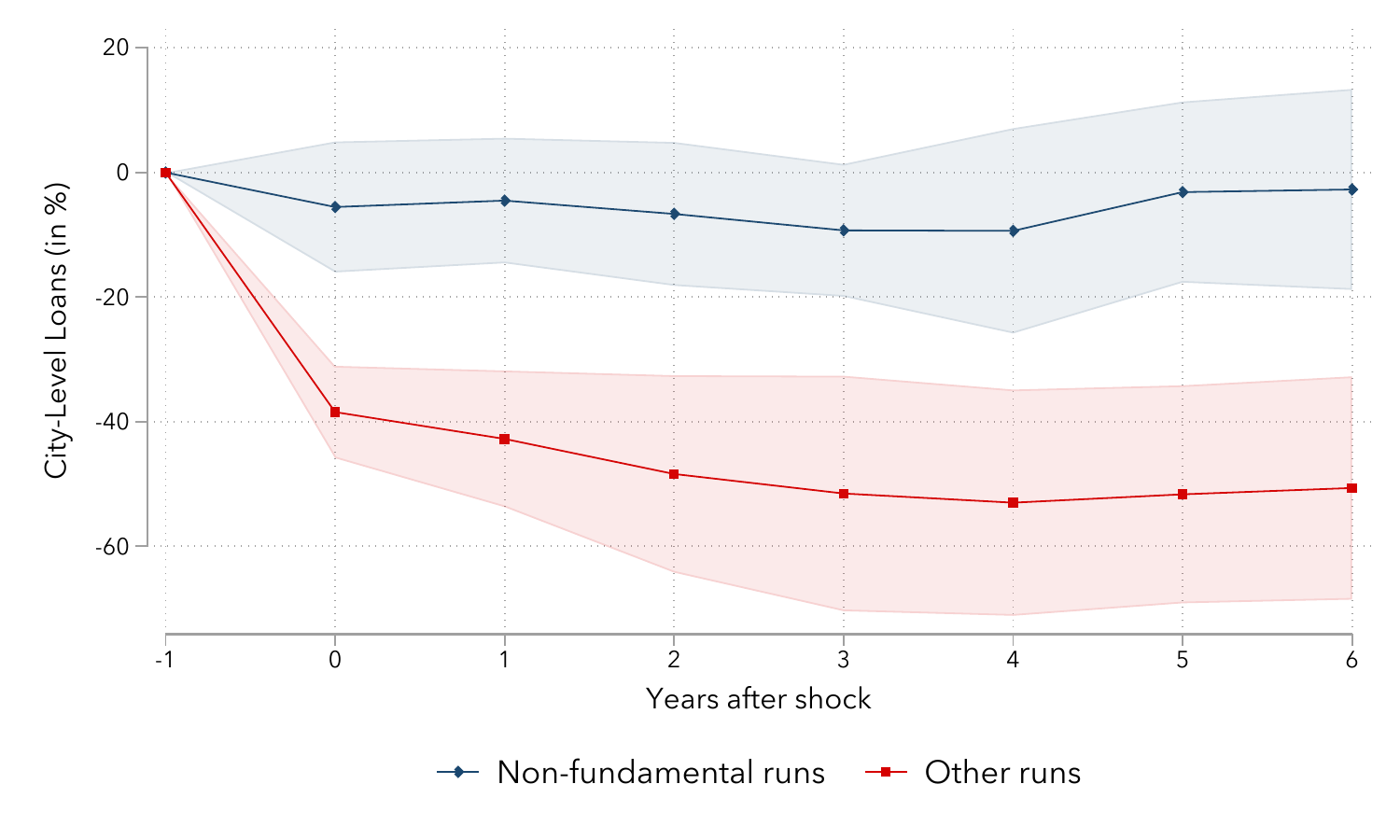}}}

\subfloat[Impact of runs with and without failure on deposits]{
{\includegraphics[width=0.49\textwidth]{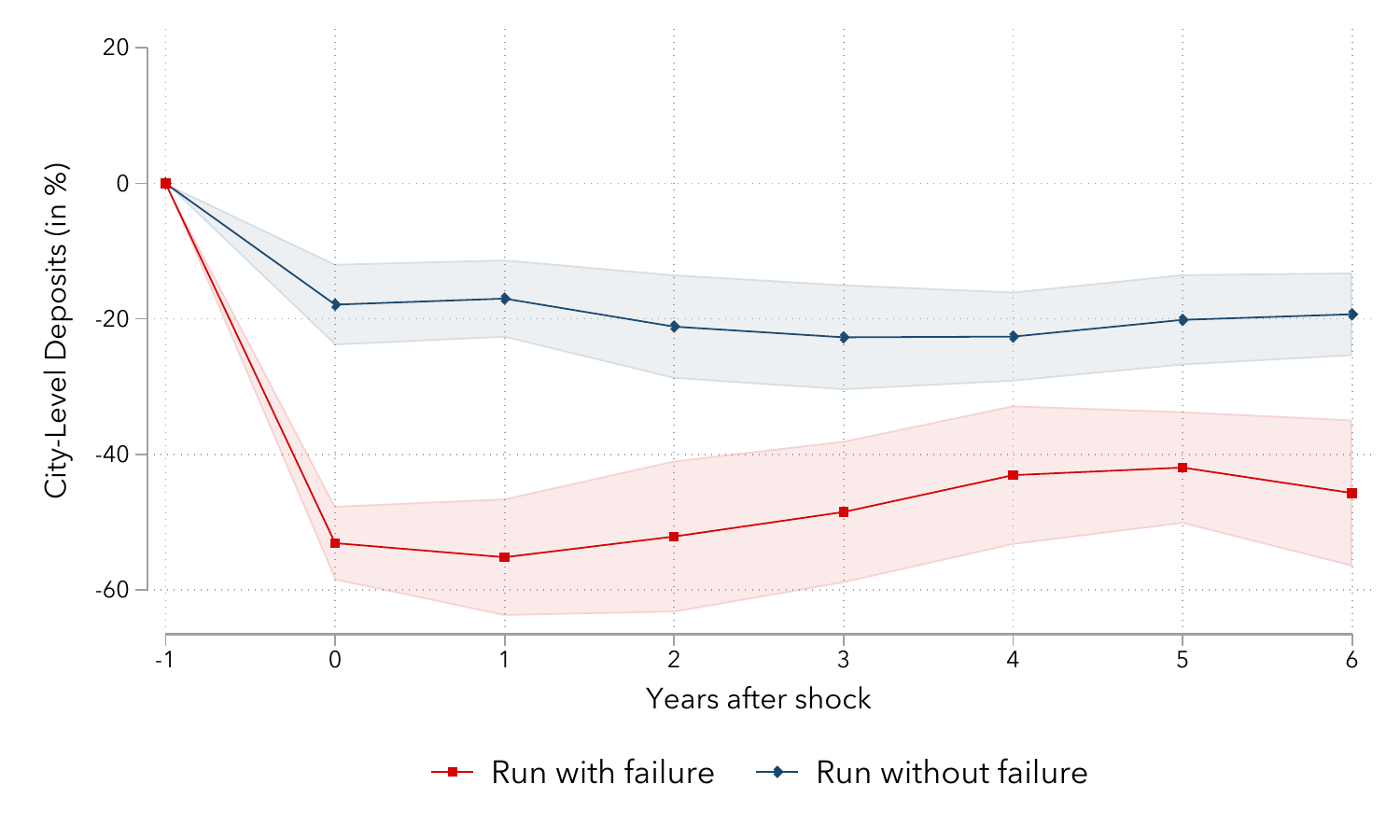}}}
\hfill
\subfloat[Impact of runs with and without failure on loans]{
{\includegraphics[width=0.49\textwidth]{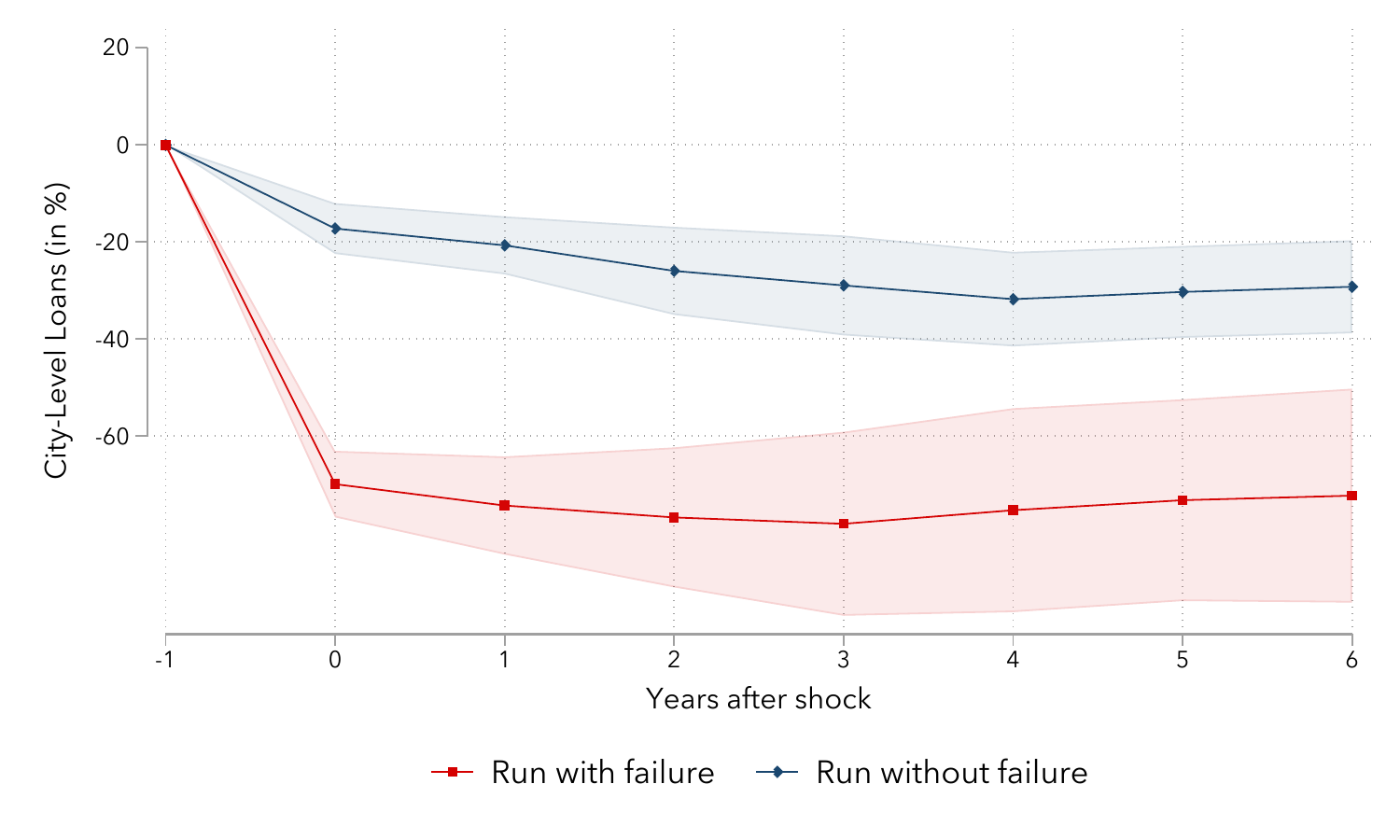}}}

\begin{minipage}{\textwidth}
\footnotesize
Notes: This figure is analogous to \Cref{fig:city_level_dynamics}, but it presents results from unweighted regressions, rather than weighting by the number of banks in a city. The figure is based on estimation of \Cref{LP1} (panels a and b), \Cref{LP_nf} (panels c and d), and \Cref{LP2} (panels e and f). The analysis is based on a city-level annual panel. Loans, deposits, and bank distress are based on aggregating the sample of national banks from 1863-1934 to the city level at an annual frequency. Error bands represent 95\% confidence intervals based on \citet{Driscoll1998} standard errors with a horizon-dependent bandwidth of $L_h = \lceil 1.5h \rceil$.
\end{minipage}
\end{figure}

\clearpage

\subsection{Additional Tables}

\begin{table}[!ht]
\tiny
\caption{\textbf{Narrative Banking Crisis Classifications and Bank Distress Events} } \label{fig:Jalil_granular}   \centering
\tiny
       \begin{minipage}{1.0\textwidth}
        \begin{center}
        {\begin{tabular}{l*{3}{c}}   \toprule  Dependent variable & Run & Failure & Suspension \\   \cmidrule(lr){2-2} \cmidrule(lr){3-3}  \cmidrule(lr){4-4}
            &\multicolumn{1}{c}{(1)}         &\multicolumn{1}{c}{(2)}         &\multicolumn{1}{c}{(3)}         \\
\hline
Major banking crisis year&        0.21\sym{***}&        0.38\sym{***}&        0.69\sym{***}\\
            &     (0.032)         &     (0.072)         &      (0.10)         \\
Major banking panic month&        0.73\sym{**} &       -0.34\sym{***}&        0.81\sym{*}  \\
            &      (0.34)         &     (0.084)         &      (0.41)         \\
May 1884 (NYC, PA, NJ)&        15.0\sym{*}  &        1.82\sym{***}&        9.00\sym{**} \\
            &      (8.05)         &      (0.45)         &      (4.28)         \\
Nov. 1890 (New York City, NY)&        3.99\sym{***}&       -0.18\sym{***}&        1.33\sym{***}\\
            &      (0.36)         &     (0.061)         &      (0.43)         \\
Dec. 1896 (IL, MN, WI)&        5.27\sym{***}&        2.53\sym{***}&        6.84\sym{***}\\
            &      (1.92)         &      (0.27)         &      (1.71)         \\
Dec. 1899 (Boston, MA; NYC, NY)&        1.44\sym{***}&        3.87\sym{***}&        5.34\sym{***}\\
            &      (0.36)         &      (1.43)         &      (0.36)         \\
Jun.-Jul. 1901 (NY: Buffalo; NYC)&        0.44\sym{***}&        1.87\sym{***}&        3.84\sym{***}\\
            &    (0.0080)         &     (0.015)         &     (0.018)         \\
Oct. 1903 (PA, MD)&        0.94         &       -0.13\sym{***}&        0.84\sym{***}\\
            &      (0.71)         &     (0.015)         &     (0.018)         \\
Dec. 1905 (Chicago, IL)&        1.94\sym{***}&       -0.13\sym{***}&        2.84\sym{***}\\
            &    (0.0080)         &     (0.015)         &     (0.018)         \\
Jan. 1908 (NYC)&        1.94\sym{***}&        3.87\sym{***}&        7.84\sym{***}\\
            &    (0.0080)         &     (0.015)         &     (0.018)         \\
Aug.-Sep. 1920 (Boston)&        3.94\sym{***}&        0.37\sym{***}&        2.84\sym{***}\\
            &    (0.0080)         &     (0.015)         &     (0.018)         \\
Nov. 1920 - Feb. 1921 (ND)&        0.94\sym{***}&        2.87\sym{***}&        7.34\sym{***}\\
            &    (0.0080)         &     (0.015)         &     (0.018)         \\
July 1926 (FL, GA)&        7.44\sym{*}  &        1.87\sym{**} &        23.8\sym{***}\\
            &      (3.93)         &      (0.71)         &      (4.29)         \\
March 1927 (FL)&        6.94\sym{***}&        0.87\sym{***}&        9.84\sym{***}\\
            &    (0.0080)         &     (0.015)         &     (0.018)         \\
Jul.-Aug. 1929 (FL)&        9.44\sym{***}&        5.87\sym{***}&        16.3\sym{***}\\
            &    (0.0080)         &     (0.015)         &     (0.018)         \\
Nov. 1930 (Nashville, TN)&        8.03\sym{***}&        3.58\sym{***}&        9.76\sym{***}\\
            &      (0.72)         &      (0.54)         &      (1.87)         \\
Nov. 1930 - Jan. 1931 (Fed districts: STL, CH, RIC, ATL, MIN)&        1.97\sym{***}&        2.24\sym{***}&        5.57\sym{***}\\
            &      (0.70)         &      (0.56)         &      (1.76)         \\
Apr. - July 1931 (Fed districts: CHI, MIN, CLE, KC)&        0.19         &        1.43\sym{**} &        0.68\sym{**} \\
            &      (0.15)         &      (0.55)         &      (0.26)         \\
June 1931 (Chicago, IL)&        26.5\sym{***}&        10.1\sym{***}&        40.5\sym{***}\\
            &      (0.16)         &      (0.60)         &      (0.30)         \\
Aug. - Oct. 1931 (Toledo, OH)&        2.94\sym{***}&       0.092         &        3.35\sym{***}\\
            &      (0.44)         &      (0.65)         &      (0.86)         \\
Sep. - Oct. 1931 (Fed districts: CHI, CLE, PHI)&        2.58\sym{***}&        2.79\sym{**} &        5.58\sym{***}\\
            &      (0.87)         &      (1.23)         &      (1.65)         \\
Sep. - Oct. 1931 (Pittsburgh/Philadelphia, PA; Chicago, IL)&        5.90\sym{***}&        8.20\sym{***}&        16.6\sym{***}\\
            &      (1.83)         &      (1.65)         &      (4.60)         \\
Oct. 1931 (West Virginia, WV)&        0.72\sym{***}&        6.49\sym{***}&        8.14\sym{***}\\
            &     (0.038)         &     (0.083)         &      (0.12)         \\
Oct. 1931 (Ohio, OH)&        0.21         &       -0.39         &        1.21         \\
            &      (0.44)         &      (0.62)         &      (0.83)         \\
June 1932 (Chicago, IL)&        14.7\sym{***}&        13.5\sym{***}&        22.1\sym{***}\\
            &     (0.038)         &     (0.083)         &      (0.12)         \\
Aug. 1932 (Idaho, ID)&        0.72\sym{***}&        4.49\sym{***}&        5.14\sym{***}\\
            &     (0.038)         &     (0.083)         &      (0.12)         \\
Nov. 1932 (Nevada, NV)&        0.72\sym{***}&       -0.51\sym{***}&        11.1\sym{***}\\
            &     (0.038)         &     (0.083)         &      (0.12)         \\
March 1933 (Detroit, MI)&       -0.28\sym{***}&        6.49\sym{***}&        90.1\sym{***}\\
            &     (0.038)         &     (0.083)         &      (0.12)         \\
\hline
\(N\)       &       41609         &       41609         &       41609         \\
Mean dep. var.&        0.10         &        0.19         &        0.28         \\
\hline\hline
\end{tabular}
}

   \end{center}
    {\footnotesize Notes: This table shows results from estimating \Cref{eq:jalil} using a monthly state-level panel with the number of distress events as the outcome variable. Major banking crisis years and major banking panic months are taken from \citet{Baron2021}, while regional banking crises are taken from \citet{Jalil2015} (1884--1929) and \citet{Wicker1996} (1930-1933). The sample covers events from 1863 through 1934. Standard errors clustered by state are reported in parentheses.   *,**, and *** indicate significance at the 10\%, 5\%, and 1\% level, respectively. }
        \end{minipage}
\end{table}

\begin{table}[!ht]
\caption{\textbf{Predictability of Bank Runs Based on Fundamentals, Runs on Other Banks, and Macroeconomic Variables} } \label{tab:pred_run_run}
  \centering
\scriptsize
       \begin{minipage}{1.0\textwidth}
        \begin{center}
        \begin{tabular}{l*{6}{c}}
        \toprule
        Dependent variable & \multicolumn{6}{c}{Run in $t+1$} \\ \cmidrule(lr){2-7}
                            &         (1)   &         (2)   &         (3)   &         (4)   &         (5)   &         (6)   \\
\cmidrule(lr){1-7} \textbf{\textit{Bank fundamentals:}}&               &               &               &               &               &               \\
\cmidrule(lr){1-7} Surplus/Equity&       -0.30** &       -0.32** &       -0.36** &       -0.45***&       -0.51***&       -0.48** \\
                    &      (0.13)   &      (0.16)   &      (0.16)   &      (0.14)   &      (0.19)   &      (0.19)   \\
Noncore funding     &       11.06***&       11.53***&       10.24***&       10.36***&       10.23***&       10.87***\\
                    &      (1.78)   &      (1.77)   &      (1.83)   &      (1.89)   &      (1.83)   &      (1.90)   \\
Surplus/Equity $\times$ Noncore funding&      -11.05***&      -12.19***&      -11.40***&      -11.09***&      -11.75***&      -11.94***\\
                    &      (2.60)   &      (2.36)   &      (2.41)   &      (2.48)   &      (2.35)   &      (2.35)   \\
Liquid Assets/Assets&       -0.65***&       -0.11   &       -0.74** &       -0.64** &       -0.67*  &       -0.25   \\
                    &      (0.24)   &      (0.19)   &      (0.36)   &      (0.28)   &      (0.34)   &      (0.19)   \\
Deposits/Assets     &        0.74*  &        0.92***&        0.43   &        0.83*  &        0.67*  &        0.83** \\
                    &      (0.40)   &      (0.32)   &      (0.34)   &      (0.46)   &      (0.34)   &      (0.34)   \\
Asset Growth (3 years)&       -0.15   &       -0.00   &       -0.01   &       -0.13   &       -0.05   &       -0.05   \\
                    &      (0.14)   &      (0.08)   &      (0.08)   &      (0.11)   &      (0.09)   &      (0.09)   \\
\cmidrule(lr){1-7} \textbf{\textit{Macro conditions:}}&               &               &               &               &               &               \\
\cmidrule(lr){1-7} NB failure rate&               &               &        0.11   &               &       -0.29   &               \\
                    &               &               &      (4.33)   &               &      (3.73)   &               \\
NB run rate         &               &               &       -4.76   &               &       -5.30   &               \\
                    &               &               &     (17.12)   &               &     (14.63)   &               \\
Stock market return &               &               &       -0.74***&               &       -0.65***&               \\
                    &               &               &      (0.26)   &               &      (0.22)   &               \\
Real GDP growth     &               &               &        0.46   &               &        0.33   &               \\
                    &               &               &      (0.76)   &               &      (0.72)   &               \\
\cmidrule(lr){1-7} \textbf{\textit{Local conditions:}}&               &               &               &               &               &               \\
\cmidrule(lr){1-7}  Run on other bank in same city&               &               &               &        3.49***&        3.45***&        3.19***\\
                    &               &               &               &      (0.97)   &      (0.95)   &      (0.86)   \\
Local business failure rate&               &               &               &        9.99***&        1.95   &        2.91   \\
                    &               &               &               &      (2.55)   &      (6.28)   &      (7.06)   \\
\cmidrule(lr){1-7} N&      290437   &      290437   &      246960   &      243908   &      243908   &      243908   \\
Mean dep. var       &         .39   &         .39   &         .38   &         .38   &         .38   &         .38   \\
Year FE             &               &  \checkmark   &               &               &               &  \checkmark   \\

        AUC                 &       0.692&       0.817&       0.749&       0.788&       0.799&       0.840 \\
 \bottomrule

        \end{tabular}
   \end{center}
    {\footnotesize Notes: This table presents estimates of \Cref{eq:run_failure_fundamentals}. \citet{Driscoll1998} standard errors in parentheses with a bandwidth of three years to allow for residual correlation within and across banks in proximate years.   *,**, and *** indicate significance at the 10\%, 5\%, and 1\% level, respectively. }
        \end{minipage}
\end{table}

\begin{table}[!ht]
\caption{\textbf{Predictability of Bank Runs without Failure Based on Fundamentals, Runs on Other Banks, and Macroeconomic Variables} } \label{tab:pred_run_run_no_fail}
  \centering
\scriptsize
       \begin{minipage}{1.0\textwidth}
        \begin{center}
        \begin{tabular}{l*{6}{c}}
        \toprule
        Dependent variable & \multicolumn{6}{c}{Run without failure in $t+1$} \\ \cmidrule(lr){2-7}
                            &         (1)   &         (2)   &         (3)   &         (4)   &         (5)   &         (6)   \\
\cmidrule(lr){1-7} \textbf{\textit{Bank fundamentals:}}&               &               &               &               &               &               \\
\cmidrule(lr){1-7} Surplus/Equity&       -0.06   &       -0.05   &       -0.08   &       -0.18** &       -0.19** &       -0.16** \\
                    &      (0.08)   &      (0.08)   &      (0.07)   &      (0.08)   &      (0.08)   &      (0.08)   \\
Noncore funding     &        1.55*  &        2.04***&        1.00   &        1.04   &        0.98   &        1.32*  \\
                    &      (0.78)   &      (0.76)   &      (0.78)   &      (0.77)   &      (0.77)   &      (0.67)   \\
Surplus/Equity $\times$ Noncore funding&        0.34   &       -0.24   &        0.33   &        0.42   &        0.13   &        0.02   \\
                    &      (1.65)   &      (1.53)   &      (1.57)   &      (1.55)   &      (1.51)   &      (1.48)   \\
Liquid Assets/Assets&       -0.40** &        0.15*  &       -0.41   &       -0.34   &       -0.35   &        0.04   \\
                    &      (0.20)   &      (0.08)   &      (0.28)   &      (0.22)   &      (0.26)   &      (0.06)   \\
Deposits/Assets     &        0.22   &        0.52** &        0.08   &        0.33   &        0.26   &        0.42** \\
                    &      (0.19)   &      (0.21)   &      (0.17)   &      (0.21)   &      (0.16)   &      (0.20)   \\
Asset Growth (3 years)&        0.03   &        0.11***&        0.09*  &        0.02   &        0.06   &        0.08** \\
                    &      (0.07)   &      (0.04)   &      (0.05)   &      (0.05)   &      (0.04)   &      (0.04)   \\
\cmidrule(lr){1-7} \textbf{\textit{Macro conditions:}}&               &               &               &               &               &               \\
\cmidrule(lr){1-7} NB failure rate&               &               &        0.83   &               &        0.54   &               \\
                    &               &               &      (2.50)   &               &      (2.05)   &               \\
NB run rate         &               &               &       -0.77   &               &       -1.44   &               \\
                    &               &               &      (9.55)   &               &      (7.79)   &               \\
Stock market return &               &               &       -0.37** &               &       -0.31** &               \\
                    &               &               &      (0.15)   &               &      (0.12)   &               \\
Real GDP growth     &               &               &        0.49   &               &        0.38   &               \\
                    &               &               &      (0.61)   &               &      (0.55)   &               \\
\cmidrule(lr){1-7} \textbf{\textit{Local conditions:}}&               &               &               &               &               &               \\
\cmidrule(lr){1-7}  Run on other bank in same city&               &               &               &        2.61***&        2.59***&        2.39***\\
                    &               &               &               &      (0.76)   &      (0.76)   &      (0.66)   \\
Local business failure rate&               &               &               &        4.17***&        0.81   &        2.03   \\
                    &               &               &               &      (1.13)   &      (3.24)   &      (3.93)   \\
\cmidrule(lr){1-7} N&      290437   &      290437   &      246960   &      243908   &      243908   &      243908   \\
Mean dep. var       &         .24   &         .24   &         .22   &         .22   &         .22   &         .22   \\
Year FE             &               &  \checkmark   &               &               &               &  \checkmark   \\

        AUC                 &       0.617&       0.798&       0.687&       0.734&       0.755&       0.818\\
 \bottomrule
        \end{tabular}
   \end{center}
    {\footnotesize Notes: This table presents estimates of \Cref{eq:run_failure_fundamentals}. \citet{Driscoll1998} standard errors in parentheses with a bandwidth of three years to allow for residual correlation within and across banks in proximate years.  *,**, and *** indicate significance at the 10\%, 5\%, and 1\% level, respectively. }
        \end{minipage}
\end{table}

\begin{table}[!ht]
\caption{\textbf{Predictability of Failure with Bank Run Based on Fundamentals, Runs on Other Banks, and Macroeconomic Variables} } \label{tab:pred_fail_w_run}
  \centering
\scriptsize
       \begin{minipage}{1.0\textwidth}
        \begin{center}
        \begin{tabular}{l*{6}{c}}
        \toprule
        Dependent variable & \multicolumn{6}{c}{Failure with run in $t+1$} \\ \cmidrule(lr){2-7}
                            &         (1)   &         (2)   &         (3)   &         (4)   &         (5)   &         (6)   \\
\cmidrule(lr){1-7} \textbf{\textit{Bank fundamentals:}}&               &               &               &               &               &               \\
\cmidrule(lr){1-7} Surplus/Equity&       -0.25***&       -0.28** &       -0.28** &       -0.28***&       -0.32** &       -0.32** \\
                    &      (0.09)   &      (0.11)   &      (0.12)   &      (0.10)   &      (0.13)   &      (0.14)   \\
Noncore funding     &        9.97***&        9.93***&        9.72***&        9.70***&        9.65***&        9.94***\\
                    &      (1.55)   &      (1.54)   &      (1.62)   &      (1.70)   &      (1.66)   &      (1.72)   \\
Surplus/Equity $\times$ Noncore funding&      -12.12***&      -12.67***&      -12.50***&      -12.12***&      -12.47***&      -12.54***\\
                    &      (1.90)   &      (1.84)   &      (1.90)   &      (1.91)   &      (1.89)   &      (1.91)   \\
Liquid Assets/Assets&       -0.24** &       -0.25*  &       -0.32** &       -0.30***&       -0.30** &       -0.27*  \\
                    &      (0.09)   &      (0.14)   &      (0.13)   &      (0.11)   &      (0.12)   &      (0.16)   \\
Deposits/Assets     &        0.51** &        0.39***&        0.35*  &        0.48*  &        0.41** &        0.41** \\
                    &      (0.23)   &      (0.14)   &      (0.18)   &      (0.25)   &      (0.19)   &      (0.16)   \\
Asset Growth (3 years)&       -0.20*  &       -0.12   &       -0.11   &       -0.15   &       -0.12   &       -0.14   \\
                    &      (0.10)   &      (0.07)   &      (0.08)   &      (0.10)   &      (0.08)   &      (0.09)   \\
\cmidrule(lr){1-7} \textbf{\textit{Macro conditions:}}&               &               &               &               &               &               \\
\cmidrule(lr){1-7} NB failure rate&               &               &       -0.98   &               &       -1.10   &               \\
                    &               &               &      (2.06)   &               &      (1.89)   &               \\
NB run rate         &               &               &       -1.92   &               &       -1.68   &               \\
                    &               &               &      (7.75)   &               &      (6.95)   &               \\
Stock market return &               &               &       -0.37** &               &       -0.34** &               \\
                    &               &               &      (0.14)   &               &      (0.13)   &               \\
Real GDP growth     &               &               &       -0.01   &               &       -0.03   &               \\
                    &               &               &      (0.28)   &               &      (0.28)   &               \\
\cmidrule(lr){1-7} \textbf{\textit{Local conditions:}}&               &               &               &               &               &               \\
\cmidrule(lr){1-7}  Run on other bank in same city&               &               &               &        0.91***&        0.89***&        0.83***\\
                    &               &               &               &      (0.33)   &      (0.31)   &      (0.31)   \\
Local business failure rate&               &               &               &        6.31***&        1.65   &        1.39   \\
                    &               &               &               &      (1.80)   &      (3.74)   &      (3.91)   \\
\cmidrule(lr){1-7} N&      290437   &      290437   &      246960   &      243908   &      243908   &      243908   \\
Mean dep. var       &         .15   &         .15   &         .16   &         .16   &         .16   &         .16   \\
Year FE             &               &  \checkmark   &               &               &               &  \checkmark   \\

        AUC                 &       0.838&       0.878&       0.865&       0.878&       0.883&       0.895\\
 \bottomrule
        \end{tabular}
   \end{center}
    {\footnotesize Notes: This table presents estimates of \Cref{eq:run_failure_fundamentals}. Driscoll-Kraay standard errors are reported in parentheses.   *,**, and *** indicate significance at the 10\%, 5\%, and 1\% level, respectively. }
        \end{minipage}
\end{table}

\begin{table}[!ht]
\caption{\textbf{Predictability of Failure Based on Fundamentals, Runs on Other Banks, and Macroeconomic Variables} } \label{tab:pred_fail}
  \centering
\scriptsize
       \begin{minipage}{1.0\textwidth}
        \begin{center}
        \begin{tabular}{l*{6}{c}}
        \toprule
        Dependent variable & \multicolumn{6}{c}{Failure in $t+1$} \\ \cmidrule(lr){2-7}
                            &         (1)   &         (2)   &         (3)   &         (4)   &         (5)   &         (6)   \\
\cmidrule(lr){1-7} \textbf{\textit{Bank fundamentals:}}&               &               &               &               &               &               \\
\cmidrule(lr){1-7} Surplus/Equity&       -1.60***&       -1.81***&       -1.85** &       -1.65***&       -1.94** &       -1.95** \\
                    &      (0.55)   &      (0.64)   &      (0.69)   &      (0.58)   &      (0.76)   &      (0.73)   \\
Noncore funding     &       65.95***&       63.46***&       66.73***&       67.34***&       66.76***&       67.39***\\
                    &     (15.99)   &     (14.67)   &     (16.16)   &     (17.32)   &     (16.58)   &     (16.30)   \\
Surplus/Equity $\times$ Noncore funding&      -92.17***&      -96.25***&     -101.92***&      -98.46***&     -102.06***&     -101.66***\\
                    &     (18.71)   &     (20.35)   &     (21.86)   &     (21.61)   &     (22.58)   &     (22.87)   \\
Liquid Assets/Assets&       -0.35   &       -2.02***&       -1.55** &       -0.96*  &       -1.54** &       -2.10***\\
                    &      (0.48)   &      (0.72)   &      (0.65)   &      (0.52)   &      (0.65)   &      (0.76)   \\
Deposits/Assets     &        3.08** &        1.45** &        1.62*  &        2.67** &        1.77** &        1.55** \\
                    &      (1.24)   &      (0.64)   &      (0.84)   &      (1.24)   &      (0.86)   &      (0.75)   \\
Asset Growth (3 years)&       -2.15** &       -1.44***&       -1.33** &       -1.81** &       -1.34** &       -1.56** \\
                    &      (0.84)   &      (0.54)   &      (0.60)   &      (0.78)   &      (0.62)   &      (0.63)   \\
\cmidrule(lr){1-7} \textbf{\textit{Macro conditions:}}&               &               &               &               &               &               \\
\cmidrule(lr){1-7} NB failure rate&               &               &       10.43   &               &       10.14   &               \\
                    &               &               &     (11.54)   &               &     (11.06)   &               \\
NB run rate         &               &               &      -53.31*  &               &      -52.69   &               \\
                    &               &               &     (30.91)   &               &     (31.55)   &               \\
Stock market return &               &               &       -2.05***&               &       -2.00***&               \\
                    &               &               &      (0.58)   &               &      (0.57)   &               \\
Real GDP growth     &               &               &       -4.41*  &               &       -4.41*  &               \\
                    &               &               &      (2.38)   &               &      (2.42)   &               \\
\cmidrule(lr){1-7} \textbf{\textit{Local conditions:}}&               &               &               &               &               &               \\
\cmidrule(lr){1-7}  Run on other bank in same city&               &               &               &        2.43***&        2.34***&        2.18***\\
                    &               &               &               &      (0.74)   &      (0.65)   &      (0.63)   \\
Local business failure rate&               &               &               &       47.23***&        6.51   &        1.35   \\
                    &               &               &               &      (7.84)   &     (16.01)   &     (15.30)   \\
\cmidrule(lr){1-7} N&      290437   &      290437   &      246960   &      243908   &      243908   &      243908   \\
Mean dep. var       &          .8   &          .8   &          .9   &         .91   &         .91   &         .91   \\
Year FE             &               &  \checkmark   &               &               &               &  \checkmark   \\

        AUC                 &       0.874&       0.903&       0.897&       0.892&       0.904&       0.911\\
 \bottomrule
        \end{tabular}
   \end{center}
    {\footnotesize Notes: This table presents estimates of \Cref{eq:run_failure_fundamentals}. \citet{Driscoll1998} standard errors in parentheses with a bandwidth of three years to allow for residual correlation within and across banks in proximate years.   *,**, and *** indicate significance at the 10\%, 5\%, and 1\% level, respectively. }
        \end{minipage}
\end{table}

\begin{table}[!ht]
\caption{\textbf{Bank Runs, Failures, and Fundamentals: Detailed Specifications} } \label{tab:fail_run_fundamentals_granular}
  \centering
\scriptsize
       \begin{minipage}{1.0\textwidth}
        \begin{center}
        {\begin{tabular}{l*{8}{c}}   \toprule  Dependent variable & \multicolumn{8}{c}{Failure in t} \\  \cmidrule(lr){2-9}
                &\multicolumn{1}{c}{(1)}         &\multicolumn{1}{c}{(2)}         &\multicolumn{1}{c}{(3)}         &\multicolumn{1}{c}{(4)}         &\multicolumn{1}{c}{(5)}         &\multicolumn{1}{c}{(6)}         &\multicolumn{1}{c}{(7)}         &\multicolumn{1}{c}{(8)}         \\
\midrule
Run             &     0.38\sym{***}&     0.54\sym{***}&     0.27\sym{***}&     0.25\sym{***}&     0.39\sym{***}&     0.24\sym{***}&     0.36\sym{***}&     0.37\sym{***}\\
                &  (0.051)         &  (0.079)         &  (0.043)         &  (0.045)         &  (0.087)         &  (0.058)         &  (0.077)         &  (0.076)         \\
Surplus/Equity  &                  &   -0.023\sym{***}&                  &                  &                  &  -0.0091\sym{***}&   -0.025\sym{***}&   -0.032\sym{**} \\
                &                  & (0.0082)         &                  &                  &                  & (0.0027)         & (0.0092)         &  (0.012)         \\
Surplus/Equity $\times$ Run&                  &    -0.48\sym{***}&                  &                  &                  &    -0.27\sym{**} &    -0.39\sym{***}&    -0.40\sym{***}\\
                &                  &   (0.12)         &                  &                  &                  &   (0.11)         &  (0.090)         &  (0.089)         \\
Noncore funding &                  &                  &     0.31\sym{***}&                  &                  &     0.15\sym{***}&     0.31\sym{***}&     0.27\sym{***}\\
                &                  &                  &  (0.099)         &                  &                  &  (0.035)         &   (0.10)         &  (0.075)         \\
Noncore funding$\times$ Run&                  &                  &     1.89\sym{***}&                  &                  &     2.94\sym{***}&     1.77\sym{***}&     1.76\sym{***}\\
                &                  &                  &   (0.31)         &                  &                  &   (0.54)         &   (0.32)         &   (0.31)         \\
OREO/Loans      &                  &                  &                  &    0.099\sym{***}&                  &    0.074\sym{***}&                  &                  \\
                &                  &                  &                  &  (0.032)         &                  &  (0.020)         &                  &                  \\
OREO/Loans$\times$ Run&                  &                  &                  &     1.93\sym{***}&                  &     1.06\sym{**} &                  &                  \\
                &                  &                  &                  &   (0.39)         &                  &   (0.41)         &                  &                  \\
Liquid/Assets   &                  &                  &                  &                  &  0.00033         & -0.00023         &    0.013         &   -0.022\sym{***}\\
                &                  &                  &                  &                  & (0.0061)         & (0.0014)         & (0.0081)         & (0.0076)         \\
Liquid/Assets $\times$ Run&                  &                  &                  &                  &   -0.011         &    0.029         &     0.15         &     0.11         \\
                &                  &                  &                  &                  &   (0.16)         &   (0.13)         &   (0.11)         &   (0.11)         \\
\midrule
Observations    &   272535         &   272535         &   272535         &    53906         &   272535         &    53905         &   272535         &   272535         \\
$R^2$           &    0.069         &    0.076         &     0.11         &     0.12         &    0.069         &     0.17         &     0.11         &     0.13         \\
State FE        &                  &                  &                  &                  &                  &\checkmark         &                  &\checkmark         \\
Year FE         &                  &                  &                  &                  &                  &\checkmark         &                  &\checkmark         \\
Sample          &1865-1934         &1865-1934         &1865-1934         &1865-1904         &1865-1934         &1865-1904         &1865-1934         &1865-1934         \\
\bottomrule
\end{tabular}
}

   \end{center}
    {\footnotesize Notes:  This table presents estimates of \Cref{eq:failure_run}. \citet{Driscoll1998} standard errors in parentheses with a bandwidth of three years to allow for residual correlation within and across banks in proximate years.   *,**, and *** indicate significance at the 10\%, 5\%, and 1\% level, respectively. }
        \end{minipage}
\end{table}

\begin{table}[!ht]
\caption{\textbf{Bank Runs, Failures, and Fundamentals: Different Time Periods}} \label{tab:fail_run_fundamentals_by_era}
  \centering
\scriptsize
       \begin{minipage}{1.0\textwidth}
        \begin{center}
        {\begin{tabular}{l*{6}{c}}   \toprule  Dependent variable & \multicolumn{6}{c}{Failure in t} \\  \cmidrule(lr){2-7}
                &\multicolumn{1}{c}{(1)}         &\multicolumn{1}{c}{(2)}         &\multicolumn{1}{c}{(3)}         &\multicolumn{1}{c}{(4)}         &\multicolumn{1}{c}{(5)}         &\multicolumn{1}{c}{(6)}         \\
\midrule
Run             &     0.37\sym{***}&     0.29\sym{***}&     0.95\sym{***}&     0.79\sym{***}&     0.41\sym{***}&     0.28         \\
                &  (0.076)         &  (0.059)         &   (0.12)         &  (0.069)         &  (0.078)         &   (0.25)         \\
Surplus/Equity  &   -0.032\sym{**} &  -0.0070\sym{***}&   -0.024\sym{***}&    -0.14\sym{***}&   -0.016\sym{***}&   -0.083\sym{*}  \\
                &  (0.012)         & (0.0016)         & (0.0075)         &  (0.023)         & (0.0053)         &  (0.036)         \\
Surplus/Equity $\times$ Run&    -0.40\sym{***}&    -0.35\sym{***}&    -0.81\sym{***}&    -0.37\sym{*}  &    -0.50\sym{***}&    -0.11         \\
                &  (0.089)         &  (0.097)         &   (0.16)         &   (0.15)         &  (0.096)         &   (0.24)         \\
Noncore funding &     0.27\sym{***}&     0.10\sym{***}&     0.21\sym{***}&     0.71\sym{**} &     0.19\sym{***}&     0.84\sym{***}\\
                &  (0.075)         &  (0.027)         &  (0.052)         &   (0.16)         &  (0.044)         &   (0.14)         \\
Noncore funding$\times$ Run&     1.76\sym{***}&     2.17\sym{***}&     0.62\sym{**} &     0.70         &     1.74\sym{***}&     1.63\sym{**} \\
                &   (0.31)         &   (0.70)         &   (0.26)         &   (0.39)         &   (0.38)         &   (0.60)         \\
Liquid/Assets   &   -0.022\sym{***}&  -0.0024\sym{*}  &   -0.012\sym{***}&   -0.051\sym{**} &   0.0035         &    0.054\sym{**} \\
                & (0.0076)         & (0.0014)         & (0.0027)         &  (0.015)         & (0.0027)         &  (0.019)         \\
Liquid/Assets $\times$ Run&     0.11         &    0.085         &    -0.82\sym{***}&    -0.59\sym{***}&    0.091         &     0.13         \\
                &   (0.11)         &   (0.13)         &   (0.18)         &  (0.081)         &   (0.12)         &   (0.30)         \\
\midrule
Observations    &   272535         &   137335         &   109611         &    25589         &   239031         &    33504         \\
$R^2$           &     0.13         &     0.13         &     0.13         &     0.16         &     0.13         &     0.12         \\
State FE        &\checkmark         &\checkmark         &\checkmark         &\checkmark         &                  &                  \\
Year FE         &\checkmark         &\checkmark         &\checkmark         &\checkmark         &                  &                  \\
Sample          &1865-1934         &1865-1913         &1914-1928         &1929-1934         &No Crisis         &Banking Crises         \\
\bottomrule
\end{tabular}
}

   \end{center}
    {\footnotesize Notes:   This table presents estimates of \Cref{eq:failure_run}. \citet{Driscoll1998} standard errors in parentheses with a bandwidth of three years to allow for residual correlation within and across banks in proximate years.   *,**, and *** indicate significance at the 10\%, 5\%, and 1\% level, respectively. }
        \end{minipage}
\end{table}

\begin{table}[!ht]
\caption{\textbf{Role of Clearinghouses in the Pass-Through of Runs to Failure} } \label{tab:fail_run_Clearinghouses}
  \centering
\footnotesize
       \begin{minipage}{1.0\textwidth}
        \begin{center}
        {\begin{tabular}{l*{4}{c}}   \toprule  Dependent variable & \multicolumn{4}{c}{Failure in t+1} \\  \cmidrule(lr){2-5}
                &\multicolumn{1}{c}{(1)}         &\multicolumn{1}{c}{(2)}         &\multicolumn{1}{c}{(3)}         &\multicolumn{1}{c}{(4)}         \\
\midrule
Run             &     0.43\sym{***}&     0.30\sym{***}&     0.42\sym{***}&     0.44\sym{***}\\
                &  (0.050)         &  (0.038)         &  (0.044)         &  (0.098)         \\
Clearinghouse City&  -0.0025\sym{**} &  0.00045         &  -0.0020\sym{*}  &  -0.0082\sym{*}  \\
                & (0.0011)         &(0.00035)         & (0.0011)         & (0.0041)         \\
Clearinghouse City $\times$ Run&    -0.14\sym{***}&   -0.098\sym{***}&    -0.17\sym{***}&   -0.092\sym{**} \\
                &  (0.035)         &  (0.036)         &  (0.050)         &  (0.036)         \\
\midrule
Observations    &   282082         &   137335         &   243762         &    38320         \\
$R^2$           &    0.071         &    0.097         &    0.092         &    0.048         \\
Mean dep. var.  &   0.0082         &   0.0030         &   0.0048         &    0.029         \\
Sample          &                  &    <1914         &No Crisis         &Banking Crisis         \\
\bottomrule
\end{tabular}
}

   \end{center}
    {\footnotesize Notes: This table presents estimates of \Cref{eq:failure_run}. \citet{Driscoll1998} standard errors in parentheses with a bandwidth of three years to allow for residual correlation within and across banks in proximate years.   *,**, and *** indicate significance at the 10\%, 5\%, and 1\% level, respectively. }
        \end{minipage}
\end{table}

\clearpage
\renewcommand{\currentappendix}{B}
\setcounter{page}{1}
\section{Model}
\label{appendix:model}

This appendix provides further details on the theoretical framework in \Cref{sec:theory}. There are two dates $t\in \{1, 2\}$. The bank holds liquid cash $C$ and illiquid loans $L$. These assets are financed by deposits $D$ and equity $E$. Deposits are demandable at $t=1$, with $D>C$. Deposits pay zero interest, but depositors who stay until $t=2$ enjoy a convenience benefit of $c$ per unit of deposits.\footnote{The convenience benefit provides a simple way to model a benefit of staying with the bank, without introducing interest payments. We assume that depositors only obtain the convenience benefit if they do not attempt to withdraw in $t=1$ and if the bank pays in full at $t=2$.} There is a continuum of risk-neutral depositors who each hold one unit of deposits. In $t=2$, the bank's loan portfolio pays off $\theta L$. The gross return $\theta$ is the fundamental of the economy.

We start by assuming that, at time $t=1$, all depositors observe the realization of $\theta$ and decide whether to keep their deposits in the bank or withdraw. We also assume that if total withdrawals $w$ exceed cash, the bank suspends and is placed into receivership at $t=1$. In receivership, assets are worth $(1-\rho) \theta L$, with $\rho \in (0,1)$ capturing the value loss of receivership. We assume sequential service. Depositors who withdraw before receivership get their full deposit, while depositors remaining in the bank after suspension get a pro-rata share of the recovery, $R(\theta) = \frac{(1-\rho)\theta L}{D-C}$.\footnote{The depositor recovery is capped at unity.} We further assume that the order of depositor arrival at the bank is randomly determined.

\subsection{Fundamental Insolvency}
The bank is fundamentally insolvent when loans plus cash are insufficient to pay all deposits:
$$C + \theta L <D.$$
This implies a fundamental solvency threshold of the fundamental:
$$\theta^S \equiv \frac{D-C}{L}.$$
For $\theta < \theta^S$, the bank is fundamentally insolvent. In this region, withdrawing is a dominant strategy.

\subsection{Complete Information: Non-Fundamental Panic Runs}

Because of the value destruction of receivership, a self-fulfilling run can cause failure when $\theta$ is below
$$ \theta^L = \frac{D-C}{(1-\rho)L} > \theta^S.$$
Thus, there is an intermediate region, $[\theta^S, \theta^L)$, where there are two equilibria. In the first equilibrium, no depositors run. In the second equilibrium, all depositors run, and the self-fulfilling run leads to suspension and failure.  Above this region, when $\theta \geq \theta^L$, the recovery rate is at least one, so staying is a (weakly) dominant strategy.

\subsection{Incomplete Information With Private Signals: Fundamental-Based Panic Runs}

Following the global games framework \citep{MorrisShin2000,Goldstein2005}, assume each depositor $i$ observes a noisy private signal of the fundamental,
$$x_i = \theta + \sigma \varepsilon_i,$$
where $\varepsilon_i$ is i.i.d.\ across depositors with a continuous density, and $\sigma > 0$ parameterizes the noise. We sketch the equilibrium in the limit as $\sigma \to 0$. In this case, there is a unique threshold $\theta^* \in (\theta^S,\theta^L)$. Below this threshold, a fundamental-based panic run occurs with probability one, and the bank fails. Above the threshold, there is no run, and the bank survives.

If a fraction $\ell$ of depositors withdraws and $\ell D > C$, each withdrawer is paid in full with probability $C/(\ell D)$, as withdrawing depositors are served in random order. Unserved withdrawers are pooled with the remaining depositors in receivership. The remaining claims, $D-C$, obtain a per unit recovery of $R(\theta) = \frac{(1-\rho)\theta L}{D - C}.$

As the signal noise goes to zero ($\sigma \to 0$), the threshold signal $x^*$ converges to a threshold fundamental, $\theta^*$, which is pinned down by the indifference of the marginal depositor. Assume the marginal depositor holds the belief that the fraction of withdrawing depositors $\ell$ is uniformly distributed on $[0,1]$. Let $\hat{\ell} \equiv C/D$ denote the withdrawal fraction that exhausts the bank's cash. We conjecture and later verify that $\theta^* \in (\theta^S, \theta^L)$. The payoff difference between staying and withdrawing, per unit of deposits, is
$$\Delta(\ell, \theta) =
\begin{cases}
c & \text{if } \ell \leq \hat{\ell}, \\[4pt]
-\dfrac{C}{\ell D}\bigl(1 - R(\theta)\bigr) & \text{if } \ell > \hat{\ell}.
\end{cases}$$
For $\ell \leq \hat{\ell}$, the bank satisfies all withdrawals from cash and, since $\theta > \theta^S$, pays remaining depositors in full at $t=2$. Thus, not withdrawing yields $1 + c$ compared to $1$ from withdrawing. For $\ell > \hat{\ell}$, the bank suspends. Not withdrawing yields $R(\theta)$, while withdrawing yields $1$ with probability $C/(\ell D)$ and $R(\theta)$ otherwise.

The indifference condition $\int_0^1 \Delta(\ell, \theta^*)\, d\ell = 0$ is thus
$$c\,\hat{\ell} = \bigl(1 - R(\theta^*)\bigr) \int_{\hat{\ell}}^{1} \frac{C}{\ell D}\, d\ell = \bigl(1 - R(\theta^*)\bigr)\, \hat{\ell}\, \ln\frac{D}{C},$$
where the last step uses that $\hat \ell=C/D$. Simplifying, we have $$1 - R(\theta^*) = c / \ln(D/C),$$ so the run threshold is
\begin{equation}
\theta^* = \frac{D - C}{(1-\rho)L}\left(1 - \frac{c}{\ln(D/C)}\right) = \theta^L \left(1 - \frac{c}{\ln(D/C)}\right).
\label{eq:theta_star}
\end{equation}
For $c > 0$, we have $\theta^* < \theta^L$. The conjecture $\theta^* > \theta^S$ holds if
\begin{equation*}
c < \rho \ln\frac{D}{C}.
\end{equation*}

The key insight is that banks with $\theta \in [\theta^S, \theta^*)$ are fundamentally solvent but fail due to a run. Banks with $\theta \in [\theta^*, \theta^L)$ are vulnerable to a self-fulfilling run in the complete-information benchmark but survive under global games selection.

The threshold $\theta^*$ in \eqref{eq:theta_star} has intuitive comparative statics. It is decreasing in the convenience benefit $c$, with $\theta^* \to \theta^L$ as $c \to 0$. It is increasing in the receivership haircut $\rho$, as greater value destruction in receivership widens the region in which coordination failure can bring down a solvent bank. The threshold is also increasing in leverage (higher $D$) and illiquidity (lower $C$).

\subsection{Incomplete Information With Public Signals: Information-Based Runs}

Next, we consider a variant of the model in which, instead of private signals, all depositors observe the same noisy public signal of the fundamental,
$$y = \theta + \epsilon,$$
where $\epsilon$ is an independent shock with mean zero and continuous cumulative distribution function $F_\epsilon$. Because the signal is public, all depositors share the same posterior beliefs, so there is again equilibrium multiplicity, as in the complete-information benchmark.

As above, if total withdrawals exceed cash, the bank suspends in $t=1$. We assume that the receiver liquidates the bank only if liquidation proceeds are insufficient to cover remaining deposits, $(1-\rho) \theta L < D-C$, or equivalently if $\theta<\theta^L$. If $\theta \geq \theta^L$, the receiver allows the bank to reopen and continue with its loan portfolio intact.

As above, we assume that, if there is a full run, each depositor is paid in full with probability $C/D$. For a depositor who runs and is not served before suspension, we assume that they receive the pro-rata recovery $R(\theta) = (1-\rho)\theta L/(D-C)$ if the bank is placed into receivership. If the bank is reopened, the depositor receives 1, but loses the convenience benefit $c$. On the other hand, a depositor who does not run receives $1 + c$ if the bank is reopened and $R(\theta)$ if it is liquidated.

We first characterize the threshold signal $\overline{y}$, below which a run is an equilibrium. Suppose all other depositors withdraw. The payoff difference between staying and withdrawing, per unit of deposits, is $c$ if $\theta \geq \theta^L$ and $-\frac{C}{D}(1 - R(\theta))$ if $\theta < \theta^L$. A run is an equilibrium if and only if withdrawing is a best response, which holds for signals below a threshold $\overline{y}$ defined by the indifference condition
\begin{equation}
c \, \Pr\!\left(\theta \geq \theta^L \mid y = \overline{y}\right)
= \frac{C}{D}\, \E\!\left[\bigl(1 - R(\theta)\bigr) \mathbf{1}[\theta < \theta^L] \,\middle|\, y = \overline{y}\right].
\label{eq:ybar0}
\end{equation}
The left-hand side is the expected convenience benefit from staying, which is the convenience benefit of having deposits and not running times the probability that the bank survives, conditional on observing $y=\overline{y}$. The right-hand side is the expected loss from staying into liquidation, which the withdrawer avoids only when served before cash is exhausted. Under a diffuse (improper) prior, and using $R(\theta) = \theta / \theta^L$, condition \eqref{eq:ybar0} is
\begin{equation}
c \, F_\epsilon\!\left(\overline{y} - \theta^L\right)
= \frac{C}{D} \int_{\overline{y} - \theta^L}^{\infty} \left(1 - \frac{\overline{y} - e}{\theta^L}\right) dF_\epsilon(e).
\label{eq:ybar_explicit0}
\end{equation}
For well-behaved $F_\epsilon$, the left-hand side of \eqref{eq:ybar_explicit0} is strictly increasing in $\overline{y}$, while the right-hand side is decreasing in $\overline{y}$. Thus, $\overline{y}$ exists and is unique. A run is an equilibrium if and only if $y \leq \overline{y}$.

If the signal is sufficiently bad, a run is the unique equilibrium. Suppose all other depositors stay. The bank then meets any individual withdrawal from cash and fails only through fundamental insolvency at $t=2$. For $\theta \geq \theta^S$, a staying depositor receives $1 + c$, while for $\theta < \theta^S$ the bank's assets $C + \theta L < D$ are divided pro rata at $t=2$, yielding $(C + \theta L)/D$ per unit of deposits. A depositor who withdraws receives $1$ in either case. Staying is no longer a best response even when other depositors stay (i.e., no-run is no longer an equilibrium), when the signal is below a threshold $\underline{y}$ defined by
\begin{equation}
c \, \Pr\!\left(\theta \geq \theta^S \mid y = \underline{y}\right)
= \E\!\left[\left(1 - \frac{C + \theta L}{D}\right) \mathbf{1}[\theta < \theta^S] \,\middle|\, y = \underline{y}\right].
\label{eq:ylow}
\end{equation}
Under parametric assumptions, we have the natural case where $\underline{y} < \overline{y}$.\footnote{This is satisfied if not running is (weakly) more attractive when others stay than when others run (strategic complementarity); that is, if the stay-minus-run payoff difference conditional on others staying is weakly greater than the stay-minus-run payoff difference conditional on others running pointwise for all $\theta$. A condition that yields this is $C/D\geq \frac{1}{2-\rho}$.}

The public signal thus partitions outcomes into three regions, similar to the complete-information benchmark. For $y > \overline{y}$, a run is no longer an equilibrium. For $y \in [\underline{y}, \overline{y}]$, both the run and no-run equilibria exist. For $y < \underline{y}$, withdrawing is strictly dominant and a run is the unique equilibrium.

The key insight of the noisy public signal case is that, because the run condition depends on the signal $y$ rather than the fundamental $\theta$, runs can occur on fundamentally strong banks in response to negative signals that are \textit{ex post} wrong. That is, there are \textit{runs without failure}. A bank with $\theta \geq \theta^L$ that draws a sufficiently negative signal can suffer a run. Under the receivership convention above, such a bank suspends and is subsequently reopened. However, again, runs can cause failure of solvent banks in the intermediate region $\theta \in [\theta^S, \theta^L)$. These banks are liquidated in receivership despite being fundamentally sound, as in the complete-information benchmark.

\subsection{Suspension, Examination, and Reopening}

Consider the public signal version of the model, as in the previous subsection. However, now suppose that if total withdrawals exceed cash, the bank, in conjunction with the receiver, can choose one of two actions after suspension. It can enter receivership, where assets are liquidated at $(1-\rho)\theta L$, as above. Alternatively, it can pay a proportional examination cost $\tau \theta L$ to have its books examined and certify that it is solvent. If assets net of the examination cost cover remaining deposits, $(1-\tau)\theta L \geq D - C$, the bank is allowed to reopen and continue to $t=2$.

We assume the bank observes the true $\theta$. The bank therefore suspends and chooses an examination whenever
$$\theta \geq \theta^{Susp} \equiv \frac{D-C}{(1-\tau)L}.$$ Otherwise, it enters receivership. Assuming that the cost of examination is lower than that of failure, $\tau < \rho$, examination is preferred to receivership whenever $\theta \geq \theta^{Susp}$. For $\theta < \theta^{Susp}$, examination and certification would fail, so the bank enters receivership. Note that $\tau \in (0, \rho)$ implies
$$\theta^{Susp} \in \left(\theta^S, \theta^L\right).$$

Suppose all other depositors withdraw. A run is an equilibrium if withdrawing is a best response, which holds for signals below a threshold $\overline{y}^\tau$ defined by an indifference condition similar to that in the previous section:
\begin{equation}
c \, \Pr\!\left(\theta \geq \theta^{Susp} \mid y = \overline{y}^\tau \right)
= \frac{C}{D}\, \E\!\left[\bigl(1 - R(\theta)\bigr) \mathbf{1}[\theta < \theta^{Susp}] \,\middle|\, y = \overline{y}^\tau \right].
\label{eq:ybar}
\end{equation}
Under the diffuse prior, and using $R(\theta) = \theta / \theta^L$, condition \eqref{eq:ybar} becomes
\begin{equation}
c \, F_\epsilon\!\left(\overline{y}^\tau - \theta^{Susp}\right)
= \frac{C}{D} \int_{\overline{y}^\tau - \theta^{Susp}}^{\infty} \left(1 - \frac{\overline{y}^\tau - e}{\theta^L}\right) dF_\epsilon(e).
\label{eq:ybar_explicit}
\end{equation}

A run is an equilibrium if and only if $y \leq \overline{y}^\tau$. Since $\theta^{Susp} < \theta^L$, the survival region is larger than in an economy without the suspension option, so $\overline{y}^\tau$ is lower than $\overline{y}$. The key result here is thus that the availability of suspension, examination, and certification (rather than fire selling or immediate receivership) shrinks the set of signals for which a run is an equilibrium. The lower the examination cost $\tau$, the lower is the cutoff $\overline{y}^\tau$.

Again, runs can occur on fundamentally strong banks in response to overly negative public signals. If true fundamentals are relatively strong, $\theta \geq \theta^{Susp}$, then, focusing on the run equilibrium, a bank that draws a signal $y \leq \overline{y}^\tau$ is subject to a run. The bank suspends and certifies solvency. The run is still costly because the bank pays the examination cost and depositors forgo liquidity during the suspension, but it is not fatal. Runs can cause failure of solvent banks only in the region $\theta \in [\theta^S, \theta^{Susp})$. A mechanism such as suspension thus reduces the set of solvent banks that can be pushed into failure by a run through two margins. First, conditional on a run, banks with $\theta \in [\theta^{Susp}, \theta^L)$ survive that would otherwise have been liquidated. Second, because survival is more likely, the run threshold $\overline{y}^\tau$ itself falls, so runs only occur for relatively worse signals.

\clearpage
\renewcommand{\currentappendix}{C}
\setcounter{page}{1}
\section{Constructing the Bank Distress Episodes Dataset}
\label{appendix:data}

\begin{table}[ht!]
\centering\small
\caption{ \textbf{Selected Keywords and Rules Used to Identify Articles Possibly Related to Bank Distress}}
\label{tab:keywords}
\begin{threeparttable}
\begin{tabular}{lp{4cm}p{6cm}}
\toprule
\textbf{Detection method} & \textbf{Required keywords} & \textbf{Notes} \\
\midrule
\multicolumn{3}{l}{\textit{Primary methods: keyword combinations}} \\
\midrule
Bank Run & \texttt{bank} \textbf{+} \texttt{run} & Exclude non-financial uses of "run" such as "trains are running". \\
\\
Bank suspension & \texttt{bank} \textbf{+} \texttt{suspen} (e.g., suspend, suspension) & Exclude non-financial suspensions, such as "suspension of production" or "rules were suspended." \\
\\
Bank receivership & \texttt{bank} \textbf{+} \texttt{receiver} or \texttt{assignee} or \texttt{assigned} & Private and state banks often used \textit{assignees} rather than \textit{receivers}. \\
\\
Bank panic & \texttt{bank} or \texttt{deposit} \textbf{+} \texttt{panic} &  \\
\\
Large withdrawals & \texttt{deposit} \textbf{+} \texttt{large} \textbf{+} \texttt{withdraw} & Catches articles describing significant withdrawals without explicitly using the word "run." \\
\midrule
\multicolumn{3}{l}{\textit{Secondary methods: high-confidence phrases}} \\
\midrule
Specific distress phrases & A specific phrase is present alongside the word \texttt{bank}. & Strong indicators of distress with low risk of false positives. The seventeen keywords include: \newline
\texttt{"heavy run"}, \texttt{"financial stringency"}, \texttt{"temporary embarrassment"}, \texttt{"heavy withdrawals"}, etc. \\
Suspension rules & 30/60/90-day notices + \texttt{deposit} & Allow for spelling variants: ``thirty days'', ``60 days'' \\
\bottomrule
\end{tabular}
\begin{tablenotes}[flushleft]\footnotesize\item
\textit{Notes:}
Matching is case-insensitive.
To avoid false positives, unrelated words are first removed from the text before searching. These include geographical features (e.g., ``river bank'', ``snow bank''), proper names (e.g., ``Albert Banks''), and common but unrelated financial terms (e.g., ``bank loans,'' ``banknote'').
Any rule involving the keyword \texttt{bank} is also repeated for the keyword \texttt{trust co}.
\end{tablenotes}
\end{threeparttable}
\end{table}

\begin{figure}[ht!]
    \centering
    \caption{\textbf{Total Number of Newspaper Articles from \textit{Chronicling America}}}
    \label{fig:ca-total-articles}
    \includegraphics[width=0.8\linewidth]{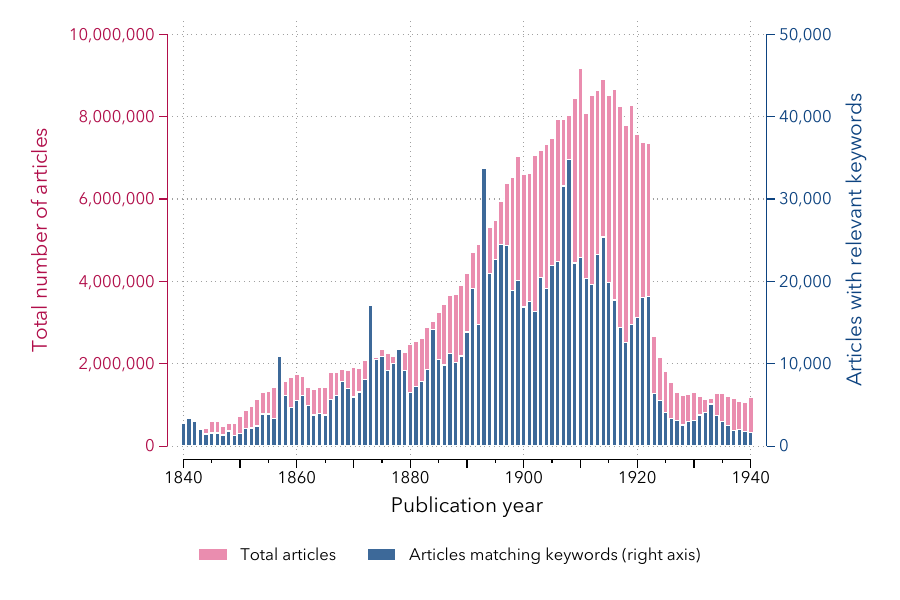}

    \begin{minipage}{\textwidth}
    \footnotesize
    Notes: The left axis depicts total articles available per year in ``Chronicling America'', while the right axis shows the number of articles with keywords related to bank distress events (see Appendix \Cref{tab:keywords}). The scale of the left axis is 200 times larger than that of the right axis. On average across the sample, 0.3\% of all articles have matching keywords.
    To identify the articles available in each newspaper page, we use the ``American Stories'' dataset \citep{AmericanStories}, which segments all pages into articles and obtains the text of each article using a simple but scalable OCR.
    \end{minipage}
\end{figure}

\begin{figure}[ht!]
    \centering
    \caption{\textbf{Number of Newspaper Articles Used in Final Sample}}
    \label{fig:ca-selected-articles}
    \includegraphics[width=0.8\linewidth]{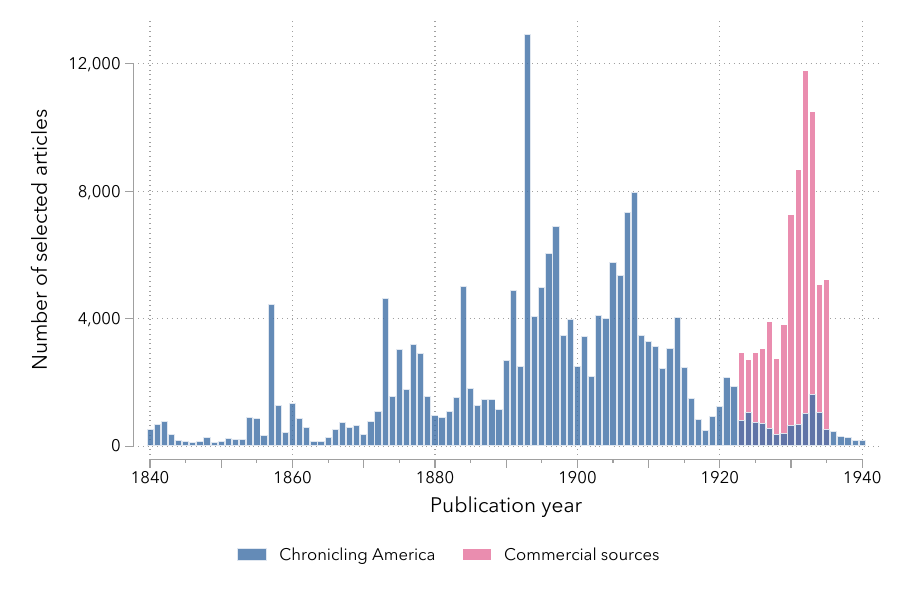}

    \begin{minipage}{\textwidth}
    \footnotesize
    Notes: This figure shows the number of newspaper articles per year included in our final sample of newspaper articles. The two requirements for inclusion are i) having at least one keyword related to bank distress events (see \Cref{tab:keywords}) and ii) being flagged by an LLM as related to a bank distress event (see \Cref{appendix:prompts}).
    This last step minimizes the number of articles about foreign bank runs, hypothetical bank runs, or unrelated uses of the keywords---for instance, ``runs to the bank of the river''---that appear in our sample. We rely on ``Chronicling America'' for most of the sample period, and complement it with commercial sources for 1923-1934.
    \end{minipage}
\end{figure}

\begin{figure}[ht!]
    \caption{\textbf{Methodology for Constructing the Bank Distress Episodes Database}}
    \centering
    \footnotesize

    \begin{subfigure}[b]{0.8\textwidth}
        \centering
        \begin{tcolorbox}[colframe=gray!30, colback=white, boxrule=1pt, arc=5pt]
            \resizebox{\linewidth}{!}{\begin{tikzpicture}[
  block/.style        ={rectangle, draw, thick, text width=2cm, align=center, minimum height=1cm, rounded corners=2mm},
  process/.style      ={rectangle, draw, thick, text width=2cm, align=center, minimum height=1cm, rounded corners=2mm},
  result/.style       ={rectangle, draw, thick, text width=2cm, align=center, minimum height=1cm, rounded corners=2mm, fill=black!5},
  arrow/.style        ={-Latex, thick, shorten >=1mm, shorten <=1mm},
  dashed_arrow/.style ={-Latex, thick, dashed, shorten >=1mm, shorten <=1mm},
  note/.style         ={font=\small},
  number_note/.style  ={font=\small, text=blue!60!black},   page_count/.style  ={font=\small, text=ForestGreen!60!black} ]

\node (CA) [result] {Scanned \\pages };

\node (AS) [block, below=10mm of CA] {Layout, \\ quick OCR};

\node (Keywords) [block, right=15mm of AS] {Keyword search};

\node (QuickLLM) [block, right=15mm of Keywords] {Quick \\  LLM};

\node (CropLLM) [process, above=10mm of QuickLLM] {Better OCR, LLM};

\node (DownCrop) [process, above=10mm of Keywords] {Download,\\Crop};

\node (DTA) [result, right=18mm of CropLLM] {314K article-events};

\draw[dashed_arrow] (CA.east |- DownCrop.west) -- (DownCrop.west)
    ;
\draw[dashed_arrow] (DownCrop.east) -- (CropLLM.west);

\draw[arrow] (CA.south) -- (AS.north)
    node[midway, left, number_note] {17M}
    ;

\draw[arrow] (AS.east) -- node[midway, above, number_note] {372M} (Keywords.west);
\draw[arrow] (Keywords.east) -- node[midway, above, number_note] {1.5M} (QuickLLM.west);

\draw[arrow] (QuickLLM.north) -- node[midway, right, number_note] {801K} (CropLLM.south);

\draw[arrow] (CropLLM.east) -- node[midway, above, number_note] {269K} (DTA.west);

\end{tikzpicture}}
        \end{tcolorbox}
        \caption{Identify events about bank distress}
        \label{fig:diagram1}
    \end{subfigure}

    \vspace{8mm}

    \begin{subfigure}[b]{0.8\textwidth}
        \centering
        \begin{tcolorbox}[colframe=gray!30, colback=white, boxrule=1pt, arc=5pt]
            \resizebox{\linewidth}{!}{\begin{tikzpicture}[
  box/.style        ={rectangle, draw, thick, text width=2.2cm, align=center, minimum height=1.2cm, rounded corners=2mm},
  result/.style     ={rectangle, draw, thick, text width=2.2cm, align=center, minimum height=1.2cm, rounded corners=2mm, fill=black!5},
  arrow/.style      ={-Latex, thick, shorten >=1mm, shorten <=1mm},
  note/.style       ={font=\small, anchor=west},
  number_note/.style={font=\small, text=blue!60!black}
]

\node (AE)  [result] {Article-events};
\node (VCN) [box, right=18mm of AE] {Validate\\state, city};
\node (VBN) [box, right=18mm of VCN] {Validate\\bank names};

\node (GBE)  [box, below=10mm of AE] {Group by\\bank-time};
\node (LLMR) [box, below=10mm of VCN] {LLM\\Ensemble};
\node (DS)   [result, below=10mm of VBN] {Dataset: \\ 20k episodes};

\draw[arrow] (AE.east) -- node[midway, above, number_note] {314K} (VCN.west);
\draw[arrow] (VCN.east) -- node[midway, above, number_note] {313K} (VBN.west);

\draw[arrow] (GBE.east) -- node[midway, above, number_note] {114K} (LLMR.west);
\draw[arrow] (LLMR.east) -- node[midway, above, number_note] {114K} (DS.west);

\draw[arrow] (VBN.east) -- ++(10mm,0)
  node[pos=0.5, above, number_note]{234K};
\draw[arrow] ($(GBE.west)+(-10mm,0)$) -- (GBE.west);

\end{tikzpicture}}
        \end{tcolorbox}
        \caption{Combine article-level events into bank distress episodes}
        \label{fig:diagram3}
    \end{subfigure}

\begin{minipage}{\textwidth}
\footnotesize
Notes: This figure summarizes the steps involved in constructing our main bank distress episodes dataset. The numbers in blue indicate how many newspaper articles---or page scans, for the first arrow---carry forward from one step to the next. \\

Panel (a) describes how we construct our intermediate dataset, where every row corresponds to a bank-level event from a given newspaper article.
    To do so, we first extract newspaper articles from each page, then we identify the articles related to bank-distress events, and lastly we obtain structured information from each article about specific events involving banks, such as runs, suspensions, failures, and reopenings. We obtain an intermediate dataset composed of 269 thousand articles discussing 314 thousand events. \\

    Panel (b) then shows how this intermediate dataset is transformed into our final dataset at the bank-episode level.
    This is done by first standardizing state, city, and bank names. We then group the articles by bank and time into episodes, and select for the next step up to 25 articles for each episode.
    This constraint is due to cost and context-window limitations, and for episodes with more than 25 articles we select them by sampling across time as well as ranking them by article quality and relevance.
    Lastly, we combine all the articles into a query and feed it to an ensemble of three LLMs for detailed episode-level information (\texttt{gpt-5-mini}, \texttt{gemini-3-flash-preview}, and \texttt{deepseek-v4-pro}). If the models disagree in their classification of the episode, we apply a majority vote with priority-based tie breaking in the order listed above. We obtain a final dataset of 19,963 episodes, with each episode containing on average 1.9 specific events. The dataset covers the years 1800-1963. Our main analysis focuses on the 1863-1934 period, which covers the National Banking Era and the pre-deposit insurance Federal Reserve Era.
\end{minipage}

    \label{fig:diagram}
\end{figure}

\begin{table}[htbp]
    \centering
    \caption{\textbf{Number of Newspaper Articles per Episode} }
    \label{tab:num_articles_per_episode}
    \footnotesize
    \begin{tabular}{lrrrrrrrr}
\toprule
Sample & N & Mean & SD & Min & P5 & Median & P95 & Max \\
\midrule
\multicolumn{9}{c}{\textbf{Panel A: All banks}} \\
\midrule
Run & 3,984 & 29.12 & 75.22 & 1 & 1 & 7 & 126 & 2205 \\
Run without failure & 2,074 & 20.17 & 75.75 & 1 & 1 & 4 & 85 & 2205 \\
Run with failure & 1,910 & 38.85 & 73.43 & 1 & 2 & 13 & 151 & 966 \\
Failure & 10,341 & 15.09 & 40.99 & 1 & 1 & 3 & 64 & 966 \\
\midrule
\multicolumn{9}{c}{\textbf{Panel B: National banks}} \\
\midrule
Run & 1,280 & 35.46 & 71.90 & 1 & 1 & 9 & 160 & 692 \\
Run without failure & 672 & 22.25 & 53.06 & 1 & 1 & 4 & 113 & 573 \\
Run with failure & 608 & 50.06 & 85.88 & 1 & 2 & 18 & 199 & 692 \\
Failure & 2,650 & 23.13 & 54.58 & 1 & 1 & 6 & 103 & 692 \\
\bottomrule
\end{tabular}

\begin{minipage}{\textwidth}
    \footnotesize
    Notes: This table reports statistics on the number of newspaper articles per episode for all bank events (panel A) and national bank events (panel B).

    \end{minipage}

\end{table}

\clearpage
\subsection{LLM Prompts Used in the Data-Constructing Process}
\label{appendix:prompts}

\begin{lstlisting}[
  caption={\textbf{Quick LLM. }Prompt used to discard obvious false positives, such as articles discussing foreign banks, hypotheticals, or other unrelated content.},
  label={prompt1},
  language={}
]
Below is the text of an old historical newspaper article that contains many OCR errors.

Return True if the article could be describing any of the following events in relation to a U.S. bank: a bank run, bank panic, bank failure, bank closing, bank reopening, deposit suspension, panic among depositors, or any similar banking crisis or issue. Otherwise, return False.

Notes:

- Again, if the article discusses foreign banks (banks located outside the United States) you must return False.
- If you believe the article might involve one of those issues listed above for U.S. banks (bank runs, etc.), return True, and False if not.
- If you are unsure, return True.

Text:

{article_text}
\end{lstlisting}

\begin{lstlisting}[
  caption={\textbf{Intermediate dataset. }Prompt used to extract structured information about each newspaper article.},
  label={prompt2},
  language={}
]
You are an expert research assistant specializing in historical U.S. banking events. Analyze the provided digitized newspaper article, keeping in mind possible OCR errors.

# Task

Identify if the article references any of these historical U.S. banking events:

- Bank run
- Bank suspension
- Bank reopening/resumption
- Bank receivership

Only consider actual events involving U.S. banks or trust companies. Exclude:

- Foreign banks (e.g., Canadian, English, Australian)
- Hypothetical, fictional, or purely general mentions of panic without concrete events

Historical synonyms to recognize include variations of:

- Runs: "run on the bank", "heavy withdrawals", "steady withdrawals", "temporary embarrassment", "invocation of the 30-day rule", "undue excitement", "drain upon cash"
- Suspensions: "failure to open", "temporarily closed"
- Reopenings: "resumption", "resumed operations"
- Receiverships: "assigned a receiver"

# Response format

{{
  "is_match": boolean,
  "events": [
    {{
      "event_type": "run" | "suspension" | "reopening" | "receivership" | "other",
      "bank_type": "national" | "state" | "private" | "unknown",
      "date": "YYYY-MM-DD" (ISO 8601, or partial like "1909-12-00", or null),
      "bank_name": string,
      "state": two-letter state code or null,
      "city": string or null,
      "cause": string describing reason(s) or null,
      "snippet": direct article excerpt or null,
      "measures": measures to address liquidity (for runs) or null
    }}
  ]
}}

# Important Guidelines

- If the article does not describe any relevant U.S. banking event, set `is_match` to False and return an empty list for events.
- Represent each distinct bank and event separately; do not combine multiple banks or events. If a given event lists multiple banks, treat it as multiple events. If the article discusses multiple events, such as a bank and subsequent closure, treat them as multiple events and list them separately.
- Handle unclear or missing information due to OCR errors:
  - Provide your best interpretation or set fields as null if uncertain.
  - **Do not invent or infer unsupported details.**
- Use provided metadata (publication date, location) to resolve date/location ambiguities:
  - For partial dates, infer missing year from the newspaper date.
  - Do not assume event location matches the newspaper location unless explicitly stated.
  - For instance, if the text mentions "the Bank of Smyrna", it should refer to Smyrna, Tennessee, if the context suggests a U.S. location (instead of Smyrna, Turkey, which is unlikely to be discussed in a U.S. newspaper from Tennessee).
- Remember: Articles might reference multiple events in various locations. List them separately.

2. Output rules
   - Match found: If the article describes one or more relevant U.S. banking events, set `is_match` to True and return a list of event objects in the events field. Do not merge multiple events into a single object. Each event must be represented as its own object in the list.  

---

Newspaper metadata:

- Name: {scan.newspaper}
- Location: {scan.city}, {scan.state}
- Date published (ISO 8601 formatted date: YYYY-MM-DD): {scan.date}

---

ARTICLE:  

```
{article_text}
```

Use the article text above to populate your response.
\end{lstlisting}

\begin{lstlisting}[
  caption={\textbf{Final dataset---Main prompt. }Prompt used to process newspaper articles related to a specific bank and distress episode, and produce a description of the overall distress episode, as well as the different events that composed it (suspensions, runs, failures, or reopenings).},
  label={prompt3},
  language={}
]
You are an expert in historical U.S. banking events. You must analyze the included list of digitized newspaper articles that discuss one or multiple events such as bank runs, bank suspensions, etc., and then classify the overall episode into one of the provided types.

# Output fields.
 
Your JSON response must follow this structure:

- `success` (boolean): True if the articles describe events that affect the specified bank. This includes (a) events that name the bank directly, and (b) broader events that demonstrably apply to the bank by jurisdiction - for example, a state-wide banking holiday in this bank's state, a city-wide panic in this bank's city, or the March 1933 federal banking holiday. Set to False only if the articles are unrelated (about a clearly different bank with no connection to the named bank's jurisdiction), or involve bank events in foreign countries outside the U.S. When False, still fill in `episode_type` (use "other"), `notes` (explain why), and any other fields you can reasonably infer.
- `bank_explicitly_mentioned` (boolean): True if at least one article names the specified bank (or an obvious OCR variant of it) directly. False if `success` is True only because the articles describe a broader event (state/city/federal-level) that affects this bank's jurisdiction.
- `confidence` (string): Your overall confidence in this classification. Must be one of: "high" (clear evidence, no ambiguity), "medium" (some ambiguity but a defensible answer), "low" (significant uncertainty, conflicting signals, or thin evidence).
- `more_data_recommended` (boolean): True if you believe additional newspaper articles about this bank would meaningfully improve the classification (e.g., the current articles are partial, only one side of an event is described, or key facts are ambiguous).
- `episode_type` (string): The type of event, as described below
- `bank_name` (string): The name of the bank. Use the provided bank name unless you detect a typo or other issue.
- `bank_name_unsure` (boolean): True if it is not clear which bank or city this event refers to.
- `bank_type` (string): The type of bank involved. Must be one of: "national", "state", "trust", "private", or "unknown". Use clues from the bank's name or the article to infer the type. For instance, names containing "National" indicate a national bank, "State" suggest a state bank, "Trust" indicate a trust company, etc. If uncertain, use unknown.
- `city` (string): The city where the bank is located. Use the provided city name unless you detect a typo or similar.
- `notes` (string): Very brief notes (1-2 short sentences MAX) on unusual nuances, ambiguities, or important caveats only. Do NOT summarize the articles, restate any structured field (episode_type, success, bank_name, etc.), or explain routine decisions. Leave empty if nothing unusual.
- `events` (array of objects): A list of events extracted from the article. Each event object must include the following fields:
  - `event_type` (string): The type of event. Must be one of: "run", "suspension", "reopening", "receivership", or "other".
  - `date` (string): An ISO 8601 formatted date (YYYY-MM-DD) string representing when the event took place, or `null` if not provided. You can return partial extractions such as the month by setting unknown fields to zero. For instance, 1909-12-00 for an unknown day in December 1909.
  - `cause` (string): The classification of why the event took place. Only applicable for bank runs and bank suspensions (see below).
  - `cause_details` (string): Short description of the immediate cause or trigger of the run or suspension, expanding on the classification listed above. Only applicable for bank runs and bank suspensions.
  - `random_run` (boolean): For `event_type="run"` only. Set True only if the article explicitly describes a specific, discrete misinformation/communication mistake that triggered withdrawals and indicates it was false, mistaken, or later corrected (e.g., a misprint, erroneous telegram/wire report, hoax/prank, false announcement that the bank failed/closed, mistaken identity with another bank, or a clearly described misunderstanding that is corrected in the text). Set False otherwise (including when the article merely says "rumors", "nervousness", or "no cause", without describing a concrete misinformation event). If the run is described as driven by (i) adverse bank-specific information (embezzlement, bad loans, insolvency, scandals), (ii) failures/distress of other banks, (iii) macro/systemic panic, or (iv) local economic shocks, then `random_run` must be False even if the article uses reassuring language. When True, the `random_run_snippet` must include the phrase(s) describing the misinformation/mistake and language indicating it was false/corrected.
  - `random_run_snippet` (string|null): 5-15 words summarizing the misinformation event (e.g., "false telegram reported bank had closed; corrected same day"). Only if `random_run` is True.
  - `measures` (string): Actions undertaken by the bank as a response to a run (if any; only applicable for bank runs).
  - `snippet` (string): A direct excerpt from an article that best demonstrates or confirms the event (including any relevant phrases or keywords), or `null` if not provided.

# Types of episode

There are nine possible episode types, based on the chronological order and the causality in which the events took place. You must write the episode type that you think took place in the `episode_type` field. To do so, follow this decision rule:

1. First determine whether a run occurred at any point.
2. Then determine whether a suspension occurred.
3. Then determine whether the bank reopened or permanently closed.
4. Use this order to select the episode type.

These are the episode types:

1. `run_only`: A bank run on this bank took place, and the bank remains open throughout the entire episode (does not suspend or fail)
2. `run_suspension_reopening`: A bank run takes place. As a consequence of the bank run, the bank suspends (either fully by suspending all payments or partially by invoking the 30/60/90 day rule). However, the bank eventually reopens and does not fail.
3. `run_suspension_closure`: A bank run takes place. As a consequence of the bank run, the bank suspends. The bank remains closed permanently and has either a receiver assigned or is taken over by another bank.
4. `run_suspension_unsure`: A bank run takes place. As a consequence of the bank run, the bank suspends. It is unsure if the bank remains closed permanently or if it reopens.
5. `suspension_closure`: A bank suspends. There is no bank run. The bank remains closed permanently and has either a receiver assigned or is taken over by another bank.
6. `suspension_reopening`: A bank suspends. There is no bank run. The bank is able to reopen successfully and continues to operate its business.
7. `suspension_run_closure`: A bank suspends. As a consequence of the suspension, depositors run and/or are agitated (i.e., they gather outside the bank) in an ill-fated attempt to regain access to their funds. The bank remains closed permanently and has either a receiver assigned or is taken over by another bank.
8. `suspension_run_reopening`: A bank suspends. As a consequence of the suspension, depositors run and/or are agitated (i.e., they gather outside the bank) in an ill-fated attempt to regain access to their funds. The bank is able to reopen successfully and continues to operate its business.
9. `other`: Other type of event or sequence. Explain in the `notes` field.


# Causes of bank runs and suspensions

1. `rumor_or_misinformation`: unfounded rumors or misinformation about the bank (false reports, misinterpreted events, malicious jokes, etc).
2. `bank_specific_adverse_info`: bank-specific adverse information affecting the solvency of the bank (embezzlement, suicides, discoveries of massive loans, or other scandals).
3. `local_banks`: information about runs and distress of other banks in the local economy
4. `local_shock`: information about the local economy that could affect the solvency of the bank (crop failures, default of local non-financial firms)
5. `correspondent`: failure or distress to correspondent banks or redemption agents.
6. `macro_news`: systemic financial panics, nationwide downturns, or other macroeconomic or political shocks (wars, currency crises, etc).
7. `government_action`: closed by government action (by receiver, examiner, banking holiday, etc). Only applicable for suspensions, not for bank runs.
8. `voluntary_liquidation`: voluntary liquidation and related events (charter expiration, etc). Only applicable for suspensions, not for bank runs.
9. `other`: other; unknown; or unable to tell.

# Guidelines and notes

You must follow these guidelines:

- If the articles are about a clearly different bank with no connection to the specified bank's jurisdiction, are irrelevant, or involve bank events in foreign countries outside the U.S., set `success` to False and explain in the `notes` field. You should still fill in the other fields as best you can. However, if the articles describe a jurisdiction-wide event (city-wide panic, state-wide banking holiday, federal banking holiday) that applies to the bank because of its location, set `success` to True, set `bank_explicitly_mentioned` to False, and capture the event(s) accordingly (for state/federal holidays, use `event_type="suspension"` with `cause="government_action"`).
- The article texts may contain significant OCR errors. Correct obvious OCR errors, especially in key fields such as bank names, dates, and locations. Document significant corrections in the notes field.
- **Do not invent or infer details** that are not reasonably supported by the text.
- Please classify the episode and the events based on what happened to the specific bank in question, including jurisdiction-wide events (e.g., a state banking holiday that applies to all banks in the state) that affect it. Do not classify based on events affecting unrelated banks discussed in the articles.
- If the event sequence matches more than one category or is ambiguous, select the most likely option and explain clearly in the `notes` field.
- Keep `notes` extremely brief (1-2 short sentences MAX). It is for unusual caveats only. Never summarize the articles or repeat structured fields. Most episodes should have an empty or near-empty `notes` field.
- If the text mentions "the Bank of Smyrna", consider that it likely refers to Smyrna, Tennessee, if the context suggests a U.S. location (instead of Smyrna, Turkey, which is unlikely to be discussed in a U.S. newspaper from Tennessee).
- The articles often talk about events in other cities, so you MUST NOT blindly use the newspaper location as the event location. You have made this mistake in the past so please avoid it!
- The `cause` field should select one of the causes of bank runs when `event_type` is a run, and one of the causes of bank suspensions when `event_type` is a suspension.
- Use any provided newspaper metadata (e.g., publication date) to help resolve ambiguities, such as when only a month and day are mentioned. For example, if the text says "January 10" without a year, you may derive the year from the newspaper date.
{receivership_guidance}
Note that historically these events might have been described using various terms, for example:

      - Bank runs: "run on the bank", "heavy withdrawals", "temporary embarrassment", "invocation of the 30 day rule", etc.
      - Bank suspensions: "failure to open", etc.
      - Bank reopenings: "resumption", "resumed operations", etc.
      - Bank receiverships: "assigned a receiver", etc.

Lastly, follow this final check before answering:

- Does `episode_type` match the events listed?
- Are event types and causes consistent?
- Does the city/bank match the actual event location?

# Bank information

- Bank: {bank}
- City: {city}
- State: {state}
{charter_date_info}{alt_name}{receivership_date_info}

# Articles

The articles are listed below:

{articles}
\end{lstlisting}

\begin{lstlisting}[
  caption={\textbf{Final dataset---Bank response to run. }Prompt used to identify the types of measures undertaken by the bank as a response to the bank run, to prevent failure},
  label={prompt4},
  language={}
]
You are an expert in historical U.S. banking events. Your task is to analyze historical newspaper articles describing bank runs, and indicate which types of measures the bank undertook in reaction to the bank run (to prevent the bank run from becoming a bank failure).

Below will be a list of newspaper articles discussing a bank distress episode that *should* include a bank run (plus other events such as suspensions, receiverships, etc.) Please answer the following questions regarding the run.

# Questions

## General questions

- `success` (boolean): True if the articles describe a bank run on the specified bank and you were able to identify response measures. Set to False if the articles are irrelevant or do not describe a run on this bank. When False, still fill in other fields as best you can.
- `no_run_described` (boolean): Set to true if you believe there was no actual run described in the articles.
- `clearinghouse_involved` (boolean): Indicate if a clearinghouse was somehow involved in the response. This can take place either through a loan, an examination, or other measures.
- `who_ran_type` (string): The main type of depositors who were part of the bank run (see below).

## Bank response measures

Indicate which of the following categories were undertaken by the bank. You can report no categories, one category, or many categories.

- `accommodation` (boolean): Accommodated withdrawals, paid out depositors as normal to prove solvency. Could also imply extending hours, opening early, hiring extra tellers, delivering money to depositors' homes, etc.
- `borrowing` (boolean): Borrowing from other banks or large institutions.
- `clearinghouse_borrowing` (boolean): Borrowing from a clearinghouse.
- `other_borrowing` (boolean): Borrowing from non-bank institutions, or from the Federal Reserve.
- `public_signal` (boolean): Send signals to display the strength of the bank and its available resources, to reassure depositors. This includes physical delivery of gold, cash, etc. in a very public way (express wagons, train, airplanes, armored trucks, etc.). Often described as "truckloads of cash", emphasizing the visual effect. It can also include posting advertisements about the strength of the bank's balance sheet, and paying depositors in a very public and visible way.
- `injection` (boolean): Equity injection or additional deposits by shareholders or other investors. Also includes mergers or takeovers, often involving additional capital.
- `partial_suspension` (boolean): Partial suspension of convertibility, either to some types of deposits (such as the invocation of the 30/60/90-day notice clauses for time deposits), or maximums per account (either in dollars or percentage).
- `full_suspension` (boolean): Full suspension of convertibility, where the bank stops paying all depositors, usually for a short period of time. Often described as the bank temporarily closing its doors, opening late, or closing early.
- `examination` (boolean): Bank books were examined by state supervisors, federal examiners (OCC, Federal Reserve), or by the local clearinghouse. The goal of this examination was to prove to depositors that the bank books were sound.
- `other` (string): Other types of measures; explain.


# Categories for `who_ran_type`

If an article mentions multiple depositors, choose the items in this list that appear first:

1. `large`: articles highlight a role for large depositors, businessmen, mercantile or commercial depositors, etc.
2. `institutional`: articles mention institutional depositors, such as county or city treasurers, trust funds, federal, universities, etc.
3. `minorities`: the articles highlight the role of immigrant groups, specific ethnicities (japanese, german, hungarian, italian, etc.), freedmen, or other demographic groups distinguishable from the general public.
4. `small`: the articles highlight a role for small depositors or retail depositors
5. `workers`: the articles highlight a role for specific professions such as farmers, miners, or factory workers.
6. `general`: the article just mention the run was driven by the general public or general depositors
7. `unspecified`: the articles only do not specify which type of depositors ran.


# Guidelines

- Only mark a category if the article describes the measure as a response to the bank run or depositor panic.
- If the article does not clearly describe a measure being undertaken, mark the category as false.
- You may mark zero, one, or multiple categories as true.
- Leave the `other` empty unless there's a specific type of measure that does not fit any of the listed categories.


# Bank information

- Bank: {bank}
- City: {city}
- State: {state}


# Articles

The articles are listed below:

{articles}
\end{lstlisting}

\begin{lstlisting}[
  caption={\textbf{Final dataset---Suspension-specific questions. }Prompt used to identify whether the suspension was caused by a local, state, or federal government action (bank holiday)},
  label={prompt5},
  language={}
]
You are an expert in historical U.S. banking events. Your task is to analyze historical newspaper articles describing bank suspensions, and determine whether the suspension was caused by a government-mandated banking holiday (i.e., the bank closed because a government authority ordered it to, rather than because the bank itself decided to suspend payments).

Below will be a list of newspaper articles discussing a bank distress episode that includes a suspension. Please answer the following questions regarding the suspension.

# Questions

- `success` (boolean): True if the articles describe a suspension affecting the specified bank - including when the suspension is due to a city-wide, state-wide, or federal banking holiday applicable to the bank's jurisdiction (even if the bank is not named individually) - and you were able to determine whether it was a government-mandated banking holiday. Set to False only if the articles are clearly unrelated to a suspension that could affect this bank. When False, still fill in the other fields as best you can.
- `is_banking_holiday` (boolean): True if the suspension was caused by a government decision (a government-mandated banking holiday), rather than the bank's own decision to suspend payments. False otherwise (e.g., the bank closed due to a run, due to insolvency, or otherwise on its own initiative).
- `holiday_type` (string or null): The level of government that ordered the holiday. Must be one of:
    - `municipal`: ordered by a city, county, or other local authority.
    - `state`: ordered by a state governor or state-level authority.
    - `federal`: ordered by the federal government (most notably the nationwide banking holiday declared by President Roosevelt in March 1933).
  Set to `null` if `is_banking_holiday` is False.

# Guidelines

- Set `is_banking_holiday` to True only if a government authority (mayor, governor, president, legislature, banking commissioner acting under a holiday declaration, etc.) ordered banks to close. The canonical examples are the March 1933 nationwide bank holiday declared by President Roosevelt, and the various state-level holidays that swept the country in late 1932 and early 1933 (e.g., Michigan in February 1933, Maryland, Ohio, etc.).
- If the bank simply chose to close on its own (e.g., in response to a run, due to insolvency, due to a clearinghouse decision, or as a voluntary suspension), then `is_banking_holiday` must be False.
- If the bank closed because of a run or because it was insolvent, but a banking holiday was *also* in effect at the time, judge whether the closure is best attributed to the holiday order or to the bank's own situation. If the articles indicate the bank would have closed regardless, set `is_banking_holiday` to False.
- For `holiday_type`, use the highest-level authority that explicitly ordered the closure as described in the articles. If a state holiday was ordered while a federal holiday was also in force, use `federal` if the federal order is what kept the bank closed.
- Do not infer a banking holiday merely because the suspension occurred in March 1933 or another holiday period; the articles should support the connection.


# Bank information

- Bank: {bank}
- City: {city}
- State: {state}


# Articles

The articles are listed below:

{articles}
\end{lstlisting}

\subsection{City Name Validation}
\label{appendix:validation:city}

In order to identify banks and track them across time---and across different newspaper articles---we first need to validate and standardize their location, as often bank names are only unique within a given city.

We start by validating banks' states and producing their two-digit FIPS code. This is quite straightforward except for former U.S. territories such as ``Colorado Territory,'' ``Indian Territory,'' ``Porto Rico,'' etc. In these cases, we try to assign the location to the successor state (e.g. ``Colorado Territory'' to ``Colorado''). When this is not possible (e.g. ``Dakota Territory'', which can be assigned to two different states) we try to search all successor states for a match.

Once the states are validated and standardized, we switch to the more complex task of validating cities\footnote{We mostly use only the word ``city'' but our data includes all populated areas, such as towns, villages, etc.}. Here we face three problems: first, the U.S. has way more cities and towns than states. Second, towns frequently changed names over time. Third, towns close to large cities could merge into the cities to become suburbs.

We address the first issue by collecting and combining several large datasets of historical U.S. cities:

\begin{enumerate}
    \item Geographic Names Information System (GNIS) Domestic Names dataset. After dropping locations other than cities and towns---ranging from ranches to national parks and tribal areas---we end up with 209,257 cities and towns.
    \item IPUMS NHGIS place point GIS files \citep{ipums}. These ``depict the locations of incorporated, unincorporated, and census-designated places for the entire U.S. from 1900 to 2015.'' After processing this data and removing certain outliers---such as towns with impossibly long names---we obtain 22,844 cities and towns.
    \item Wikipedia articles on U.S. populated areas. We downloaded the entire list of Wikipedia articles, selected those about current and past U.S. populated areas, and compiled a list of those areas including their location and population across time. This yields 25,431 cities and towns.
    \item Stanford's Center for Spatial and Textual Analysis \citep{CESTA2017}. This is a curated dataset of 8,848 cities and towns, mostly compiled from the U.S. census and state-level sources.
    \item Manually collected datasets. We further rely on the work of Jacob Alperin-Sheriff who manually digitized 20,844 cities from the U.S. decennial census, and of James Feigenbaum, who digitized a partly overlapping set of 2,442 cities also from the census.
\end{enumerate}

When combining these cities, we perform minimal standardizations on city names, such as replacing ``centre'' with ``center'' and ``borough'' with ``boro''. This minimizes the possibility of multiple spellings across the different datasets.

Lastly, to deal with city name changes across time as well as conversions into suburbs, we manually create a crosswalk between about 1,000 city names and a chosen canonical representation. Thus, ``Wolfboro, NH'' becomes ``Wolfeboro, NH'' and ``Morton Park, IL'' becomes a suburb of Chicago.

\subsection{Bank Name Validation}
\label{appendix:validation:bank}

Newspapers often mentioned banks with multiple abbreviations and conventions.
For instance, the ``Enterprise National Bank of Allegheny'' could be referred to as ``The Enterprise National Bank,'' ``Enterprise National,'' ``Enterprise NB,'' ``The Enterprise Natl Bank at Allegheny,'' and so on.

To address this naming heterogeneity, and standardize bank names across newspaper articles---and when merging with other datasets---we started an ancillary project, \texttt{bank-census}, consisting of several well-documented lists of banks, plus a Python package that performs numerous heuristics in order to standardize the bank names obtained from newspapers and match against the valid list of banks.

The six datasets we rely on are:

\begin{enumerate}
    \item Census of antebellum state banks \citep{WeberJEH}. This contains a manually collected list of all U.S. state banks from 1782 to 1861.
    \item Census of national banks between 1863 and 1940 \citep{CLV2026}. This contains the names and main events pertaining to all national banks up to 1940.
    \item Federal Financial Institutions Examination Council (FFIEC) bank attribute tables. These tables contain the names and main events about all commercial banks between approximately 1940 and today, plus some pre-1940 events for banks that still existed by that date.
    \item Census of all banks between 1870 and 1900 \citep{Jaremski_Fishback_2018}. This dataset includes not only state banks but also private banks, not included in the previous sources.
    \item Census of all banks regulated by the Department of Financial Services of the State of New York between 1784 and 2025 \citep{nysdfs}.
    \item Lastly, a manually collected list of banks. When combined, the previous sources are quite comprehensive but have some gaps. For instance, they exclude state banks that operated exclusively between 1862 and 1869, as well as certain state banks, private banks, and trusts that operated  between 1901 and 1940. To address this, we manually verify banks that appear in newspaper articles but not in any of the previous sources, and validate them against contemporaneous records such as Rand McNally, the annual reports of state bank examiners, etc.
\end{enumerate}

We use these datasets to create a validated list of U.S. banks, which are the main input to our \texttt{bank-census} Python package.

This package receives tuples of (state, city, bank), cleans up and standardizes these names, and then matches them against the validated list of banks.
If a direct match fails, the package attempts several fallback approaches to find the correct institution. These include replacing common abbreviations, fixing typos, etc. As a last resort, we manually constructed a crosswalk linking about 2,000 names to their standardized versions.

\clearpage
\subsection{Cross-Validation of CLV Bank Distress Events Against Ground Truth Datasets}

\label{app:cross-validation}

This section presents cross-validation exercises against four ground truth datasets:
\begin{enumerate}
\item Cross-validation against a ground truth dataset of 287
\unskip runs, suspensions, and reopenings based on \citet{Wicker1996,Wicker2006};
\item Cross-validation against runs and failures discussed in \cite{Rockoff2021};
\item Cross-validation against \textit{Bradstreet's} list of bank suspensions in 1893;
\item Cross-validation against a sample of 271
\unskip bank runs hand-collected from \url{newspapers.com} by undergraduate research assistants.
\end{enumerate}
The information in these ground truths was not used to build the CLV bank distress events dataset, so they provide out-of-sample tests of the accuracy and coverage of our dataset on bank runs and other bank distress events. Overall, for national banks, we find that our novel bank distress events dataset matches about 95\% of the runs and other events across these four ``ground truth'' datasets. The events that are not matched are most often driven by these events not being captured or found in our corpus of newspaper articles, rather than false negatives caused by LLM errors.

\subsubsection*{\#1 Cross-Validation Against Events Discussed in \citet{Wicker1996,Wicker2006}}

We manually construct a ``ground truth'' of bank runs, suspensions, and reopenings based on bank-level events discussed in two prominent studies \citet{Wicker1996} (\textit{Banking Panics of the Great Depression}) and \citet{Wicker2006} (\textit{Banking Panics of the Gilded Age}). These two studies are particularly useful, because they are based on archival research of newspapers to document the crises and near-crises of 1873, 1884, 1890, 1893, 1907, and 1930-33.

To build the ground truth, we first hand-collect all bank events mentioned in these two studies. We then manually code all runs, suspensions, and reopenings associated with the corresponding episode. Note that because these studies generally do not provide the full set of events within a given episode (e.g., they do not always say whether a suspension was temporary or permanent, or whether a run resulted in failure), to map out all events corresponding to an episode we hand-collect additional evidence from local newspapers, national financial press such as the \textit{Commercial and Financial Chronicle}, banking publications such as \textit{Banker's Magazine}, publications from the Office of the Comptroller of the Currency, case law, and other studies such as \citet{Sprague1910,FriedmanSchwartz,GortonTallman2016,Rockoff2021,Anderson2024}. We document the rationale for our classification of each event, usually relying on at least two separate sources. Thus, while the initial sample of bank events is based on \citet{Wicker1996,Wicker2006}, our ground truth is more comprehensive than the information contained in these studies because it contains the full set of events for a given bank distress episode.

\begin{table}[htbp]
    \centering
    \caption{\textbf{Cross-Validation against Ground Truth Events Based on \citet{Wicker1996,Wicker2006}}}
    \label{tab:gt_validation}
    \begin{tabular}{lrrrrrrrr}
\toprule
 & \multicolumn{4}{c}{All Banks} & \multicolumn{4}{c}{National Banks} \\
\cmidrule(lr){2-5}\cmidrule(lr){6-9}
 & & \multicolumn{3}{c}{Match within\ldots} & & \multicolumn{3}{c}{Match within\ldots} \\
\cmidrule(lr){3-5}\cmidrule(lr){7-9}
 & GT & month & 12 months & Rate & GT & month & 12 months & Rate \\
\midrule
All events & 287 & 241 & 250 & 87.1\% & 98 & 90 & 93 & 94.9\% \\
\addlinespace[2pt]
\midrule
\addlinespace[2pt]
\quad Runs & 100 & 84 & 85 & 85.0\% & 34 & 31 & 32 & 94.1\% \\
\quad Suspensions & 122 & 109 & 112 & 91.8\% & 41 & 39 & 40 & 97.6\% \\
\quad Reopenings & 65 & 48 & 53 & 81.5\% & 23 & 20 & 21 & 91.3\% \\
\bottomrule
\end{tabular}

\begin{minipage}{\textwidth}
    \footnotesize
    Notes: This table reports a cross-validation exercise of the CLV bank distress events database against a hand-collected ``ground truth'' based on events discussed in \citet{Wicker1996} and \citet{Wicker2006}. See text for details on the construction of the ground truth. GT refers to the number of events in the ground truth. The next columns under ``Match within...'' report the number of the GT events that are matched within one month or twelve months in the CLV bank distress events database. The rate refers to the match rate within 12 months.

    \end{minipage}

\end{table}

We then ask how many events in this ground truth based on \citet{Wicker1996,Wicker2006} are in our CLV bank distress events dataset. \Cref{tab:gt_validation} reports the results from this cross-validation exercise separately for all banks and national banks. For this exercise, we require a match across bank, event type, and date. We consider matches of an event in the same month and within twelve months. The ground truth based on Wicker's events contains 287
\unskip events for all banks, of which our dataset includes 250
\unskip (match rate: 87.1
\unskip\%). For all banks, the ground truth contains 100
\unskip runs, and we match 85
\unskip (match rate: 85.0
\unskip\%).

Among national banks---for which we have a comprehensive census throughout the sample and which are central to our analysis---there are 98
\unskip ground truth events, of which we match 93
\unskip (match rate: 94.9
\unskip\%). There are 34
\unskip runs on national banks in the ground truth, and our novel database matches 32
\unskip of these (match rate: 94.1
\unskip\%). \Cref{tab:wicker_comparison} reports all 34
\unskip runs on national banks from the ground truth, showing that most are matched on the exact same day.

 \begin{table}[htbp]
  \centering
  \caption{\textbf{Cross-Validation Against National Bank Runs Discussed in \citet{Wicker1996,Wicker2006}}}
  \label{tab:wicker_comparison}
  \footnotesize
  \scalebox{.9}{
  \begin{tabular}{llll}
\toprule
Date of run & Bank Name & City, State & Matched \\
\midrule
Sep. 19, 1873 & First NB & Philadelphia, PA & Exact \\
Sep. 19, 1873 & Fourth NB & New York, NY & Exact \\
Sep. 20, 1873 & NB of the Commonwealth & New York, NY & same month \\
Sep. 23, 1873 & Commercial NB & Petersburg, VA & Exact \\
Sep. 23, 1873 & First NB & Petersburg, VA & No match \\
Sep. 23, 1873 & Merchants NB & Petersburg, VA & Exact \\
Sep. 24, 1873 & First NB & Memphis, TN & Exact \\
Sep. 25, 1873 & Merchants \& Planters NB & Augusta, GA & Exact \\
Sep. 26, 1873 & Union NB & Chicago, IL & same month \\
May 14, 1884 & Metropolitan NB & New York, NY & Exact \\
May 14, 1884 & Second NB & New York, NY & same month \\
Mar. 27, 1893 & First NB & Nashville, TN & Exact \\
May 6, 1893 & Chemical NB & Chicago, IL & same month \\
May 11, 1893 & Capital NB & Indianapolis, IN & Exact \\
May 11, 1893 & Columbia NB & Chicago, IL & Exact \\
May 22, 1893 & NB of Deposit & New York, NY & No match \\
June 21, 1893 & Consolidated NB & San Diego, CA & Exact \\
July 14, 1893 & Kansas City NB & Kansas City, MO & same month \\
July 15, 1893 & Missouri NB & Kansas City, MO & Exact \\
July 22, 1893 & Kentucky NB & Louisville, KY & Exact \\
July 24, 1893 & Louisville City NB & Louisville, KY & Exact \\
July 25, 1893 & Fourth NB & Louisville, KY & Exact \\
July 25, 1893 & Merchants NB & Louisville, KY & Exact \\
July 29, 1893 & Commercial NB & Portland, OR & Exact \\
Aug. 4, 1893 & National German American Bank & St Paul, MN & Exact \\
Aug. 10, 1893 & American NB & Nashville, TN & Exact \\
Aug. 10, 1893 & Fourth NB & Nashville, TN & same month \\
Aug. 17, 1893 & First NB & Dubuque, IA & Exact \\
Oct. 16, 1907 & Mercantile NB & New York, NY & same month \\
Oct. 16, 1907 & NB of North America & New York, NY & 12 months \\
Oct. 16, 1907 & New Amsterdam NB & New York, NY & same month \\
Nov. 10, 1930 & Holston Union NB & Knoxville, TN & Exact \\
Nov. 14, 1930 & Tennessee Hermitage NB & Nashville, TN & Exact \\
Nov. 17, 1930 & NB of Kentucky & Louisville, KY & Exact \\
\midrule
\multicolumn{3}{l}{Total events} & 34 \\
\multicolumn{3}{l}{Matched on exact date} & 23 \\
\multicolumn{3}{l}{Matched within 1 month} & 31 \\
\multicolumn{3}{l}{Matched within 12 months} & 32 \\
\bottomrule
\end{tabular}
}

\begin{minipage}{\textwidth}
    \footnotesize
    Notes: This table lists the 34
\unskip national bank runs mentioned in \citet{Wicker1996} and \citet{Wicker2006}. The last column lists whether the bank run is in the CLV bank distress events database.
   \end{minipage}

  \end{table}

Inspecting events we do not match, there are two main reasons they are not included in the CLV bank distress events data. First, prior to around 1900, the bank census we built has imperfect coverage of non-national banks, especially private banks, so our procedure does not find events for these banks in our corpus of newspaper articles. Note that this is not an issue for national banks, as we have a complete census for these banks. Second, some events are not covered in the corpus of newspapers we obtain from \textit{Chronicling America} and commercial sources. This explains the two runs on national banks we miss at the 1-month level. For these cases, we could find newspaper articles on these events from other historic newspaper sources, but we did not include these to avoid contaminating the cross-validation exercise. We did not find cases where the failure to match was driven by the LLMs incorrectly interpreting the newspaper evidence. Instead, the issue is usually not having original newspaper coverage of an event.

\subsubsection*{\#2 Cross-Validation Against Runs and Failures in \citet{Rockoff2021}}

\cite{Rockoff2021} studies the failures and near-failures of financial institutions during a set of major U.S. financial crises. His Table 1 provides a list of these failures and near-failures. \Cref{tab:rockoff_comparison} reproduces the 19
\unskip events for our sample period, 1863-1934. It shows that the CLV bank distress events database captures all of these events, save one. The event that is not matched involves a trust, rather than a commercial bank. The reason it is not matched is that it is not in our bank census.

\begin{table}[htbp]
    \centering
    \caption{\textbf{Cross-Validation Against Failures and Near Failures in \citet{Rockoff2021}} }
    \label{tab:rockoff_comparison}
    \footnotesize
    \scalebox{.9}{
    \begin{tabular}{lllllll}
\toprule
Year & Bank & City & State & Bank Type & CLV Episode & Match \\
\midrule
1873&Jay Cooke \& Co.&Philadelphia&PA&private&run+susp+fail&Matched\\
1873&Kenyon, Cox \& Co.&New York&NY&private&susp+fail&Matched\\
1873&New York Warehouse \& Security Co.&New York&NY&private&susp+fail&Matched\\
1884&Grant \& Ward&New York&NY&savings&susp+fail&Matched\\
1884&Marine NB&New York&NY&national&run+susp+fail&Matched\\
1884&Metropolitan NB&New York&NY&national&run+susp+reopen&Matched\\
1884&Second NB&New York&NY&national&run+only&Matched\\
1890&Charles M. Whitney \& Co.&New York&NY&private&susp+fail&Matched\\
1890&Decker, Howell \& Co.&New York&NY&broker&susp+reopen&Matched\\
1893&Capital NB&Indianapolis&IN&national&run+susp+reopen&Matched\\
1893&Chemical NB&Chicago&IL&national&run+susp+fail&Matched\\
1893&Columbia NB&Chicago&IL&national&run+susp+fail&Matched\\
1893&Herman Schaffner \& Co.&Chicago&IL&private&susp+fail&Matched\\
1893&United States Loan and Trust Co.&Chicago&IL&&&Not matched\\
1893&Wisconsin Marine \& Fire Insurance Co.&Milwaukee&WI&state&run+susp+reopen&Matched\\
1907&Knickerbocker Trust Co.&New York&NY&trust&run+susp+reopen&Matched\\
1907&Mercantile NB&New York&NY&national&run+susp+fail&Matched\\
1930&Bank of United States&New York&NY&state&run+susp+fail&Matched\\
1930&Caldwell \& Co.&Nashville&TN&private&susp+fail&Matched\\
\midrule
\multicolumn{6}{l}{Matched} & 18 \\
\multicolumn{6}{l}{Not matched} & 1 \\
\bottomrule
\end{tabular}
}

    \begin{minipage}{\textwidth}
      \footnotesize
      Notes: This table lists the bank runs and failures mentioned in
  \citet{Rockoff2021}. The last column lists whether the event is in our database.
    \end{minipage}

  \end{table}

\subsubsection*{\#3 Cross-Validation Against 1893 Suspensions from \textit{Bradstreet's}}

Following the Panic of 1893, the trade journal \textit{Bradstreet's} published a list of suspended national, state, savings, and private banks in its November 11, 1893 issue.\footnote{This source is also used by  \cite{Carlson2005} and \cite{Wicker2006}.} This provides a ground truth of bank suspensions to which we can compare our CLV bank distress events database. \Cref{tab:bradstreets_summary} shows the comparison for all suspended national banks. The table shows that of the 123
\unskip national bank suspensions in 1893, our database based on newspaper events matches 115
\unskip (match rate: 93.5
\unskip\%).

\begin{table}[htbp]
    \centering
    \caption{\textbf{Cross-Validation Against 1893 National Bank Suspensions Reported in \emph{Bradstreet's}}}
    \label{tab:bradstreets_summary}
    \begin{tabular}{lrrr}
\toprule
 & & \multicolumn{2}{c}{Match\ldots} \\
\cmidrule(lr){3-4}
Sample & GT & Number & Rate \\
\midrule
National bank suspensions & 123 & 115 & 93.5\% \\
\bottomrule
\end{tabular}

        \begin{minipage}{\textwidth}
      \footnotesize
      Notes: The table summarizes a cross-validation exercise of our bank runs database against national bank suspensions in 1893 as recorded in \textit{Bradstreet's}. The GT column (``Ground Truth'') refers to the number of national bank suspensions in \textit{Bradstreet's} November 11, 1893 issue. The next columns under ``Match'' report the number of these GT events that are matched within one month in our bank distress events database. The rate refers to the corresponding match rate.

    \end{minipage}
\end{table}

\subsubsection*{\#4 Humans vs LLMs: Cross-Validation Against Hand-Collected Dataset of Runs}

At the outset of this project, we hired eight MIT undergraduate RAs for one month to create a hand-collected database of bank runs. We gave each RA an account with \url{newspapers.com} to search for bank runs using a list of keywords.\footnote{The keywords were: bank run, large deposit withdrawal, bank panic, banking panic, bank suspension, and suspend payments.} This provided us with a sample of 271
\unskip bank runs from 1863 through 1934. We manually checked the RAs' list of runs against the original articles to remove false positives and, in a few instances, improve the bank name, city, and state.

This exercise serves two purposes. First, the cross-validation against events based on \cite{Wicker1996,Wicker2006} in exercise \#1 above may skew toward relatively salient runs and runs that occur during crises. The RA-collected sample provides a somewhat more random ground truth sample of runs that are mentioned in newspapers, during and outside of crisis years. Second, the underlying set of articles available on \url{newspapers.com} is not the same as in \textit{Chronicling America}, our main source throughout most of the sample, so this exercise also provides a sense of whether we are missing many runs because they are not captured in \textit{Chronicling America}.

\begin{table}[!ht]
    \centering
    \caption{\textbf{Cross-Validation Against Runs Hand-Collected by a Team of RAs} }
    \label{tab:ra_comparison}

    \begin{tabular}{lrrrr}
\toprule
 & & \multicolumn{3}{c}{Match within\ldots} \\
\cmidrule(lr){3-5}
Sample & GT & month & 12 months & Rate \\
\midrule
All RA runs & 271 & 214 & 233 & 86.0\% \\
National bank runs & 87 & 73 & 82 & 94.3\% \\
\bottomrule
\end{tabular}

    \begin{minipage}{\textwidth}
      \footnotesize
      Notes: The table summarizes a cross-validation exercise of the CLV bank runs database against a hand-collected ``ground truth'' of runs collected by a team of RAs from searching \url{newspapers.com} for bank runs that we subsequently verified. See text for details on the construction of the ground truth. GT refers to the number of events in the ground truth. The next columns under “Match within...” report the number of these GT events that are matched within one month or twelve months in the CLV bank distress events database. The rate refers to the match rate
within 12 months.

    \end{minipage}
\end{table}

\Cref{tab:ra_comparison} summarizes the cross-validation against this RA-collected dataset of runs. Of the 271
\unskip runs in this hand-collected dataset, the CLV bank distress events dataset captures 233
\unskip for a match rate of 86.0
\unskip\%. For national banks, there are 87
\unskip runs in the hand-collected dataset, of which we match 82
\unskip, for a match rate of 94.3
\unskip\%.

The 5
\unskip national bank runs that are not matched are instructive. In all cases, they are caused by a difference in the corpus of news articles in \textit{Chronicling America} and \url{newspapers.com}. For example, in one case, an article on \url{newspapers.com} mentioned a ``slight run,'' but this article is not in the \textit{Chronicling America} corpus. In another case, our dataset identifies the bank failing on the same day, but none of the articles in our corpus discuss a run or large withdrawals. In none of these cases was the failure to match driven by the LLMs misinterpreting the news articles.

\clearpage
\subsection{Illustrative Examples of Newspaper Articles}
\label{appendix:sec:examples}

\begin{figure}[ht]
  \centering
  \caption{Example of the Layout and OCR Steps}
  \label{fig:example-layout-ocr}
  \begin{subfigure}[t]{0.49\textwidth}
    \caption{\textbf{Layout:} Page segmentation}
    \label{fig:example-layout}
    \includegraphics[width=\linewidth]{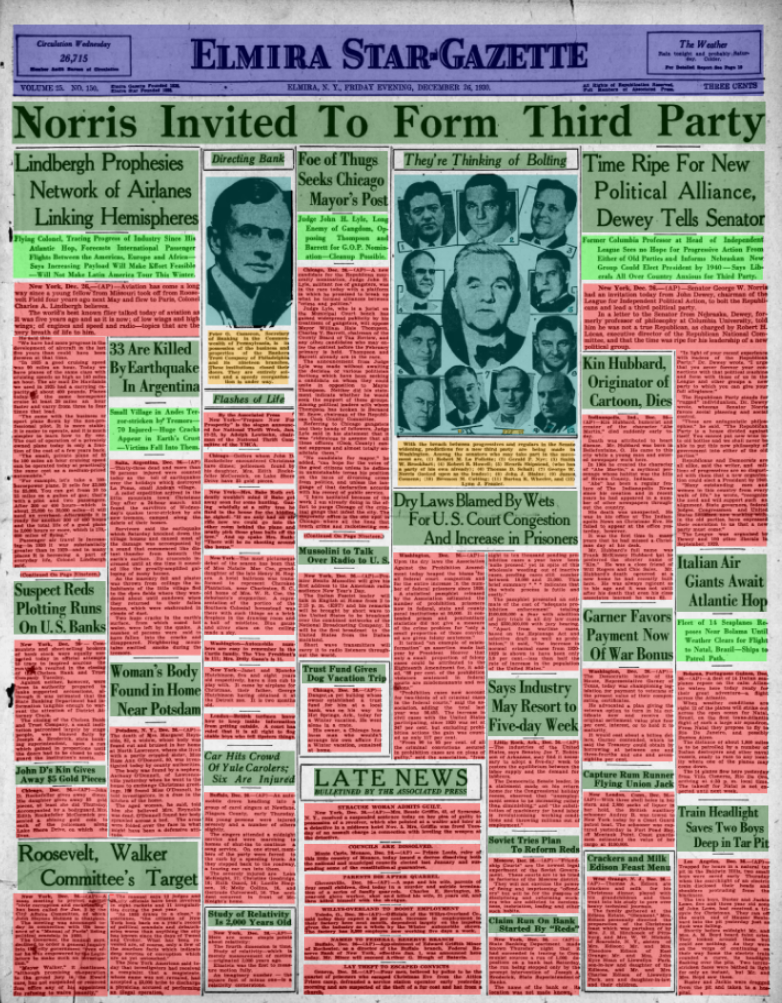}
  \end{subfigure}\hfill
  \begin{subfigure}[t]{0.49\textwidth}
    \caption{\textbf{OCR:} Text extraction}
    \label{fig:example-ocr}
    \includegraphics[width=\linewidth]{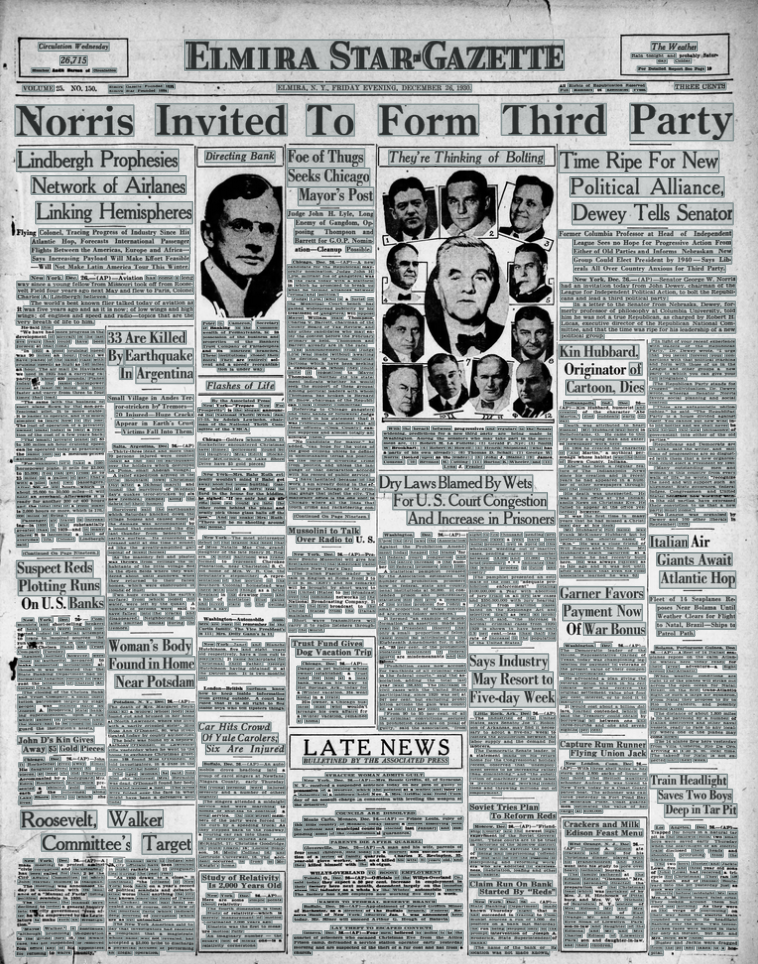}
  \end{subfigure}
  \begin{minipage}{\textwidth}
    \footnotesize
    Notes: Panel (a) shows the output of the page segmentation step, where each page is split into multiple articles, which in turn can be segmented further into headers, main text, images, etc. For most pages, this was done through code available in the \textit{American Stories} project \citep{AmericanStories}, but we also used as fallbacks the \textit{Arcanum} newspaper segmentation model and the open source \textit{DocLayout-YOLO} model. Panel (b) shows the OCR step, where we identify the text in each page. It was implemented either via the \textit{Textract} product by AWS or through the OCR tools embedded in \textit{Gemini 2.0} by Google (we used this product as a fallback in case the quality of the OCR obtained by \textit{Textract} was low).
    \end{minipage}
\end{figure}

\begin{figure}[ht]
  \centering
  \caption{Examples of False Positives Flagged by Keywords and Excluded by LLMs}
  \label{fig:example-false-positives}

  \begin{subfigure}[t]{0.49\textwidth}
    \caption{Unrelated event}
    \label{fig:example-false-positives1}
    \includegraphics[width=\linewidth]{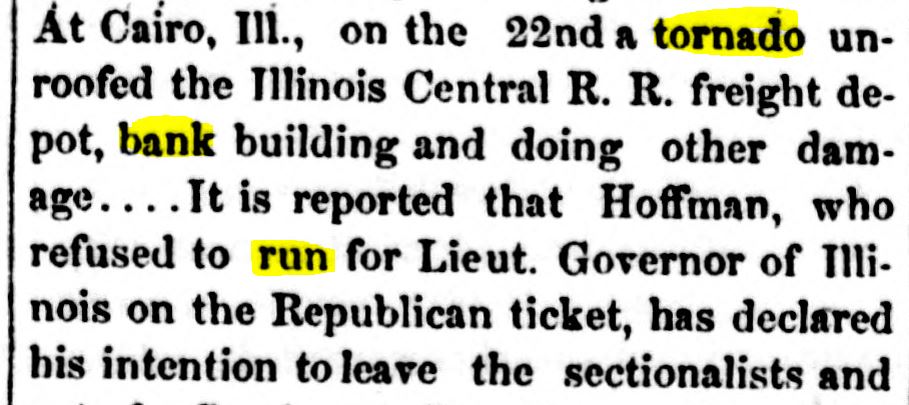}
  \end{subfigure}\hfill
  \begin{subfigure}[t]{0.49\textwidth}
    \caption{Foreign bank run}
    \label{fig:example-false-positives2}
    \includegraphics[width=\linewidth]{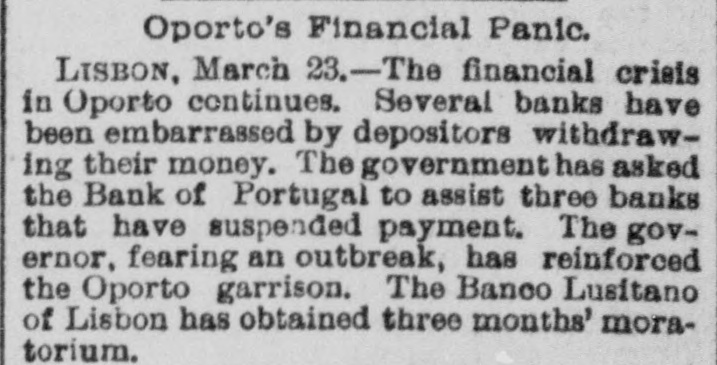}
  \end{subfigure}

  \begin{subfigure}[t]{0.49\textwidth}
    \caption{Fictional run}
    \label{fig:example-false-positives3}
    \includegraphics[width=\linewidth]{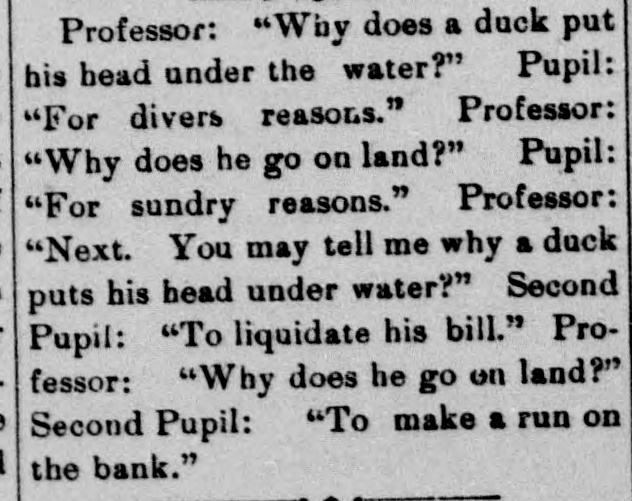}
  \end{subfigure}\hfill
  \begin{subfigure}[t]{0.49\textwidth}
    \caption{Only tangentially related}
    \label{fig:example-false-positives4}
    \includegraphics[width=\linewidth]{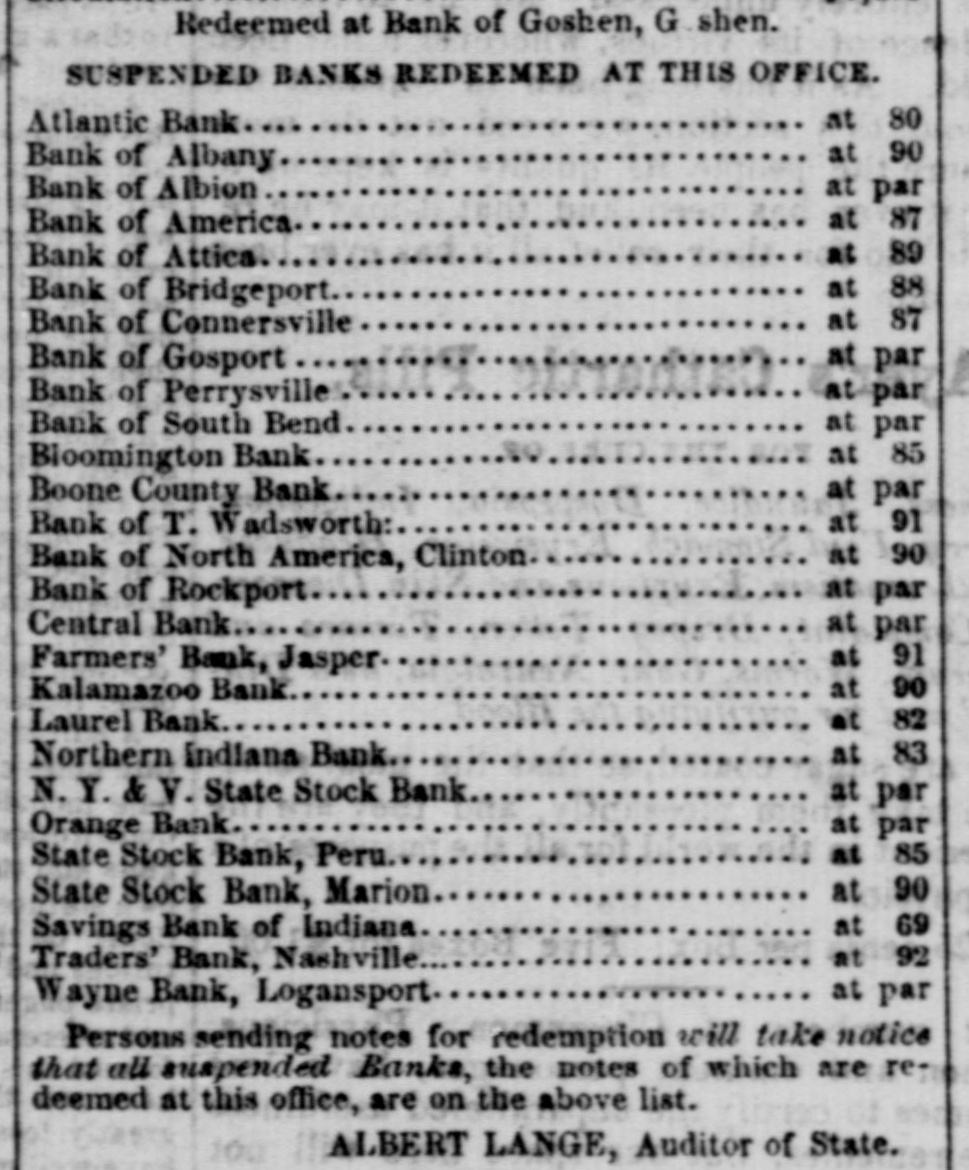}
  \end{subfigure}

  \begin{minipage}{\textwidth}
    \footnotesize
    Notes: These four examples illustrate cases where the keyword detection step flagged an article as potentially relevant for bank distress, but the LLM step detected that the article was, in fact, not relevant.
    \end{minipage}
\end{figure}

\begin{figure}[ht]
  \centering
  \caption{Examples of Articles Discussing Bank Runs or Suspensions}
  \label{fig:examples}

  \begin{subfigure}[t]{0.43\textwidth}
    \caption{Jay Cooke's Failure (1873)}
    \label{fig:example1}
    \includegraphics[width=\linewidth]{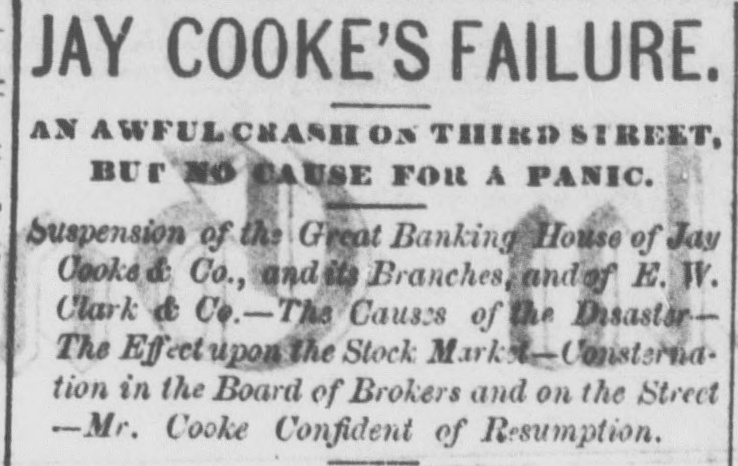}
  \end{subfigure}\hfill
  \begin{subfigure}[t]{0.43\textwidth}
    \caption{Run without failure (1854)}
    \label{fig:example2}
    \includegraphics[width=\linewidth]{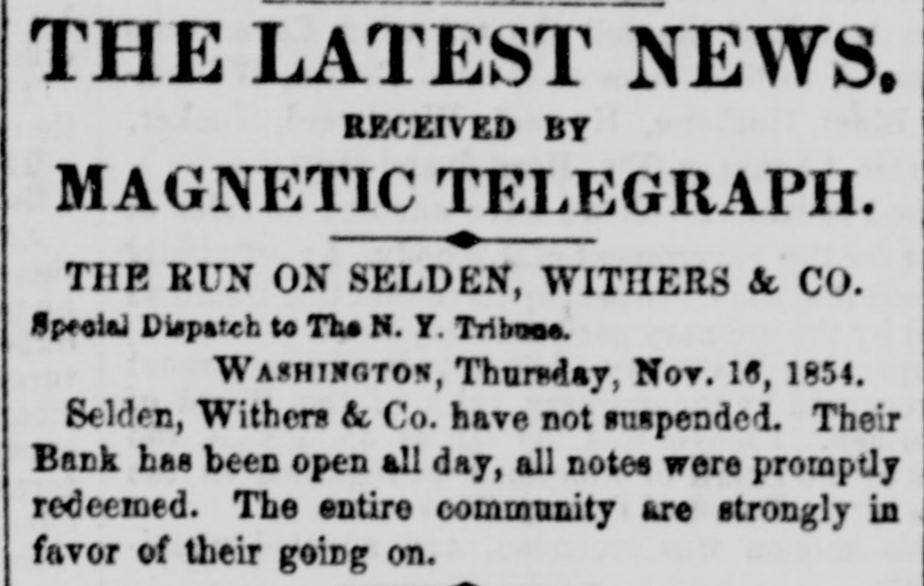}
  \end{subfigure}

  \vspace{5mm}

  \begin{subfigure}[t]{0.43\textwidth}
    \caption{Bank failure (1907)}
    \label{fig:example3}
    \includegraphics[width=\linewidth]{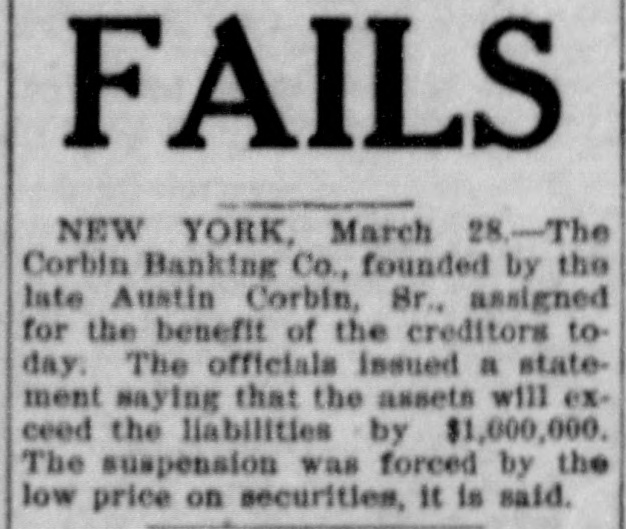}
  \end{subfigure}\hfill
  \begin{subfigure}[t]{0.43\textwidth}
    \caption{Sunspot run (1929)}
    \label{fig:example4}
    \includegraphics[width=\linewidth]{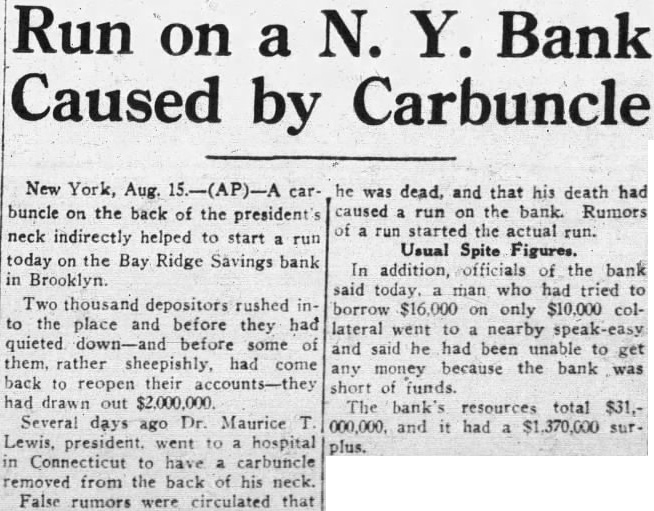}
  \end{subfigure}

  \vspace{5mm}

  \begin{subfigure}[t]{0.43\textwidth}
    \caption{Bank suspensions (1907)}
    \label{fig:example5}
    \includegraphics[width=\linewidth]{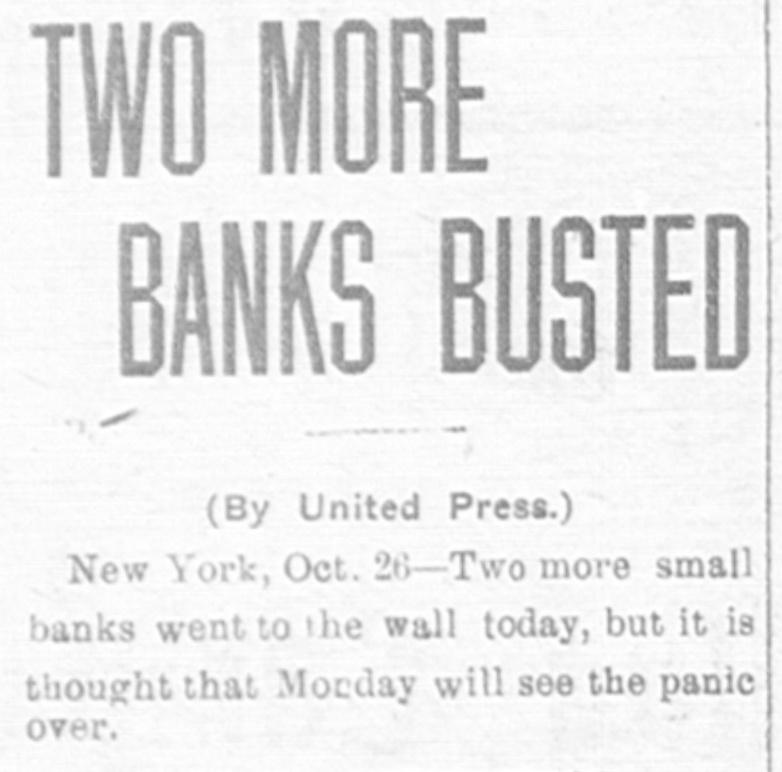}
  \end{subfigure}\hfill
  \begin{subfigure}[t]{0.43\textwidth}
    \caption{Run followed by failure (1928)}
    \label{fig:example6}
    \includegraphics[width=\linewidth]{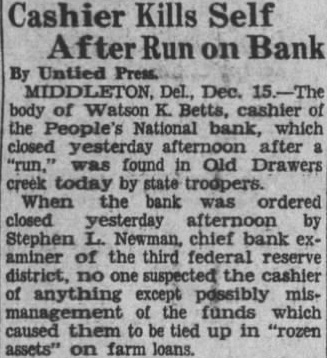}
  \end{subfigure}

  \begin{minipage}{\textwidth}
    \footnotesize
    Notes: Examples of bank runs, suspensions, and failures.
    \end{minipage}
\end{figure}

\begin{figure}[ht]
  \centering
  \caption{Examples of Articles Illustrating Runs Exemplifying Various Theories}
  \label{fig:run_examples_theories}

  \begin{subfigure}[t]{0.4\textwidth}
    \caption{Asymmetric information run: The presence of a famous boxer at a bank attracted a large crowd, which was misinterpreted by passersby as evidence of trouble at the bank, triggering a bank run (Merchants National Bank, February 24, 1910)}
    \label{fig:ex_asym_info2}
    \includegraphics[width=\linewidth]{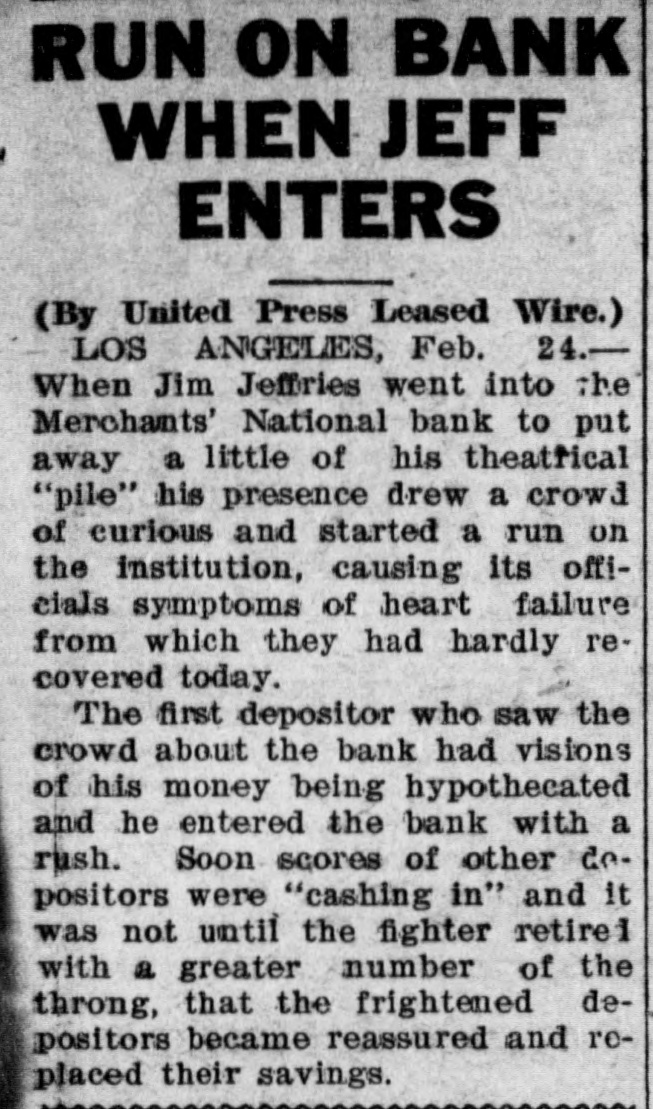}
  \end{subfigure}
~
  \begin{subfigure}[t]{0.45\textwidth}
    \caption{Non-fundamental run: Run based on false rumor in newspaper (First National Bank of Minneapolis, October 1, 1873)
}
    \label{fig:false_rumor1}
    \includegraphics[width=\linewidth]{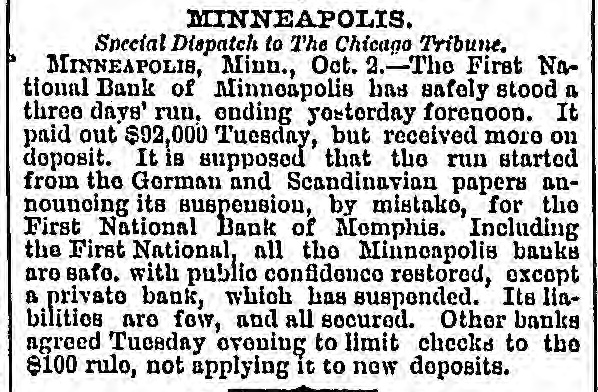}
  \end{subfigure}

  \begin{subfigure}[t]{0.45\textwidth}
    \caption{Non-fundamental run: Run based on false rumor by depositor (Bank of Wheeling, March 11, 1884)
}
    \label{fig:false_rumor2}
    \includegraphics[width=\linewidth]{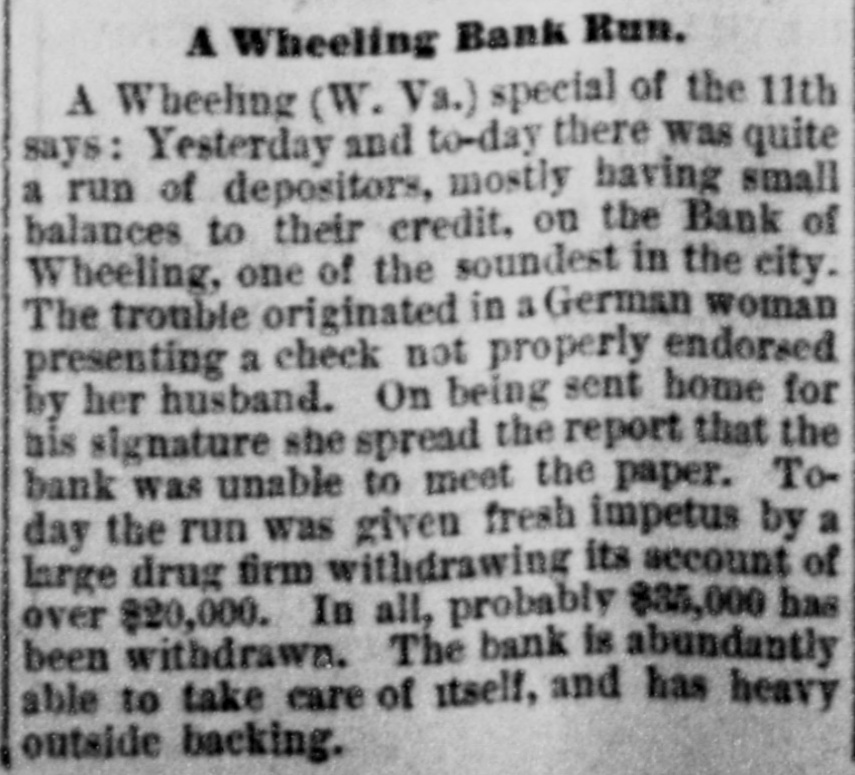}
  \end{subfigure}
~
  \begin{subfigure}[t]{0.45\textwidth}
    \caption{Non-fundamental run: Run based on joke article misinterpreted by excited depositors (Norwalk Savings Bank, May 26, 1884)
}
    \label{fig:false_rumor3}
    \includegraphics[width=\linewidth]{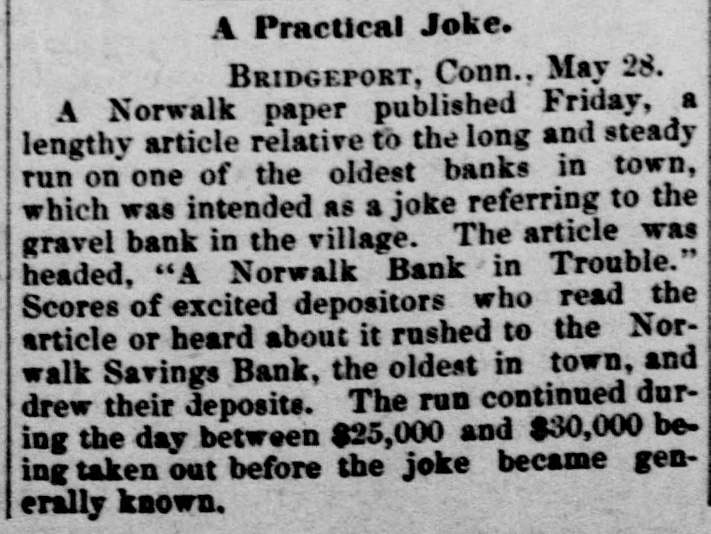}
  \end{subfigure}

\end{figure}

\clearpage
\renewcommand{\currentappendix}{D}
\setcounter{page}{1}
\clearpage
\section{Details on Other Data}
\normalsize
\label{appendix:other_data}

\subsection{Bradstreet's Trade at a Glance}
\label{app:bradstreets}
\begin{figure}[ht]
  \centering
  \caption{\textbf{Bradstreet's Trade at a Glance: Example from January 3, 1931}}
  \label{fig:Bradstreets_example}

    \includegraphics[scale=.7]{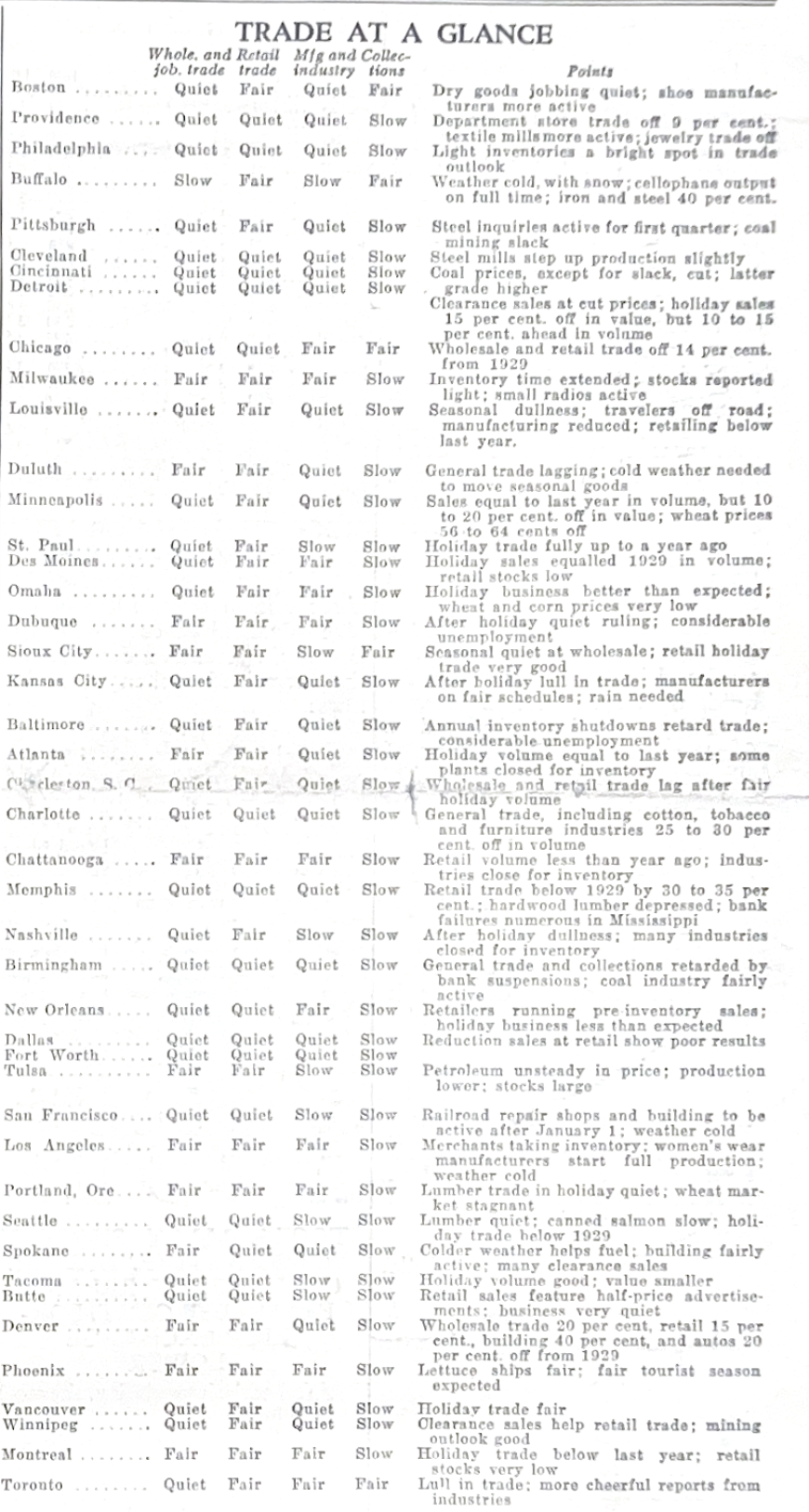}

\end{figure}

\paragraph{Source.} The weekly business and financial journal \textit{Bradstreet's} published a summary of economic conditions at the city level for major North American cities called Trade at a Glance (TAAG). \Cref{fig:Bradstreets_example} provides an example of the raw data. Building on \cite{CLV_Pandemic_2022}, we digitize these tables for all years they are available, from 1917 to March 1933. From March 1933 through 1935, \textit{Bradstreet's} provides short (one or two-paragraph) descriptions of business conditions, from which we extract key phrases. Data from March 1933 onward is available at a monthly frequency. Given the sample break, we verify that results are robust to excluding the data after March 1933.

\paragraph{Conversion to numeral score for TAAG tables.} For 1917-1933, the TAAG tables provide a short description of conditions in various sectors in the city. We convert the text into an integer value from 1 (bad conditions), 2 (fair conditions), and 3 (good conditions). The exact classification is given in \Cref{tab:TAAG_class}.

\paragraph{Conversion to numeral score for TAAG paragraphs.} For 1933-1935, we search the text for whether a sector is described with any of the keywords given in \Cref{tab:TAAG_class}. We then apply the same classification.

\paragraph{Sample.} Throughout the 1917 to 1935 period, Bradstreet's covers 121
distinct cities. However, some cities only enter the sample occasionally. We therefore focus on a sample of 30
cities with consistent coverage throughout the sample. Because even these cities can occasionally enter and exit the sample we fill missing values forward. For our analysis at the city level, we also take a seven-week moving average to smooth out the discrete changes in the index. We verify that the main results are not sensitive to forward-filling and smoothing.

\paragraph{Comparison with industrial production.} To validate the usefulness of our local economic activity index based on \textit{Bradstreet's} TAAG, we compare the manufacturing index to industrial production (IP) estimates from \cite{MironRomer1990}. Since IP is monthly at the national level, we take a cross-city average of the TAAG manufacturing index in each month. Moreover, since IP contains a trend, while TAAG refers to cyclical conditions, we consider the year-on-year change in log industrial production and the deviation of log industrial production from the one-sided three-year moving average.

\begin{figure}[ht]
  \centering
  \caption{\textbf{Comparing Bradstreet's Trade at a Glance Manufacturing Index and Industrial Production}}
  \label{fig:TAAG_IP}

    \includegraphics[scale=1.0]{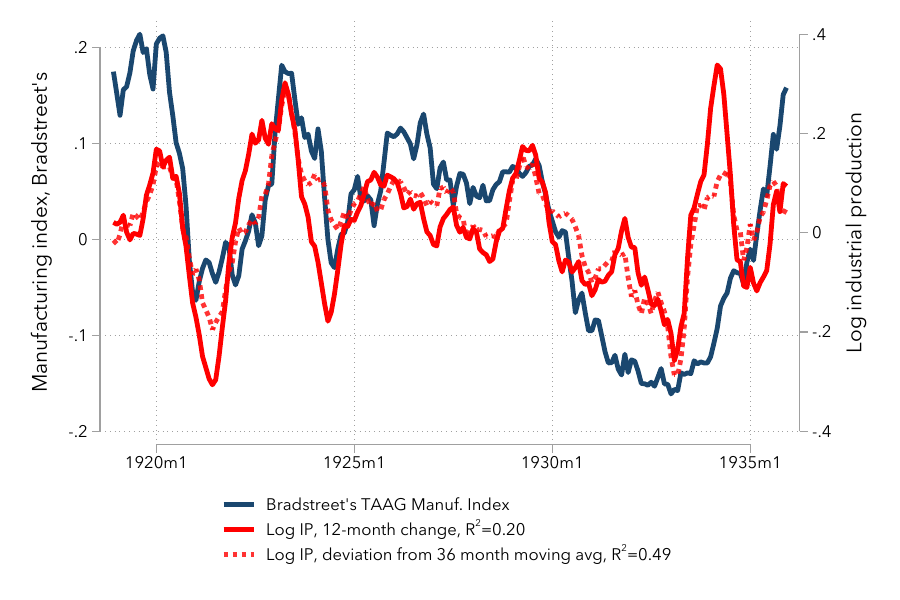}

    \begin{minipage}{\textwidth}
    \footnotesize
    Notes: This figure compares the national monthly average of our manufacturing index constructed from digitized \textit{Bradstreet's} Trade at a Glance tables with industrial production. Industrial production is detrended using the year-on-year log change or the deviation from a three-year moving average.
    \end{minipage}
\end{figure}

\Cref{fig:TAAG_IP} reports the comparison. The correlation is quite high. For example, the $R^2$ between the TAAG manufacturing index and IP's deviation from its three-year moving average is 0.49. Moreover, the TAAG manufacturing index clearly captures the 1920-21 recession and the downturn and recovery in the Great Depression.

\begin{table}[tbp] \centering
\newcolumntype{R}{>{\raggedleft\arraybackslash}X}
\newcolumntype{L}{>{\raggedright\arraybackslash}X}
\newcolumntype{C}{>{\centering\arraybackslash}X}

\caption{Bradstreet's TAAG Classification}
\label{tab:TAAG_class}
\begin{tabularx}{\linewidth}{@{}lCC@{}}

\toprule
{Bad (cat=1)}&{Fair (cat=2)}&{Good (cat=3)} \tabularnewline
\midrule \addlinespace[\belowrulesep]
75 percent&better&active \tabularnewline
backward&better outlook&brisk \tabularnewline
blackening&cautions&enlarging \tabularnewline
contracted&fair&fairly active \tabularnewline
curtailed&fair activity&gaining \tabularnewline
curtailing&improved&good \tabularnewline
disturbed&improving&improvement \tabularnewline
dull&irregular&increased \tabularnewline
hampered&normal&increasing \tabularnewline
hesitating&readjusting&more active \tabularnewline
inactive&shifting&picking up \tabularnewline
interrupted&spotty&very active \tabularnewline
lagging&steady& \tabularnewline
less active&unsettled& \tabularnewline
paralyzed&varying& \tabularnewline
poor&& \tabularnewline
quiet&& \tabularnewline
quieter&& \tabularnewline
reduced&& \tabularnewline
restricted&& \tabularnewline
retarded&& \tabularnewline
slackened&& \tabularnewline
slow&& \tabularnewline
slower&& \tabularnewline
slowing up&& \tabularnewline
strikes hurtful&& \tabularnewline
suspended&& \tabularnewline
very dull&& \tabularnewline
very quiet&& \tabularnewline
waiting&& \tabularnewline
\bottomrule

\end{tabularx}
\end{table}

\clearpage
\renewcommand{\currentappendix}{E}
\setcounter{page}{1}
\clearpage

\section{Case Studies}

\label{appendix:case_studies}

This appendix provides case studies of bank runs, suspensions and failures, thus showcasing and providing further validation of our new bank distress dataset. Each case study contains a link to the episode page with primary sources at \href{https://finhist.com/bank-runs}{finhist.com/bank-runs}. Some episodes are well-documented in existing work. For these, this appendix provides new facts and easily accessible primary contemporary sources that validate the database. Other episodes are not well covered in existing scholarly work, providing new material for understanding the nature and causes of runs.

\subsection{Prominent Runs and Failures during Major Crises}

\subsubsection{The Panic of 1873}

\begin{figure}[ht]
  \centering
  \caption{Panic of 1873}
  \label{fig:maps-1873}
  \begin{subfigure}[t]{0.49\textwidth}
    \caption{August 1873}
    \label{fig:map-1873-8}
    \includegraphics[width=\linewidth]{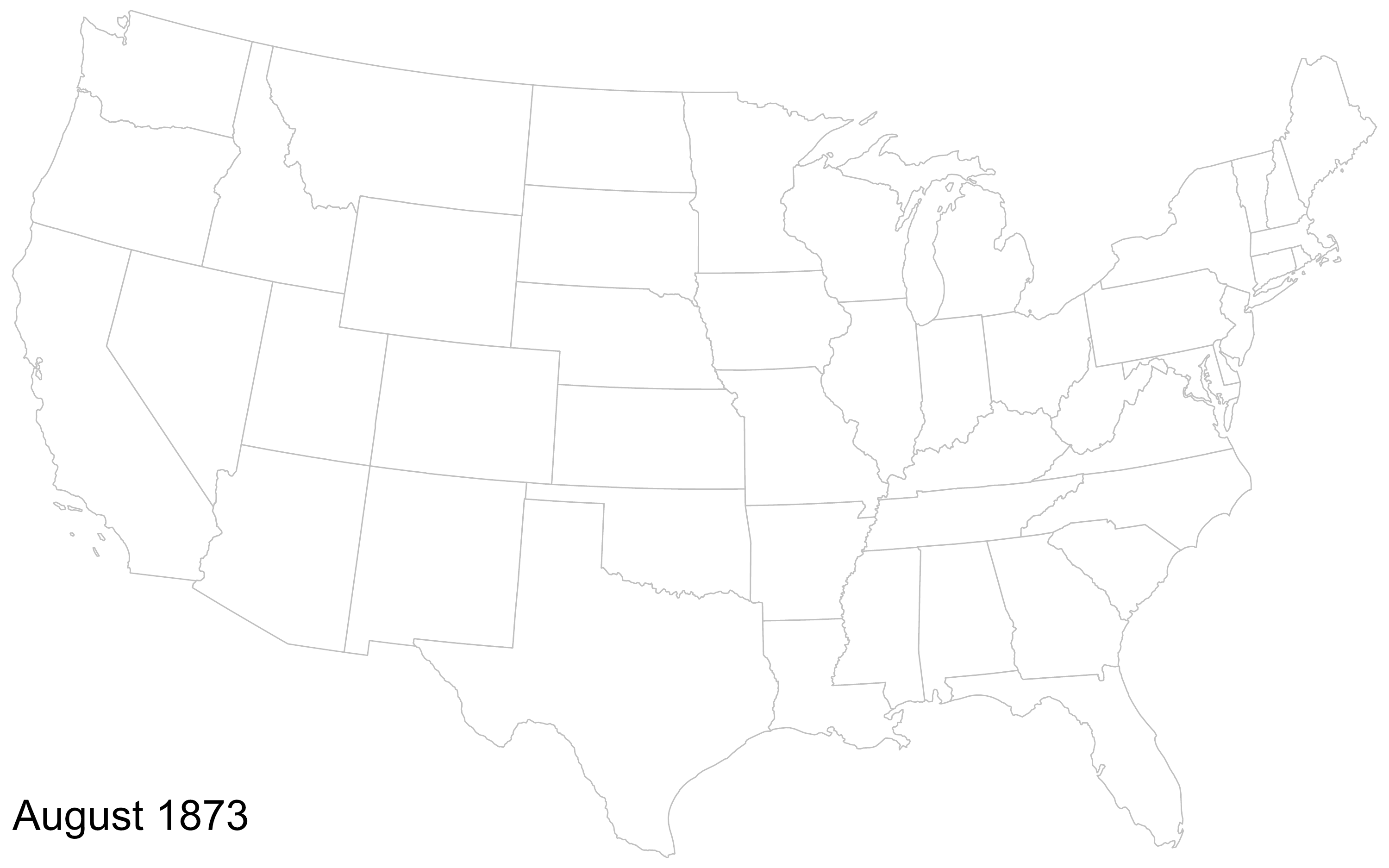}
  \end{subfigure} \hfill
  \begin{subfigure}[t]{0.49\textwidth}
    \caption{September 1873}
    \label{fig:map-1873-9}
    \includegraphics[width=\linewidth]{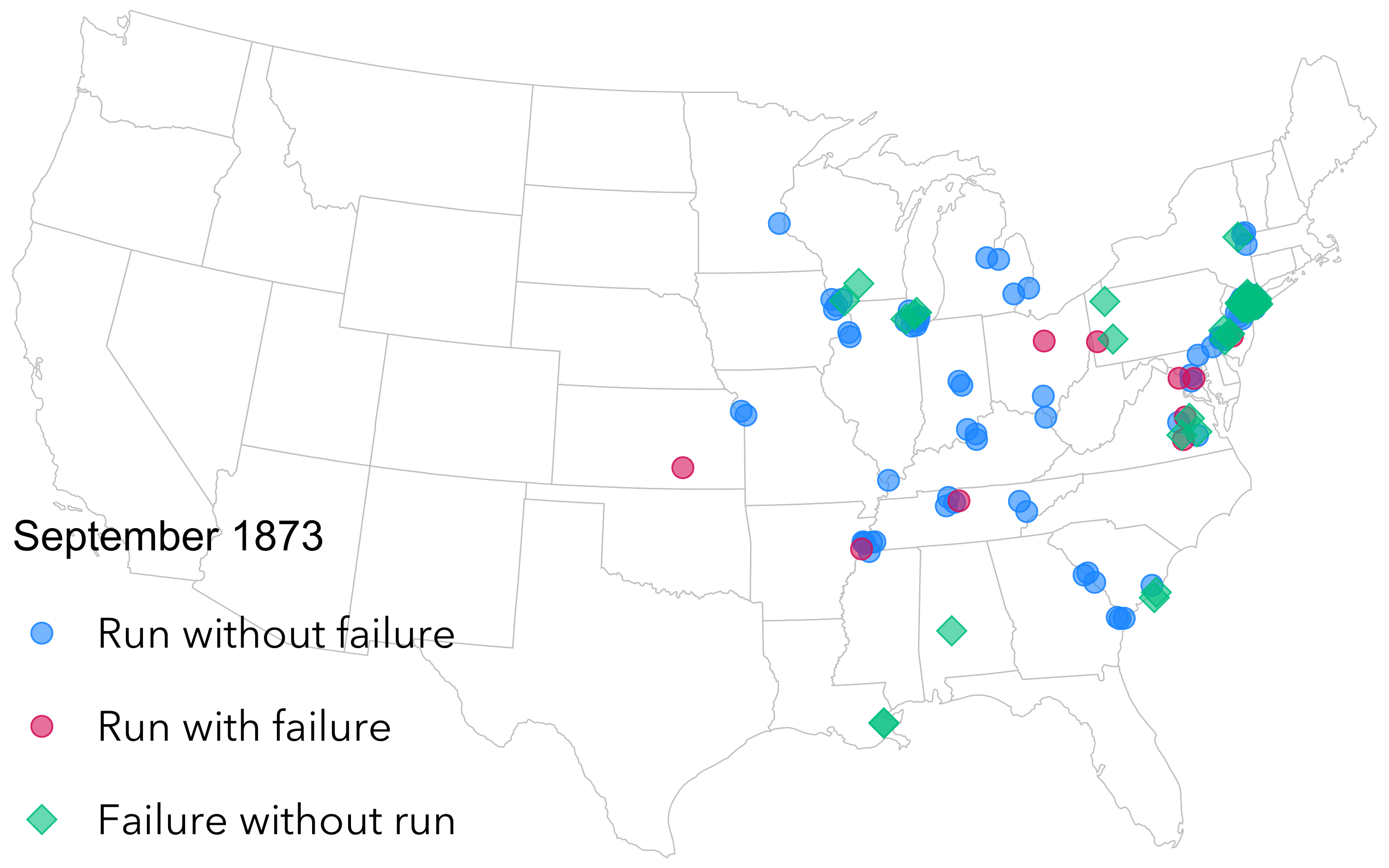}
  \end{subfigure}

  \vspace{0.5em}

  \begin{subfigure}[t]{0.49\textwidth}
    \caption{October 1873}
    \label{fig:map-1873-10}
    \includegraphics[width=\linewidth]{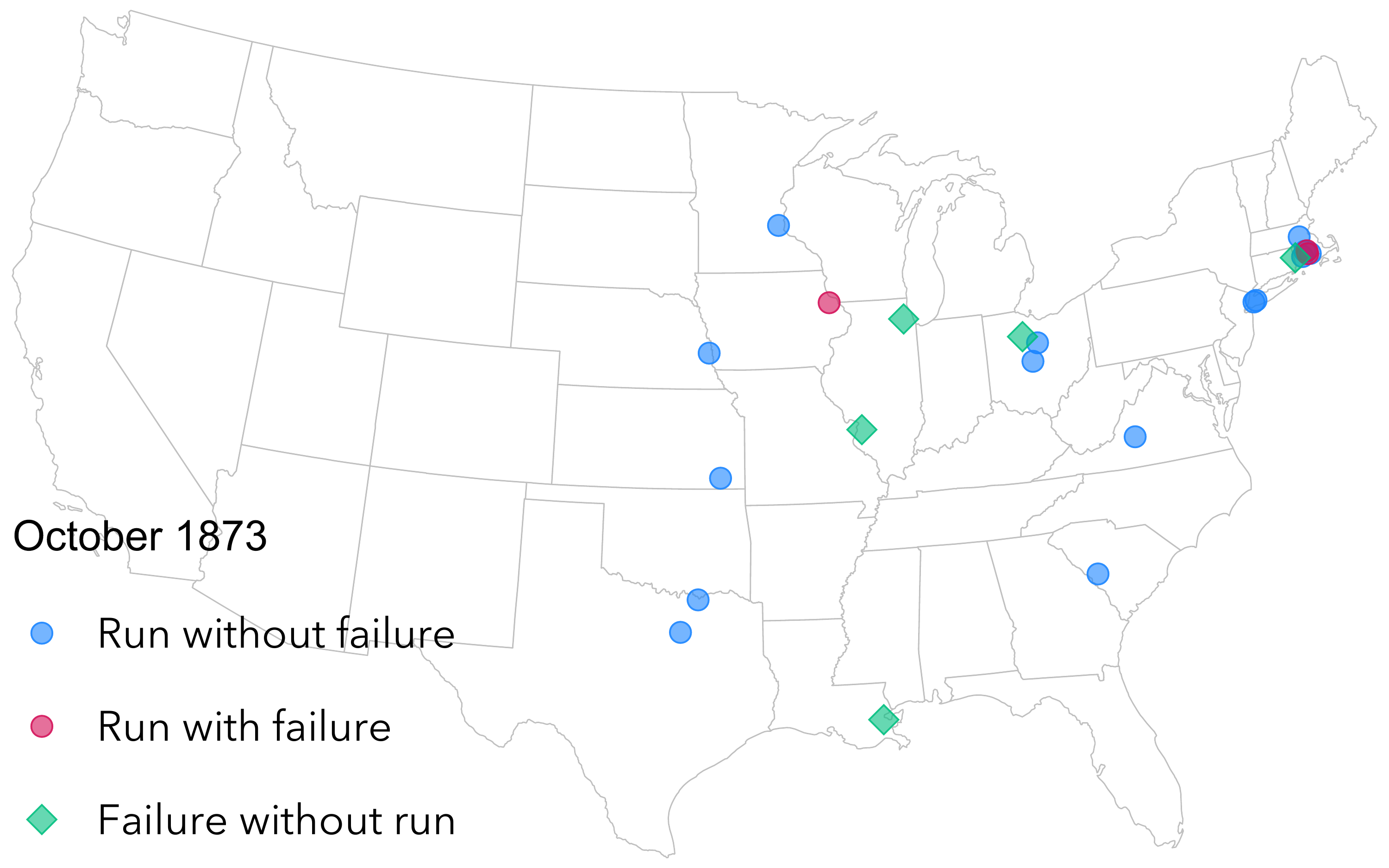}
  \end{subfigure}\hfill
  \begin{subfigure}[t]{0.49\textwidth}
    \caption{November 1873}
    \label{fig:map-1873-11}
    \includegraphics[width=\linewidth]{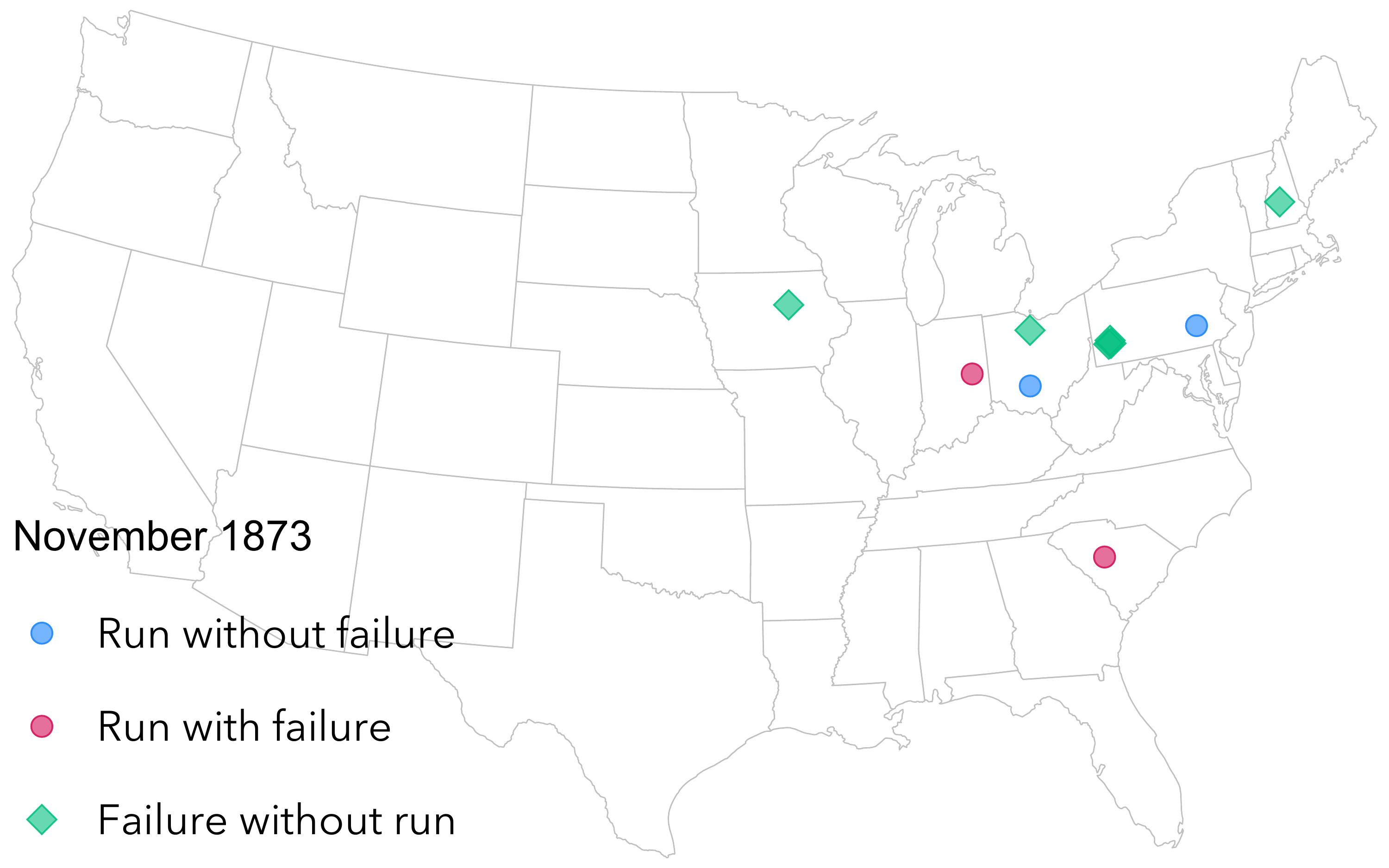}
  \end{subfigure}

  \begin{minipage}{\textwidth}
    \footnotesize
    \textit{Notes:} Each panel shows the location of bank distress episodes in a given month during the Panic of 1873. Blue circles indicate banks subject to a run that did not subsequently fail. Red circles indicate banks subject to a run that subsequently failed. Green diamonds indicate banks that failed without a reported run.
  \end{minipage}
\end{figure}

The Panic of 1873 followed a credit and investment boom in railroads after the Civil War. Railways borrowed heavily, and projects outpaced demand for new capacity. \cite{OCCAnnualReport1873} reports the money market became ``overloaded with debt,'' much of which was sold abroad. When foreign demand dried up following a stock market crash in Austria, prices for stocks and bonds fell. New York banks were saddled with assets from struggling railroad businesses, leading to runs and a reduction in credit supply to railroads. In turn, this led to widespread bankruptcies of railroad companies \citep{Richardson2015}. The associated downturn became known as the Long Depression. NBER dates the business cycle peak in October 1873. \Cref{fig:maps-1873} provides maps of runs and failures for the key months up to and during the panic.

The first hint of trouble on Wall Street was the suspension on July 19, 1873 of the \href{https://finhist.com/bank-runs/episodes/8778914790883.html}{Brooklyn Trust Company (New York, NY)}. We code this as a Suspension $\to$ Reopening. The suspension followed the death of Brooklyn Trust's president and discovery of defalcations.  The directors suspended preemptively in anticipation of a run, with a statement that included the following explanation: ``The unexpected action of the bank, however, yesterday afternoon, through which the company made its exchanges, declining any longer to clear for us, and which action will, in our opinion, necessarily cause a run to be made upon us which we cannot at once meet, has induced us to determine upon a general suspension of business as a matter of justice, not only to the depositors of the company, but also to the stockholders.''\footnote{\textit{The New York Herald}, July 20, 1873.} Reports claimed it had loan losses between \$225,000 and \$300,000. It had exposures to several railroad companies. \cite{Wicker2006} notes the episode was an early warning sign of financial fragility before the September panic, but it did not have any immediate effects. It was reopened on August 9, 1873 after stockholders invested additional capital and directors provided a guarantee fund.

The \href{https://finhist.com/bank-runs/episodes/2064105190885.html}{New York Warehouse \& Security Company (New York, NY)}, a private bank, failed shortly before the full outbreak of the panic on Wall Street (Suspension $\to$ Closure). \cite{OCCAnnualReport1873} stated that its failure marked the beginning of the panic (p. 26). Newspapers report it suspending on September 8, 1873. The bank had large and concentrated exposure to railroad firms, especially Missouri, Kansas, and Texas Railway, with which it had two overlapping directors. These positions were illiquid and likely impaired. It suspended when it could not refinance its short-term paper in the deteriorating money market. Its suspension ``added to the general demoralization'' on Wall Street.\footnote{\textit{The Sun}, September 9, 1873.}

The full-blown panic phase occurred with the run and failure of \href{https://finhist.com/bank-runs/episodes/9028931590885.html}{Jay Cooke \& Company (Philadelphia, PA)} (Run $\to$ Suspension $\to$ Closure). Cooke was a prominent private bank headquartered in Philadelphia with branches in New York City and Washington, D.C. It had been the government's chief financier during the Civil War. It had heavy exposure to Northern Pacific Railroad, whose bonds it was trying to market. It experienced a large deposit drain on September 18, 1873, forcing it to suspend on the same day. The suspension was a ``financial thunderbolt'' that had ``a great effect upon the Stock Exchange, and was the great topic of discussion in financial circles.''\footnote{\textit{New-York Tribune}, September 19, 1873.} It was the most important bank that failed in the Panic of 1873 according to \cite{FriedmanSchwartz} and the immediate trigger of the panic \citep{Rockoff2021}. Reports describe a sharp fall in stocks, and an increase in money market rates and gold prices in response. On September 20, the New York Stock Exchange closed its doors for ten days, the first such closing in its history. A receiver was appointed at Jay Cooke \& Co. on January 15, 1874.

The troubles of Jay Cooke \& Company led to the run on \href{https://finhist.com/bank-runs/episodes/100885.html}{First National Bank of Philadelphia, PA}. The main stockholders were members of the firm of Jay Cooke \& Co. It was mismanaged and had large advances exceeding regulations to Cooke, which led to its failure.

The failure of Jay Cooke \& Company also led to the run, suspension, and closure of the \href{https://finhist.com/bank-runs/episodes/2600885.html}{First National Bank of Washington, D.C.} on September 19. Its president was Henry D. Cooke, brother of Jay Cooke, Republican political insider, and first territorial governor of the District of Columbia. Like the FNB of Philadelphia, it had exposures to Jay Cooke \& Co. exceeding National Currency Act legal lending limits. One report notes the bank had lent Jay Cooke \& Co. \$600,000 with ``no security, except \$100,000 in Northern Pacific bonds.''\footnote{\textit{Nashville Union and American}, August 1, 1874.} The bank had taken Northern Pacific bonds at 83 percent of par value plus interest, but in August 1874 they were reported to be worth only 30 percent of par. Former President Andrew Johnson was reported to have a deposit account with the bank. President U.S. Grant also had an account that was overdrawn, and he later paid the debt to the receiver. Ultimately, the receivership paid out 100\% to depositors, so this failure was among the 19\% of national bank failures without losses to depositors from 1863 to 1934 \citep{CLV2026}. According to \cite{OCCAnnualReport1873}, the failure of national banks in 1873 ``may be attributed to the criminal mismanagement of their officers, or to the neglect or violation of the act on the part of their directors,'' so this bank was closed, despite possibly being solvent, for legal violations.\footnote{\cite{Conti2025} notes that the Panic of 1873 led to the use of forbearance as a new tool, whereby the OCC sometimes decided not to exercise its available authority to initiate receiverships for all legal violations, such as a failure to maintain required reserves. In this sense, failure involved some supervisory discretion.}

In New York, other prominent runs, suspensions, and/or failures after Jay Cooke's suspension included: the temporary three-month suspension of the private bank \href{https://finhist.com/bank-runs/episodes/8076409690885.html}{Fisk \& Hatch} of New York, which had large railroad securities holdings; the run on \href{https://finhist.com/bank-runs/episodes/29000885.html}{Fourth National Bank of New York}, which received NY clearinghouse loan certificate support, and did not suspend or fail; the temporary suspension of \href{https://finhist.com/bank-runs/episodes/1479944390885.html}{National Trust Company of New York} to avoid liquidating assets at a loss, while ``calling in'' loans ``as rapidly as possible,'' contributing to a credit crunch for borrowers.\footnote{\textit{New-York Tribune}, September 23, 1873.}

The largest national bank failure in the Panic of 1873 by assets and capital stock was that of the \href{https://finhist.com/bank-runs/episodes/137200885.html}{National Bank of the Commonwealth (New York, NY)} (Run $\to$ Suspension $\to$ Closure). The bank was subject to a run on September 20, 1873 after the failure of Jay Cooke \& Co and the ensuing panic. It was closed by the OCC and put into receivership on September 22, 1873. The Comptroller later stated that a receiver had been appointed for violation of the March 1869 certified-check act, which authorized appointment of a receiver when a national bank certified checks for drawers without sufficient funds on deposit \citep{OCCAnnualReport1873}.\footnote{The \textit{Commercial and Financial Chronicle} \citep[cited in][]{Sprague1910} wrote ``The failure of the Commonwealth Bank was occasioned by permitting a banking firm to overdraw their account some \$200,000, and the bank has also gone into a receiver's hands. If the accounts be true, it appears that the depositors both in the bank and trust company will be almost or wholly free from loss whenever the market becomes settled so that securities can be sold at a fair price.''}

The evidence on the solvency of the National Bank of the Commonwealth is mixed. The OCC classified 20\% of assets as doubtful, 19.2\% as worthless, and said it failed because of losses (``Injudicious banking and depreciation of securities''). This suggests the assets were impaired, though these figures are better than for the average failed bank \citep[see][]{CLV2026}. One newspaper reported its capital was ``impaired 38.5 per cent.'' However, the bank later paid out 100\% to depositors. The bank president also claimed it was solvent.\footnote{``...president of the Bank, is very much interested in the matter and has been here to urge that the receiver should be withdrawn. He said the action of the officers of the bank was hasty; that they asked for a receiver without due deliberation when the panic came on, and that the bank was and is in a solvent condition. Controller Knox, who went to New-York on Saturday, will have a consultation with the receiver and be ready to decide the question when he returns.'' \textit{New-York Tribune}, October 7, 1873.}  The bank directors sent an application to the Comptroller to revive the bank by ``reducing or filling up the capital,'' but this was denied by a vote of the New York Clearinghouse.\footnote{``The Controller directed the receiver to submit his report to the Loan Committee of the Clearing-house. That Committee accordingly met at noon on Monday last... After due consideration of the condition of the bank, it was voted that its affairs ought to be wound up. The vote was unanimous.'' \textit{New-York Tribune}, October 7, 1873.} Overall, the bank was not insolvent to depositors ex post, but its capital was impaired and it was not regarded as viable by the receiver or Clearinghouse committee. The president George Ellis was arrested in June 1874 for misappropriation of funds.\footnote{``George Ellis, formerly president of the National Bank of the Commonwealth, which failed in the panic of last fall, was arrested... on an indictment in the United States circuit court for having misapplied \$53,000 of its money in December, 1872, to buy stock of the bank, which he afterwards (as alleged) deposited with the Security National Bank to obtain credit on, drawing against the value of the stock so deposited. It is also charged that false entries were made to cover these transactions'' \textit{Evening Star}, July 3, 1874.}

The failure of Jay Cooke and other institutions was followed by an increase in withdrawals, especially by the country correspondents of the New York banks \citep[][p. 26]{OCCAnnualReport1873}. This led the New York Clearinghouse Association to meet September 20, where it decided to issue clearinghouse loan certificates. Banks could borrow certificates up to 75\% of approved securities or bills receivable deposited in order to settle balances at the clearinghouse. The clearinghouse also pooled reserves by treating legal tenders of the associated banks as a ``common fund, held for mutual aid and protection'' \citep[][p. 27]{OCCAnnualReport1873}. New York banks then partially suspended currency payments on September 24 in New York, but continued making payments to interior banks \citep{Wicker2006}. Suspension followed in other large cities, generally lasting 40 days \citep[][p. 27]{OCCAnnualReport1873}. The decisive action by the clearinghouse to pool member reserves helped contain the acute panic phase of the crisis.\footnote{While loan certificates were issued by the NYCH in subsequent panics, reserve pooling was not repeated. Thus, the Panic of 1873 was a high mark for the NYCH's willingness to act collectively as a lender of last resort \citep{Wicker2006}.}

Runs also occurred elsewhere in the country, but there were relatively few failures. \cite{Wicker2006} notes that the panic next spread down the eastern seaboard to Petersburg, VA, Richmond, VA, and Augusta, GA. For instance, in Petersburg, the \href{https://finhist.com/bank-runs/episodes/154800885.html}{Merchants National Bank}, \href{https://finhist.com/bank-runs/episodes/137800885.html}{First National Bank of Petersburg}, and \href{https://finhist.com/bank-runs/episodes/3284103390885.html}{Planters \& Mechanics Bank} all suspended on September 23, 1873. The OCC classified the failure of both the Merchants NB and First NB as caused by fraud. The president of Merchants, who was involved in speculating in common railroad stock and ``brought himself and all the firms with which he was connected into debt'' ``by a princely style of living,'' was charged with embezzlement and supposedly entered false entries of board minutes to ratify his borrowings.\footnote{\textit{The New York Herald}, November 27, 1873.} At suspension, the OCC assessed 27.5\% of Merchants' assets as doubtful and 35.1\% as worthless; depositors ultimately recovered only 34\%. This illustrates how runs sometimes revealed deeper fraud and insolvency. In a related example, the suspension of all banks in New Orleans ultimately revealed the poor condition of the  \href{https://finhist.com/bank-runs/episodes/182500886.html}{New Orleans National Banking Association}, which was put into receivership.

In Chicago, banks were reported to have adopted the ``thirty and sixty day rule,'' thereby not redeeming larger deposits on demand and only paying out up to \$100. The \href{https://finhist.com/bank-runs/episodes/23600885.html}{Third National Bank of Chicago} (Run $\to$ Suspension $\to$ Reopening) experienced a run starting on September 22 and suspended on September 27 as part of a suspension coordinated by the Chicago clearinghouse. One newspaper noted ``it is believed to be a perfectly sound institution, and the suspension is thought to be merely temporary.''\footnote{\textit{The Rock Island Daily Argus}, September 29, 1873} It reopened on October 8 with ``an entire absence of anything like a run.'' Upon reopening it ``received more than it paid out during the day, a fact which is taken as proof not alone of confidence in the bank.''\footnote{\textit{Nashville Union and American}, October 9, 1873.}

The \href{https://finhist.com/bank-runs/episodes/69800885.html}{Union National Bank} of Chicago (Run $\to$ Suspension $\to$ Reopening) suspended on September 30, 1873. \cite{Wicker2006} states that the importance of the suspension in spreading fear throughout the Midwest ``cannot be overstated'' (p. 24). The president of Union National stated the bank was solvent. Due to continued pressure from country banks, the bank announced it would go into voluntary liquidation. However, by October it was reported it ``contemplates resumption'' and it reopened on October 16. The example illustrates that a solvent bank was able to avoid failure due to a pure but severe liquidity problem. The panic nevertheless produced a large credit contraction in Chicago \citep{Wicker2006}.

\subsubsection{The Incipient Panic of 1884}

\begin{figure}[ht]
  \centering
  \caption{Incipient Panic of 1884}
  \label{fig:maps-1884}
  \begin{subfigure}[t]{0.49\textwidth}
    \caption{April 1884}
    \label{fig:map-1884-4}
    \includegraphics[width=\linewidth]{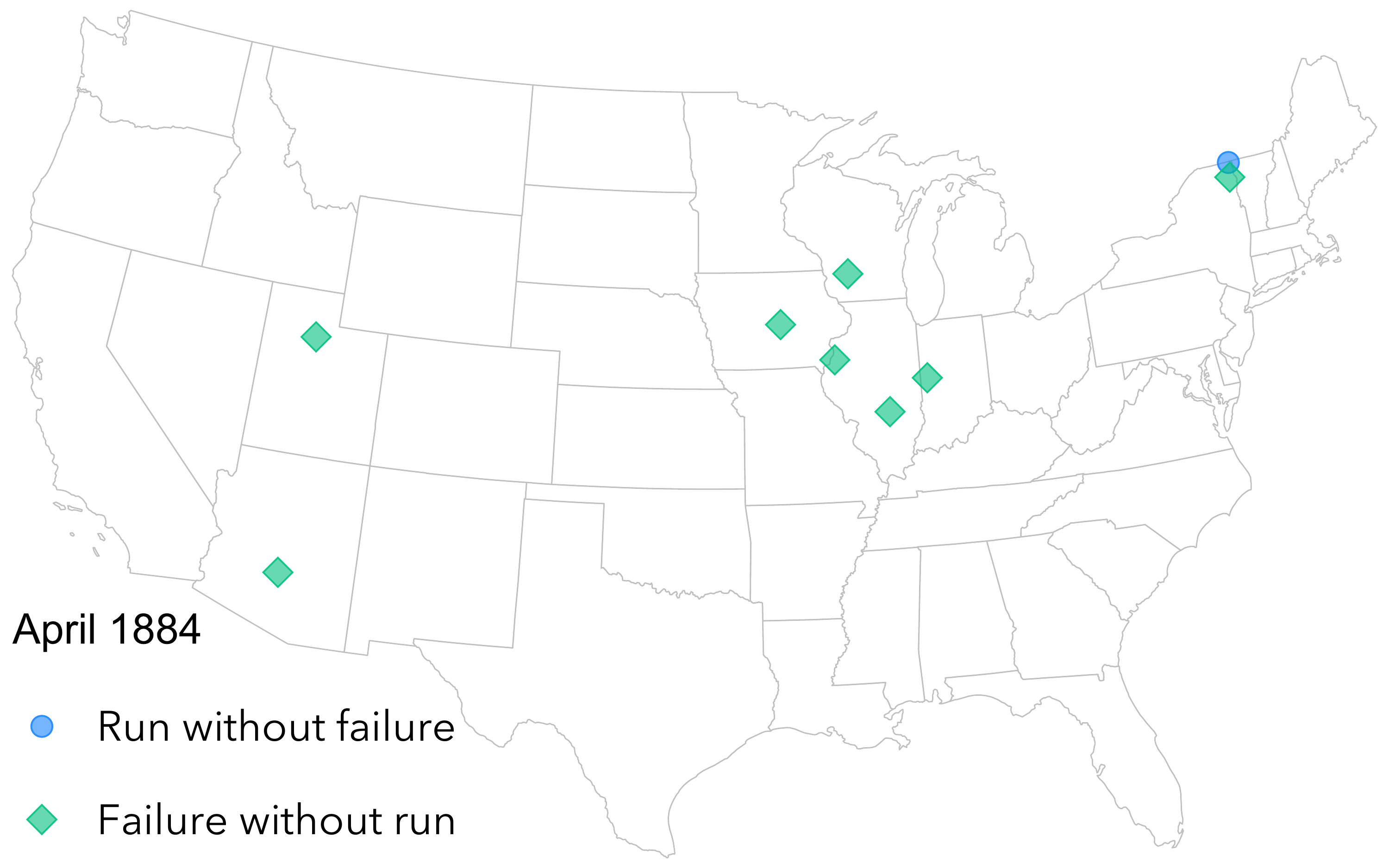}
  \end{subfigure}\hfill
  \begin{subfigure}[t]{0.49\textwidth}
    \caption{May 1884}
    \label{fig:map-1884-5}
    \includegraphics[width=\linewidth]{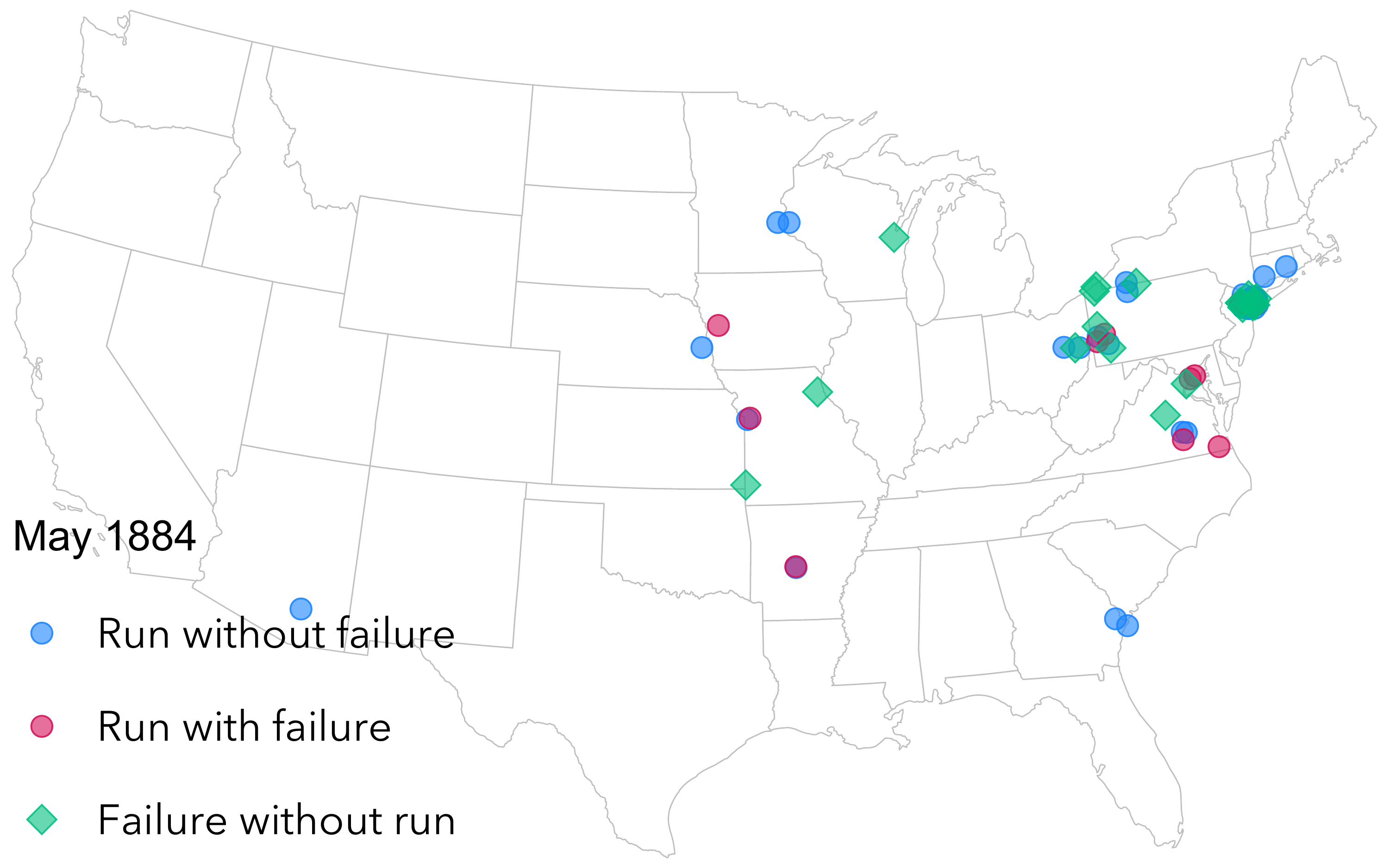}
  \end{subfigure}

  \vspace{0.5em}

  \begin{subfigure}[t]{0.49\textwidth}
    \caption{June 1884}
    \label{fig:map-1884-6}
    \includegraphics[width=\linewidth]{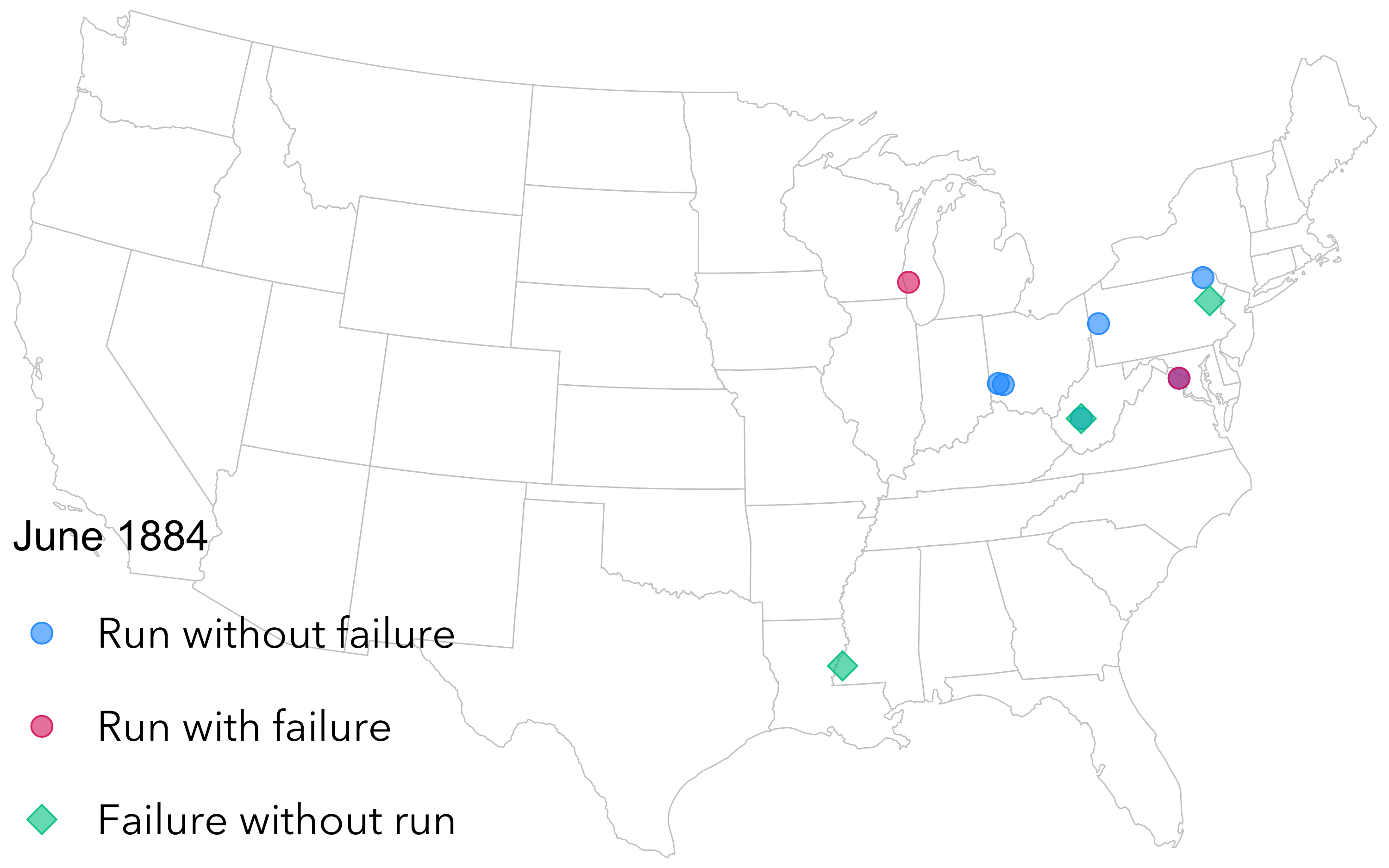}
  \end{subfigure}\hfill
  \begin{subfigure}[t]{0.49\textwidth}
    \caption{July 1884}
    \label{fig:map-1884-7}
    \includegraphics[width=\linewidth]{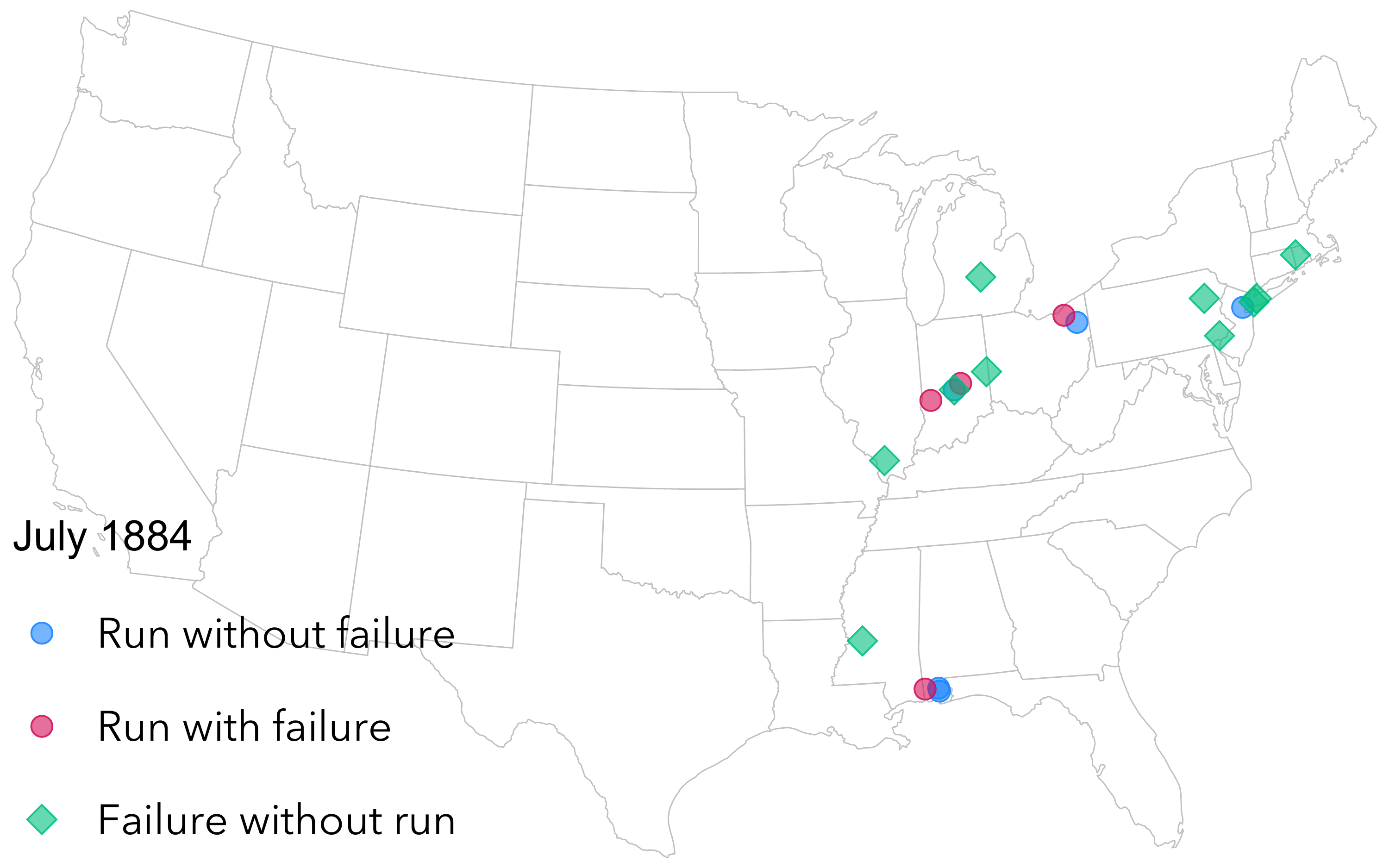}
  \end{subfigure}

  \begin{minipage}{\textwidth}
    \footnotesize
    \textit{Notes:} Each panel shows the location of bank distress episodes in a given month during the Panic of 1884. Blue circles indicate banks subject to a run that did not subsequently fail. Red circles indicate banks subject to a run that subsequently failed. Green diamonds indicate banks that failed without a reported run.   \end{minipage}
\end{figure}

The Panic of 1884 was mild \citep{Wicker2006,Rockoff2021}. According to \citet{Wicker2006}, there was no general loss of depositor confidence or suspension of cash payment, so it does not rise to a full-scale banking panic but rather qualifies as an ``incipient'' panic. The panic was centered on Wall Street. It provides an example of how clearinghouse support and bank examination helped separate weak from solvent institutions. \Cref{fig:maps-1884} provides a map of runs and failures during key months up to, during, and after the panic. Bank distress is visibly less widespread than during 1873 or 1893. The panic occurred in the context of a recession that, according to the NBER, had started already two years earlier in March 1882 and ended in May 1885.

The panic was triggered by the failure of the \href{https://finhist.com/bank-runs/episodes/121501013.html}{Marine National Bank} of New York City on May 6, 1884 when it was unable to pay a \$500,000 debt to the NY clearinghouse. A newspaper reported ``a visit to the Bank discovered some fifty disconsolate depositors standing in the pouring rain around the closed doors.''\footnote{\textit{The Dallas Daily Herald}, May 7, 1884.} The bank had losses from large overdrafts by the firm Grant \& Ward, which experienced losses following the recent decline in the stock market.\footnote{Grant refers to the President and General. Though Grant lent his name to Grant \& Ward, \cite{Chernow2017} portrays Grant as financially naive and unaware of Ward's fraud. Ward had run a Ponzi scheme and was deeply insolvent \citep{Sprague1910}. Marine's president Fisk was also a partner at Grant \& Ward.} The overdrafts amounted to legal violations by Marine NB, as the bank certified checks for Grant \& Ward before receiving a deposit \citep{OCCAnnualReport1884}. Newspapers also refer to heavy real estate advances by Marine NB. It was put into receivership on May 13. The national bank examiner described the bank as ``hopelessly insolvent'' \citep{Conti2025}. The depositor recovery rate was 83.5\%, consistent with the examiner's assessment. The OCC gave ``fraud'' as the cause of failure, and assessed 13.9\% of assets doubtful and 26.7\% worthless. Both Marine NB's president Fisk and Ward were sentenced to prison for 10 years \citep{Wicker2006}. Newspapers report the stock market ``became panicky'' following the failure.\footnote{\textit{Evening Star}, May 6, 1884.}

On May 14, the \href{https://finhist.com/bank-runs/episodes/6201013.html}{Second National Bank} of New York was subject to a run when it was revealed that its president John Eno had misused \$3 million in bank funds to speculate and suffered losses. We classify this as a ``run only.'' Eno's wealthy father, Amos Eno, came to the rescue to cover the loss, which allowed the bank to avoid suspending. John Eno resigned as president. Despite assurances that the loss had been covered, depositors ran on the bank. After the Comptroller telegraphed the OCC bank examiner to investigate the bank, the examiner reported: ``There is a run on the Second National Bank. I have secured guarantees for all deficiencies, and money will be supplied the bank until the run ceases. The capital is left, with a small surplus.''\footnote{\textit{The Indianapolis Journal}, May 15, 1884.} The bank posted the examiner's solvency report: ``The certificate of the examiner that the bank was solvent was posted in both front windows as early as 11 o’clock, and attracted much attention.''\footnote{\textit{New-York Tribune}, May 15, 1884.} The run was ``finally broken'' by May 15,\footnote{\textit{New-York Tribune}, May 16, 1884.} and the bank was ``completely secured from its difficulties by depositors who brought in large amounts of ready money to the relief of the bank.''\footnote{\textit{The Daily Enterprise}, May 16, 1884.}  This case provides an example of a run following news of a loss, where failure was prevented by recapitalization, accommodating withdrawals, and information provision to signal solvency by a bank examiner.\footnote{\cite{Conti2025} discusses this as an example of valuable information provision by banking supervision. The OCC could not provide liquidity, so its main tools of supervision were information production and provision, forbearance or receivership, and chartering.}

\href{https://finhist.com/bank-runs/episodes/112101013.html}{Metropolitan National Bank} of New York City (Run $\to$ Suspension $\to$ Reopening) provides an example of a run where failure was averted through suspension, clearinghouse lending, examination, and information provision. The run and suspension nevertheless appear to have done lasting damage to the bank's reputation.

Metropolitan, a major correspondent bank, experienced a run on May 14, 1884 following reports that it had made loans to a Wall Street firm run by Metropolitan NB's president Seney's two sons and son-in-law, along with Southern railroad investments subject to large losses. The bank suspended the same day: ``when the cashier of the Metropolitan slammed down his slide and ordered the doors closed a wild scene of tumultuous confusion and excitement followed.''\footnote{\textit{The Canton Advocate}, May 22, 1884.}

In anticipation of large withdrawals by country correspondent banks, the New York Clearinghouse called a meeting on May 14 to address the crisis \citep{OCCAnnualReport1884}. The NYCH took two key steps. First, it resolved to issue loan certificates up to 75\% of the value of bills or securities pledged.\footnote{Unlike in 1873, however, the NYCH did not pool legal tenders in a common fund.} Second, it appointed a committee to examine Metropolitan NB to verify its solvency. Its securities were assessed to provide good collateral, allowing the bank to borrow against them. This allowed the bank to resume on May 15: ``A committee of the Clearing House investigated the affairs of the bank and it was decided that its assets would justify the resumption of business. It will, therefore, open its doors to-day.''\footnote{\textit{New-York Tribune}, May 15, 1884.} The reopening calmed the panic; ``people were gradually returning to their senses.''\footnote{\textit{The Canton Advocate}, May 22, 1884.} The OCC examiner also confirmed that the bank was sound, saying ``It was a terrible blow to this institution, and a great mistake was made when its doors were closed, and it will take time for it to recover its former standing, but it is sound and can pay every dollar of its liabilities to-day.''\footnote{\textit{New-York Tribune}, May 25, 1884.} However, the bank ultimately went into voluntary liquidation in November 1884, after it lost nearly 90\% of its deposits over the summer of 1884 due to a ``lack of confidence'' following the suspension \citep[][p. 37]{OCCAnnualReport1884}.

\subsubsection{The Incipient Panic of 1890}

\begin{figure}[ht]
  \centering
  \caption{Incipient Panic of 1890}
  \label{fig:maps-1890}
  \begin{subfigure}[t]{0.49\textwidth}
    \caption{September 1890}
    \label{fig:map-1890-9}
    \includegraphics[width=\linewidth]{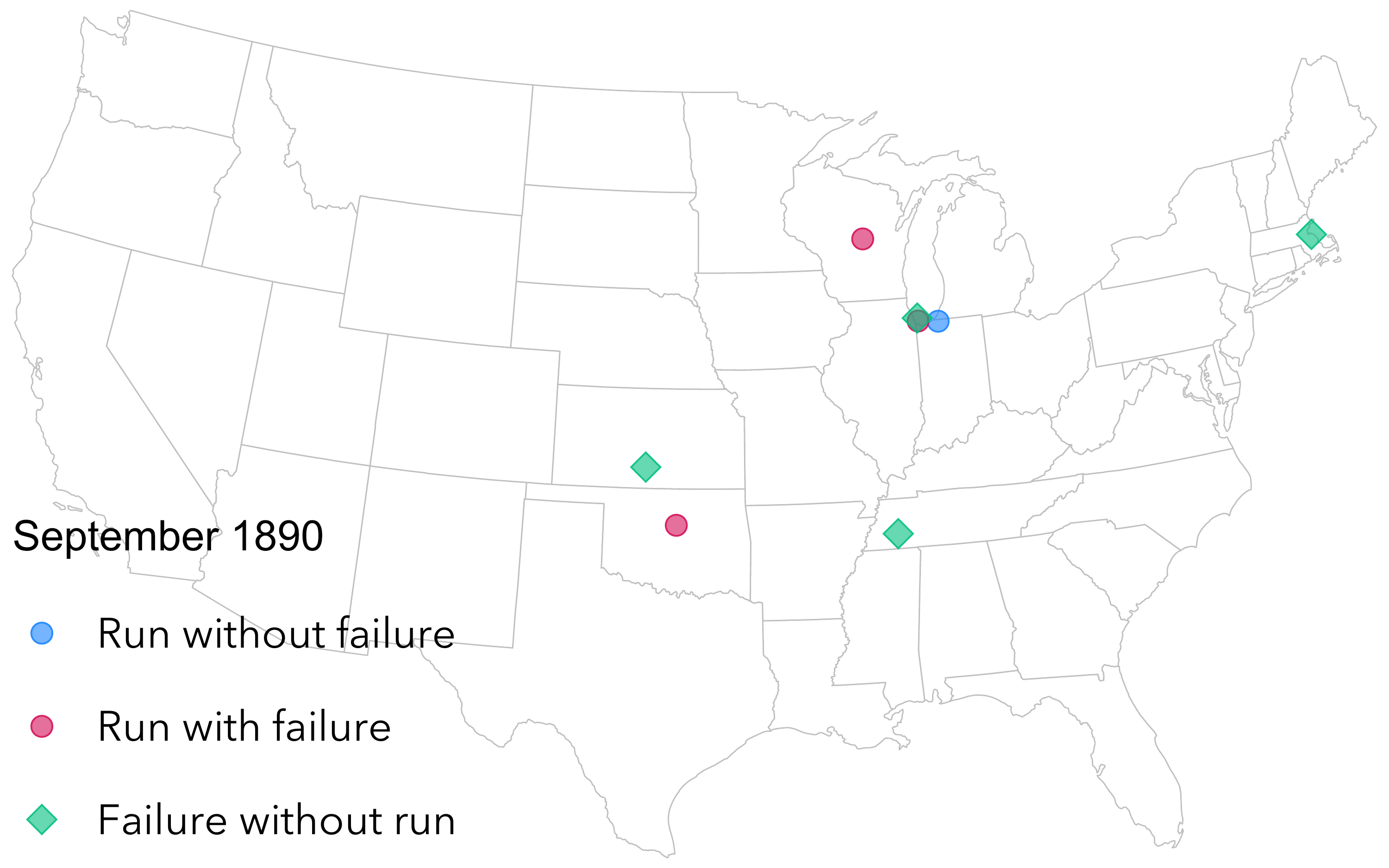}
  \end{subfigure}\hfill
  \begin{subfigure}[t]{0.49\textwidth}
    \caption{October 1890}
    \label{fig:map-1890-10}
    \includegraphics[width=\linewidth]{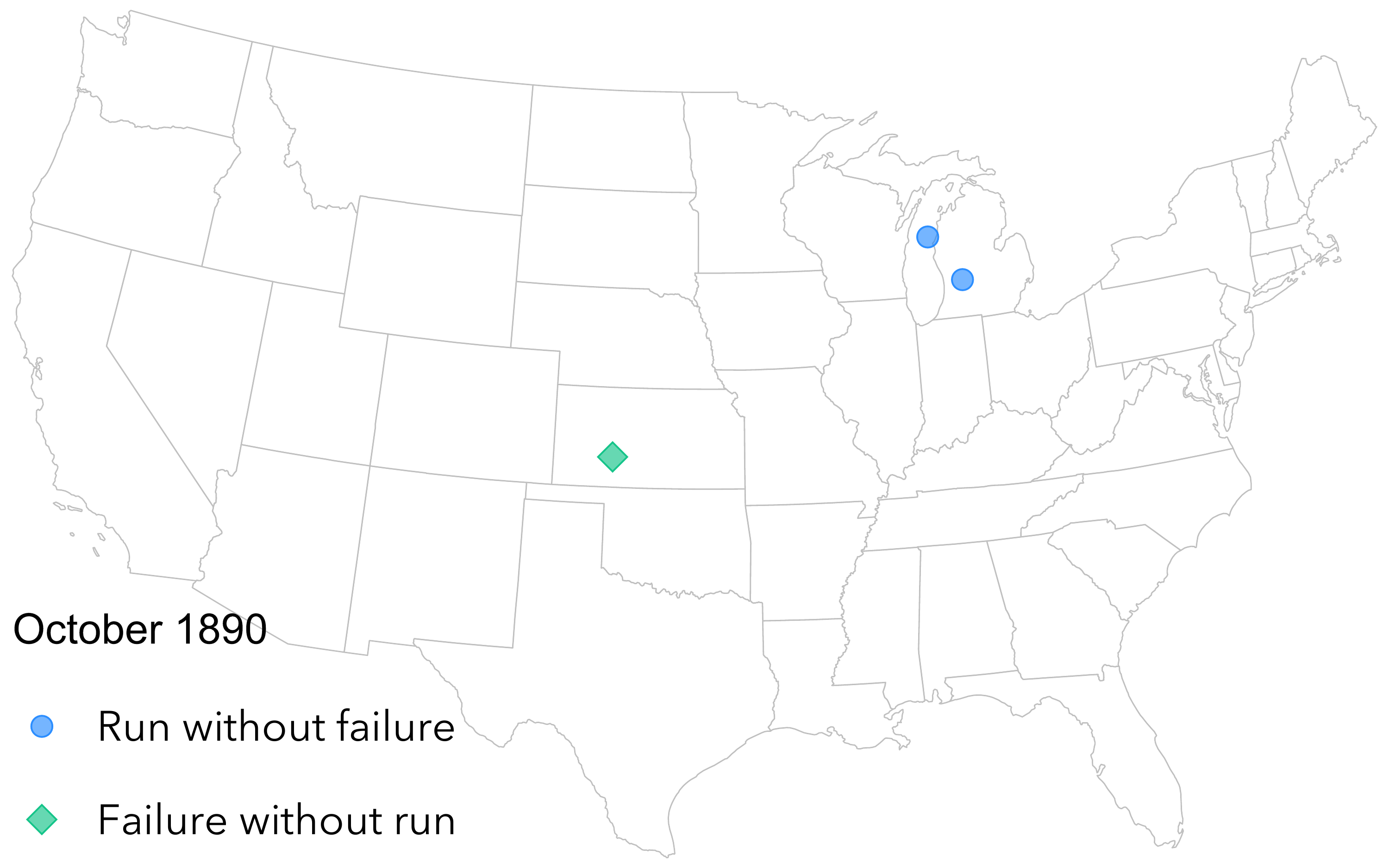}
  \end{subfigure}

  \vspace{0.5em}

  \begin{subfigure}[t]{0.49\textwidth}
    \caption{November 1890}
    \label{fig:map-1890-11}
    \includegraphics[width=\linewidth]{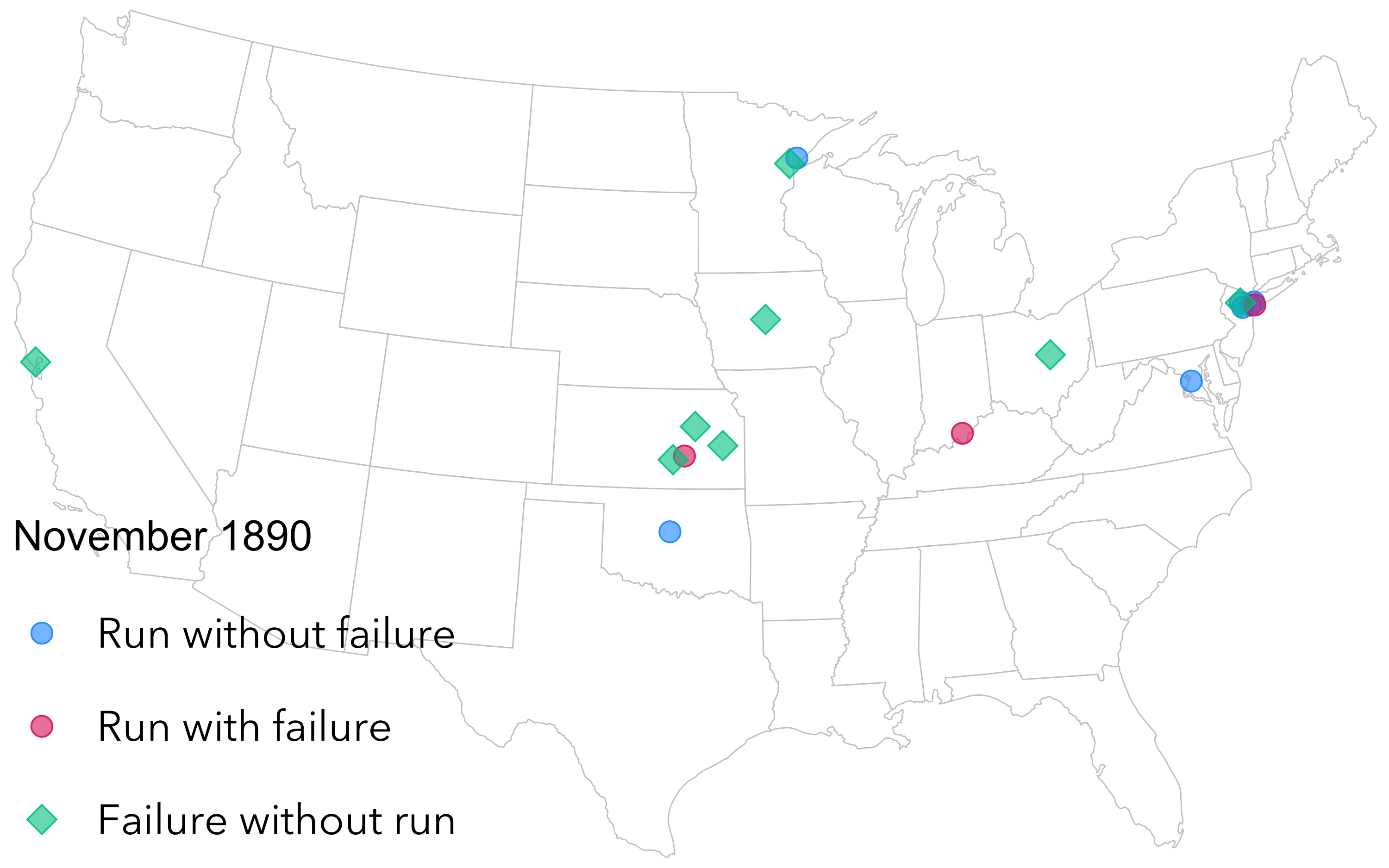}
  \end{subfigure}\hfill
  \begin{subfigure}[t]{0.49\textwidth}
    \caption{December 1890}
    \label{fig:map-1890-12}
    \includegraphics[width=\linewidth]{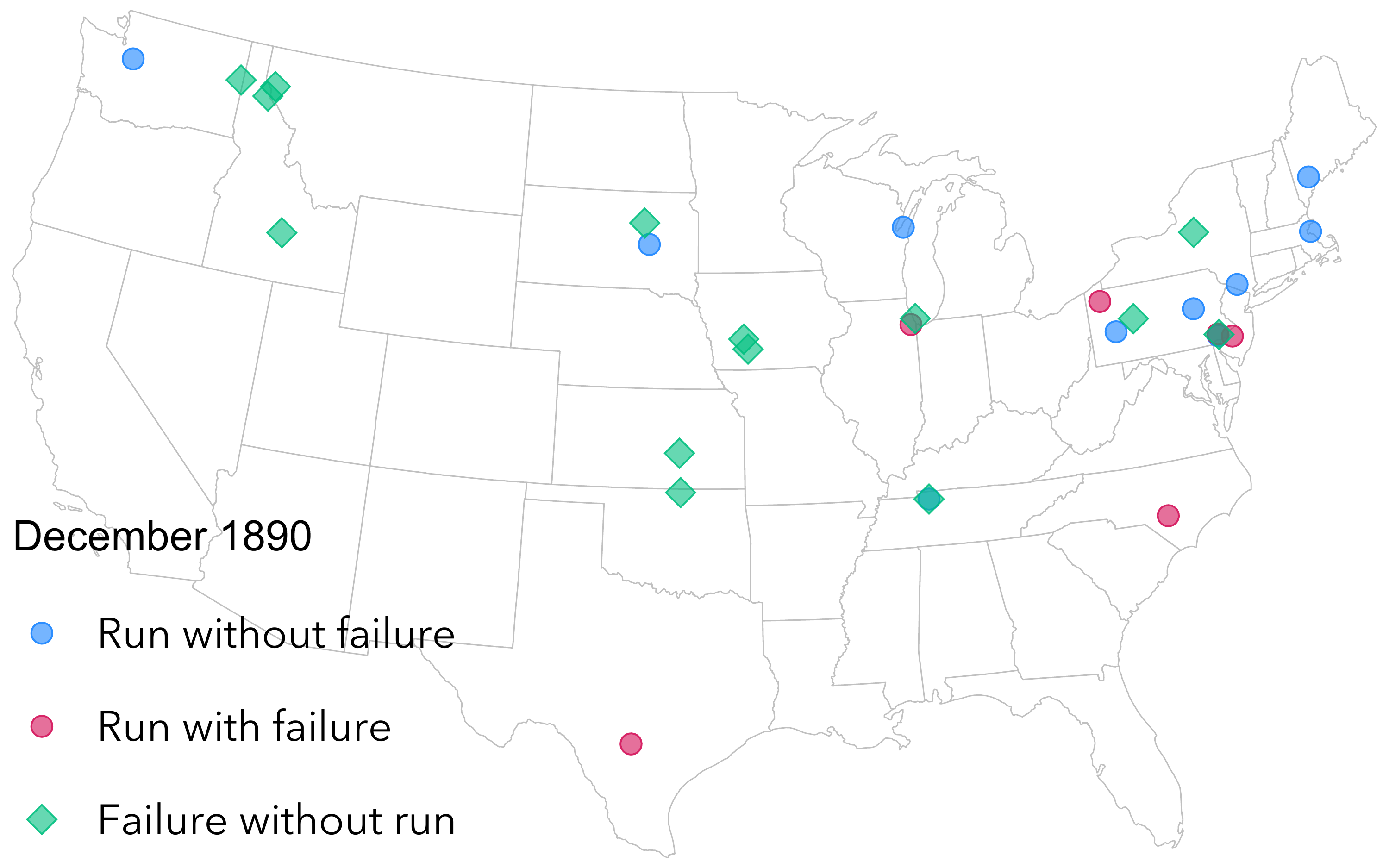}
  \end{subfigure}

  \begin{minipage}{\textwidth}
    \footnotesize
    \textit{Notes:} Each panel shows the location of bank distress episodes in a given month during the Panic of 1890. Blue circles indicate banks subject to a run that did not subsequently fail. Red circles indicate banks subject to a run that subsequently failed. Green diamonds indicate banks that failed without a reported run.
  \end{minipage}
\end{figure}

Like 1884, the panic in November of 1890 was also limited. \cite{Wicker2006} refers to it as a ``banking disturbance'' and argues it might have evolved into a full-scale panic were it not for the intervention of the New York Clearinghouse. \cite{OCCAnnualReport1890} and \cite{Sprague1910} refer to it as an episode of ``monetary stringency.'' It is important to distinguish these money market crises from crises of bank fundamentals tied to severe real economic distress. \Cref{fig:maps-1890} confirms the distress was relatively limited.

The key event in 1890 was the suspension of the Wall Street broker \href{https://finhist.com/bank-runs/episodes/8119717491091.html}{Decker, Howell \& Company} in New York on November 12, 1890 (Suspension $\to$ Reopening). Stock price declines lowered the value of its stock holdings, which included the North American Company, a large utility and railway conglomerate, whose price had fallen 44\% \citep{Wicker2006}. Tight money market conditions meant the firm could not roll over large daily borrowings. Joseph Decker stated: ``Our trouble has come upon us chiefly from the troubles in the money market, the recent stringency in money, the want of confidence which has been felt in Wall Street since August last.''\footnote{\textit{New-York Tribune}, November 12, 1890.} The failure of Decker, Howell \& Co. led to a collapse in stock prices.

The assignee stated that Decker was solvent but held illiquid assets: ``The liabilities are about \$10,000,000 and the assets, at the present market price, largely exceed that sum. The liabilities are due almost entirely to banks and bankers on loans made and are all well secured. The cause of the suspension was the inability of the firm to borrow the necessary amount of cash required in the day's business. The firm's transactions were very large, it being necessary to borrow several millions daily. The firm had an abundance of collateral today, and it was not for lack of security, but the inability to make it available that caused the crash. As the securities are of a special line, there may be a disposition among the creditors to sacrifice them on the market, but such a course would be suicidal, The character of the securities show that their price on the market is far below their actual value, and if creditors have the good judgment to hold their securities, they will be amply protected.''\footnote{\textit{St. Paul Daily Globe}, November 12, 1890.}

Later, in December, a less optimistic report suggested the firm was nearly insolvent or insolvent at market prices: ``The schedules in assignment of Decker, Howell \& Co., bankers, show liabilities of \$9.430,331, nominal assets \$35,181,982, and actual assets \$8,760,357.''\footnote{\textit{The Seattle Post-Intelligencer}, December 6, 1890.} However, by January 1891, newspapers reported that it would resume and repay creditors in full.

Decker, Howell, and Co.'s troubles spilled over to its bank, the \href{https://finhist.com/bank-runs/episodes/7883195591091.html}{Bank of North America}, to which it had a \$1.4 million overdraft. This led the New York Clearinghouse to call an emergency meeting on November 11 to provide loan certificates to \href{https://finhist.com/bank-runs/episodes/7883195591091.html}{Bank of North America}, Mechanics and Trader's Bank, and \href{https://finhist.com/bank-runs/episodes/976831691091.html}{North River Bank}. Moreover, with the leadership of J.P. Morgan, nine banks provided \$100,000 to the Bank of North America and Mechanics and Trader's to cover their deficit at the NYCH. \cite{Wicker2006} argues that the NYCH intervention averted an incipient panic. The \href{https://finhist.com/bank-runs/episodes/976831691091.html}{North River Bank} ultimately failed in a run when it became known it received clearinghouse support: ``On account of the mention of the North River bank as one of the defaulters at the clearing house yesterday the depositors started a quiet but steady run on it.''\footnote{\textit{The Anaconda Standard}, November 13, 1890.}

\subsubsection{Panic of 1893}

\begin{figure}[ht]
  \centering
  \caption{Panic of 1893}
  \label{fig:maps-1893}
  \begin{subfigure}[t]{0.45\textwidth}
    \caption{April 1893}
    \label{fig:map-1893-4}
    \includegraphics[width=\linewidth]{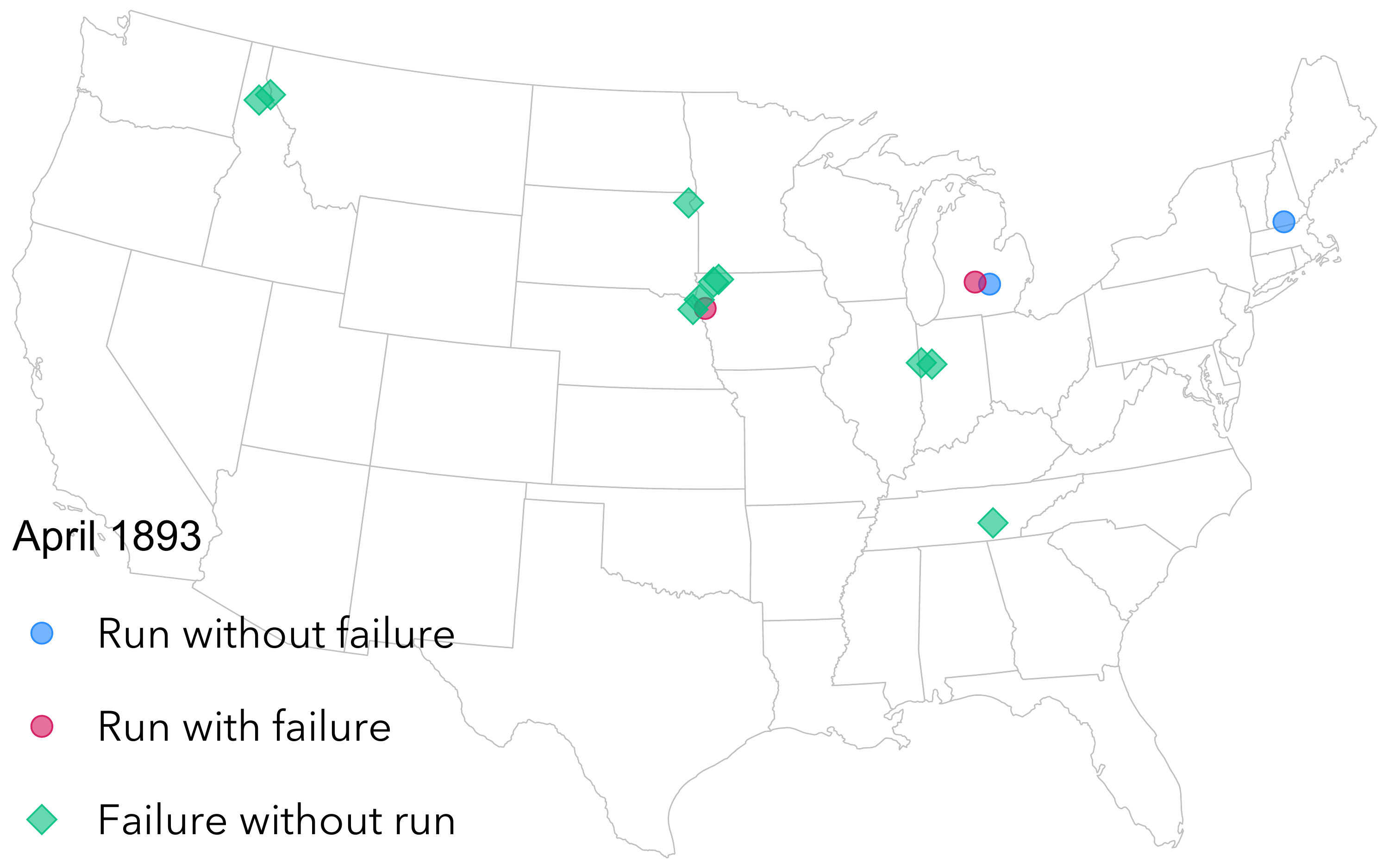}
  \end{subfigure}\hfill
  \begin{subfigure}[t]{0.45\textwidth}
    \caption{May 1893}
    \label{fig:map-1893-5}
    \includegraphics[width=\linewidth]{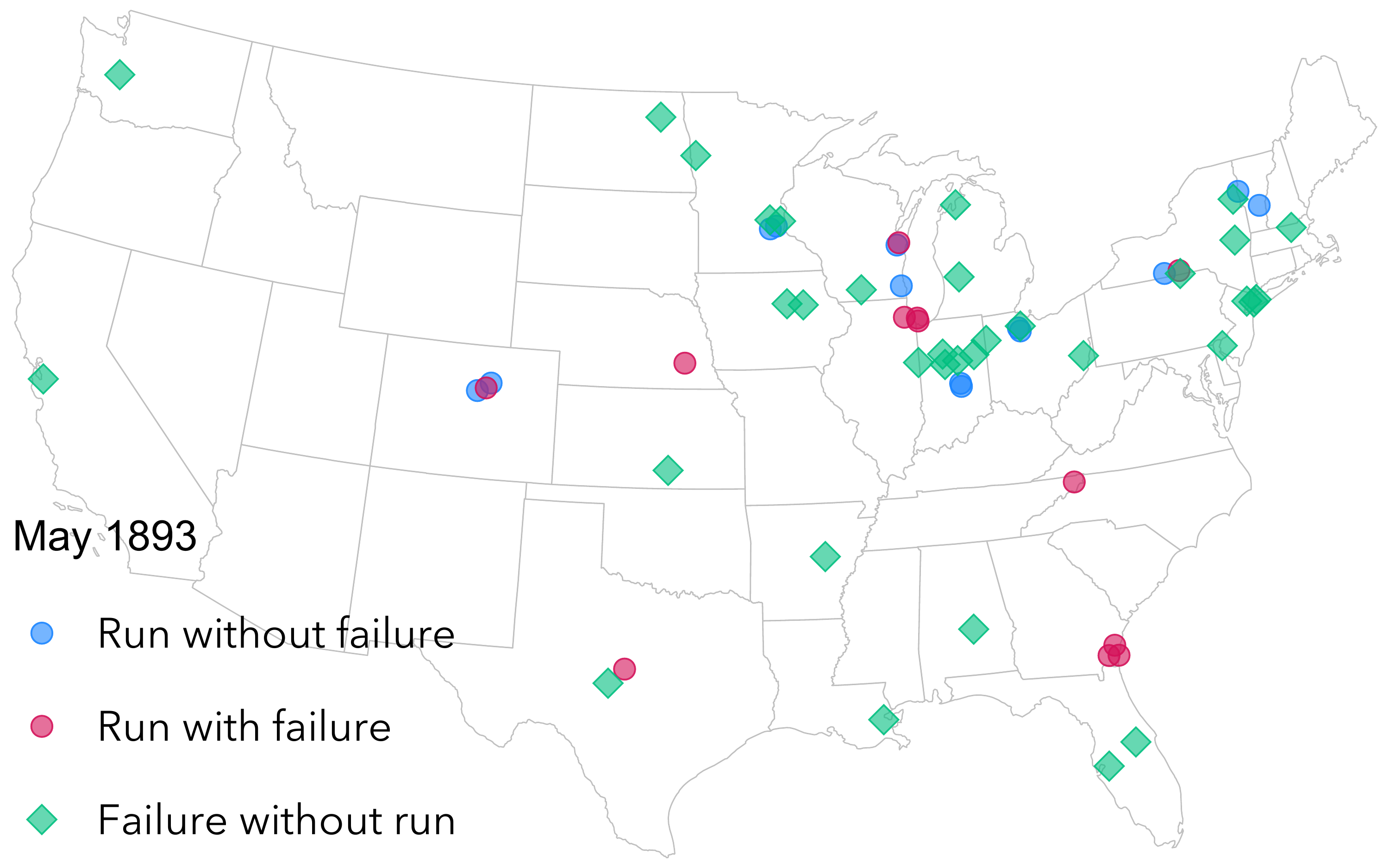}
  \end{subfigure}

  \vspace{0.5em}

  \begin{subfigure}[t]{0.45\textwidth}
    \caption{June 1893}
    \label{fig:map-1893-6}
    \includegraphics[width=\linewidth]{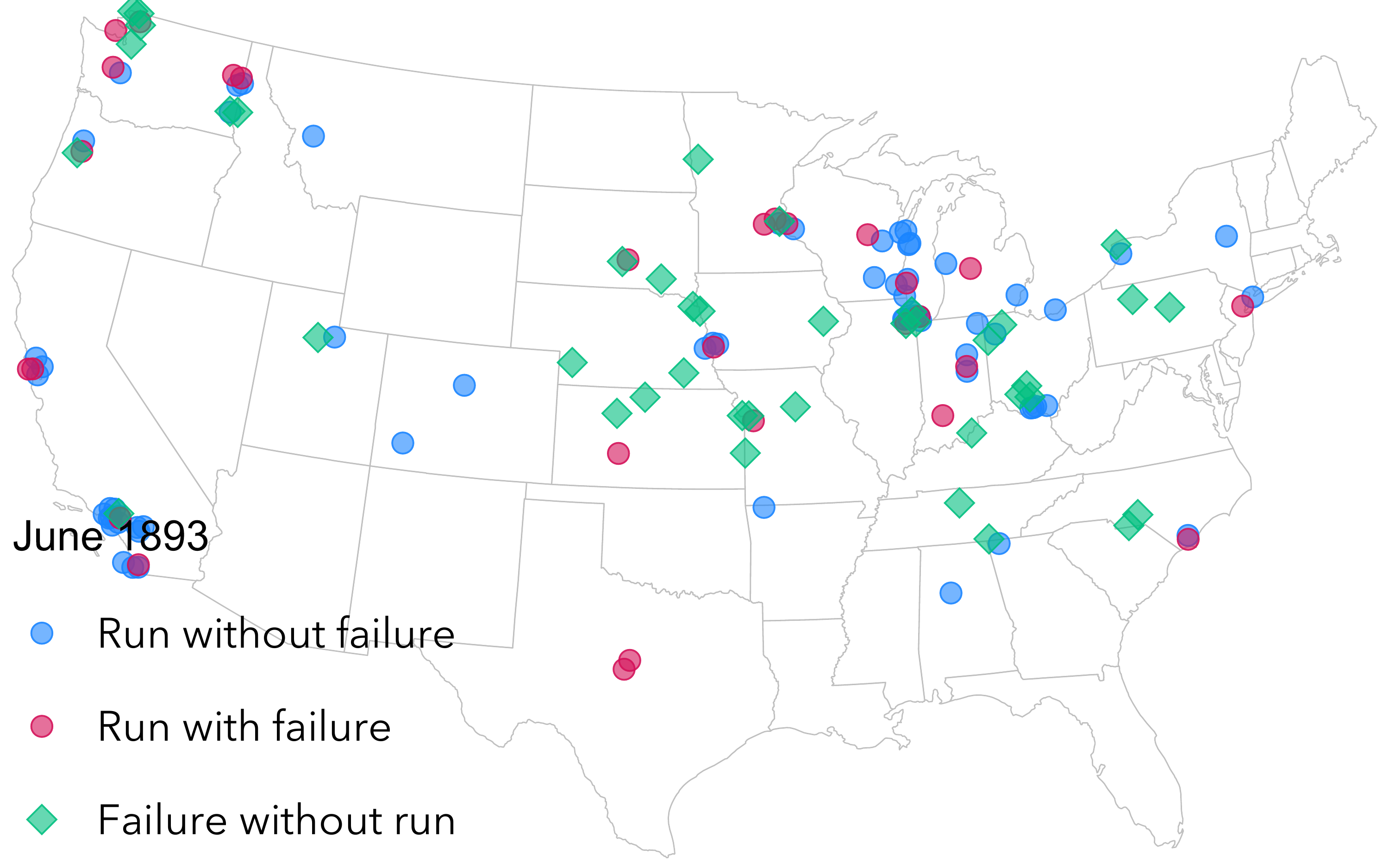}
  \end{subfigure}\hfill
  \begin{subfigure}[t]{0.45\textwidth}
    \caption{July 1893}
    \label{fig:map-1893-7}
    \includegraphics[width=\linewidth]{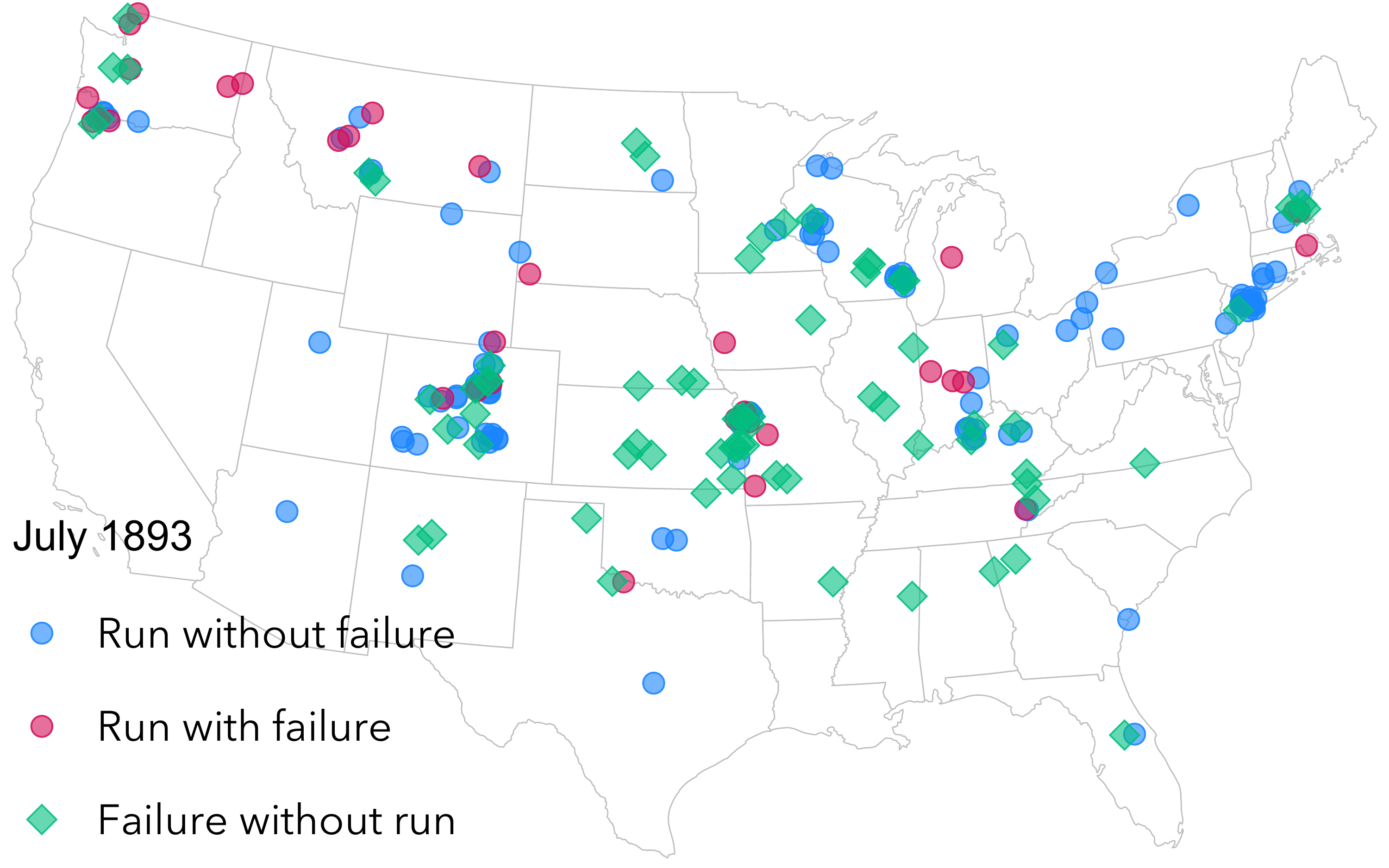}
  \end{subfigure}

  \vspace{0.5em}

  \begin{subfigure}[t]{0.45\textwidth}
    \caption{August 1893}
    \label{fig:map-1893-8}
    \includegraphics[width=\linewidth]{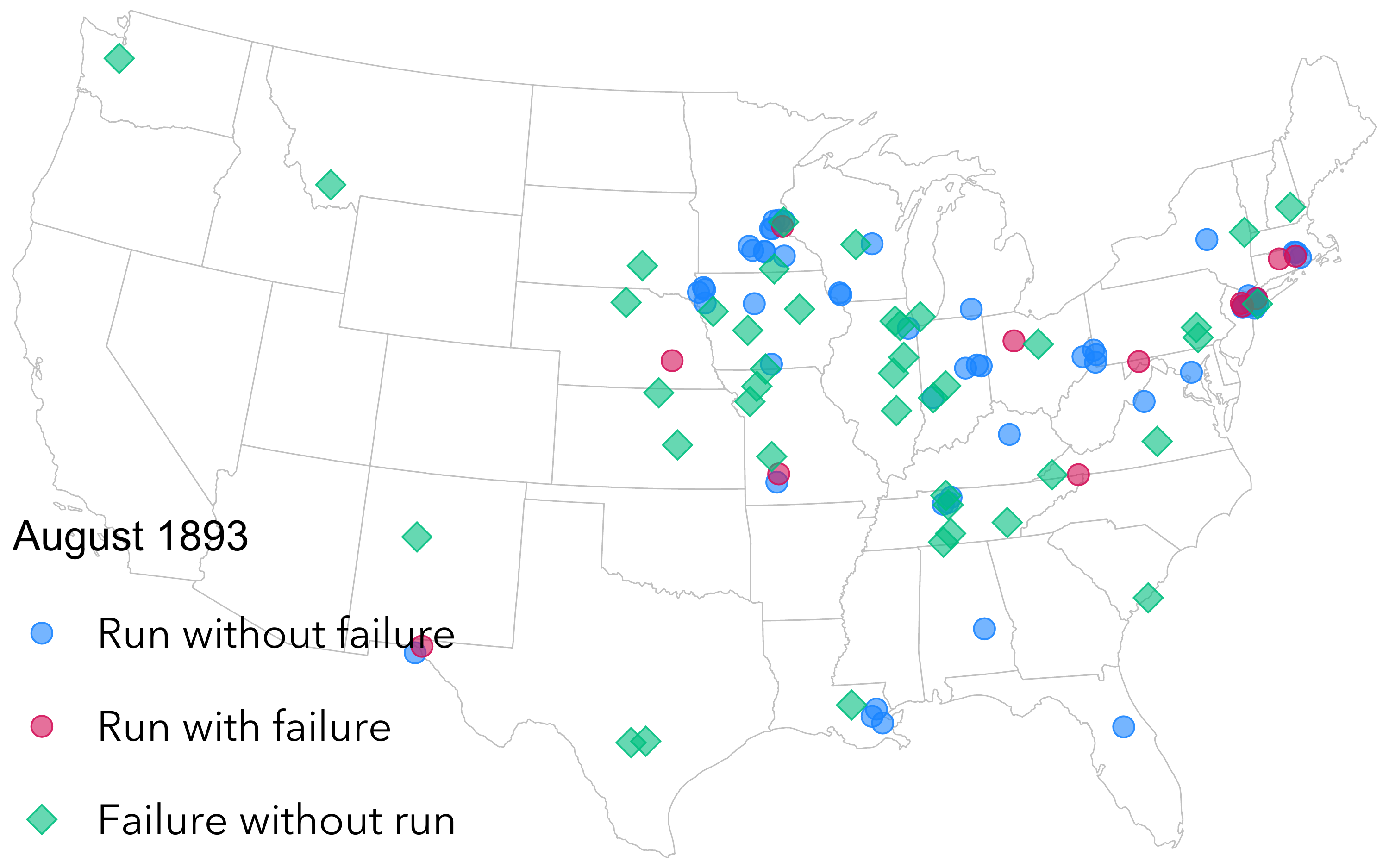}
  \end{subfigure}\hfill
  \begin{subfigure}[t]{0.45\textwidth}
    \caption{September 1893}
    \label{fig:map-1893-9}
    \includegraphics[width=\linewidth]{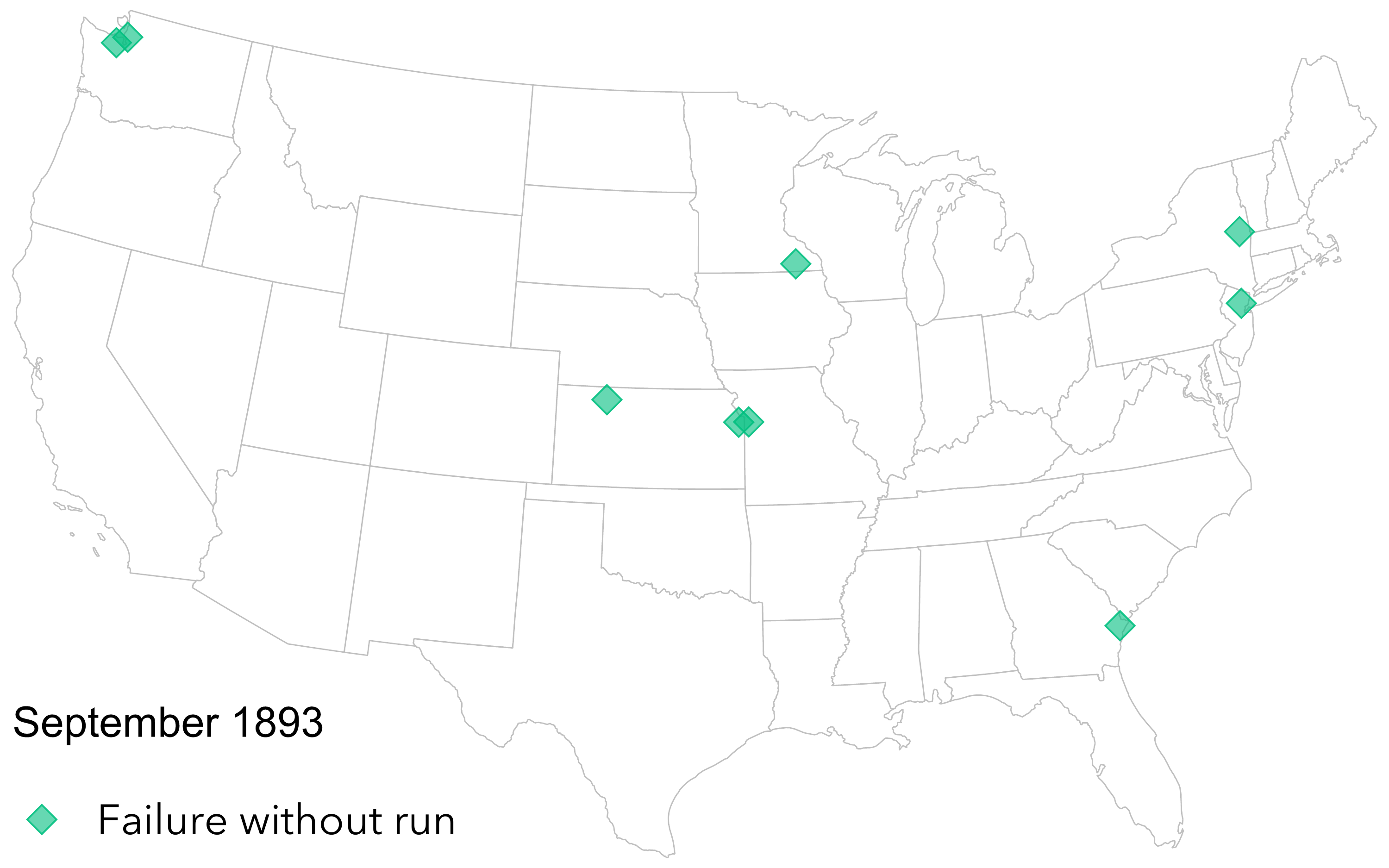}
  \end{subfigure}
  \begin{minipage}{\textwidth}
    \footnotesize
    \textit{Notes:} Each panel shows the location of bank distress episodes in a given month during the Panic of 1893. Blue circles indicate banks subject to a run that did not subsequently fail. Red circles indicate banks subject to a run that subsequently failed. Green diamonds indicate banks that failed without a reported run.
  \end{minipage}
\end{figure}

The Panic of 1893 was the most severe banking crisis of the National Banking Era. As seen in \Cref{fig:events}, it featured the highest rate of runs, as well as the highest rate of bank failures in this era \citep{CLV2026}. The crisis was preceded by a railroad investment and lending boom. The rate of bank failures started to rise gradually over 1890-92. The literature identifies two important precursors to the crisis \citep[e.g.,][]{Carlson2005}. The first was distress among nonfinancial firms, including railroads and silver mines, and a broader economic downturn that the NBER dates to have started in January of 1893. Prominent nonfinancial failures include the Philadelphia and Reading Railroad, which failed in February, and the National Cordage Company, which failed in early May. Second, a decline in the Treasury’s gold reserves raised concerns that the U.S. might abandon gold convertibility. Banking unrest began in May, accelerated in June, and peaked in July, as seen in \Cref{fig:maps-1893}. \cite{Wicker2006} stresses that this crisis was the only panic of the National Banking Era to start in the interior rather than in New York. In this sense, it was more similar to the Great Depression than the other crises of the National Banking Era.

There was a small localized flurry of failures and runs in Nashville in late March, which, to our knowledge, is not discussed in previous work. The immediate trigger was the failure of \href{https://finhist.com/bank-runs/episodes/322801119.html}{Commercial National Bank} (Nashville, TN). The bank had large exposures to Dobbins \& Dazey, a cotton firm that failed, leading to ``a heavy loss.''\footnote{\textit{The Morning Call}, March 26, 1893.} The OCC cause of failure was ``Fraudulent management and injudicious banking,'' the share of assets assessed worthless was 40.3\%, and the depositor recovery rate was 71.5\%. The cashier was later arrested for false statements and embezzlement after suspension.

The failure of Commercial NB led to predictions of failures of other local banks by ``curbstone loungers,''\footnote{\textit{The Indianapolis Journal}, March 28, 1893.} leading to depositor uneasiness and runs on other local banks. In response, larger banks prepared by visibly displaying ``great stacks of currency'' and paying withdrawals.\footnote{\textit{The Morning News}, March 28, 1893.}

Existing accounts identify \href{https://finhist.com/bank-runs/episodes/466601121.html}{Chemical National Bank} of Chicago as a key early failure of the Panic of 1893 (Run $\to$ Suspension $\to$ Closure). The bank was subject to a run on Saturday May 6 and Monday May 8 and suspended on May 9. This episode illustrates how solvency considerations were key both in terms of the decision to suspend and the decision to place the bank into receivership.

The bank requested assistance from the Chicago Clearinghouse. However, the Clearinghouse refused: ``being unable to satisfy the committee as to its solvency, the committee decided to render no assistance. There was nothing left to do but to suspend.''\footnote{\textit{New-York Tribune}, May 9, 1893.}  Ironically, the bank had an exhibit at the Chicago World's Fair. The run revealed deeper problems in its loan book and management. The OCC classified the cause of failure as ``Fraudulent management, excessive loans to officers and directors, and excessive loans to others.'' However, the bank's president tried to frame the issue as a liquidity problem: ``The reason of the suspension, were the extraordinary heavy runs of Saturday and Monday.''\footnote{\textit{The Wilmington Daily Republican}, May 10, 1893} He went on to say: ``One thing is certain neither our customers nor any other bank is going to get hurt. Our loans may be slow, but they are sound.''\footnote{\textit{New-York Tribune}, May 9, 1893.}

The bank was put into receivership on July 20, 1893: ``After long, patient and earnest efforts, the committee of the stockholders of the Chemical National Bank of Chicago has concluded that resumption is not practicable at this time. At first the problem seemed easy, but the many financial disasters throughout the country have excited the public and produced unusual caution, and made it very difficult to collect outstanding bills or raise money by usual processes.''\footnote{\textit{New-York Tribune}, July 21, 1893. Ultimately, however, the depositors did receive 100\% payout in nominal terms, suggesting future returns on the bank's assets were better than expected. } There is evidence the failure had broader contagion effects: ``The collapse of the Chemical National Bank of Chicago… led to rumors affecting the credit of other financial institutions of that city.''\footnote{\textit{St. Paul Daily Globe}, May 10, 1893.}

The failure of \href{https://finhist.com/bank-runs/episodes/466601121.html}{Chemical National Bank} had immediate consequences through the correspondent network. On May 10, \href{https://finhist.com/bank-runs/episodes/415801121.html}{Capital National Bank} in Indianapolis (Run $\to$ Suspension $\to$ Reopening) was subject to a run, as its funds were tied up in Chemical NB; this relationship ``was known and caused distrust and heavy drafts.''\footnote{\textit{The Salt Lake Herald}, May 12, 1893.} The bank was viewed as relatively risky: ``The Capital National has been conducted on what are termed liberal lines. Soon after its organization it refused to be bound by the rigid rules of the clearing house, the result of which was its withdrawal from that institution.''\footnote{\textit{The Indianapolis Journal}, May 12, 1893.} It also had exposure to a firm that had recently gone into receivership.  

Capital NB suspended on May 11. By May 26, the Indianapolis Journal reported that Examiner Young had informed the Comptroller that no receiver was needed and that his report would show the bank was in condition to resume. The examiner went on to say: ``Any man that would sell his claim on that bank for a shave would be a fool.''\footnote{\textit{The Indianapolis Journal}, May 26, 1893.} On June 19, the OCC examiner ``turned the institution over to the officials'' and the bank reopened:
\begin{quote}
``There were hardly half a dozen depositors at the doors at that hour and the first man to withdraw his deposit was a laboring man. Like others who followed him, he explained that he had confidence in the bank but needed the money which had been tied up. This was the general explanation of those who took out the sums they had in the bank. Within ten minutes after the opening the line of depositors increased to fifteen, and about that time men began to come in to make deposits and show their confidence in the institution and its new management. Many of these were South Meridian-street merchants. Other depositors came, and, after sizing up the situation, went away without asking for their money. On the paying teller's desk were huge heaps of greenbacks and coin. Whenever the crowd threatened to become large a second paying teller was put to work, and in this manner the line was thinned out at a rapid rate. The light run lasted about an hour. At noon Examiner Young stated that the bank had paid out about \$50,000 and had received in deposits about \$40,000... The day's business, he said, showed an excellent financial condition in the city.''\footnote{\textit{The Indianapolis Journal}, June 20, 1893.} \end{quote}
This episode illustrates the role of the correspondent network in transmitting bank distress, bank examination and suspension in assessing solvency and avoiding failure, and the use of conspicuous displays of cash to ensure confidence and play to depositor psychology.

Another key failure in May was \href{https://finhist.com/bank-runs/episodes/367701121.html}{Columbia National Bank} in Chicago, which was subject to a run on May 11 (Run $\to$ Suspension $\to$ Closure). It was subject to a ``constant run'' since the announcement of the Chemical NB suspension. It was considered vulnerable and not well-regarded: ``Rumors that it was in trouble had been current nearly a week,'' and it ``was not in esteem among the other city banks.''\footnote{\textit{Deseret Evening News}, May 11, 1893.} One paper reported it had poor collections on loans. It was not a Chicago clearinghouse member, as it had been refused entry a year earlier after it was investigated. This provides an example of how other banks are often informed of a bank's vulnerable condition. Columbia NB was forced to suspend when its clearing bank refused to clear for it after it learned it would be short of funds. The OCC cause of failure was ``Fraudulent management and injudicious banking,'' and the examiner's asset classification at failure was poor: 32.8\% ``good,'' 43.2\% ``doubtful,'' and 24.0\% ``worthless.'' The depositor recovery rate was 81\%.  The failure had contagion effects on small country banks: ``Some eighteen or twenty small banks and firms of bankers in Illinois, Ohio, Michigan and Indiana, connected with this Chicago bank have also suspended creating quite a financial flurry in those states. There were about forty banks connected with the Columbian [sic] bank combination.''\footnote{\textit{The Somerset Reporter}, May 18, 1893.}

By late May, there were also clusters of failures in New York. For example, \href{https://finhist.com/bank-runs/episodes/410501121.html}{Elmira National Bank} (Elmira, NY) failed following large exposures from overdrafts by Elmira's mayor, Col. David C. Robinson. The bank examiner assessed the bank was ``hopelessly insolvent.''\footnote{\textit{New-York Tribune}, May 27, 1893.} The failure caused a run on the \href{https://finhist.com/bank-runs/episodes/50010971121.html}{Elmira Savings Bank} (Run $\to$ Suspension $\to$ Reopening), which invoked the 30/60 day withdrawal rule for larger withdrawals in a partial suspension. This provides another example of how a bank failure triggered a run (without failure) on another local bank.

In June, there was an increase in runs, especially in the Midwest and West. The panic spread through city-level waves in cities such as Omaha, Detroit, Kansas City, Iowa, Minneapolis, Spokane, Denver, and the Pacific Northwest. There were runs on Chicago savings banks following the failure of \href{https://finhist.com/bank-runs/episodes/2422241791122.html}{Herman Schaffner \& Company} (Chicago) on June 3, a private bank that specialized in commercial paper. The firm was exposed to the deteriorating market for commercial paper. The immediate trigger for failure was the disappearance and apparent suicide of senior partner Herman Schaffner.

Schaffner's failure led to the runs ``on almost if not every one of the savings banks of Chicago,'' especially by alarmed Jewish residents.\footnote{\textit{The Morning Call}, June 6, 1893.} The \href{https://finhist.com/bank-runs/episodes/2001671122.html}{Illinois Trust \& Savings Bank}, which had ``a larger line of this class of deposits than any other bank in the city,'' saw a ``run only.'' In response, they added tellers and accommodated withdrawals to signal strength: ``It was after 2 o'clock this morning before the Illinois Trust and Savings Bank closed its doors, after paying the last depositors that remained in line at that hour. This performance, it is said, broke the record, it being the first time in this country, so far as is known, that a bank has remained open after midnight in order to meet a run.''\footnote{\textit{The Morning News}, June 7, 1893.} The \href{https://finhist.com/bank-runs/episodes/6626048291122.html?utm_source=chatgpt.com}{Dime Savings Bank} in Chicago invoked the savings bank's sixty-day notice-of-withdrawal rule and paid out only a portion of deposits in cash.

The runs on Chicago savings banks in June 1893 provide an example of a bank failure leading small depositors to update their belief that the deposits of other banks might be at risk. In response, confused depositors failed to discriminate between healthy and weak banks. However, such runs alone need not spell death for a bank.

On June 15, the NY Clearinghouse preemptively issued clearinghouse certificates.\footnote{\textit{Freeland Tribune}, June 22, 1893.} The Philadelphia and Boston clearinghouses followed \citep{Wicker2006}. The peak of the panic occurred in July, with runs and suspensions in the West and interior. From July 18 to 20, there was a cluster of runs in Denver, Colorado. By late July, there was a cluster of events in New York savings banks.  By August, the panic was still ongoing but tapering off, as seen in \Cref{fig:maps-1893}.

\clearpage

\subsubsection*{Panic of 1907}

\begin{figure}[ht]
  \centering
  \caption{Panic of 1907}
  \label{fig:maps-1907}
  \begin{subfigure}[t]{0.49\textwidth}
    \caption{September 1907}
    \label{fig:map-1907-9}
    \includegraphics[width=\linewidth]{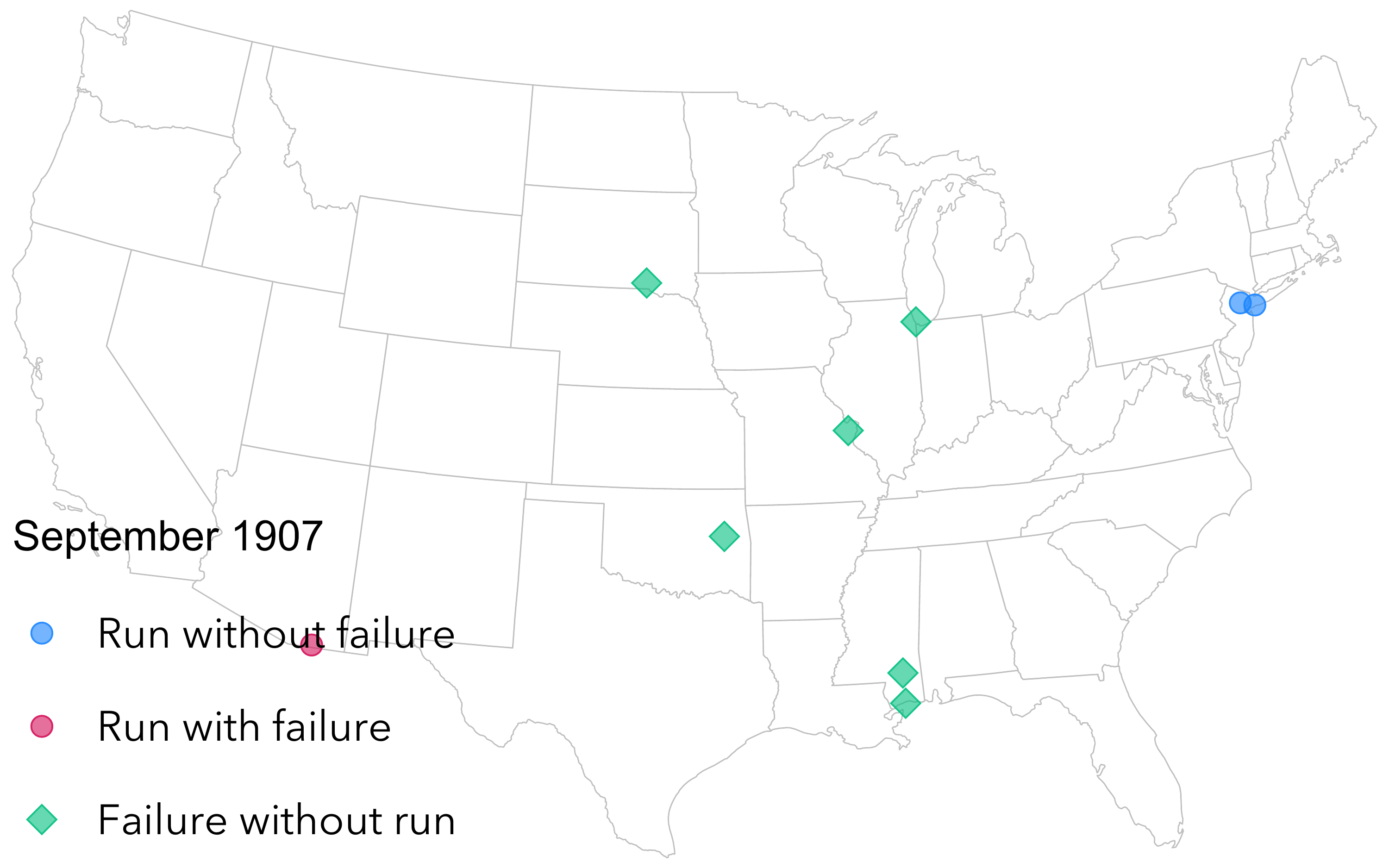}
  \end{subfigure}\hfill
  \begin{subfigure}[t]{0.49\textwidth}
    \caption{October 1907}
    \label{fig:map-1907-10}
    \includegraphics[width=\linewidth]{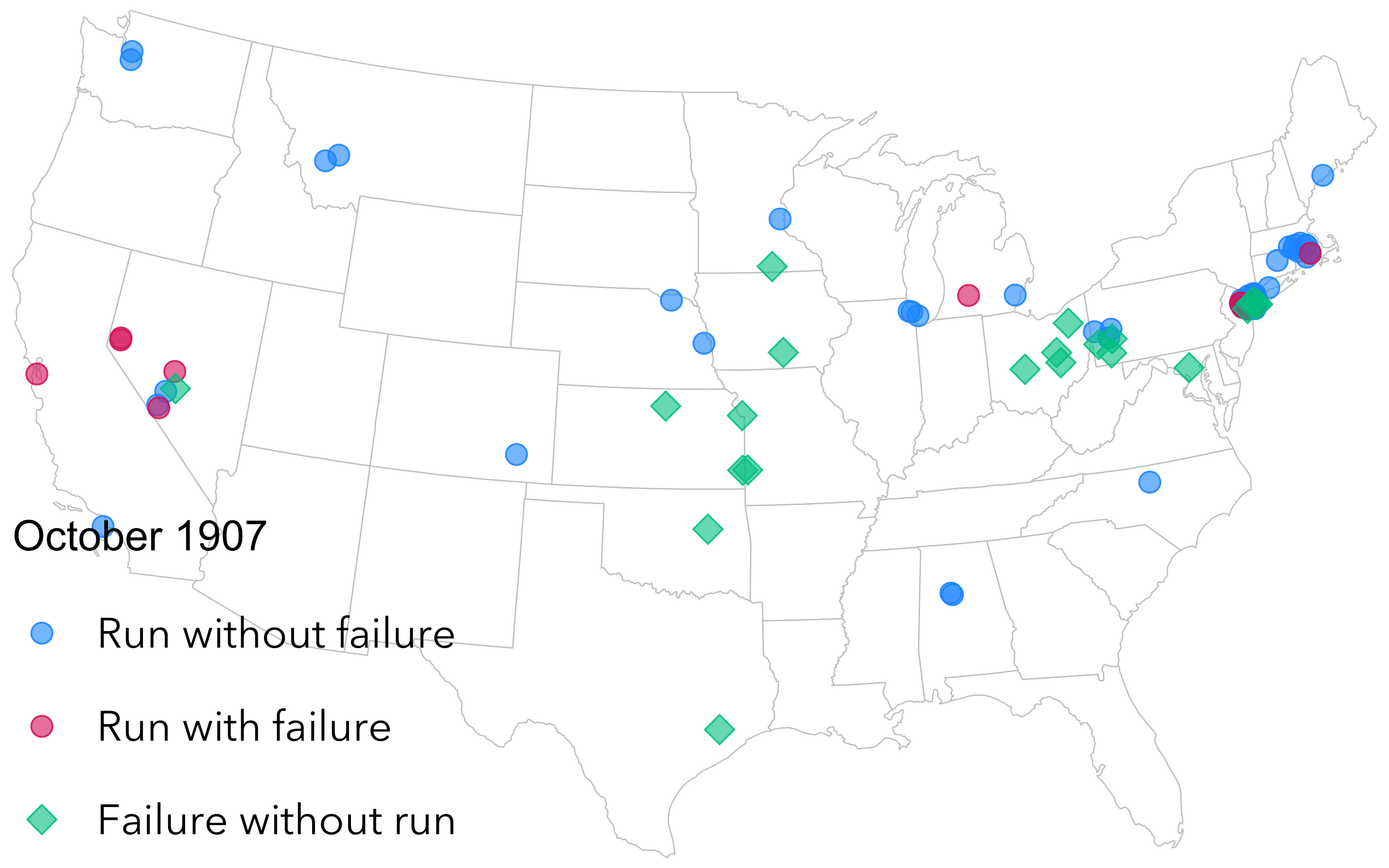}
  \end{subfigure}

  \vspace{0.5em}

  \begin{subfigure}[t]{0.49\textwidth}
    \caption{November 1907}
    \label{fig:map-1907-11}
    \includegraphics[width=\linewidth]{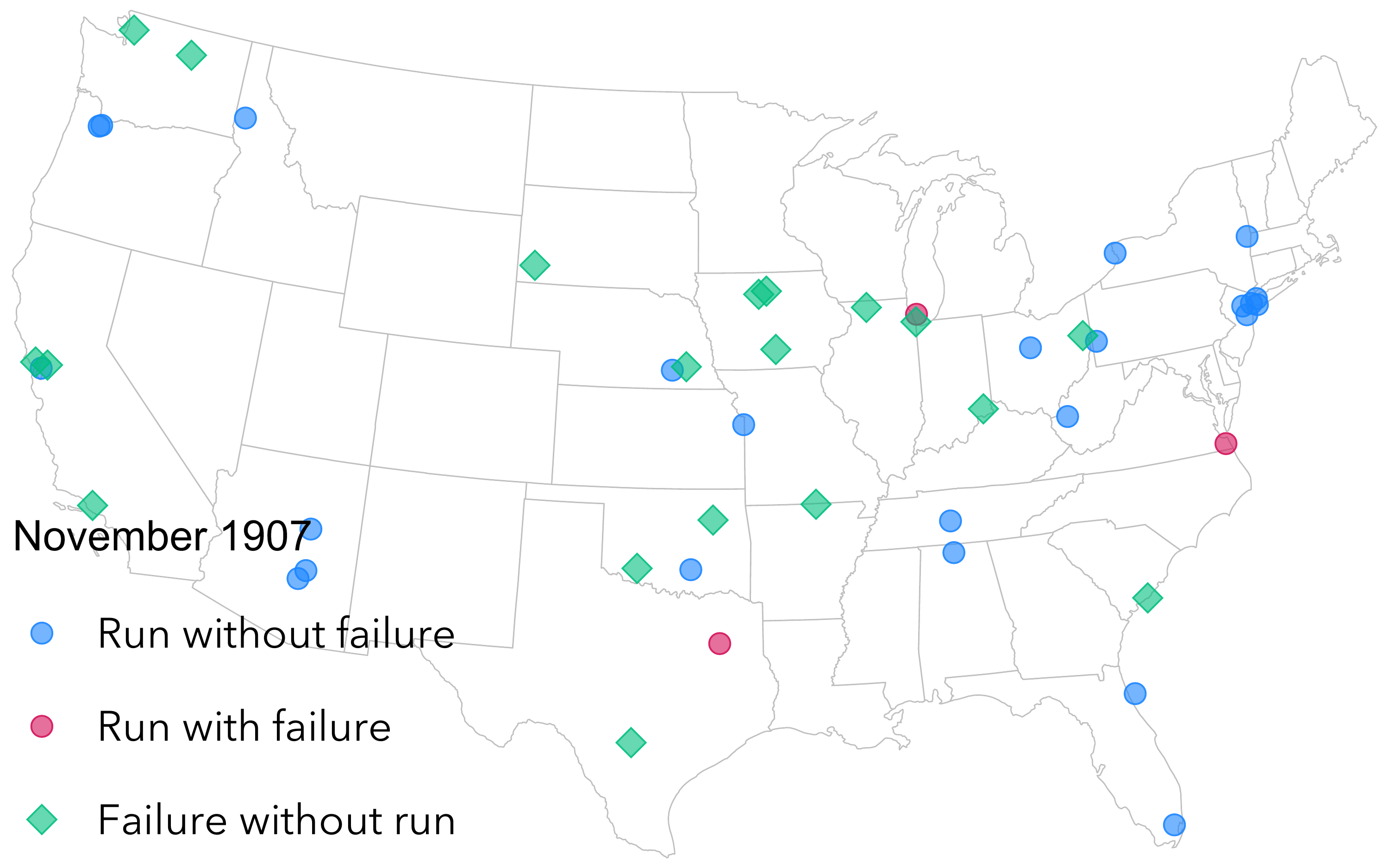}
  \end{subfigure}\hfill
  \begin{subfigure}[t]{0.49\textwidth}
    \caption{December 1907}
    \label{fig:map-1907-12}
    \includegraphics[width=\linewidth]{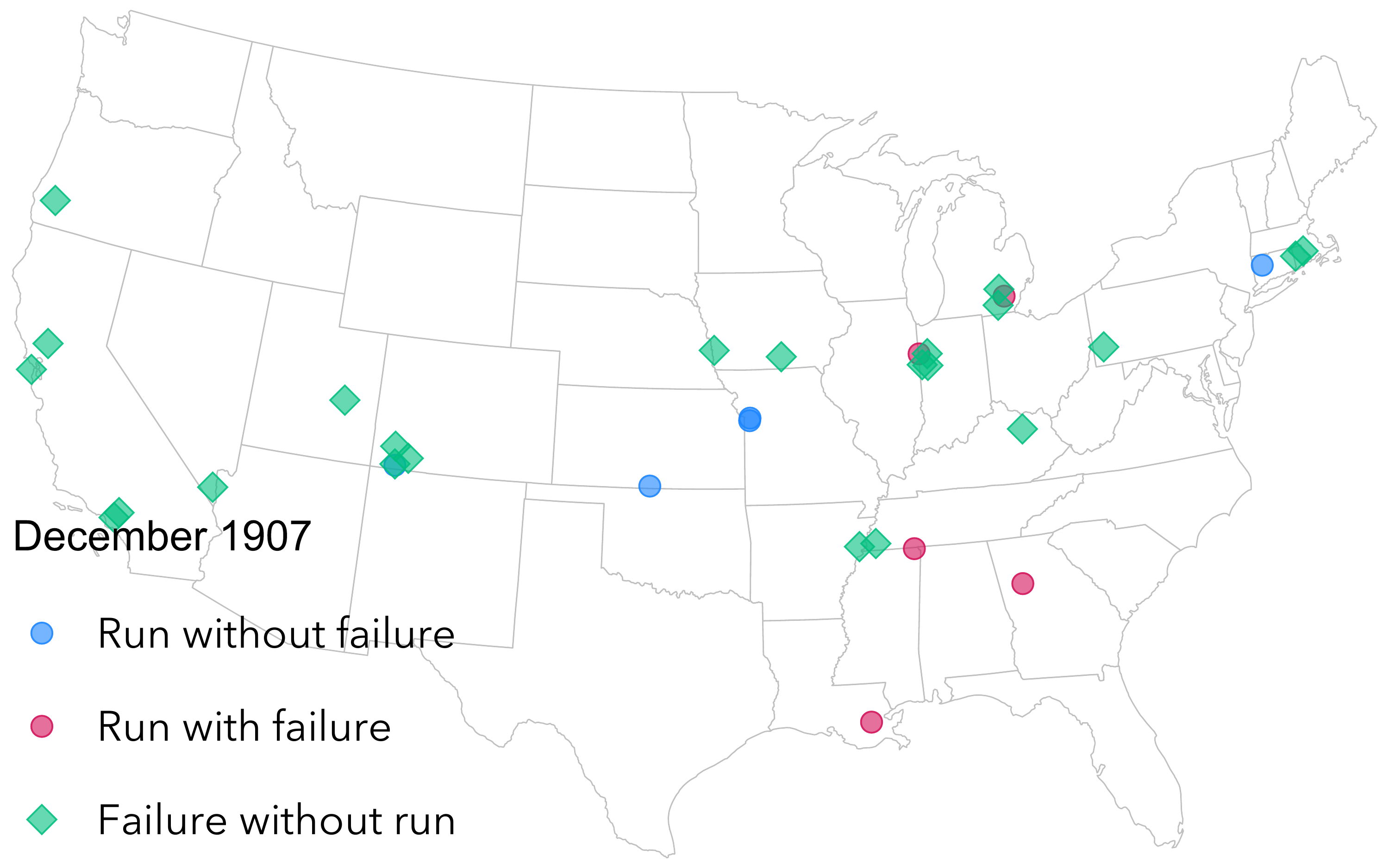}
  \end{subfigure}

  \begin{minipage}{\textwidth}
    \footnotesize
    \textit{Notes:} Each panel shows the location of bank distress episodes in a given month during the Panic of 1907. Blue circles indicate banks subject to a run that did not subsequently fail. Red circles indicate banks subject to a run that subsequently failed. Green diamonds indicate banks that failed without a reported run.
  \end{minipage}
\end{figure}

The Panic of 1907, which occurred in October, was the last panic of the National Banking Era. \Cref{fig:maps-1907} shows it was centered in New York, but it rippled to the interior. The real economy was already in a downturn before the start of the panic, with the NBER dating the previous business cycle peak to May 1907. The stock market was also declining in 1907, having peaked in December 1906 following a boom from 1904 to 1906. Moreover, credit spreads were widening well before October of 1907, and the money market was tight \citep{TallmanMoen1990,MishkinWhite2002}. The trough of the NBER recession was reached in June 1908.

The immediate trigger of the panic in October was the failed attempt by a group of speculators---F. Augustus Heinze, his brother Otto Heinze, and Charles W. Morse---to corner the stock of United Copper Company. The failed speculation raised concerns about the solvency of banks to which the speculators were connected. This led to a run on the \href{https://finhist.com/bank-runs/episodes/106701294.html}{Mercantile National Bank} on October 16, and at least seven other banks \citep{Wicker2006}. Mercantile's president was F.A. Heinze. In response to the run, the NY clearinghouse examined its books and found it solvent. It then provided liquidity to the bank on the condition that Heinze and all the board of directors resign \citep{TallmanMoen1990}. One report notes that ``nine clearing house banks had promised to contribute \$200,000 each to tide the Mercantile Bank over any distress.''\footnote{\textit{Evening Star}, October 19, 1907.}

Banks connected to Morse were also subject to runs \citep{TallmanMoen1990}. The NYCH managed to calm the run on banks connected to the failed speculators by examining member banks, certifying their solvency, forcing out implicated management, and providing liquidity support through a \$10 million fund \citep{TallmanMoen1990,Frydman2015}. However, the panic spread to trust companies. Trusts, which had grown rapidly before the panic, were not part of the NY clearinghouse. Trusts did not have the same mechanisms as banks for dealing with runs through information provision, interbank cooperation, and suspension.

News that \href{https://finhist.com/bank-runs/episodes/1019971294.html}{Knickerbocker Trust Company} president Charles Barney was connected to Charles W. Morse had already prompted withdrawals by October 18, and on October 21 Knickerbocker’s board dismissed Barney. The National Bank of Commerce also announced that it would no longer act as Knickerbocker’s clearing agent. These events were interpreted as a sign of deeper problems.  On October 22, Knickerbocker ``temporarily suspended payment at its downtown office this afternoon, after paying out \$8,000,000.''\footnote{\textit{Perth Amboy Evening News}, October 22, 1907.} Knickerbocker was forced to suspend after emergency assistance was denied by the NYCH. J.P. Morgan also did not provide assistance because Benjamin Strong, whom he asked to examine Knickerbocker, could not determine the trust's solvency in the limited time.\footnote{Knickerbocker ultimately reopened in March 1908 following the infusion of new capital.}

The revelation that the NYCH would not aid trusts led to more runs. \href{https://finhist.com/bank-runs/episodes/1021271294.html}{Trust Company of America} was subject to a run which we date to October 23 when ``several hundred depositors were waiting for the bank to open this morning.''\footnote{
\textit{Hattiesburg Daily News}, October 24, 1907.} Another prominent run ``of large proportions'' occurred on the \href{https://finhist.com/bank-runs/episodes/1011671294.html}{Lincoln Trust Company}.\footnote{\textit{The Richmond Palladium and Sun-Telegram}, October 24, 1907.}

The Panic of 1907 is famous for the intervention of J.P. Morgan, who led a rescue of the trusts, creating a rescue fund from loans by large financial and industrial firms to provide liquidity to trusts. ``Wall St sighed with relief today when it learned that money would be supplied `up to the limit' to support the Trust Company of America and the Lincoln Trust company, where runs have been on... A formal statement was issued from the Morgan mansion this morning to the effect that an examination shows that on the present basis of values, the assets of the two companies are sufficient to pay depositors in full. ''\footnote{\textit{The Seattle Star}, November 6, 1907.} A committee examined the books of the Trust Company of America and Lincoln Trust, determining they were solvent. These efforts calmed the run. Given the rivalry between trusts and banks, the willingness of Morgan and other financiers to lend is consistent with their belief that the institutions were solvent \citep{MoenTallman2000}. Meanwhile, on October 26, after some delay, the NY clearinghouse met to issue clearinghouse loan certificates, temporarily easing liquidity pressure.

The Panic of 1907 illustrates that private sector arrangements can often resolve runs on solvent institutions. That process is reasonably smooth when such institutional arrangements are in place, as in the case of NY banks and the clearinghouse. It is more frictional when they are not, as in the case of the trusts, as they were outside of the NY clearinghouse, the clearinghouse did not assume responsibility for the stability of the broader system, and the trusts were not well organized on their own \citep{Wicker2006}. Overall, the severity of the Panic of 1907 was down to institutional failure \citep{Wicker2006}.

The aftermath of the Panic of 1907 is also worth discussing. In January 1908, the NY clearinghouse pushed banks to repay clearinghouse loan certificates \citep{CFC8Feb1908}. This refocused attention on banks that had required support during the panic, leading to renewed withdrawals because of lingering distrust of some banks. Two banks that were associated with Morse and had survived runs with clearinghouse support in October 1907 (\href{https://finhist.com/bank-runs/episodes/458101297.html}{National Bank of North America} and \href{https://finhist.com/bank-runs/episodes/578301294.html}{New Amsterdam National Bank}) both closed in January 1908 after funding pressure.  In both bank receiverships, the depositor recovery rate was 100\%, consistent with the banks being solvent during the panic.

\end{document}